\documentclass[12pt,a4paper]{report}
\usepackage{epsf,epsfig,color}
\usepackage[small,bf]{caption} 
\usepackage{latexsym,amssymb,amsmath,mathrsfs,amsfonts,psfrag}
\usepackage{setspace,cite}
\usepackage{fancyhdr}
\usepackage{a4}
\usepackage{graphics, graphpap,array}
\begin{document}
\newpage
\pagenumbering{arabic}
\setcounter{page}{1} 
\newcommand{\quark}{\langle \bar q q\rangle}
\newcommand{\mixed}{\langle \bar q \sigma g_sG q\rangle}
\newcommand{\squark}{\langle \bar s s\rangle}
\newcommand{\smixed}{\langle \bar s \sigma g_sG s\rangle}
\newcommand{\gluon}{\left\langle \frac{\alpha_s}{\pi}\,G^2\right\rangle}
\newcommand{\deriv}{\stackrel{\leftrightarrow}{D}}
\newcommand{\derleft}{\stackrel{\leftarrow}{D}}
\newcommand{\derright}{\stackrel{\rightarrow}{D}}
\newcommand{\ub}{\bar{u}}
\newcommand{\ds}{\displaystyle}
\makeatletter
\def\slash#1{\rlap/{\mkern-1mu {#1}}} 
\newcommand{\etapb}{\eta^{(\prime)}}
\newcommand{\etap}{\eta^{\prime}}
\newcommand{\bra}[1]{\langle #1|}
\newcommand{\ket}[1]{|#1 \rangle}
\newif\ifContLineOne
\newif\ifContLineTwo
\newif\ifContLineThree
\def\conC#1{\vbox{\ialign{##\crcr
  \ifContLineThree\hrulefill\else\vphantom{\hrulefill}\fi\crcr
  \noalign{\kern3.2pt\nointerlineskip}
  \ifContLineTwo\hrulefill\else\vphantom{\hrulefill}\fi\crcr
  \noalign{\kern3.2pt\nointerlineskip}
  \ifContLineOne\hrulefill\else\vphantom{\hrulefill}\fi\crcr
  \noalign{\nointerlineskip}
  $\hfil\textstyle{\vbox to 14pt{}#1}\hfil$\crcr}}}
\def\DrawLeg#1#2{
  \kern-.2pt              
  \dimen2 =#1             
  \advance\dimen2 by 2pt  
  \dimen3 = 10.6pt        
  \dimen4 =3.6pt          
  \advance\dimen3 by -\dimen2 
  \multiply\dimen4 by #2
  \advance\dimen3 by \dimen4
  \raise\dimen2 \hbox{\vrule height\dimen3 width .4pt} 
  \kern-.2pt}             
\def\begC#1#2{\setbox0 =\hbox{$\textstyle{#2}$}
  \dimen0=.5\wd0 \dimen1=\ht0
  \conC{\hskip\dimen0}
  \count255=#1
  \ifnum\count255 =1 \ContLineOnetrue\else
  \ifnum\count255 =2 \ContLineTwotrue\else
  \ifnum\count255 =3 \ContLineThreetrue\fi\fi\fi
  \DrawLeg{\dimen1}{\count255}
  \conC{\hskip\dimen0}
  \kern-\dimen0\kern-\dimen0 \box0}
\def\endC#1#2{\setbox0 =\hbox{$\textstyle{#2}$}
  \dimen0=.5\wd0 \dimen1=\ht0
  \conC{\hskip\dimen0}
  \count255=#1
  \ifnum\count255 =1 \ContLineOnefalse\else
  \ifnum\count255 =2 \ContLineTwofalse\else
  \ifnum\count255 =3 \ContLineThreefalse\fi\fi\fi
  \DrawLeg{\dimen1}{\count255}
  \conC{\hskip\dimen0}
  \kern-\dimen0\kern-\dimen0 \box0}
\def\backskip{\vskip -3.6pt}  
\def\bbackskip{\vskip -7.2pt}
\thispagestyle{empty}
\topskip 1cm
\begin{titlepage}
  \vfill
  \begin{center}
    {\Huge\bf Meson Distribution Amplitudes}\\
       \vspace{1.cm}
        {\Large\bf Applications To Weak Radiative $B$ Decays}\\ {\Large\bf And $B$ Transition Form Factors}\\
    \vspace{4.cm}
    \doublespacing
    {\large A thesis presented for the degree of\\ Doctor of Philosophy\\ by\\}
    \vspace{1.cm}
    {\LARGE\bf Gareth W. Jones}\\
    \vspace{1.5cm}
        {\large  September 2007}\\
        \vspace{1.5cm}
    {\large Institute for Particle Physics Phenomenology\\ University of Durham\\}
    \singlespacing
    \vfill
  \scalebox{0.2}{\includegraphics{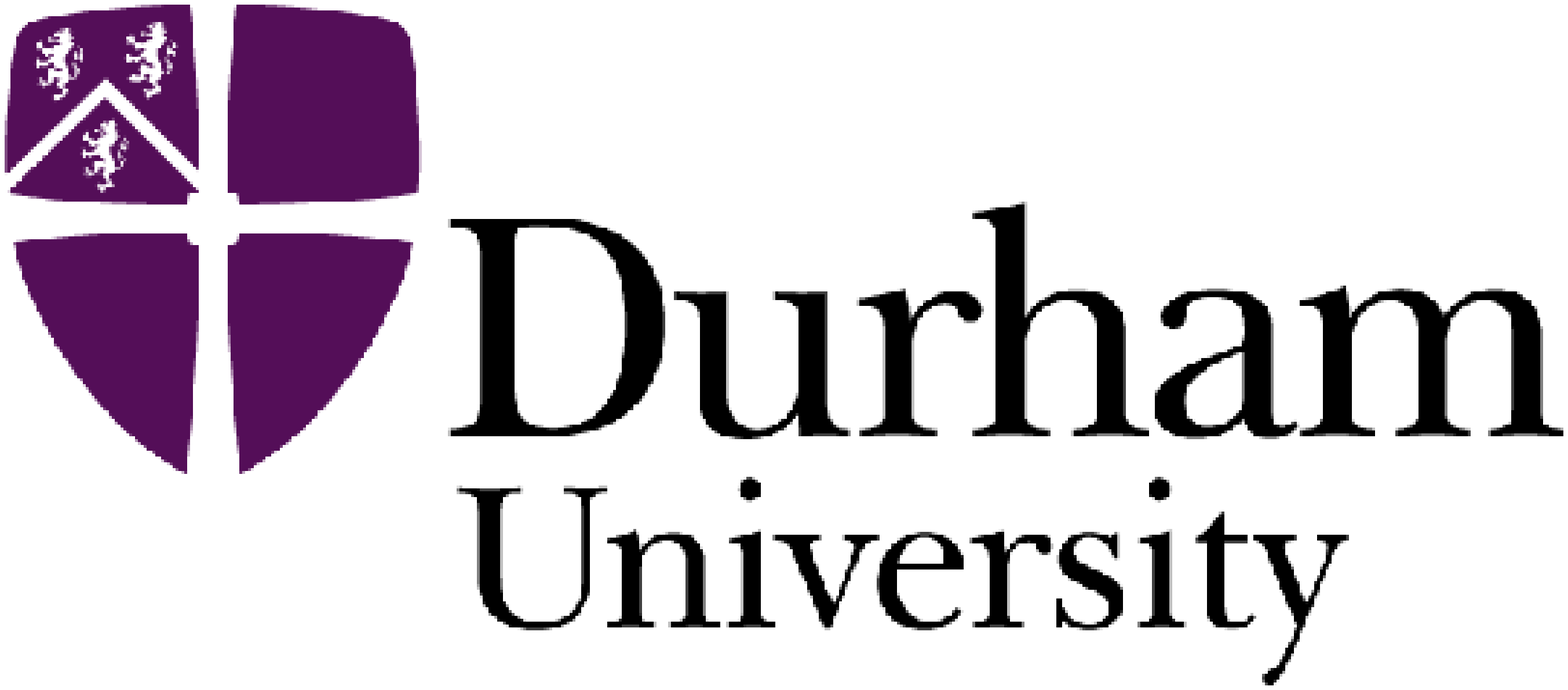}}
  \end{center}
\end{titlepage}
\onehalfspacing

\begin{center}
  {\Huge\bf Abstract\\}
\end{center}
\pagestyle{empty}
This thesis examines the applications and determinations of meson light-cone distribution amplitudes, which enter the theoretical description of exclusive processes at large momentum transfer. The investigation of such processes, in the context of $B$ physics, provides one with a rich and extensive way of determining the Standard Model parameters of the CKM matrix, which are essential in describing CP violation, and searching for tell-tale signs of new  physics beyond the Standard Model.

We investigate the twist-2 and twist-3 distribution amplitudes of vector mesons and fully examine $\rm SU(3)_F$-breaking effects and include leading G-parity violating terms. We use the conformal expansion allowing the distribution amplitudes to be described by a set of non-perturbative hadronic parameters which is reduced by invoking the QCD equation of motion to find various interrelations between the distribution amplitudes.  Numerical values of the leading non-perturbative hadronic parameters are determined from QCD sum rules.

The new distribution amplitude results find direct application in the radiative $B$ decays to light vector mesons $B \rightarrow V \gamma$. We examine the phenomenologically most important observables in this decay mode using the formalism of QCD factorisation in which the distribution amplitudes play a vital role. We also include long-distance photon emission and soft quark loop effects, which formally lie outside the QCD factorisation formalism. The analysis  encompasses all the relevant modes, that is $B_{u,d} \to \rho, \omega, K^*$ and  $B_{s} \to \phi, \bar{K}^*$.

We also calculate the $B \to \etapb$ transition form factor using QCD sum rules on the light-cone. The method relies on the collinear factorisation of the QCD dynamics into a perturbatively calculable hard-scattering kernel and the non-perturbative universal distribution amplitudes. We include the singlet contribution originating from the $\rm U(1)_A$ anomaly and bring the calculation consistently within the $\eta$-$\etap$ mixing framework.

\onehalfspacing
\pagestyle{empty}
\pagenumbering{roman}
\setcounter{page}{1} \pagestyle{plain}


\chapter*{Acknowledgements}

First and foremost, I would like to thank my supervisor Patricia Ball for all her help and guidance over the last three years. It has been a great opportunity to work with her, and a fantastic learning experience. I must also thank Roman Zwicky for always finding the time to quell my confusions, and with whom it was a pleasure to collaborate. Also, I thank Angelique Talbot for all her friendly discussions, and I wish Aoife Bharucha all the best with her future projects.

I also thank my office mates Ciaran Williams, Karina Williams, Kemal Ozeren, Martyn Gigg and Stefan Hoeche, and the many other friends who have made my time in Durham and the IPPP so enjoyable.

To those whose support cannot be appreciated enough; I must thank my parents. I thank my brother too for all the discussions and debates we had over coffee, and finally, I must also thank my grandparents. 

This work was supported by a PPARC studentship which is gratefully acknowledged.

\chapter*{Declaration}
I declare that no material presented in this thesis has previously been submitted
for a degree at this or any other university. The research described in this thesis has been carried out in collaboration with Prof.~Patricia Ball and Dr.~Roman Zwicky and has been published as follows:
\begin{itemize}
  \item{``$B \to V \gamma$ beyond QCD factorisation,''\newline
  P.~Ball, G.~W.~Jones and R.~Zwicky, Phys.\ Rev.\  D {\bf 75} (2007) 054004,\newline [arXiv:hep-ph/0612081].}
    \item{``Twist-3 distribution amplitudes of $K^*$ and $\phi$ mesons,''\newline
P.~Ball and G.~W.~Jones, JHEP {\bf 03} (2007) 069, \newline [arXiv:hep-ph/0702100].}
    \item{``$B \to \etapb$ Form Factors in QCD,''\newline
P.~Ball and G.~W.~Jones, JHEP {\bf 08} (2007) 025, \newline  arXiv:0706.3628 [hep-ph].}
\end{itemize}
The copyright of this thesis rests with the author.  No quotation from it should be published without their prior written consent and information derived from it should be acknowledged. 
\begin{flushright}
garethwarrenjones@gmail.com
\end{flushright}
\tableofcontents
\listoffigures
\listoftables
\chapter*{Introduction}\label{chapter0_intro}\addcontentsline{toc}{chapter}{\protect\numberline{Introduction\hspace{-96pt}}} 
One only has to ask the question ``why?'' a handful of times before one reaches the answer ``I don't know'', regardless of the topic considered and regardless of the person asked.  It is safe to say, however, almost all questions of the structure of matter at the smallest of distances leads one directly to, or at least through, the field of modern particle physics. The beginnings of our understanding of the physical world harks back to the dawn of scientific reasoning in the ancient world; logic and reasoning were applied with the aim of describing the behaviour of physical systems in terms of simple universal axioms, a philosophy which still holds strong today. Through experimentation and the language of mathematics the scientific method has driven back the edge of ignorance to frontiers unimaginable to those physicists of 100 years ago, let alone the natural philosophers of millennia ago. The present ``coal face'' is known as the Standard Model \cite{weak, QCD} which describes three of the four known forces of nature -- electromagnetism, and the weak and strong nuclear forces -- in one unifying framework. 

Frustratingly, the Standard Model does not explain many of the things which it encompasses; it does not provide an origin for CP violation but only gives a parameterisation, nor does it explain why there are three generations of quarks and leptons, or their hierarchy of masses.  All attempts to bring gravity into the fold have so far failed, however, whatever theory lies beyond must yield the Standard Model as some limiting case.

The Standard Model has been scrutinised relentlessly since its inception. Remarkably, nearly without fail it has held its ground over the entire breadth of its theoretical reach and so the task of finding new ways to probe its structure requires ever more the creativity and ingenuity of both theorists and experimentalists alike.  Novel experimental signatures, against which to pit theory, must be used to maximum potential. From a theoretical standpoint there are still many challenges to be met, especially in preparation for the next generation of collider experiments now just round the corner. Particularly, the control and reduction of the theoretical uncertainty of Standard Model predictions is of paramount importance as only then can one hope to be in a position to discern signs of new physics from that of the Standard Model background.

Some of its most challenging tests of the Standard Model fall in the field of heavy-flavour physics, within which $B$ physics has proven itself to be rich and fertile. Today it is an area of high activity with many success stories, including the recent measurement of the $B_s^0$-$\bar{B}_s^0$ mass difference $\Delta m_s$ at the Tevatron \cite{Abulencia:2006ze}. Moreover, two dedicated ``$B$-factories'', Belle at KEK \cite{:2000cg} and \textsc{BaBar} at SLAC \cite{Aubert:2001tu}, have measured a range of observables, such as branching fractions and CP asymmetries, of a vast number of $B$ decay modes. Looking to the future, the $B$ physics community eagerly await the forthcoming LHCb experiment, and beyond that so-called ``superflavour factories'' \cite{superB} have been championed with the aim of probing rare $B$ decays to extract CP violation parameters to much higher levels of accuracy. It is imperative to find tests of the Standard Model  which may be observed in these up-and-coming experiments \cite{Gershon:2006mt} and promising modes include the rare decays $B\to V \gamma$ and $B\to K \mu^+\mu^-$. 

The strict pattern of CP violation of the Standard Model finds its origin in the Cabbibo-Kobayashi-Maskawa (CKM) matrix \cite{Cabibbo:1963yz,Kobayashi:1973fv}. CP violation was discovered in $B$ physics via the decay mode $B_d^0\to J/\psi K_S^0$ and found to be large, in contrast to $K$ decays where the violation is tiny. The possible largeness of CP violation in $B$ decays offers promising ways to detect new physics indirectly via CP violating observables testing the CKM paradigm. 

Theoretically, central to the description of $B$ decays is the disentanglement of the weak decay process from strong interaction effects leading to a low-energy effective Hamiltonian in which the physics at a scale $\mathcal{O}(M_W)$ is well under control. Achieving this goal for the wide range of $B$ decays of interest has only been possible through huge calculational effort; the availability in the literature of Wilson coefficients at next-to-leading-order, and in some cases next-to-next-to-leading-order, is testament to this.  Furthermore, the theoretical description of the matrix elements of effective $B$ decay operators has been hugely improved through QCD factorisation methods. We discuss and make use of one such framework, namely that introduced by Beneke, Buchalla, Neubert and Sachrajda \cite{Beneke:1999br, Beneke:2000ry,Beneke:2001ev}. The so-called BBNS approach showed, to leading-order in a $1/m_b$ expansion, that the $\alpha_s$ corrections beyond naive-factorisation of a large class of non-leptonic $B$ decay matrix elements are calculable in terms of $B$ transition form factors and meson light-cone distribution amplitudes. Armed with the corresponding amplitudes the phenomenologist may construct observables, such as branching ratios, CP asymmetries and isospin symmetries, which may then be compared to experiment. The predictive power of the QCD factorisation framework is jeopardised by a poor understanding of both these non-perturbative QCD quantities and the impact of the generally unknown power-suppressed contributions $\mathcal{O}(1/m_b)$; this in part motivates the work of this thesis.

In this thesis we investigate $\rm SU(3)_F$-breaking effects in vector meson distribution amplitudes which are crucial in differentiating between the particles $\rho$, $K^*$ and $\phi$. The leading non-perturbative DA parameters are determined via the method of QCD sum rules introduced by Shifman, Vainshtein and Zakharov  \cite{Shifman:1978bz, Shifman:1978by, Shifman:1978bx}. The method provides a prescription for the systematic calculation of non-perturbative QCD parameters,  albeit with an irreducible error $\sim 20-30\%$, and constitutes an extremely useful theoretical tool. 

The sum rule results have a direct application in the QCD factorisation description of $B$ decays to $\rho$, $K^*$ and $\phi$ mesons. In particular, radiative $B$ decays to vector mesons $B\to V\gamma$, are an excellent example of a process potentially sensitive to new physics contributions, as at leading order the decays are mediated at loop level in the Standard Model. We perform a phenomenological analysis of these decays using the QCD factorisation framework of Bosch and Buchalla \cite{Bosch:2001gv,Bosch:2002bw} including leading power-suppressed corrections for which the updated non-perturbative distribution amplitude parameters find use.  The impact of the power-suppressed corrections on the key decay observables is discussed and leads to a better understanding of the theoretical uncertainty of the QCD factorisation predictions. 

Also, we calculate important contributions to the $B\to\etapb$ transition form factors via a variant sum rule approach, known as light-cone sum rules, for which distribution amplitudes play a crucial role. The result of the analysis elucidates a major source of theoretical uncertainty of the $B\to \etapb$ form factor. The result impacts $B\to K^* \etapb$, for example, where the experimental data and QCD factorisation predictions of the branching ratios are inconsistent.

The thesis is structured as follows: 
\begin{itemize}
\item{Chapter~\ref{chapter1_basics} introduces some of the fundamentals of the Standard Model and its application to $B$ physics. We define the QCD Lagrangian and the CKM matrix, introduce CP violation in Standard Model $B$ decays, and briefly discuss the structure of the $\Delta B =1$ weak effective Hamiltonian. }
\item{Chapter~\ref{chapter2_DAs} covers the definitions of the light-cone distribution amplitudes of the light vector mesons $\rho$, $K^*$ and $\phi$. We determine their structure up to twist-3 accuracy and using the conformal expansion and QCD equations of motion express the distribution amplitudes in terms of a finite set of non-perturbative parameters. We extend previous determinations in order to fully differentiate between the three particles by including all G-parity violating contributions and $\rm{SU}(3)_F$-breaking effects. }
\item{Chapter~\ref{chapter3_SR}  discusses the QCD sum rule method and its extension light-cone sum rules.  The methods allow, amongst other things, the determination of the non-perturbative distribution amplitude parameters and transition form factors respectively, and are very widely applicable in and beyond $B$ physics.}
\item{In Chapter~\ref{chapter4_det} we apply QCD sum rules to determine the leading non-perturbative distribution parameters defined in Chapter~\ref{chapter2_DAs}. Consistency requires the inclusion of all G-parity violating contributions and $\rm{SU}(3)_F$-breaking effects to the sum rules, and we extend previous determinations by including higher-order strange quark mass effects and $\mathcal{O}(\alpha_s)$ contributions to the quark condensates. We analyse the resulting sum rules and provide updated numerical results for all parameters. The results of this section find immediate application in QCD factorisation and light-cone sum rule descriptions of processes involving these vector mesons.}
\item{In Chapter~\ref{chapter5_eta} we calculate the gluonic flavour-singlet contribution to the semileptonic $B\to \etapb$ transition form factor in the framework of light-cone sum rules. In doing so we discuss pseudoscalar meson and two-gluon distribution amplitudes. The new contribution is combined with the previous determination of the quark contribution, to complete the theoretical treatment of these form factors. The $\etapb$ system is complicated due to large mixing effects via the $\rm U(1)_A$ anomaly. We introduce the phenomenological framework of $\eta$-$\etap$ mixing and connect it to the form factor calculation in a consistent manner. The results of this chapter find immediate application in the QCD factorisation description of $B\to \etapb$ transitions, which in turn, in principle, allow a determination of the CKM matrix element $|V_{ub}|$ from $B\to\etapb l \nu$.}
\item{Chapter~\ref{chapter6_QCDF}  introduces the framework of QCD factorisation, which is an important application of meson distribution amplitudes and transition form factors. We briefly discuss the BBNS approach and then go on to discuss the leading contributions to QCD factorisation in the context of $B\to V \gamma$ decays.}
\item{In Chapter~\ref{chapter7_rad} we investigate the impact of the relevant, power-suppressed contributions to $B\to V \gamma$ beyond the QCD factorisation formula. We include long-distance photon emission from weak annihilation diagrams and soft gluon emission from quark loops. The non-perturbative distribution amplitude parameters determined in Chapter~\ref{chapter4_det} find use in a light-cone sum rule estimation of the latter. The key observables are the branching ratios, isospin asymmetries and the indirect time-dependent CP asymmetry $S(V\gamma)$  which, as has been know for some time, forms the basis of a ``null test'' of the Standard Model. Assuming no new physics contributions, we extract the ratio of CKM matrix parameters $\left|V_{td}/V_{td}\right|$ to a competitive degree of accuracy.}
\item{We summarise and conclude in Chapter~\ref{chapter8_conc}.}
\end{itemize}
The material of Chapters~\ref{chapter2_DAs} and \ref{chapter4_det} follows Ref.~\cite{Ball:2007rt} and the material of Chapters~\ref{chapter5_eta} and \ref{chapter7_rad} follows Refs.~\cite{Ball:2007hb} and \cite{Ball:2006eu}, respectively. Some of the more bulky equations, and material not necessary in the general flow of reading the thesis, are given in two appendices.

\pagestyle{plain}
\pagenumbering{arabic}
\onehalfspacing
\chapter{Fundamentals Of $B$ Physics}\label{chapter1_basics}
In this chapter we begin with the basics of the Standard Model and then go on to  discuss two concepts which are central to the investigations of $B$ physics, and those of this thesis:
\begin{itemize}
\item{CP violation in the flavour sector, which follows a strict pattern in the Standard Model and can readily be sensitive to new physics;}
\item{the $\Delta B=1$ effective weak Hamiltonian, which we briefly discuss as it is the starting point of many phenomenological studies in $B$ physics.}
\end{itemize}

\section{The Standard Model}
The \textit{Standard Model} (SM) \cite{weak,QCD} is a model of great scope and predictive power. Despite its successes, however, we know it to be incomplete; for example, the recent discovery of neutrino oscillation and the lack of conclusive evidence for the Higgs particle providing two areas of intense theoretical and experimental effort.  The SM describes three of the four known fundamental forces of nature;  the strong force, the weak force and electromagnetism. \textit{Quantum Chromodynamics} (QCD) is a Yang-Mills gauge theory based on the gauge group $\rm SU(3)$ and describes the fundamental interactions of the strong interaction as  interactions between quarks and gluons \cite{Yang:1954ek,Gell-Mann:1964nj,FGM,Fritzsch:1973pi}. The basic QCD Lagrangian is
\begin{equation}
\mathcal{L}_{\rm QCD}=\sum_q \bar{q}^i \left(i \gamma_\mu \left(D^\mu\right)_{ij}-m_q \delta_{ij}\right)q^j -\frac{1}{4} G^a_{\mu\nu} G^{a \mu\nu}\,,
\label{basics_eq1}
\end{equation}
with
\begin{equation}
(D_\mu)_{ij} = \delta_{ij} \partial_\mu -i g_s (t^a)_{ij} A^a_\mu\,, \qquad G^a_{\mu\nu}=\partial_\mu A_\nu^a-\partial_\nu A_\mu^a + g_s f^{abc}A^b_\mu A^c_\nu\,,
\label{basics_eq2}
\end{equation}
where the sum is over all quark flavours $q$, $i,j=\{1,2,3\}$ are colour indices, the $t^d$ are the $3\times3$ colour matrices with $d=\{1,\dots,8\}$ and $f^{abc}$ are the structure constants.  $G^a_{\mu\nu}$ is the gluonic field strength tensor, and $A^a_\mu$ is the gluon field. We will make use of the notation $(G_{\mu\nu})_{ij}=G^a_{\mu\nu} (t^a)_{ij}$ and the relation $g_s^2=4 \pi \alpha_s$ (and $e^2=4 \pi \alpha_{\rm QED}$). The Lagrangian can alternatively be defined with the replacement $g_s\to-g_s$ and the sign convention matters for the applications in Chapters~\ref{chapter4_det} and \ref{chapter7_rad}.

The non-Abelian nature of QCD leads to the possibility of gluon self-interaction and the celebrated \textit{asymptotic freedom} property of QCD \cite{FGM,Fritzsch:1973pi,Politzer:1973fx,Gross:1973id1}. The coupling tends to zero, giving a theory of free quarks, at asymptotically high energy. On the other hand, at low energy, or large distances, the coupling increases.  At energies for which $\alpha_s\gtrsim1$ perturbation theory is not applicable, and one has to resort to \textit{non-perturbative} methods to determine the effects of QCD. Despite the simplicity of the QCD Lagrangian (\ref{basics_eq1}) an accurate determination of non-perturbative QCD from first principles, and hence  \textit{confinement}, poses a major challenge. One such method, based on ideas of Wilson \cite{Wilson:1974sk}, is that of Lattice QCD, which aims to calculate the QCD action computationally on a grid of discretised spacetime points.  An altogether different, and less rigourous, method is that of QCD sum rules, which encodes non-perturbative effects in terms of non-vanishing vacuum expectation values of operators with the quantum numbers of the vacuum. This method is central to the work in this thesis, and shall be discussed in Chapter~\ref{chapter3_SR}.

The electroweak force is the unification of the weak nuclear force and electromagnetism given by the \textit{Glashow-Salam-Weinberg model}. The model is based on the gauge group $\rm SU(2)_L \otimes U(1)_Y$, which is broken by  \textit{spontaneous symmetry breaking} to yield $\rm U(1)_Q$ - the gauge group corresponding to \textit{Quantum Elecrodynamics} (QED).  The weak interaction is mediated by three massive gauge bosons $W^\pm$ and $Z^0$ and occurs between quarks and leptons. The quarks and leptons are arranged, within the three generations, into left-handed doublets and right-handed singlets under $\text{SU(2)}_{\rm L}$
\begin{eqnarray}
 Q_{\rm L}=  \left(\begin{array}{c} 
               U\\ 
                D \\ 
         \end{array}\right)_{\rm L}\,,\quad
          E_{\rm L}=  \left(\begin{array}{c} 
               \nu_l \\ 
                 l^- \\ 
         \end{array}\right)_{\rm L} \,;\quad
         U_{\rm R}\,,D_{\rm R}\,,l^-_{\rm R}\,,
         \label{basics_eq5}
\end{eqnarray}
where the \textit{weak eigenstates} $U=\{u,c,t\}$, $D=\{d,s,b\}$ and $l^-=\{e^-,\mu^-,\tau^-\}$ are the up-type quarks,  down-type quarks and charged leptons respectively. The subscript L (R) represents the left (right)-handed projectors $q_{\rm L (R)}=\frac{1}{2}(1\mp \gamma_5)q$ which reflect the chiral nature of the weak  interaction. The neutrinos are massless in the SM, and the right handed neutrino does not exist.  The electroweak interactions of the quarks are described by the following Lagrangian, which consists of a \textit{charged current} ($CC$) and a \textit{neutral current} ($NC$)
\begin{eqnarray}
   \mathcal{L}^{\textrm{ew}} &=& \mathcal{L}_{CC} +
   \mathcal{L}_{NC}\,,\nonumber \\
    &=& \frac{g}{\sqrt{2}}\left[J_\mu^+W^{+\mu} + J_\mu^-W^{-\mu}\right]\,,\nonumber \\
    & +&
   e\,\left[J^{\textrm{em}}_\mu A^\mu  \right]+ \frac{g}{\cos{\theta_W}}\left[ \left(J^3_\mu - \sin^2{\theta_W}J^{em}_\mu\right) Z^\mu   \right]\,.
\end{eqnarray}
The neutral current part of the Lagrangian is made up of the electromagnetic  current $J^{\textrm{em}}_\mu$ and neutral weak current $J^3_\mu$:
\begin{equation}
   J^{\textrm{em}}_\mu = Q_U\,\bar{U}_{\rm L}\gamma_\mu U_{\rm L}+Q_D\,\bar{D}_{\rm L}\gamma_\mu D_{\rm L}\,, \qquad
   J^3_\mu= \frac{1}{2}(\bar{U}_{\rm L} \gamma_\mu U_{\rm L}-\bar{D}_{\rm L}\gamma_\mu D_{\rm L})\,,
\end{equation}
where $Q_{U(D)}=2/3\,(-1/3)$ is the electric charge of the $U$ $(D)$ quarks, $\theta_W$ is the weak mixing angle and $g$ is the electroweak coupling related to the electromagnetic coupling by $e=g \sin \theta_W$. Rotating to the basis of \textit{mass eigenstates} modifies the charged current in the quark sector to
\begin{equation}
   J^+_\mu =  \bar{U}_{\rm L}^m \gamma_\mu \,\hat{V}_{\rm{CKM}}\, D_{\rm L}^m\,,
\end{equation}
where $\hat{V}_{\mathrm{CKM}}$ is the \textit{Cabbibo-Kobayashi-Maskawa} matrix \cite{Cabibbo:1963yz,Kobayashi:1973fv} and the superscript $m$ denotes mass eigenstates. The CKM matrix is $3\times3$ (for three quark generations), unitary, and its off-diagonal entries allow for transitions between the quark generations.  There are no flavour-changing neutral-currents (FCNC) at tree-level in the SM as the neutral currents $ J^{\textrm{em}}_\mu$ and $J^3_\mu$ are invariant under the transformation to the mass eigenbasis, which is known as  the \textit{Glashow-Iliopoulos-Maiani (GIM) mechanism} \cite{Glashow:1970gm}. The entries of the CKM matrix are written as
\begin{equation}
   \hat{V}_{\mathrm{CKM}} = \left(\begin{array}{ccc}
                    V_{ud} & V_{us} & V_{ub} \\
                    V_{cd} & V_{cs} & V_{cb} \\
                    V_{td} & V_{ts} & V_{tb}  
              \end{array}\right)\,,
\label{basics_eq6}
\end{equation}
and are fundamental parameters of the SM that have to be determined from experiment. Evidently, the matrix has $n^2=9$ parameters $n (n-1)/2=3$ of which are rotation angles due to its unitarity. The six quark fields in Eq.~(\ref{basics_eq5}) can be re-phased, up to an overall phase, leaving the Lagrangian invariant and therefore $9-5-3=1$ phase remains giving rise to complex entries -- complex coupling constants. This is the origin of \textit{CP violation} in the quark sector of the weak interaction. The leptonic sector is described by an analogous mixing matrix which, in the absence of neutrino masses, is given by the unit matrix because all phases can be rotated away.

\begin{figure}[h]
$$  \epsfxsize=0.5\textwidth\epsffile{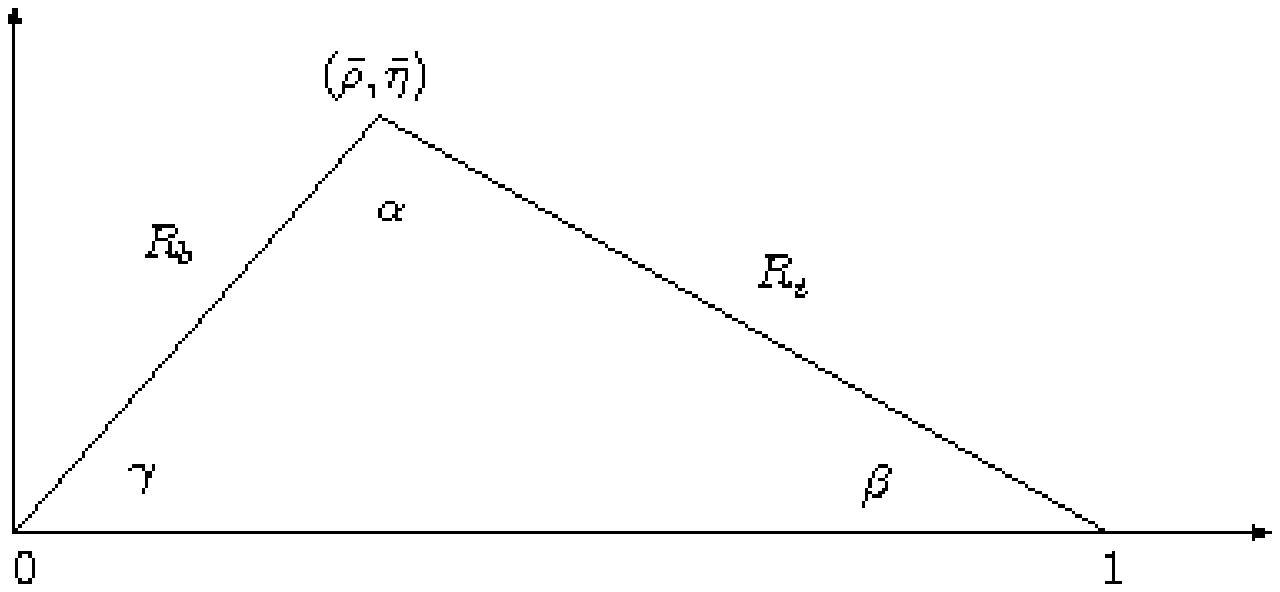}$$
\caption[The Unitary Triangle.]{\small The Unitary Triangle. The determination of the sides $R_b$ and $R_t$ and the angles $\alpha$, $\beta$ and $\gamma$ lead to stringent tests of the Standard Model.} 
\label{basics_fig1}
$$  \epsfxsize=0.5\textwidth\epsffile{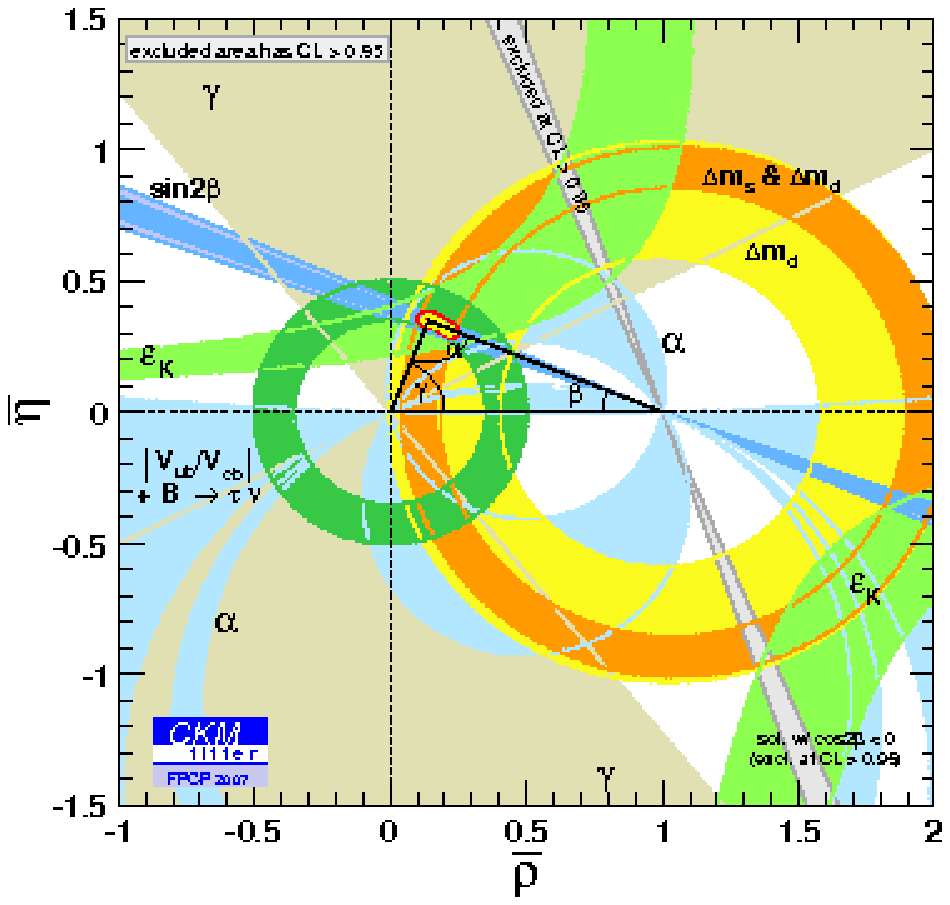}$$
\caption[Constraints on the angles and sides of the Unitarity Triangle.]{\small Constraints on the angles $\alpha$, $\beta$, and $\gamma$ and sides $R_b$ and $R_t$ of the Unitarity Triangle as imposed from numerous experimental sources. Complied by the CKM fitter group \cite{global}.} 
\label{basics_fig2}
\end{figure}

The CKM matrix (\ref{basics_eq6}) is often parameterised to incorporate the constraints of unitarity.\footnote{The ``standard'' parameterisation of the CKM matrix is in terms of the three mixing angles $\theta_{ij}$ $(i,j=1,2,3)$ and the CP violating phase $\delta$ \cite{Yao:2006px}.} A very useful and convenient parameterisation is the \textit{Wolfenstein parameterisation} \cite{Wolfenstein:1983yz} which, along with unitarity, incorporates the experimental observations $|V_{us}|\ll 1$, $|V_{cb}|\sim|V_{us}|^2$ and $|V_{ub}| \ll |V_{cb}|$. It is an expansion in $\lambda=|V_{us}| \approx 0.22$, and as such is only approximately unitary at a given order in $\lambda$:
\begin{equation}
  \hat{V}_\mathrm{CKM}=\left(
  \begin{array}{ccc} 
     1-\frac{\lambda^2}{2} & \lambda & A\lambda^3 (\rho-i\eta) \\
     -\lambda & 1-\frac{\lambda^2}{2} & A\lambda^2 \\
     A \lambda^3 (1-\rho-i\eta) & -A\lambda^2 & 1
   \end{array}\right)+\mathcal{O}(\lambda^4)\,.
   \label{basics_eq7}
\end{equation}
The matrix is given in terms of the four parameters ($A,\lambda,\rho,\eta$); $A$ and $\rho^2+\eta^2$ are order unity and the hierarchy of sizes of elements can be infered from the powers of $\lambda$. The smallness of $V_{cb}$ and $V_{ub}$ are responsible for the relatively long lifetime of $B$ mesons (and baryons), which facilitates their experimental detection. The unitarity of the CKM matrix gives six equations that equal zero and can be represented as triangles in the complex plane. The most widely used of these relations in $B$ physics  is
\begin{equation}
V_{ud}V_{ub}^* + V_{cd}V_{cb}^* + V_{td}V_{tb}^* = 0\,,
\label{basics_eq8}
\end{equation}
which is invariant under phase transformations and is an observable. The above relation is divided by $V_{cd}V_{cb}^*$ to give a triangle in the complex plane with a base of unit length and upper apex at the point $(\bar\rho,\bar\eta)$\footnote{The following rescaling proves convenient to the definition of the UT: $\rho\to\bar\rho=\rho\,(1-\lambda^2/2)$ and $\eta\to\bar\eta=\eta\,(1-\lambda^2/2)$.} known as \textit{The Unitary Triangle} (UT), see Figs.~\ref{basics_fig1} and \ref{basics_fig2}. The sides of the UT are given by
\begin{eqnarray}
R_b &\equiv& \frac{|V_{ud}V_{ub}^*|}{|V_{cd}V_{cb}^*|} = \sqrt{\bar\rho^2+ \bar\eta^2}=
\left(1-\frac{\lambda^2}{2}\right)\frac{1}{\lambda}
\left|\frac{V_{ub}}{V_{cb}}\right|\,,\label{basics_eq88}\\ 
R_t &\equiv& \frac{|V_{td}V_{tb}^*|}{|V_{cd}V_{cb}^*|} = \sqrt{(1-\bar\rho)^2+ \bar\eta^2}\,.
\label{basics_eq9}
\end{eqnarray}
The angles are given by
\begin{equation}
\alpha\equiv\arg\left(-V_{td}V_{ub}V^*_{tb}V_{ud}^*\right)\,,\quad\beta\equiv\arg\left(-V_{cd}V_{tb}V^*_{cb}V_{td}^*\right)\,,\quad\gamma\equiv\arg\left(-V_{ud}V_{cb}V^*_{ub}V_{cd}^*\right)\,.
\end{equation}
The (over) determination of the sides and angles of the UT is a major quest in understanding the SM.   To achieve this goal one must construct decay observables, which can then be matched to experimental results in order to extract values for the desired CKM (or equivalently UT) parameters. Such observables include branching ratios, which may appear simply proportional to a CKM matrix element, and CP asymmetries, which encode the effects of the SM predictions of CP violation, and can also be measured experimentally.

\section{CP Violation In $B$ Decays}
Does  the CKM matrix (\ref{basics_eq6}) account for the CP violation observed in nature? Examining CP violation in $B$ decays allows one to probe the structure of the CKM matrix and is a very promising way to detect the effects of new physics, which many not be expressed through other decay observables. Consequently, the CP properties of  FCNC processes, which are characterised by their potential sensitivity to new physics effects, have been under intense theoretical and experimental investigation for many years. Prime examples of such processes include $B^0$-$\bar B^0$ mixing (see for example Ref.~\cite{Ball:2006xx}) and radiative $B$ decays, see Chapter~\ref{chapter7_rad}.

The idea that the weak interaction may violate parity was first suggested many years ago by Lee and Yang \cite{Lee:1956qn}, and quickly confirmed in the $\beta$ decay of $^{60}$Co by Wu \textit{et al.} \cite{Wu:1957my}. The violation of the combined CP symmetry was first observed in the context of $K$ decays in 1964 \cite{Christenson:1964fg} and it was not until 2001 that it was first observed outside the $K$ system in $B^0_d \to J/ \psi \,K^0_S$ decays \cite{Aubert:2001nu,Abe:2001xe}; in both cases the CKM paradigm was upheld.  Recently discoveries in $B$ physics include the measurement by CDF of the mass difference $\Delta m_s$ \cite{Abulencia:2006ze}. Some of the most important sources of information about the UT from $B$ physics include: the determination of $\sin 2 \beta$ from the ``gold-plated'' decay $B\to J/\psi \,K_S$; the  extraction of $\alpha$ from non-leptonic $B$ decays such as $B \to \pi\pi$; the extraction of $|V_{td}|/|V_{ts}|$ from $B$ mixing and radiative $B$ decays, such as $B\to V \gamma$; and the determination of $|V_{ub}|$ from $B\to\pi l\nu$.

The $B^0_q$-$\bar B^0_q$ systems, where $q=\{d,s\}$, exhibit the phenomenon of particle-antiparticle mixing, which, in the SM is mediated by so-called \textit{box diagrams} whose amplitudes are $\sim G_F^2$ and therefore very small.  We do not go into any detail about the theory of neutral state mixing and we restrict ourselves to only the formulas required in this thesis; for more information see Refs.~\cite{Buras:1998ra,CPV}. State mixing causes, for example, an initially pure beam of $B^0$ mesons to evolve into a time-dependent linear combination of $B^0$ and $\bar B^0$ mesons. There are four main quantities that describe the $B^0_q$-$\bar B^0_q$ system and its decays: the width difference $\Delta \Gamma_q$, the mass difference $\Delta m_q$, the CP violating mixing phase $\phi_q$ and $\lambda_f$ (not to be confused with the Wolfenstein CKM parameter $\lambda\approx 0.22$). One begins by writing the  heavy (H) or light (L) eigenstates of evolution in terms of the flavour states:
\begin{equation}
\ket{B_{\rm H}}=p \ket{B^0}-q \ket{\bar{B}^0}\,,\qquad \ket{B_{\rm L}}=p \ket{B^0}+q \ket{\bar{B}^0}\,,
\label{basics_eq10}
\end{equation}
with $|p|^2+|q|^2=1$. The ratio $q/p$ is given in terms of the $B^0_q$-$\bar B^0_q$  mixing matrix $M_{12}^q$,  by
\begin{equation}
\left.\frac{q}{p}\right|_{q} = \sqrt{\frac{(M_{12}^{q})^*}{M_{12}^{q}}} = e^{-i\phi_{q}}\,,
\label{basics_eq11}
\end{equation}
under the condition $\Delta \Gamma_q \ll \Delta m_q$.  Experimentally, there is no evidence for \textit{mixing-indiced} CP violation in the $B^0_q$-$\bar B^0_q$ systems, i.e. $\left|q/p\right|_{d,s}\approx 1$ \cite{Barberio:2007cr}.   The CP violating mixing phase is given by $ \phi_q={\rm arg}\left[M_{12}^q\right]$ which in the SM and the Wolfenstein parametrisation of the CKM matrix can be written in terms of the UT angles as
\begin{equation}
\phi_d \equiv {\rm arg}[(V_{td}^* V_{tb})^2] = 2 \beta\,,\qquad 
\phi_s \equiv {\rm arg}[(V_{ts}^* V_{tb})^2] = -2 \lambda
\left|\frac{V_{ub}}{V_{cb}}\right|  \sin\gamma\,.
\label{basics_eq12}
\end{equation}
Besides mixing-induced CP violation  there also exists \textit{direct} and \textit{indirect} CP violation for $B$ and $\bar B$ decays to a common CP eigenstate $f$. The corresponding time-dependent CP asymmetry is given by
\begin{eqnarray}
A_{CP}(t) &=& \frac{\Gamma(\bar B^0_q(t)\to f) - \Gamma( B^0_q(t)\to   \bar   f)}{\Gamma(\bar B^0_q(t)\to  f) + \Gamma( B^0_q(t)\to \bar     f)} \nonumber\\
&=&\underbrace{S(f)}_{\rm indirect} \sin(\Delta m_q\, t )-\underbrace{C(f)}_{\rm direct}\cos(\Delta m_q\, t)\,,
\label{basics_eq13}
\end{eqnarray}
where we have neglected the width difference $\Delta\Gamma_q=2 {\rm Re}\left[M^q_{12} \Gamma^{q*}_{12}\right]/|M^q_{12}|$. The oscillation frequency is set by the mass difference between the heavy and light states
\begin{equation}
\Delta m_q = m_H^q-m_L^q=2|M_{12}^{q}|\,,
\label{basics_eq14}
\end{equation}
and the current world averages are \cite{Barberio:2007cr}:  
\begin{equation}
\Delta m_d =0.507\pm 0.004 \,{\rm ps}^{-1}\,,\qquad\Delta m_s=17.77\pm \overbrace{0.10}^{stat.}\pm \overbrace{0.07}^{sys.} {\rm ps}^{-1}\,.
\label{basics_eq15}
\end{equation}
Finally, if  we define the observable quantity 
\begin{equation}
\lambda_f =\frac{q}{p} \frac{\bar A}{A}\,,
\label{basics_eq16}
\end{equation}
where $A$ denotes the decay amplitude, then the two CP asymmetries can be written as
\begin{equation}
C(f)=\frac{1-|\lambda_f|^2}{1+|\lambda_f|^2}\,,\qquad S(f)=\frac{2 \,{\rm Im}\left[ \lambda_f\right]}{1+|\lambda_f|^2}\,.
\label{basics_eq17}
\end{equation}

\section{Effective Field Theories Of Weak Decays}\label{basics_eftowd}
A very widely used tool in the theoretical description of $B$ decay processes is the framework of \textit{effective field theories} \cite{Gilman:1979bc,Buras:1998ra}. The framework simplifies the dynamics of the weak decay by relying on an \textit{operator product expansion}  (OPE) \cite{Wilson:1969zs} of the weak vertices to separate the short and long distance physics. The OPE yields a concise \textit{effective Hamiltonian} $\mathcal{H}^{eff}$ built from a set of local effective operators $Q_i$ multiplied by renormalisation-scale dependent perturbatively calculable \textit{Wilson coefficient functions} $C_i(\mu)$:
\begin{equation}
\left<\mathcal{H}\right> \stackrel{\rm OPE}{\longrightarrow}\left<\mathcal{H}^{eff}\right> \sim \sum_i C_i(\mu) \left<Q_i\right>+\mathcal{O}(k^2/M_W^2)\,,\
\label{basics_eq18}
\end{equation}
where $k$ is the momentum flowing through the $W$ boson propagator. The separation of energy scales stems naturally from the fact that the weak decay of the $B$ meson is governed by physics originating at well separated scales: $m_t,\,M_W\gg m_{b,c}\gg \Lambda_{\rm QCD} \gg m_{u,d,s}$. It is the interplay of weak and strong effects that complicates the treatment of these decays, and must be dealt with appropriately.  By taking into account radiative corrections to tree-level and penguin diagrams, ultimately one obtains the effective Hamiltonian in terms of the set of all relevant local operators, which is closed under renormalisation. The full $\Delta B=1$ effective Hamiltonian is, for a final state containing a $D$ quark
\begin{equation}
\mathcal{H}^{eff}=\frac{G_f}{\sqrt{2}}\sum_{U=u,\,c}\lambda_U^{(D)} \left[C_1 Q_1^U+C_2 Q_2^U+C_{7\gamma} Q_{7\gamma}+C_{8g} Q_{8g}+\sum_{i=3,\dots,10} C_i Q_i\right]\,,
\label{basics_eq20}
\end{equation}
where make use of the standard short-hand notation for the product of CKM matrix elements $\lambda_U^{(D)}\equiv V^*_{UD} V_{Ub}$.  The form of Eq.~(\ref{basics_eq20}) is chosen by assuming the unitarity of the CKM matrix (\ref{basics_eq8}) to explicitly remove the dependence of the top quark CKM matrix elements which originate from penguin loops.  The effective operators are
\begin{eqnarray}
\lefteqn{\bf{Current-Current\footnote{The literature is not consistent concerning the labelling of the two operators $Q_{1,2}$ and one should be aware that the practice of swapping of these two operators is commonplace. We use the convention that the larger Wilson coefficient belongs to $Q_2$; that is, $Q_1$ is the new operator.}:}}\hspace{4cm}\nonumber\\
Q^U_1 &=& (\bar D_i U_j)_{V-A}(\bar U_j b_i)_{V-A}\,,
\qquad Q^U_2 = (\bar DU)_{V-A}(\bar Ub)_{V-A}\,,\nonumber\\
\lefteqn{\rm\bf{QCD~Penguin:}}\hspace{4cm}\nonumber\\
Q_3 &=& (\bar Db)_{V-A} \sum_q (\bar qq)_{V-A}\,,
\qquad Q_4 = (\bar D_i b_j)_{V-A} \sum_q (\bar q_j q_i)_{V-A}\,,\nonumber\\
Q_5 &=& (\bar Db)_{V-A} \sum_q (\bar qq)_{V+A}\,, 
\qquad Q_6 = (\bar D_i b_j)_{V-A} \sum_q (\bar q_j q_i)_{V+A}\,,\nonumber\\
\lefteqn{\rm\bf{Electroweak~Penguin:}}\hspace{4cm}\nonumber\\
Q_7 &=& (\bar Db)_{V-A} \sum_q \frac{3}{2} e_q (\bar qq)_{V+A}\,,
\qquad Q_8 = (\bar D_i b_j)_{V-A} \sum_q  \frac{3}{2} e_q (\bar q_j q_i)_{V+A}\,,\nonumber\\
Q_9 &=& (\bar Db)_{V-A} \sum_q \frac{3}{2} e_q  (\bar qq)_{V-A}\,, 
\qquad Q_{10} = (\bar D_i b_j)_{V-A} \sum_q  \frac{3}{2} e_q (\bar q_j q_i)_{V-A}\,,\nonumber\\
\lefteqn{\rm\bf{Electromagnetic~Dipole:}}\hspace{4cm}\nonumber\\
Q_{7\gamma} &=& \frac{e}{8\pi^2}m_b\, 
        \bar D\sigma^{\mu\nu}(1+\gamma_5)F_{\mu\nu}\,b
         + \frac{e}{8\pi^2}m_D\, 
        \bar D\sigma^{\mu\nu}(1-\gamma_5)F_{\mu\nu}\,b\,, \nonumber\\
\lefteqn{\rm\bf{Chromomagnetic~Dipole:}}\hspace{4cm}\nonumber\\
Q_{8g} &=& \frac{g_s}{8\pi^2}m_b\, 
        \bar D\sigma^{\mu\nu}(1+\gamma_5)G_{\mu\nu}\, b
        + \frac{g_s}{8\pi^2}m_D\, 
        \bar D\sigma^{\mu\nu}(1-\gamma_5)G_{\mu\nu}\, b\,,
        \label{basics_eq21}
\end{eqnarray}
where $e_q$ is the electric charge of the quark $q$ in units of $|e|$ and $F_{\mu\nu}$ is the photonic field strength tensor. The Wilson coefficients entering the effective Hamiltonian are essentially effective coupling constants of the local effective operators. One can view the renormalisation of the matrix elements as an equivalent renormalisation of their Wilson coefficients. One makes use of renormalisation-group techniques to sum the potentially large logarithms $\sim \ln M_W^2/\mu^2$ that appear naturally in the evolution from weak scales $\mathcal{O}(M_W)$ to hadronic scales, such as $\mu\sim m_b$. The operators (\ref{basics_eq21}) mix with each other under evolution and from the renormalisation-scale invariance of $\mathcal{H}^{eff}$ one finds
\begin{equation}
\mu \frac{d}{d \mu} C_i (\mu)=\gamma_{ji}(\mu)\, C_j(\mu)\,, 
\label{basics_eq22}
\end{equation}
where $\hat \gamma$ is the \textit{anomalous dimension} matrix, which can be given as an expansion in the strong coupling via the renormalisation constant $\hat Z$
\begin{equation}
\gamma_{ji}(\mu)= Z^{-1}_{ik}\frac{d Z_{kj}}{d \ln \mu}\,,\qquad \hat{\gamma}=\left(\frac{\alpha_s}{4\pi}\right)\hat{\gamma}^{(0)}+\left(\frac{\alpha_s}{4\pi}\right)^2\hat{\gamma}^{(1)}+\mathcal{O}(\alpha_s^3)\,.
\label{basics_eq23}
\end{equation}
Solving  Eq.~(\ref{basics_eq22}) yields the evolution of the Wilson coefficients via the evolution matrix $\hat U(\mu,\mu_0)$
\begin{equation}
 C_i(\mu)= U_{ij}(\mu,\mu_0)\, C_j(\mu_0)\,,\qquad\hat{U}(\mu,\mu_0)=\exp \int^{g(\mu)}_{g(\mu_0)}dg^\prime\,\frac{\hat{\gamma}^{T}(g^\prime)}{\beta(g^\prime)}\,,
\label{basics_eq24}
\end{equation}
where $\beta(g)$ is the QCD $\beta$-function. To leading order one has
\begin{equation}
\hat{U}^{\rm LO}(\mu,\mu_0)=\left(\frac{\alpha_s(\mu_0)}{\alpha_s(\mu)}\right)^{\frac{\hat{\gamma}^{(0)T}}{2 \beta_0}}= \hat V\left[\left(\frac{\alpha_s(\mu_0)}{\alpha_s(\mu)}\right)^{\frac{\overrightarrow{\gamma}^{\left(0\right)}}{2 \beta_0}}\right]_D\hat V^{-1}\,,
\label{basics_eq25}
\end{equation}
where $V$ is the matrix that diagonalises $\hat{\gamma}^{(0)T}$ and $\overrightarrow{\gamma}^{\left(0\right)}$ is a vector of the eigenvalues of the leading order anomalous dimension matrix $\hat{\gamma}^{(0)}=\hat V\hat{\gamma}^{(0)T}_D\hat V^{-1}$. At NLO we have
\begin{equation}
C_i(\mu)=C_i^{(0)}(\mu)+\frac{\alpha_s(\mu)}{4\pi}C_i^{(1)}(\mu)\,,
\label{basics_eq26}
\end{equation}
and the evolution is a bit more complicated:
\begin{equation}
\hat{U}^{\rm NLO}(\mu,\mu_0)=\left[1+\frac{\alpha_s(\mu)}{4\pi}\hat{J}\right]\hat{U}^{\rm LO}(\mu,\mu_0)\left[1-\frac{\alpha_s(\mu_0)}{4\pi}\hat{J}\right]\,,
\label{basics_eq27}
\end{equation}
with
\begin{equation}
\hat{J}=V \hat{S}V^{-1}\,,\qquad S_{ij}=\delta_{ij}\gamma_{i}^{(0)}\frac{\beta_1}{2\beta_0^2}-\frac{G_{ij}}{2\beta_0+\gamma_i^{(0)}-\gamma_j^{(0)}}\,,\qquad\hat{G}=V^{-1} \hat{\gamma}^{(1)T}V\,.
\label{basics_eq28}
\end{equation}
To NLO the required $\beta$-function coefficients are $\beta_1=\frac{34}{3}N_c^2-\frac{10}{3}N_c N_f -2 C_F N_f$ and $\beta_0=\frac{11}{3}N_c  -\frac{2}{3} N_f$ with $N_f$ is the number of active flavours, $C_F=(N_c^2-1)/(2N_c)$ and $N_c$ the number of colours. Care must be taken in evolving through ``thresholds'' where the number of active flavours $N_f$ changes; the evolution must then be taken in stages, as a change in $N_f$ changes the $\beta$-function coefficients and the anomalous dimension matrices. If there is a flavour threshold $\mu_{\rm th}$  between $\mu_0$ and $\mu$, which changes the number of active flavours from $N_f$ to $N_f+1$, then one has to make the replacement
\begin{equation}
\hat U(\mu,\mu_0)\to \left.\hat U(\mu,\mu_{\rm th})\right|_{N_f +1}\cdot \left.\hat U(\mu_{\rm th},\mu_0) \right|_{N_f}.
\label{basics_eq29}
\end{equation} 
The effective Hamiltonian, combined with the renormalisation-group improvement of the perturbative series forms an exceptionally powerful framework. The matrix elements of the local operators $\left<Q_i\right>$ are the subject of QCD factorisation theorems, such as that discussed in Chapter~\ref{chapter6_QCDF}, which allow the calculation of $B$ decay amplitudes.  From these amplitudes one can construct observables such as branching fractions, CP asymmetries, and isospin asymmetries which can be investigated phenomenologically.  
\chapter{Vector Meson Light-Cone Distribution Amplitudes}\label{chapter2_DAs}
In this chapter we discuss light vector meson light-cone distribution amplitudes and via the (approximate) conformal symmetry of QCD present expressions for the distribution amplitudes up to twist-3. The method introduces a set of non-perturbative parameters which is reduced in size by invoking the QCD equations of motion to relate the two-particle twist-3 distribution amplitudes to the three-particle twist-3 and two-particle twist-2 distribution amplitudes. In our analysis we include all $\rm SU(3)_F$-breaking effects and G-parity violating terms thus allowing one to fully differentiate between $\rho$, $K^*$ and $\phi$ mesons. Moreover, a non-zero quark mass induces a mixing between twist-2 and twist-3 parameters under a change of renormalisation scale $\mu$. To simplify notation we explicitly consider the $K^*$ meson, with quark composition $s\bar{q}$ where $q=\{u,d\}$.\footnote{The notation in this thesis, $K^*$ being a $(s\bar q)$   bound state, is in contrast to the standard labelling, according to  which $K^{*0}=(d\bar s)$ and $\bar K^{*0}=(s\bar d)$. This is the standard notation used for light-cone distribution amplitudes where $K^*$ always contains an $s$ quark, and $\bar K^*$ an $\bar s$  quark. This distinction is relevant because of  a sign change of G-odd matrix elements under $(s\bar  q)\leftrightarrow (q\bar s)$. This notation also applies to calculations of form factors and other matrix  elements which involve light-cone distribution amplitudes.}

There are two main applications of meson distribution amplitudes that motivate their study:
\begin{itemize}
\item{they are directly applicable to the theoretical description of exclusive decay processes via QCD factorisation theorems, which require the distribution amplitudes as a non-perturbative input, see Chapter~\ref{chapter6_QCDF}.}
\item{they are also applicable to the determination of transition form factors from the light-cone sum rule approach and as such are indirectly applicable to the same QCD factorisation theorems for which the transition form factors are also required,  see Chapters~\ref{chapter3_SR} and \ref{chapter5_eta}.}
\end{itemize}
In Chapter~\ref{chapter4_det} we calculate,  from QCD sum rules,  numerical values for the leading twist-2 and twist-3 distribution amplitude  parameters defined here. Standard notations used, such as the light-cone coordinates, are given in Appendix~\ref{appendixA}. The material covered in this chapter partially follows that of Ref.~\cite{Ball:2007rt}.

\section{Introduction}
Hadronic light-cone distribution amplitudes (DAs) of light mesons were first discussed in the ground-breaking papers of Brodsky, Lepage, and others, see Refs.~\cite{Chernyak:1977fk, Chernyak:1980dk, Lepage:1980fj, Efremov:1979qk, Efremov:1978rn,Chernyak:1977as, Lepage:1979zb,Chernyak:1980dj} and play an essential role in the QCD description of hard exclusive processes \cite{Chernyak:1981zz,Brodsky:1989pv}.  The amplitudes that describe such processes factorise in the asymptotic limit $Q^2\sim 1/x^2 \to \infty$ -- where $Q^2$ is the momentum transfer and $x$ the transverse separation of the partons -- and are dominated by contributions from near the light-cone. The factorisation is given by the convolution of a hard-scattering kernel, calculable in perturbation theory, and process-independent, universal, non-perturbative DAs. 

The study of hadronic DAs has a long history. The simplest and first to be investigated were the twist-2 DA of the $\pi$ \cite{Lepage:1980fj,Efremov:1979qk,Chernyak:1977as,Lepage:1979zb}. Higher twist DAs of the $\pi$, alongside those of the other pseudoscalar mesons followed \cite{Ball:1998je}.   For vector mesons, the leading-twist DAs of the $\rho$ were first investigated by Chernyak and Zhitnitsky in Ref.~\cite{Chernyak:1983ej} and later in Refs.~\cite{Ali:1993vd, Ball:1996tb}. The formalism of higher twist-3 and twist-4 contributions, including meson mass corrections, was investigated by Ball \textit{et al.} in Refs.~\cite{Ball:1998sk,Ball:1998ff,Ball:2006wn,Ball:2007zt}. 

The DAs of the $K^*$  ($K$) differ to those of the $\rho$ ($\pi$) due to the non-zero strange quark mass which yields $\rm SU(3)_F$-breaking and G-parity violating corrections from a number of different sources.\footnote{Perfect $\rm SU(3)_F$ symmetry is realised for equal $u,d,$ and $s$ quark masses.} The study of the various contributions span many publications:
\begin{itemize}
\item{explicit quark mass corrections to DAs and evolution equations are generated by the QCD equations of motion (EOM) and only affect higher twist DAs. The contributions for vector mesons were calculated in Ref.~\cite{Ball:1998sk} up to twist-3, and those to the evolution equations for vector mesons in Ref.~\cite{Ball:2007rt} and flavour-octet pseudoscalar mesons Ref.~\cite{Ball:2006wn}.}
\item{G-parity violating contributions, which are proportional to $m_s-m_q$ and hence vanish for equal quark masses, i.e. for $\rho$ and $\phi$, were investigated for twist-2 DAs in Refs.~\cite{Chernyak:1983ej,Ball:1998sk,Ball:2003sc,Braun:2004vf,Ball:2005vx,Ball:2006fz} and for twist-3 DAs in Ref.~\cite{Ball:2007rt}.}
\item{$\rm SU(3)_F$-breaking of non-perturbative hadronic parameters entering the DAs.
The effects for the twist-2 parameters are known from
Refs.~\cite{Chernyak:1983ej,Ball:1998sk,Ball:2003sc}, twist-3 from Ref.~\cite{Ball:2007rt} and twist-4 from Ref.~\cite{Ball:2007zt}. The twist-3 vector meson parameters are discussed in Chapter~\ref{chapter4_det} where we include all these effects in a determination of numerical values using QCD sum rules.}
\end{itemize}
The objects which define the DAs are vacuum-to-meson matrix elements of non-local operators at strictly light-like separations $z^2=0$ \cite{Chernyak:1983ej}. Two examples we shall encounter are
\begin{eqnarray}
\bra{0}\bar{q}(z)\Gamma [z,-z] s(-z) \ket{K^*(p,\lambda)}\,,\qquad \bra{0}\bar{q}(z) [z,v z] g_s G_{\mu \nu}(vz)\Gamma [v z,-z] s(-z) \ket{K^*(p,\lambda)}\,,\nonumber\\
\label{das_eq1}
\end{eqnarray}
where $\Gamma$ is a general Dirac matrix, $\lambda=\{\parallel,\perp\}$ is the polarisation of the $K^*$ meson and the quark fields are taken at symmetric separation for simplicity.\footnote{The Dirac matrices  $\Gamma =\{ \sigma_{\mu \nu},\,i \gamma_5,\, \bf 1\}$ give rise to so-called \textit{chiral-odd} distributions because they are chirality-violating. Likewise, distributions generated from $\Gamma =\{ \gamma_{\mu},\,\gamma_{\mu} \gamma_5\}$ are \textit{chiral-even}.}  The first (second) matrix element above corresponds to a two- (three-) particle Fock state. To render the matrix element gauge invariant the path-ordered gauge factor is included
\begin{equation}
[x,y]=\textrm{P}\, \exp \left[ i g_s \int^1_0 dt\, (x-y)_\mu A^\mu (t x+(1-t)y)\right].
\end{equation}
For convenience we work in the \textit{fixed-point} gauge\footnote{also known as the \textit{Fock-Schwinger} gauge.}
\begin{equation}
(x-x_0)^\mu A_\mu^a (x)=0\,,
\label{das_eq2}
\end{equation}
and by choosing $x_0=0$ we have $[x,-x]=1$. The gauge factor will be implied unless otherwise stated. The DAs are dimensionless functions of  the collinear momentum fractions of a fixed number of constituents within a meson, at zero transverse separation. For two-particle DAs the constituent strange quark and antiquark  ($\bar{q}$) share $u$ and $\bar{u}=1-u$ of the meson momentum $p$ respectively.  For three-particle DAs we have $\underline{\alpha} = (\alpha_1, \alpha_2, \alpha_3 )$ corresponding to the momentum fractions carried by the strange quark, antiquark ($\bar{q}$) and gluon, respectively.  For a minimum number of constituents, the DAs are related to the \textit{Bethe-Salpeter wavefunction} $\phi_{BS}$ by integration over the transverse momenta
\begin{equation}
\phi(u, \mu) \sim \int^{|k_{\perp}|<\mu}d^2 k_{\perp}\,\phi_{BS}(u,k_{\perp})\,,
\label{das_eq3}
\end{equation}
where  $\mu$ is the renormalisation scale.  The price to pay for integrating out $k_{\perp}$ below $\mu$ is a renormalisation-scale dependence of the DAs governed by renormalisation-group equations.  The DAs have to be evaluated at the scale $\mu^2 \sim x^{-2}$ i.e. of the order of the deviation from the light-cone \cite{Balitsky:1987bk}.

Non-local operators that appear at finite $Q^2$ or mass scales are expanded near the light-cone $x^2 \neq 0$ as an OPE in terms of the renormalised  non-local operators on the light-cone - the \textit{light-cone expansion} \cite{Balitsky:1987bk}.\footnote{The expansion is facilitated by using light-cone coordinates which are given in Appendix~\ref{appendixA}.} After taking matrix elements the resulting Lorentz-invariant amplitudes are matched to the definitions of the DAs with the coefficient functions of the expansion taken at tree-level, to leading logarithmic accuracy.  

The structure of vector meson DAs follows the same pattern as the nucleon structure functions and can be classified in the same way \cite{Jaffe:1991ra}. They are described by separate DAs for each polarisation and thus there are more vector meson DAs than pseudoscalar DAs. 

Lastly, we briefly mention some other DAs. Flavour-singlet pseudoscalar meson DAs
are complicated by the $\rm U(1)_A$ anomaly of QCD and are discussed in Chapter~\ref{chapter5_eta} in the context of the $B \to \etapb$ transition form factor \cite{Ball:2007hb}. Much work has been done concerning the DAs of heavy mesons, such as the $B$ meson \cite{Szczepaniak:1990dt ,Braun:2003wx}; indeed, the DAs of $B$ mesons enter the QCD factorisation framework of radiative and non-leptonic $B$ decays, as discussed in Chapter~\ref{chapter6_QCDF}, and a variant light-cone sum rule method devised in Ref.~\cite{Khodjamirian:2006st}. There also exist DAs of the photon which describe its ``soft'' hadronic components, along with the usual ``hard'' electromagnetic components \cite{Ball:2002ps}. The photonic DAs can be important in, for example, $B \to V \gamma$ decays \cite{Ball:2006eu} as investigated in Chapter~\ref{chapter7_rad}, and $B \to \gamma e \nu$ \cite{Descotes-Genon:2002mw,Ball:2003fq}. Finally, the field of baryon DAs is also active and many of the tools and concepts we cover in this thesis find application there, see for example Ref.~\cite{Braun:2006hn} for a review.

\section{The Conformal Expansion}
The standard determination of meson DAs  proceeds by making use of the conformal symmetry of massless QCD at tree-level.  The conformal expansion is analogous to the partial wave expansion of wave functions in quantum mechanics in spherical harmonics $\psi(r,\theta,\phi) \to R(r) \sum_{m,l} Y^l_m (\theta,\phi)$. The expansion uncovers a simple multiplicative renormalisation at leading-order, and as such different partial waves, with different \textit{conformal spin}, do not mix under  a change of renormalisation scale.  At next-to-leading-order this is not the case, because strictly speaking the conformal symmetry of a quantum theory requires its $\beta$ function to vanish. Proximity to the conformal limit in QCD is therefore governed by the value of the strong coupling constant, becoming true as $\alpha_s \to 0$ and we pass to the free theory.\footnote{It must be noted that mass terms break the conformal expansion immediately at the classical level. This does not upset the conformal expansion, however. See Ref.~\cite{Braun:2003rp} for details.} Using the QCD equations of motion we can elucidate this mixing order-by-order in the conformal expansion. 

The application of conformal symmetry to exclusive processes has recieved a lot of attention in the literature, see Refs.~\cite{Brodsky:1980ny,Brodsky:1985ve,Ohrndorf:1981qv,Braun:1989iv,Makeenko:1980bh}. The main benefit of the conformal expansion is the systematic separation of the longitudinal and transverse degrees of freedom in meson DAs. The former correspond to the longitudinal momentum fractions and is given by irreducible representations of the relevant symmetry group, SL(2,$\mathbb R$).  The latter are integrated out to yield a renormalisation-scale dependence of the DAs, described by renormalisation-group equations.  Here we focus on the most important points, see Ref.~\cite{Braun:2003rp} for a detailed review. 

\subsection{Conformal Group}
The conformal group is defined as all transformations that change only  the scale of the metric and as such preserve angles and leave the light-cone invariant $g_{\mu\nu}^\prime(x^\prime)=\omega(x) g_{\mu\nu}(x)$; the spacetime interval $ds^2=g_{\mu\nu}(x) \,dx_\mu dx_\nu$ is conserved up to scaling. These transformations form a generalisation of the Poincar\'e group. The full conformal algebra in 4 dimensions includes fifteen generators
\begin{eqnarray}
\textbf{P}_\mu &\to& \rm 4~ Translations, \nonumber\\
\textbf{M}_{\mu \nu}&\to& \rm 6 ~Lorentz ~rotations, \nonumber\\
\textbf{D} &\to& \rm 1 ~Dilatation,\nonumber\\
\textbf{K}_\mu &\to& \rm 4 ~Special ~conformal ~translations.
\label{das_eq4}
\end{eqnarray}
Our hadronic picture is of partons moving collinearly in, say the $p_{\mu}$ direction, existing near the light-cone. We therefore restrict the \textit{fundamental fields} of the conformal group to the light-cone
$\Phi(x) \to \Phi(\alpha z)$, where $\alpha $ is a real number, and we assume fields to be eigenstates of the spin operator 
\begin{equation}
\Sigma^{\mu\nu} \psi = \frac{i}{2}\sigma^{\mu\nu} \psi\,,
\label{das_eq5}
\end{equation}
so as to have a fixed Lorentz-spin projection $s$ in the $z_{\mu}$ (``plus'') direction $\Sigma_{+-} \Phi(\alpha z) = s \,\Phi(\alpha z)$. For leading-twist operators this is automatically satisfied and for higher-twist operators projections are used to separate different spin states, as we shall discuss shortly. The full conformal symmetry (\ref{das_eq4}) is now modified  and it turns out that the resulting group of transformations form the special linear group SL(2,$\mathbb R$), or so-called \textit{collinear conformal group}, given by just four generators. They are written in standard form by constructing the following linear combinations
\begin{eqnarray}
\textbf{L}_+= \textbf{L}_1+i \textbf{L}_2=-i \textbf{P}_+\,,& \qquad& \textbf{L}_-= \textbf{L}_1-i \textbf{L}_2= \frac{i}{2} \textbf{K}_-\,,\nonumber\\
\textbf{L}_0= \frac{i}{2}(\textbf{D}+\textbf{M}_{+-})\,,& \qquad& \textbf{E}=\frac{i}{2}(\textbf{D}-\textbf{M}_{+-})\,.
\end{eqnarray}
which leads to the familiar relations
\begin{equation}
[\textbf{L}_0,\,\textbf{L}_\mp]=\mp \textbf{L}_\mp\,, \qquad [\textbf{L}_-,\,\textbf{L}_+]=-2 \textbf{L}_0\,.
\end{equation}
The operators act on the fundamental fields as
\begin{eqnarray}
\left[\textbf{L}_+,\Phi(\alpha z) \right]&=&-\partial_{\alpha} \Phi(\alpha n)\,,
\label{das_eq6}\\
\left[\textbf{L}_{-},\Phi(\alpha z)\right]&=&(\alpha^2 \partial_{\alpha} +2 j \alpha)\Phi(\alpha n)\,,
\label{lower}\\
\left[\textbf{L}_0,\Phi(\alpha z) \right]&=&(\alpha \partial_{\alpha}+j) \Phi(\alpha n)\,,
\label{das_eq7}\\
\left[\textbf{E},\Phi(\alpha z)\right]&=&\frac{1}{2} (l-s) \Phi(\alpha n)\,,
\label{das_eq8}
\end{eqnarray}
where $t=l-s$ is the \textit{twist},\footnote{strictly it is the \textit{collinear twist} which is defined as ``dimension minus spin projection on the positive direction''. There also exists \textit{geometric twist} which is defined as ``dimension minus spin''.} $l$ is the canonical mass dimension,\footnote{For example, $l=3/2$ for quarks and $l=2$ for gluons.} $s$ the Lorentz-spin projection, and $j=\frac{1}{2}(l+s)$ the \textit{conformal spin} of the field $\Phi$. The conformal spin specifies the representation of the collinear conformal group. The operator $\textbf{E}$ commutes with all $\textbf{L}_i$ and therefore twist is a good quantum number for each conformal field. The Casimir operator commutes with all $\textbf{L}_i$ and is given by
\begin{equation}
\sum_{i=0,1,2}[\textbf{L}_i,[\textbf{L}_i,\Phi(\alpha z)]]=j(j-1)\Phi(\alpha z) = \textbf{L}^2 \Phi(\alpha z)\,.
\end{equation}
At the origin of the light-cone $\alpha=0$ and the field $\Phi(0)$ is killed by the lowering operator $\textbf{L}_-$ and as such has the minimum spin projection $j_{\rm min}$ of states of conformal spin $j$. One can define a \textit{conformal operator} $\mathbb{O}_n=\Phi(0)$ by requiring that it transforms just as the fundamental field, Eqs.~(\ref{lower} - \ref{das_eq8}), and is killed by the lowering operator $\textbf{L}_-$. The raising operator $\textbf{L}_+$ can be repeatedly applied to $\Phi(0)$ to give
\begin{equation}
\mathbb{O}_{n,n+k}=\underbrace{\left[ \textbf{L}_+,...,\left[\textbf{L}_+,\left[\textbf{L}_+\right.\right.\right.}_{k},\left.\left.\left.\Phi(0)\right]\right]\right]=\left(-i \partial_+\right)^k \mathbb{O}_n\,,
\end{equation}
where $\mathbb{O}_{n,n}=\mathbb{O}_n$ and the subscript $n$ defines the \textit{conformal tower} of states, of conformal spin $j_{\textrm{min}}<j_{\textrm{min}}+k<\infty$, generated by the collinear conformal algebra. This is an infinite dimensional representation of the collinear conformal group.

\subsection{States of Definite Spin}
Now the main language of the collinear conformal group is defined, if one can relate the fundamental fields $\Phi$ to the operators of hard processes in QCD one can export all the machinery above and immediately reap the benefits. To this end,  consider the non-local two-particle operator at light-like separation (\ref{das_eq1}) and expand at small distances
\begin{equation}
\bar{q}(z)\Gamma s(-z)= \sum_k \frac{1}{k!}\, \bar{q}(0) (\deriv \cdot z)^k \Gamma s(0)\,,
\label{das_eq9}
\end{equation}
where $\deriv_\mu= \derright_\mu- \derleft_\mu$. The question is, how does one express these local operators in terms of conformal operators and thus separate all the different twist contributions?  To proceed one decomposes the quark fields into definite Lorentz-spin components using the projection operators
\begin{equation}
\Pi_+ = \frac{1}{2} \frac{\slash{p}\slash{z}}{ p\cdot z}\,, \qquad \Pi_- = \frac{1}{2} \frac{\slash{z}\slash{p}}{ p\cdot z}\,, \qquad \Pi_+ +\Pi_- =1\,.
\end{equation}
which project onto the ``plus'' and ``minus'' components of the spinor respectively. Using the generator of the spin rotations of a spinor field (\ref{das_eq5}) one can show
\begin{equation}
\underbrace{\psi_+ = \Pi_+ \psi}_{ \shortstack{\footnotesize{$s=+1/2$} \\ \footnotesize{$j=1$}}}\,, \qquad \underbrace{\psi_- = \Pi_- \psi}_{ \shortstack{\footnotesize{$s=-1/2$} \\ \footnotesize{$j=1/2$}}}\,.
\end{equation}
The composite operators (\ref{das_eq9}) have  conformal spin projection $j=j_q+j_s+k$  where the subscripts correspond to the separate quark fields. The composite operators are ordered by increasing twist
\begin{equation}
\underbrace{\bar{q}_{+} \Gamma s_{+}}_{ \rm t=2}\,, \qquad \underbrace{\bar{q}_{+} \Gamma s_{-} + \bar{q}_{-} \Gamma s_{+}}_{ \rm t=3}\,, \qquad \underbrace{\bar{q}_{-} \Gamma s_{-}}_{ \rm t=4}\,. 
\label{das_eq10}
\end{equation}
It can be shown that the corresponding local conformal operators are
\begin{eqnarray}
\mathbb{Q}_n^{\rm t=2}(x) &=& (i\partial_+)^n \left[\bar{q}(x) \gamma_+ C^{3/2}_{n} \left(\deriv_+ / \partial_+\right) s(x) \right]\,,
\label{das_eq11}\\
\mathbb{Q}_n^{\rm t=3}(x) &=& (i\partial_+)^n \left[\bar{q}(x)\gamma_+ \gamma_\perp \gamma_-P^{(1,0)}_{n} \left(\deriv_+ / \partial_+\right) s(x) \right]\,,\label{das_eq12}\\
\mathbb{Q}_n^{\rm t=4}(x) &=& (i\partial_+)^n \left[\bar{q}(x) \gamma_- C^{1/2}_{n} \left(\deriv_+ / \partial_+\right) s(x) \right]\,,
\label{das_eq13}
\end{eqnarray}
where $\partial_+ =\derleft_+ +\derright_+$, $C^m_n(x)$ are \textit{Gegenbauer polynomials} and $P^{(r,s)}_n(x)$ are \textit{Jacobi polynomials}. There is another twist-3 operator corresponding to Eq.~(\ref{das_eq12}) with the replacement $P^{(1,0)}_{n}(x) \to P^{(0,1)}_{n}(x)$. One can now connect firmly to QCD with the specific example of the leading-twist operator; consider again Eq.~(\ref{das_eq1}) and specify the twist-2 Dirac matrix $\Gamma \to \gamma_\mu$ (projected onto $z_\mu$) and define the DA as
 \begin{equation}
 \bra{0} \bar{q}(z)\gamma_z s(-z) \ket{K^*(p,\lambda)}= f_{K^*}^\parallel m_{K^*} (e^{(\lambda)}\cdot z) \int^1_0 du\, e^{-i [\bar{u} z+ u(-z)]\cdot  p} \phi^{\parallel}_{2;K^*}(u,\mu)\,.
 \label{das_eq14}
\end{equation}
Then, using Eq.~(\ref{das_eq9}) and comparing the result to Eq.~(\ref{das_eq11}), one finds
\begin{equation}
\int^1_0 du\,  C^{3/2}_{n}(\xi)\,\phi^{\parallel}_{2;K^*}(u,\mu) = \langle\langle \mathbb{Q}_n^{\rm t=2}\rangle\rangle\,,
\label{das_eq15}
\end{equation}
where we introduce the shorthand $\xi = u-\bar{u} = 2u-1$, and $\langle\langle \mathbb{Q}_n^{\rm t=2}\rangle\rangle$ are the reduced matrix elements of the operator $\mathbb{Q}_n^{\rm t=2}$. The Gegenbauer polynomials form a complete set of orthogonal functions over the weight function $6 u (1-u)$ in the interval $0<u<1$
\begin{equation}
\int^1_0 du\,  u \bar u \,C^{3/2}_{n}(\xi)\, C^{3/2}_{m}(\xi)=\delta_{m n}\frac{(n+1)(n+2)}{4(2n+3)}\,,
\end{equation}
and so one can invert (\ref{das_eq15}) to find
\begin{equation}
\phi^{\parallel}_{2;K^*}(u,\mu) = 6 u (1-u) \left\{1+ \sum_{n=1}^{\infty} a_n^{\parallel}(\mu) \,C^{3/2}_{n}(\xi)\right\}\,,
\label{das_eq16}
\end{equation}
where the \textit{Gegenbauer coefficients} $a_n^{\parallel}(\mu)$ are related to the reduced matrix elements, Eq.~(\ref{das_eq15}), as
\begin{equation}
a_n^{\parallel}(\mu) =\frac{2(2n+3)}{3(n+1)(n+2)}\langle\langle \mathbb{Q}_n^{\rm t=2}\rangle\rangle\,, \qquad a_0^{\parallel} = 1\,.
\label{das_eq17}
\end{equation}
The result is that the conformal symmetry has separated the longitudinal degrees of freedom -- as contained in the orthogonal Gegenbauer polynomials which are function of the momentum fraction $u$ for the  twist-2 distribution -- from the transverse degrees of freedom, which now show up as the renormalisation-scale dependence of the Gegenbauer coefficients $a_n^{\parallel}$. The Gegenbauer coefficients  contain the non-perturbative information of the DA and, for the leading-twist DA, are of conformal spin $n+2$.  The higher-twist two-particle DAs are expanded analogously in $P^{(1,0)}_{n}$ and $P^{(0,1)}_{n}$ for twist-3 and $C^{1/2}_{n}$ for twist-4, see Eqs. (\ref{das_eq12})  and (\ref{das_eq13}) respectively.  The explicit expression for the DA of an $m$-particle state with the lowest possible conformal spin $j=j_1+\dots+j_m$, the so-called \textit{asymptotic distribution amplitude}, is given by
\begin{equation}
\phi_{as}(\alpha_1,\alpha_2,\cdots,\alpha_m) =
\frac{\Gamma(2j_1+\cdots +2j_m)}{\Gamma(2j_1)\cdots \Gamma(2j_m)}\,
\alpha_1^{2j_1-1}\alpha_2^{2j_2-1}\ldots \alpha_m^{2j_m-1}\,,
\label{das_eq18}
\end{equation}
where $\sum_{k=1}^m \alpha_k =1$ \cite{Braun:1989iv}. For the twist-2 two-particle DA considered $j=1$ and we recover the weight function $6 u (1-u)$. Analogously, for a multi-particle DA, states higher in conformal spin are multiplied by polynomials orthogonal over the weight function Eq.~(\ref{das_eq18}). The matrix element Eq.~(\ref{das_eq14}) with the chiral-odd Dirac matrix $\Gamma \to \sigma_{z p}$ starts at twist-2 also, and gives rise to the second two-particle twist-2 DA
\begin{equation}
\phi^{\perp}_{2;K^*}(u,\mu) = 6 u (1-u)\left\{1+ \sum_{n=1}^{\infty} a_n^{\perp}(\mu) \,C^{3/2}_{n}(\xi)\right\}\,.
\label{das_eq19}
\end{equation}
The expansion of DAs in terms of an infinite sum of partial waves, as in Eq. (\ref{das_eq19}), is very general, and is valid at the level of operators. In practice the concept of \textit{G-parity} allows one to classify which Gegenbauer coefficients contribute for specific matrix elements of those operators. The G-parity operator $\cal{G}$ is defined as $\mathcal{G}=\mathcal{C} e^{-i\pi T_2}$ where $T_2$ is the isospin generator of the $2$ axis and $\cal{C}$ is the charge conjugation operator.  G-parity is the generalisation of charge conjugation to particle multiplets, for example, $\mathcal{G}\ket{\pi^{\pm,0}}=-\ket{\pi^{\pm,0}}$ and is conserved in QCD. It effectively swaps quarks for anti-quarks and therefore for equal mass quarks $u \leftrightarrow \bar{u}$ and consequently, for $\pi,\,\rho,\,\omega$ and $\phi$, the odd Gegenbauer coefficients $a_{2n+1}$ vanish. 

\section{Two-Particle Twist-2 Distribution Amplitudes}\label{das_sec1.3}
As mentioned in the last section, there are two two-particle matrix elements that begin at twist-2 \cite{Ball:1998sk}:\footnote{The vacuum-vector meson matrix elements vanish for $\Gamma =i \gamma_5$ because it is impossible to construct a pseudoscalar quantity from the three available 4-vectors $p_\mu,\,z_\mu$ and $e^{(\lambda)}_\mu$.}
\begin{eqnarray}
\lefteqn{
\bra{0} \bar q(x) \gamma_\mu s(-x) \ket{K^*(P,\lambda)} = }\hspace*{1cm}
\nonumber\\
&&{}f_{K^*}^\parallel m_{K^*} \left\{ \frac{e^{(\lambda)}x}{P\cdot x}\,P_\mu \int_0^1 du\, e^{i\xi P\cdot x} \left[ \phi_{2;K^*}^\parallel(u) +   \frac{1}{4}\, m_{K^*}^2 x^2  \phi^\parallel_{4;K^*}(u)\right]\right.\nonumber\\
&&{} + \left( e^{(\lambda)}_\mu -
P_\mu\,\frac{e^{(\lambda)}x}{P\cdot x}\right) \int_0^1 du\,e^{i\xi  P\cdot x}\,\phi_{3;K^*}^\perp(u) \nonumber\\
&&\left. - \frac{1}{2}\,x_\mu \,\frac{e^{(\lambda)}x}{(P\cdot x)^2} \, m_{K^*}^2 \int_0^1 du\,e^{i\xi P\cdot x}\,
\left[ \psi_{4;K^*}^\parallel(u) + \phi_{2;K^*}^\parallel(u) - 2   \phi_{3;K^*}^\perp(u)\right]\right\}\,,
  \label{das_eq20}\\[10pt]
\lefteqn{
\bra{0} \bar q(x) \sigma_{\mu\nu} s(-x) \ket{ K^*(P,\lambda)} =}\hspace*{1cm} \nonumber\\
&&{}i f_{K^*}^\perp \left\{ (e_\mu^{(\lambda)} P_\nu - e_\nu^{(\lambda)}   P_\mu) \int_0^1 du \, e^{i\xi P\cdot x} \left[ \phi_{2;K^*}^\perp(u) +   \frac{1}{4}\, m_{K^*}^2 x^2  \phi^\perp_{4;K^*}(u)\right]\right. \nonumber\\
&&{}
+ (P_\mu x_\nu - P_\nu x_\mu)\,\frac{e^{(\lambda)}x}{(P\cdot x)^2}\,m_{K^*}^2 \int_0^1 du \,
e^{i\xi P\cdot x} \left[ \phi_{3;K^*}^\parallel(u) -   \frac{1}{2}\phi_{2;K^*}^\perp(u) -   \frac{1}{2}\psi_{4;K^*}^\perp(u)\right] \nonumber\\
&&{}\left.
+ \frac{1}{2}\,(e_\mu^{(\lambda)} x_\nu - e_\nu^{(\lambda)}x_\mu)\, \frac{m_{K^*}^2}{P\cdot x} \int_0^1 du \, e^{i\xi P\cdot x} \left[ \psi_{4;K^*}^\perp(u)-\phi_{2;K^*}^\perp(u)  \right]\right\}\,.
\label{das_eq21}
\end{eqnarray}
All other DAs in the above relations are of twist-3 or -4 and all terms in the light-cone expansion of twist-5 and higher are neglected. The twist-4 DAs are shown for completeness. The normalisation of all DAs is given by
\begin{equation}
\int_0^1 du\, \phi(u) = 1\,.
\end{equation}
The conformal expansions of the leading-twist DAs $\phi_{2;K^*}^\parallel$ and $\phi_{2;K^*}^\perp$ are given by Eqs.~(\ref{das_eq16}) and (\ref{das_eq19}) respectively. In the local limit $x^\mu \to 0$ the matrix elements (\ref{das_eq20}) and (\ref{das_eq21}) reduce to the longitudinal $f^{\parallel}_{K^*}$ and transverse $f^{\perp}_{K^*}$ decay constants;
\begin{eqnarray}
\bra{0} \bar q(0) \gamma_\mu s(0) \ket{K^*(P,\lambda)} &=&  e_{\mu}^{(\lambda)}  m_{K^*} f^{\parallel}_{K^*}\,,
\label{das_eq22}\\
\bra{0} \bar q(0) \sigma_{\mu\nu} s(0) \ket{ K^*(P,\lambda)}&=& i \left(e_{\mu}^{(\lambda)} P_\nu-e_{\nu}^{(\lambda)}
P_{\mu}\right)f^{\perp}_{K^*}(\mu)\,.
\label{das_eq23}
\end{eqnarray}
Note that $f^\perp_{K^*}(\mu)$ is scale-dependent because the tensor current is not conserved. Numerical values of the decay constants are discussed in Chapter~\ref{chapter4_det}. The above DAs are related to those defined in Refs.~\cite{Ball:1998sk,Ball:1998ff} by
\begin{equation}
\addtolength{\arraycolsep}{5pt}
\begin{array}[b]{r@{\ =\ }l@{\qquad}r@{\ =\ }l@{\qquad}r@{\ =\ }l}
\phi_{2;K^*}^{\parallel(\perp)} & \phi_{\parallel(\perp)}\,, &
\phi_{3;K^*}^\parallel & h_\parallel^{(t)}\,, & \psi_{4;K^*}^\parallel &
g_3\,,\\[10pt]
\phi_{4;K^*}^{\parallel(\perp)} & {\mathbb A}_{(T)}\,, &
\phi_{3;K^*}^\perp & g_\perp^{(v)}\,, & \psi_{4;K^*}^\perp &
h_3\,.
\end{array}
\addtolength{\arraycolsep}{-5pt}
\label{das_eq24}
\end{equation}

\section{Two-Particle Twist-3 Distribution Amplitudes}\label{das_sec1.4}
The two-particle twist-3 DAs $\phi_{3;K^*}^{\perp,\parallel}$ have already been defined in
Eqs.~(\ref{das_eq20}) and (\ref{das_eq21}). There are two more two-particle DAs,
$\psi_{3;K^*}^{\perp,\parallel}$, defined as:\footnote{In the notations of Ref.~\cite{Ball:1998sk}, $\psi_{3;K^*}^\perp =   \{1-(f_{K^*}^\perp/f_{K^*}^\parallel) (m_s+m_q)/m_{K^*}\} g_\perp^{(a)}$, $\psi_{3;K^*}^\parallel =   \{1-(f_{K^*}^\parallel/f_{K^*}^\perp) (m_s+m_q)/m_{K^*}\} h_\parallel^{(s)}$.}
\begin{eqnarray}
\bra{0}\bar q(z) \gamma_{\mu} \gamma_{5}s(-z)\ket{K^*(P,\lambda)}
&=& \frac{1}{2}\,f_{K^*}^\parallel m_{K^*} \epsilon_{\mu}^{\phantom{\mu}\nu \alpha \beta}
e^{(\lambda)}_{\nu} p_{\alpha} z_{\beta}\int_{0}^{1} \!du\, e^{i \xi p \cdot z} \psi_{3;K^*}^\perp(u)\,,\\
\bra{0}\bar q(z)s(-z)\ket{K^*(P,\lambda)} & = & {} -i f_{K^*}^\perp(e^{(\lambda)} \cdot z) m_{K^*}^{2}
\int_{0}^{1} \!du\, e^{i \xi p \cdot z} \psi_{3;K^*}^\parallel(u)\,.
\end{eqnarray}
The normalisation is given by
\begin{equation}
\int_0^1 du\, \psi_{3;K^*}^{\parallel(\perp)}(u) = 1-\frac{f_{K^*}^{\parallel(\perp)}}{f_{K^*}^{\perp(\parallel)}}\,\frac{m_s+m_q}{m_{K^*}}\,,
\end{equation}
which differs from Ref.~\cite{Ball:1998sk}, where all DAs were normalised to 1; here we keep the full dependence on the quark masses. 

\section{Three-Particle Twist-3 Distribution Amplitudes}\label{das_sec1.5}
There are also three three-particle DAs of twist-3:
\begin{eqnarray}
\bra{0}\bar q(z) g_s \widetilde{G}_{\beta z}(vz)\gamma_{z}\gamma_5 s(-z) \ket{K^*(P,\lambda)} & = & f_{K^*}^\parallel m_{K^*} (p\cdot z)^2
e^{(\lambda)}_{\perp\beta}
\widetilde\Phi_{3;K}^\parallel(v,p\cdot z)+\dots\,,\\
\bra{0}\bar q(z) g_s G_{\beta z}(vz)i\gamma_{z} s(-z) \ket{K^*(P,\lambda)} & = & 
f_{K^*}^\parallel m_{K^*} (p\cdot z)^2
e^{(\lambda)}_{\perp\beta} \Phi_{3;K}^\parallel(v,p\cdot z)+\dots\,,\nonumber\\
\bra{0}
\bar q(z) g_s G_{z\beta}(vz)\sigma_{z\beta} s(-z) \ket{K^*(P,\lambda)} & = & f_{K^*}^\perp m_{K^*}^2
(e^{(\lambda)}\cdot z)(p\cdot z) \Phi_{3;K^*}^\perp(v,p\cdot z)+\dots\,,\nonumber
\label{das_eq25}
\end{eqnarray}
where the dots denote terms of higher twist and we use the short-hand notation
\begin{equation}
{\cal F}(v,p\cdot z) = \int {\cal D}\underline{\alpha}\,e^{-ip\cdot z(\alpha_2-\alpha_1+v\alpha_3)} {\cal F}(\underline{\alpha}) 
\end{equation}
with ${\cal F}(\underline{\alpha})$ being a three-particle DA  and the integration measure ${\cal D}\underline{\alpha}$ is defined as
\begin{equation}
\int{\cal D}\underline{\alpha} \equiv \int_0^1 d\alpha_1 d\alpha_2 d\alpha_3\, \delta(1-\sum \alpha_i)\,.
\label{das_eq26}
\end{equation}
The twist-3 three-particle DAs correspond to the light-cone projection $\gamma_z G_{z\perp}$ and $\sigma_{\perp z} G_{\perp z}$, respectively, which picks up the $s=\frac{1}{2}$ component of the
quark fields and the $s=1$ component of the gluonic field strength tensor. According to Eq.~(\ref{das_eq18}), the (normalised) asymptotic DA is then given by $360\,\alpha_1\alpha_2\alpha_3^2$. To NLO in the conformal expansion, each three-particle twist-3 DA involves three hadronic parameters, which we label in the following way: $\zeta,\kappa$ are LO and $\omega,\lambda$ NLO parameters.  $\zeta$ and $\omega$ are G-parity conserving, whereas 
$\kappa$ and $\lambda$ violate G-parity and hence vanish for mesons with quarks of equal mass, i.e.\  $\rho$, $\omega$ and $\phi$. We then have
\begin{eqnarray}
\Phi_{3;K^*}^\parallel(\underline{\alpha})
 & = & 360\alpha_1\alpha_2\alpha_3^2 \left\{
\kappa_{3K^*}^\parallel + \omega_{3K^*}^\parallel (\alpha_1-\alpha_2) +
\lambda_{3K^*}^\parallel \frac{1}{2}\,(7\alpha_3 -
3)\right\},\nonumber\\
\widetilde\Phi_{3;K^*}^\parallel(\underline{\alpha}) & = & 
360\alpha_1\alpha_2\alpha_3^2 \left\{
\zeta_{3K^*}^\parallel + \widetilde\lambda_{3K^*}^\parallel (\alpha_1-\alpha_2) +
\widetilde\omega_{3K^*}^\parallel \frac{1}{2}\,(7\alpha_3 -
3)\right\},\nonumber\\
\Phi_{3;K^*}^\perp(\underline{\alpha}) 
& = & 360\alpha_1\alpha_2\alpha_3^2 \left\{
\kappa_{3K^*}^\perp + \omega_{3K^*}^\perp (\alpha_1-\alpha_2) +
\lambda_{3K^*}^\perp \frac{1}{2}\,(7\alpha_3 -
3)\right\}\,.
\label{das_eq27}
\end{eqnarray}
The parameters defined above are related to those of Ref.~\cite{Ball:1998sk} by:
\begin{eqnarray}
\zeta_{3}^A = \zeta^\parallel_{3},&\qquad & \zeta_{3}^V =\omega_{3}^\parallel/14,  \nonumber\\[10pt]
\zeta_{3}^T = \omega_{3}^\perp/14,&\qquad &\zeta_3^\parallel\,\omega^A_{1,0} =
\widetilde\omega_{3}^\parallel\,.
\label{das_eq28}
\end{eqnarray}
G-parity breaking terms were not considered in Ref.~\cite{Ball:1998sk}. For equal mass quarks, $\Phi^{\perp,\parallel}_{3;K^*}$ are antisymmetric under $\alpha_1\leftrightarrow \alpha_2$, whereas $\widetilde\Phi^{\parallel}_{3;K^*}$ is symmetric. All these parameters can be defined in terms of matrix elements of local twist-3 operators. For chiral-odd operators, for instance, one has
\begin{eqnarray}
\bra{0}\bar q \sigma_{z\xi} g_sG_{z\xi}
s\ket{K^*(P,\lambda)} & = & f_{K^*}^\perp m_{K^*}^2 
(e^{(\lambda)}\cdot z)(p\cdot z)\kappa_{3K^*}^\perp\,,\nonumber\\
\bra{0}\bar q \sigma_{z\xi}  [iD_z,g_sG_{z\xi}] s
- \frac{3}{7}\, i\partial_z \bar q \sigma_{z\xi} 
g_sG_{z\xi} s \ket{K^*(P,\lambda)} 
& = & f_{K^*}^\perp m_{K^*}^2 (e^{(\lambda)}\cdot z)(p\cdot z)^2
\,\frac{3}{28}\, \lambda_{3K^*}^\perp\,,
\nonumber\\
\bra{0} \bar q i \!\stackrel{\leftarrow}{D}_z\!
\sigma_{z\xi}  g_sG_{z\xi} s - \bar q
\sigma_{z\xi}  g_sG_{z\xi} i\! \stackrel{\rightarrow}{D}_z\! s
 \ket{K^*(P,\lambda)}  & = & f_{K^*}^\perp m_{K^*}^2 (e^{(\lambda)}\cdot z)(p\cdot z)^2
\,\frac{1}{14}\, \omega_{3K^*}^\perp\,;\nonumber
\label{das_eq29}
\end{eqnarray}
the formulas for chiral-even operators are analogous.  In Chapter~\ref{chapter4_det} we calculate numerical values for all the parameters in Eq.~(\ref{das_eq27}) from QCD sum rules.

\section{Relations Between Distribution Amplitudes}
The QCD EOM are a crucial ingredient in simplifying the kinematic contributions of different operators, a task which is facilitated by the fact that they are preserved to all orders in the conformal expansion. The EOM relate via integral equations the two-particle twist-3 DAs, defined in Sections~\ref{das_sec1.3} and \ref{das_sec1.4}, to the two-particle twist-2 DAs, Eqs.~(\ref{das_eq16}) and (\ref{das_eq19}), and three-particle twist-3 DAs, defined in Section~\ref{das_sec1.5}. We do not quote the EOM themselves for which we refer the reader to the literature. The framework for the procedure was developed in Ref.~\cite{Braun:1989iv} based on deriving the EOM for non-local light-ray operators \cite{Balitsky:1987bk}. The operator relations are then sandwiched between the vacuum and meson states and the definitions of the DAs used to convert them into integral equations, making use of partial integration to remove explicit dependence on co-ordinates and momentum 4-vectors. The resulting expressions are then solved order-by-order in the conformal expansion, see Ref.~\cite{Braun:2003rp} for an overview. The EOM contain mass dependent contributions $\propto m_s\pm m_q$ that were calculated in Ref.~\cite{Ball:1998sk}. In the present analysis,  G-parity breaking terms of the three-particle twist-3 DAs are included, which then, via the EOM, impact on the two-particle DAs \cite{Ball:2007rt}. The resulting integral equations are:
\begin{eqnarray}
\psi_{3;K^*}^\parallel(u) & = & \ub\int_0^u dv\,\frac{1}{\bar v}\,
\Upsilon(v) + u \int_u^1 dv\,\frac{1}{v}\, \Upsilon(v)\,,\nonumber\\
\phi_{3;K^*}^\parallel(u) 
& = & \frac{1}{2}\,\xi\left[\int_0^u dv\,\frac{1}{\bar v}\,
\Upsilon(v) - \int_u^1 dv\,\frac{1}{v}\, \Upsilon(v)\right] + 
\frac{f_{K^*}^\parallel}{f_{K^*}^\perp}\,\frac{m_s+m_q}{m_{K^*}}\,
\phi_{2;K^*}^\parallel(u)\nonumber\\
&& {}+\frac{d}{du}\,\int_0^u d\alpha_1\int_0^{\bar u} d\alpha_2
\frac{1}{\alpha_3}
\,{\Phi}^\perp_{3;K^*}(\underline{\alpha})
\end{eqnarray}
with
\begin{eqnarray}
\Upsilon(u) & = & 2\phi_{2;K^*}^\perp(u) - 
\frac{f_{K^*}^\parallel}{f_{K^*}^\perp}\,\frac{m_s+m_q}{m_{K^*}} \left[
1 - \frac{1}{2}\,
\xi\frac{d}{du}\right]\phi_{2;K^*}^\parallel(u) -
\frac{1}{2}\,\frac{f_{K^*}^\parallel}{f_{K^*}^\perp}\,
\frac{m_s-m_q}{m_{K^*}}\,\frac{d}{du}\,\phi_{2;K^*}^\parallel(u)\nonumber\\
& & {}+\frac{d}{du}\,\int_0^u\!\!d\alpha_1\int_0^{\bar
u}\!\! d\alpha_2
\,\frac{1}{\alpha_3}\left(\alpha_1\,\frac{d}{d\alpha_1} +
\alpha_2\,\frac{d}{d\alpha_2}\, - 1\right) 
\Phi_{3;K^*}^\perp(\underline{\alpha})
\end{eqnarray}
and
\begin{eqnarray}
\psi_{3;K^*}^\perp(u)& = & \ub\int_0^u dv\,\frac{1}{\bar v}\,
  \Omega(v) + u \int_u^1 dv\,\frac{1}{v}\, \Omega(v)\,,\nonumber\\
\phi_{3;K^*}^\perp(u) 
& = & \frac{1}{4}\left[\int_0^u dv\,\frac{1}{\bar v}\,
\Omega(v) + \int_u^1 dv\,\frac{1}{v}\, \Omega(v)\right] + 
\frac{f_{K^*}^\perp}{f_{K^*}^\parallel}\,\frac{m_s+m_q}{m_{K^*}}\,
\phi_{2;K^*}^\perp(u)\nonumber\\
&& {}+\frac{d}{du}\,\int_0^u d\alpha_1\int_0^{\bar u} d\alpha_2
\frac{1}{\alpha_3}
\,{\Phi}^\parallel_{3;K^*}(\underline{\alpha})\nonumber\\
&&{} + \int_0^u d\alpha_1\int_0^{\bar u}
d\alpha_2\,\frac{1}{\alpha_3}\left( \frac{d}{d\alpha_1} +
\frac{d}{d\alpha_2} \right) 
\widetilde{\Phi}_{3;K^*}^\parallel(\underline{\alpha})
\end{eqnarray}
with
\begin{eqnarray}
\Omega(u)  &=&  2\phi_{2;K^*}^\parallel(u) + 
\frac{f_{K^*}^\perp}{f_{K^*}^\parallel}\,\frac{m_s+m_q}{m_{K^*}}\, 
\xi\,\frac{d}{du}\,\phi_{2;K^*}^\perp(u)-
\frac{f_{K^*}^\perp}{f_{K^*}^\parallel}\,\frac{m_s-m_q}{m_{K^*}}\, 
\frac{d}{du}\,\phi_{2;K^*}^\perp(u)\nonumber\\
& & {}+2\frac{d}{du}\,\int_0^u\!\!d\alpha_1
\int_0^{\bar u}\!\! d\alpha_2
\,\frac{1}{\alpha_3}\left(\alpha_1\,\frac{d}{d\alpha_1} +
\alpha_2\,\frac{d}{d\alpha_2}\right)
\Phi_{3;K^*}^\parallel(\underline{\alpha})\nonumber\\
&&{} +2\,\frac{d}{du}\,\int_0^u\!\!d\alpha_1
\int_0^{\bar u}\!\! d\alpha_2
\,\frac{1}{\alpha_3}\left(\alpha_1\,\frac{d}{d\alpha_1} -
\alpha_2\,\frac{d}{d\alpha_2}\right) 
\widetilde\Phi_{3;K^*}^\parallel(\underline{\alpha})\,.
\end{eqnarray}
Using Eq.~(\ref{das_eq27}), and the corresponding relations for twist-2 DAs, one obtains expressions for the twist-3 two-particle DAs, which are valid to NLO in the conformal expansion. As discussed in Ref.~\cite{Ball:1998sk}, the structure of this expansion is complicated by the fact that  these DAs do not correspond to a fixed Lorentz-spin projection $s$ of the quark fields.  The resulting expansion is in $C^{3/2}_n(\xi)$ for $\psi^{\perp,\parallel}_{3;K^*}$ and $C^{1/2}_n(\xi)$ for $\phi^{\perp,\parallel}_{3;K^*}$:
\begin{eqnarray}
\phi_{3;K^*}^\parallel(u) & = & 3\xi^2 + \frac{3}{2}\,\xi(3\xi^2-1)
a_1^\perp + \frac{3}{2}\,\xi^2 (5\xi^2-3)a_2^\perp\nonumber\\
&&{} + \left(\frac{15}{2}\,\kappa_{3K^*}^\perp -
\frac{3}{4}\,\lambda_{3K^*}^\perp\right) \xi (5\xi^2-3) +
\frac{5}{8}\,\omega_{3K^*}^\perp (3-30\xi^2 + 35\xi^4) \nonumber\\
&&{}+\frac{3}{2}\,\frac{m_s+m_q}{m_{K^*}}\,
\frac{f^\parallel_{K^*}}{f_{K^*}^\perp}
\left\{ 1 + 8\xi a_1^\parallel + 3 (7-30 u\ub )a_2^\parallel + \xi \ln
\ub (1+3 a_1^\parallel + 6 a_2^\parallel)\right.\nonumber\\
&&{}\left.\hspace*{3.6cm} - \xi \ln u (1-3
a_1^\parallel + 6 a_2^\parallel)\right\}\nonumber\\
&&{}-\frac{3}{2}\,\frac{m_s-m_q}{m_{K^*}}\,
\frac{f^\parallel_{K^*}}{f_{K^*}^\perp}\,
\xi \left\{ 2 + 9\xi a_1^\parallel + 2 (11-30 u\ub )a_2^\parallel + \ln
\ub (1+3 a_1^\parallel + 6 a_2^\parallel)\right.\nonumber\\
&&{}\hspace*{3.9cm}\left. + \ln u (1-3
a_1^\parallel + 6 a_2^\parallel)\right\},
\label{das_eq30}
\end{eqnarray}
\begin{eqnarray}
\psi_{3;K^*}^\parallel(u) & = & 6 u\bar u \left\{ 1 + \left(
\frac{a_1^\perp}{3} + \frac{5}{3} \kappa_{3K^*}^\perp\!\right)
C_1^{3/2}(\xi) + \left( \frac{a_2^\perp}{6}  + \frac{5}{18}
\omega_{3K^*}^\perp\right) C_2^{3/2}(\xi) -
\frac{1}{20}\lambda_{3K^*}^\perp C_3^{3/2}(\xi)\right\}\nonumber\\
&&{}+ 3\,
\frac{m_s+m_q}{m_{K^*}}\,\frac{f_{K^*}^\parallel}{f_{K^*}^\perp}\left\{
u\bar u (1 + 2 \xi a_1^\parallel + 3 (7-5u\bar u) a_2^\parallel) + \bar
u \ln \bar u (1+3 a_1^\parallel + 6 a_2^\parallel)\right.\nonumber\\
&&{}\left.\hspace*{3.2cm} + u \ln u 
(1-3 a_1^\parallel + 6 a_2^\parallel)\right\}\nonumber\\
&&- 3\,
\frac{m_s-m_q}{m_{K^*}}\,\frac{f_{K^*}^\parallel}{f_{K^*}^\perp}\left\{
u\bar u (9 a_1^\parallel + 10\xi a_2^\parallel) + \bar
u \ln \bar u (1+3 a_1^\parallel + 6 a_2^\parallel)\right.\nonumber\\
&&{}\left.\hspace*{3.2cm} - u \ln u 
(1-3 a_1^\parallel + 6 a_2^\parallel)\right\},
\label{das_eq31}
\end{eqnarray}
\begin{eqnarray}
\psi_{3;K^*}^\perp(u) & = & 6 u\bar u \left\{ 1 + \left(
\frac{1}{3} a_1^\parallel + \frac{20}{9} \kappa_{3K^*}^\parallel\right)
C_1^{3/2}(\xi) 
\right.\nonumber\\
&&\left.\hspace*{0pt} + \left( \frac{1}{6} a_2^\parallel +
\frac{10}{9}\zeta_{3K^*}^\parallel + \frac{5}{12}\,
\omega_{3K^*}^\parallel-\frac{5}{24}\,
\widetilde\omega_{3K^*}^\parallel\!\right)
C_2^{3/2}(\xi) +
\left(\frac{1}{4}\widetilde\lambda_{3K^*}^\parallel - \frac{1}{8}
\lambda_{3K^*}^\parallel\right) C_3^{3/2}(\xi)\right\}\nonumber\\
&&{}+ 6\,
\frac{m_s+m_q}{m_{K^*}}\,\frac{f_{K^*}^\perp}{f_{K^*}^\parallel}\left\{
u\bar u (2 + 3 \xi a_1^\perp + 2 (11-10u\bar u) a_2^\perp) + \bar
u \ln \bar u (1+3 a_1^\perp + 6 a_2^\perp)\right.\nonumber\\
&&{}\left.\hspace*{3.2cm} + u \ln u 
(1-3 a_1^\perp + 6 a_2^\perp)\right\}\nonumber\\
&&- 6\,
\frac{m_s-m_q}{m_{K^*}}\,\frac{f_{K^*}^\perp}{f_{K^*}^\parallel}\left\{
u\bar u (9 a_1^\perp + 10\xi a_2^\perp) + \bar
u \ln \bar u (1+3 a_1^\perp + 6 a_2^\perp)\right.\nonumber\\
&&{}\left.\hspace*{3.2cm} - u \ln u 
(1-3 a_1^\perp + 6 a_2^\perp)\right\},
\label{das_eq32}
\end{eqnarray}
\begin{eqnarray}
\phi_{3;K^*}^\perp(u) & = & \frac{3}{4}\,(1+\xi^2) + \frac{3}{2}\,\xi^3
a_1^\parallel + \left\{ \frac{3}{7}\,a_2^\parallel + 5
\zeta_{3K^*}^\parallel \right\} (3\xi^2-1) + \left\{5\kappa_{3K^*}^\parallel
- \frac{15}{16}\,\lambda_{3K^*}^\parallel \right.\nonumber\\
&&{} \left. +
\frac{15}{8}\,\widetilde{\lambda}_{3K^*}^\parallel \right\} \xi
(5\xi^2-3)+ \left\{ \frac{9}{112}\,a_2^\parallel +
\frac{15}{32}\,\omega_{3K^*}^\parallel -
\frac{15}{64}\,\widetilde\omega_{3K^*}^\parallel \right\} (35\xi^4-30
\xi^2+3)\nonumber\\
&&{}+\frac{3}{2}\,\frac{m_s+m_q}{m_{K^*}}\,\frac{f_{K^*}^\perp}{
f_{K^*}^\parallel} \left\{ 2 + 9 \xi a_1^\perp + 2 (11-30 u \ub)
    a_2^\perp\right.\nonumber\\
&&{}\left.\hspace*{3.5cm} + 
(1-3 a_1^\perp + 6 a_2^\perp) \ln u + (1+3 a_1^\perp + 6 a_2^\perp) 
\ln \ub \right\}\nonumber\\
&&{}-\frac{3}{2}\,\frac{m_s-m_q}{m_{K^*}}\,\frac{f_{K^*}^\perp}{
f_{K^*}^\parallel} \left\{ 2\xi + 9 (1-2u \ub) a_1^\perp + 2 \xi (11-20 u \ub)
    a_2^\perp\right.\nonumber\\
&&{}\left.\hspace*{3.5cm} + 
(1+3 a_1^\perp + 6 a_2^\perp) \ln \ub - (1-3 a_1^\perp + 6 a_2^\perp) 
\ln u \right\}\,.
\label{das_eq33}
\end{eqnarray}
These expressions  supersede those given in Ref.~\cite{Ball:1998sk} where G-parity violating terms in $\kappa_3$ and $\lambda_3$ were not included. The DAs given above now contain a minimum number of parameters which can be determined from one's favourite method, such as QCD sum rules or Lattice QCD.  We briefly mention that the matrix elements of QCD operator identities can also be used to relate twist-2 and twist-4 DA parameters to each other; such investigations were performed for the G-partiy violating twist-2  $K^*$ parameters $a_1^{\parallel,\perp} (K^*)$ in Refs.~\cite{Braun:2004vf,Ball:2006fz}.

\section{Evolution Equations}\label{evolution}
The scale dependence of the leading-twist DAs of Eqs.~(\ref{das_eq16}) and (\ref{das_eq19}) can be investigated using perturbation theory. The resulting renormalisation-group equation is the \textit{Efremov-Radyushkin-Brodsky-Lepage (ER-BL) evolution equation} \cite{Lepage:1980fj,Efremov:1979qk,Efremov:1978rn,Lepage:1979zb}
\begin{equation}
\mu^2 \frac{d}{d \mu^2}\, \phi(u,\mu)=\int^1_0dv\,V(u,v;\alpha_s(\mu))\,\phi(v,\mu)\,,
\label{das_eq34}
\end{equation}
which completely specifies $\phi(u,\mu)$ given $\phi(u,\mu_0)$. The kernel is given by an expansion in $\alpha_s$
\begin{equation}
V(u,v;\alpha_s(\mu)) = \frac{\alpha_s(\mu)}{4 \pi} V^{(0)}(u,v)+ \left(\frac{\alpha_s(\mu)}{4 \pi}\right)^2 V^{(1)}(u,v)+\dots \,.
\end{equation}
The evolution equation (\ref{das_eq34}) can be solved readily at leading-order using the conformal expansion \cite{Brodsky:1980ny, Ohrndorf:1981qv, Makeenko:1980bh, Brodsky:1980XX}. This amounts to finding its eigenfunctions, which we already know to be Gegenbauer polynomials, and using this fact it can be shown that the leading-order kernel can be written as
\begin{equation}
V^{(0)}(u,v)=-6u \bar{u} \sum^\infty_{n=0} \frac{4(2n+3)}{(n+1)(n+2)}\gamma_n^{(0)}C_n^{3/2}(2u-1)C_n^{3/2}(2v-1)\,,
\end{equation}
giving the LO anomalous dimensions of the Gegenbauer coefficients $\gamma_n$. The renormalisation is hence multiplicative at leading-order
\begin{equation}
a^{\rm LO}_n(\mu^2)=a_n(\mu^2_0)L^{\gamma^{(0)}_n /(2 \beta_0)}\,,
\end{equation}
where $L=\alpha_s(\mu^2)/\alpha_s(\mu^2_0)$. The one-loop anomalous dimensions of the twist-2 Gegenbauer coefficients are \cite{Gross:1973id1,Shifman:1980dk}
\begin{eqnarray}
\gamma^{\parallel(0)}_{(n)}&=&8 C_F \left(\sum_{k=1}^{n+1}\frac{1}{k} -\frac{3}{4}-\frac{1}{2(n+1)(n+2)}\right)\,,
\label{das_eq35}\\
\gamma^{\perp(0)}_{(n)}&=&8 C_F\left(\sum_{k=1}^{n+1} \frac{1}{k}-\frac{3}{4}\right)\,.
\label{das_eq36}
\end{eqnarray}
At next-to-leading-order the scale dependence is more complicated \cite{Mueller:1993hg,Mueller:1994cn}
\begin{equation}
a_n^{\rm  NLO}(\mu^2)=a_n(\mu^2_0) E_n^{\rm NLO}+\frac{\alpha_s}{4 \pi}\sum_{k=0}^{n-2} a_k(\mu^2_0)L^{
\gamma^{(0)}_k /(2 \beta_0)} d_{nk}^{(1)}\,,
\label{evo}
\end{equation}
where
\begin{equation}
E_n^{\rm  NLO}=L^{\gamma^{(0)}_k /(2 \beta_0)}\left\{1+\frac{\gamma_n^{(1)} \beta_0 - \gamma_n^{(0)} \beta_1}{8 \pi \beta_0^2}\left[\alpha_s(\mu^2)-\alpha_s(\mu^2_0)\right]\right\}\,.
\end{equation}
$\gamma_n^{(1)}$ are the diagonal two-loop anomalous dimensions which are available for the vector current \cite{Floratos}, and the tensor current \cite{Haya}. The mixing coefficients $d^{(1)}_{nk}$ are given in closed form in Refs.~\cite{Mueller:1993hg,Mueller:1994cn} where the formulas are valid for arbitrary currents by substitution of the corresponding one-loop anomalous dimension. For the lowest moments $n=\{0,1,2\}$ one has
\begin{eqnarray}
\gamma_0^{\parallel(1)} =0\,, \qquad 
\gamma_1^{\parallel(1)} = \frac{23110}{243} - \frac{512}{81}\, N_f\,,\qquad  
\gamma_2^{\parallel(1)} =  \frac{34072}{243}-\frac{830}{81}\, N_f\,,\\[10pt]
\gamma_0^{\perp(1)} =\frac{724}{9} - \frac{104}{27}\,N_f\,, \qquad 
\gamma_1^{\perp(1)} = 124 - 8 N_f\,,\qquad  
\gamma_2^{\perp(1)} =  \frac{38044}{243}-\frac{904}{81}\, N_f\,,
\end{eqnarray}  
and
\begin{equation}
  d^{\parallel(1)}_{20}  = 
  \frac{35}{9}\,\frac{20-3\beta_0}{50-9\beta_0} \left(1-L^{50/(9\beta_0)-1}\right)\,, \quad  d^{\perp(1)}_{20}  =   \frac{28}{9}\,\frac{16-3\beta_0}{40-9\beta_0}  \left(1-L^{40/(9\beta_0)-1}\right)\,.
\end{equation}
It is evident from Eqs.~(\ref{das_eq35}) and  (\ref{das_eq36}) that the anomalous dimensions of the Gegenbauer coefficients increase logarithmically  with conformal spin. This implies that as $\mu \to \infty$ the coefficients higher in conformal spin are damped and the DA approaches its asymptotic form (\ref{das_eq18})
\begin{equation}
\lim_{\mu \to \infty} \phi_{2;K^*}(u,\mu^2) \to 6 u \bar{u}\,.
\end{equation}
This limit offers a great simplification in that if it can be verified, experimentally of otherwise, that a given process is well described by the asymptotic form of the DA at hadronic scales, there are no non-perturbative parameters to be determined. The convergence of the conformal expansion in general has to be verified case by case and there is no \textit{a priori} reason why it should do so. In practice one has to truncate the expansion at some order in conformal spin, usually $n=2$, and as this constitutes an approximation it thus introduces a model dependent assumption. In Ref.~\cite{Ball:2005ei} for example,  these issues are discussed and an alternative method suggested.

The scale dependence of the three-particle twist-3 DAs can in principle be deduced from evolution equations for the non-local operators in (\ref{das_eq25}) using the techniques of Ref.~\cite{Balitsky:1987bk}. The evolution of the parameters in (\ref{das_eq27}) could then be found by projecting out the desired conformal spin, as  for the leading-twist DA. Another approach is to consult the literature of results for the corresponding nucleon structure functions \cite{andim}. The three-particle twist-3 parameters $\zeta_{3K^*}^\parallel$, $\kappa_{3K^*}^{\perp,\parallel}$, $\omega_{3K^*}^\perp$ and $\lambda_{3K^*}^\perp$ renormalise multiplicatively in the chiral limit, and the others mix with each other. For non-zero strange quark mass, there is additional mixing with twist-2 parameters with the mass corrections featuring as $m_s \pm m_q$ depending on the G-parity of the parameter. Here, we write down explicitly only the renormalisation-group improved relations for the above  5 parameters.   The relations can be written in compact form as
\begin{equation}
P_i(\mu^2) = L^{(\gamma_P)_i/\beta_0}\, P_i(\mu_0^2) + \sum_{j=1}^3
C_{ij} \left( L^{(\gamma_Q)_{ij}/\beta_0} -
L^{(\gamma_P)_i/\beta_0}\right) Q_{ij}(\mu_0^2)\,,
\label{das_eq37}
\end{equation}
where the parameters are given by:
\begin{eqnarray}
P & = & \{f_{K^*}^\parallel \zeta_{3K^*}^\parallel,\,f_{K^*}^\parallel
\kappa_{3K^*}^\parallel,\,f_{K^*}^\perp \kappa_{3K^*}^\perp,\,f_{K^*}^\perp
\omega_{3K^*}^\perp,\,f_{K^*}^\perp \lambda_{3K^*}^\perp\}\,,\nonumber\\
Q_{1(2)} &=& \frac{f_{K^*}^\perp}{m_{K^*}}\,\{
m_s\pm m_q,\,(m_s\mp m_q) a_1^\perp,\,(m_s\pm m_q)a_2^\perp\}\,,\nonumber\\
Q_{3,5} &=& \frac{f_{K^*}^\parallel}{m_{K^*}}\,\{
m_s-m_q,\,(m_s+m_q) a_1^\parallel,\,(m_s-m_q) a_2^\parallel\}\,,\nonumber\\
Q_{4}& =& \frac{f_{K^*}^\parallel}{m_{K^*}}\,\{
m_s+m_q,\,(m_s-m_q) a_1^\parallel,\,(m_s+m_q) a_2^\parallel\}\,,
\nonumber\\
\gamma_P & = & \left\{
\frac{77}{9},\,\frac{77}{9},\,\frac{55}{9},\,\frac{73}{9},\,
\frac{104}{9}\right\}, \nonumber\\
(\gamma_Q)_{1,2} & = & \left\{ \frac{16}{3},\, 8,\,
\frac{88}{9} \right\},\qquad
(\gamma_Q)_{3,4,5}  =  \left\{ 4,\, \frac{68}{9},\,
\frac{86}{9} \right\}\,,
\end{eqnarray}
\begin{eqnarray}
C & = & \left(
\begin{array}{r@{\hskip10pt}r@{\hskip10pt}r}
\ds \frac{2}{29} & \ds \frac{6}{25} & \ds 0\\[10pt]
\ds -\frac{2}{29} & \ds-\frac{6}{25} & 0\\[10pt]
\ds-\frac{4}{19} & \ds \frac{12}{65} & 0\\[10pt]
\ds \frac{14}{37} & \ds-\frac{42}{25} & \ds\frac{12}{13}\\[10pt]
\ds -\frac{1}{85} & \ds -\frac{1}{5} & \ds-\frac{4}{15}
\end{array} \right)\,.
\end{eqnarray}
We refrain from delving into a full discussion here of the mixing including the remaining parameters $\lambda_3^\parallel$, $\omega_3^\parallel$, and guide the reader to Appendix A of Ref.~\cite{Ball:2007rt} for details. 
\chapter{QCD Sum Rules}\label{chapter3_SR}
The original QCD sum rule approach was introduced by the revolutionary work of Shifman, Vainshtein and Zakharov in Refs.~\cite{Shifman:1978bz, Shifman:1978by, Shifman:1978bx}, and has proven itself to be one  of the most effective tools for determining non-perturbative parameters of low-lying hadronic states. It does so in terms of a finite number of  universal non-perturbative input parameters, and as such has great predictive power. The approach has been massively successful in ascertaining  a wide range of phenomena of non-perturbative origin.  QCD sum rules are particularly advantageous for $B$ physics because the presence of an  intrinsic heavy-quark mass scale $m_b$ provides the necessary conditions required for the application of the short-distance OPE or light-cone  expansion of relevant correlation functions, from which the relevant quantities can be extracted. The heavy-quark limit $m_b\to\infty$  is not necessary and sum rules can be derived in full QCD for finite $m_b$. Despite its successes, the  method is limited by an inherent  irreducible systematic uncertainty of $20 - 30\%$. However, such is the relative ease of the QCD sum rule method, as compared to, for example, Lattice QCD, that its place in the tool-box of the QCD practitioner is ensured.

Firstly we  discuss step-by-step the methodology of the original QCD sum rule approach. Secondly we discuss its modification to accommodate non-local correlation functions which aides the extraction of DA parameters of beyond leading-order in conformal spin. Thirdly, we outline an extension of the original
approach; light-cone sum rules. All three methods find application in this thesis:
\begin{itemize}
\item{in Chapter~\ref{chapter4_det} we make use of the non-local formalism to extract numerical values for the leading twist-2 and twist-3 DA parameters defined in the Chapter~\ref{chapter2_DAs}.}
\item{In Chapter~\ref{chapter5_eta} we calculate important contributions to the semileptonic $B\to \etapb$ transition form factors in light-cone sum rules.}
\item{An example of the original method is presented in the last section of this chapter, section \ref{example}, where we calculate the $\alpha_s$ corrections to the gluon condensate contribution from a local correlation function.}
\end{itemize}
 We focus solely on the points required for future chapters. For more information on sum rules, see for example Refs.~\cite{Colangelo:2000dp,Narison:1989aq,Shifman:1992xu,Reinders:1984sr}.

\section{SVZ Sum Rules}
The original sum rule method, which we refer to as \textit{SVZ sum rules}, parameterises unknown non-perturbative QCD vacuum effects in terms of the so-called universal \textit{vacuum condensates}. These quantities are vacuum expectation values of local operators $O_i$ that vanish in perturbation theory by definition and are ordered by their dimension $D$.

The calculation of a QCD sum rule starts from the calculation in QCD of a suitable correlation function in which the mesons are represented by interpolating currents possessing the correct quantum numbers. The method proceeds by equating two different representations of the correlation function. The first is obtained by performing a short-distance OPE, the result of which is matched to a second representation, in terms of a dispersion relation over physical hadronic states, leading to a sum rule from which various properties of the hadronic states can be extracted.

The SVZ sum rules find an important application in determining the universal hadronic parameters that appear in meson DAs.  Indeed, some of the first SVZ sum rule calculations were performed to extract decay constants $f_{\pi, \rho}$ \cite{Shifman:1978bx}, and Gegenbauer coefficients  $a_n$ \cite{Zhitnitsky:1982dd,Zhitnitsky:1983dd,Zhitnitsky:1985dd} of light meson DAs. The method can only be applied to parameters of the lowest few orders in conformal spin; parameters higher in conformal spin must be determined from other methods because the sum rules become unreliable.  One such method is that of \textit{non-local} condensates, see for example Ref.~\cite{Mikhailov:1986be}.

\subsection{Correlator}
The following \textit{two-point correlation function} describes the propagation of a quark-antiquark pair
\begin{equation}
\Pi(q) = i \int d^4 x \,e^{i q \cdot x} \bra{0} T J_{1}(x) J_{2}(0) \ket{0}\,,
\label{sr_eq1}
\end{equation}
where $q$ is the incoming momentum and possible Lorentz indices are omitted for simplicity. The local currents $J_i$  are chosen to have the correct quantum numbers and particle content corresponding to the particular hadronic parameters under investigation.  The physical picture of a hadron is of quarks, antiquarks and gluons confined within a typical hadronic size $R$  which is large when compared to the scale associated with perturbative effects. If one can show, however, that the correlation function is dominated by small spacial distances and time intervals
\begin{equation}
|\vec{x}| \sim x^0 \sim 1/\sqrt{Q} \ll R\,,
\label{sr_eq2}
\end{equation}
for a certain momentum configuration, then one has ensured the small size of the strong coupling $\alpha_s$ and hence the use of perturbation theory in our calculation. One begins by noting that after contracting any Lorentz indices which may appear in the currents $J_i$ in Eq.~(\ref{sr_eq1}), the correlation function can only depend on the interval $x^2=x^2_0-\vec{x}^2$. By taking the Fourier transform, completing the square, and shifting the variable $x$ one finds
\begin{equation}
\Pi(q^2) = i \int d\kappa \,\int d^4 x \,e^{i \kappa x^2}\,e^{i Q^2/4 \kappa} f(\kappa)\,.
\label{sr_eq3}
\end{equation}
The integral is dominated by the region where the arguments of the exponential vary slowly. This condition requires  $\kappa \sim 1/x^2$ and $\kappa \sim Q^2$ which are both  fulfilled for $x^2 \sim 1/Q^2$; for large momentum transfer the quarks propagate near the light-cone.\footnote{An expansion round $x^2 \to 0$ is the basis of QCD sum rules on the light-cone -- see Section \ref{LCSR}.} To find the true short-distance dominance one needs to dig a little further and by choosing the Lorentz frame $q_0=0$ one finds $\vec{x}^{2} \sim 1/Q^2$ as required (\ref{sr_eq2}). In the case of light quarks one needs the momentum transfer to the quarks to be large, $Q^2 \equiv -q^2\gg \Lambda^2_{\rm QCD}$. In the case of heavy quarks a large energy scale is introduced through the quark mass, for example $m_b$, which then serves to set the characteristic distances for the correlation function $|\vec{x}|\sim x_0\sim 1/(2 m_b)$; one is thus automatically in the perturbative regime.

\subsection{Short-Distance OPE}
The first of the two representations of the correlation function is obtained by performing the QCD calculation, valid for $Q^2 \equiv -q^2 \gg  \Lambda^2_{\rm QCD}$, using the short-distance OPE
\begin{equation}
\Pi (q^2) \stackrel{\rm{OPE}}{\longrightarrow}\sum_{i} C_i(q^2) \left< O_i\right> \equiv \Pi^{\rm{OPE}}\,,
\label{sr_eq4}
\end{equation}
where the non-perturbative long distance effects of QCD are encoded in the condensates $\left< O_i\right>$ and the short-distance effects are included in the Wilson coefficient functions $C_i$ which are calculable in perturbation theory. Both the condensates and their coefficients are in general renormalisation scale dependent. Perturbative corrections to the condensates are calculated when necessary.  The perturbation theory contribution to Eq.~(\ref{sr_eq4}) has $D=0$ and corresponds to the unit operator $\left<O_{\rm PT}\right>=\bf{1}$. The condensates play the role of power-corrections and are suppressed by inverse powers of the hard scale as $(Q^2)^{-D/2}$. In the asymptotic limit $Q^2\to\infty$ only the unit operator survives, corresponding to asymptotic freedom. 

\subsection{Condensates}
The condensates represent the effects of non-perturbative QCD and they cannot be determined from first principles due to the unknown nature of the QCD vacuum. The determination of the condensates is an industry in itself. The light quark condensate $\bra{0} \bar{q} q\ket{0}$ has been known for a long time \cite{Gell-Mann:1968rz} and it drives the breakdown of the chiral symmetry of the light quarks $q=\{u,d\}$ and its value can be extracted from experiment:
\begin{equation}
m_\pi^2 f_\pi^2 \approx-(m_u+m_d)\left<\bar q q\right>\,,
\end{equation}
where we use the notation $\left<O_i\right>\equiv\bra{0}O_i\ket{0}$. To define other condensates, one notes that the only vacuum expectation values of operators that can survive are those which are Lorentz invariant, spin zero, colour and flavour-singlets i.e. possess the quantum numbers of the vacuum. The complete set of condensates $\left<O_i\right>$ that contribute with $D\le 6$ are
\begin{eqnarray}
\underbrace{\left<\textbf{1}\right>}_{D=0}\,, \hspace{1cm}&\underbrace{m_q\left< \bar{q} q\right>}_{D=4}\,,& \hspace{1cm} \underbrace{\left<\frac{\alpha_s}{\pi}G^2\right>}_{D=4}\,,\nonumber \\
\underbrace{m_q\left<  \bar{q}\sigma g_s G q \right>}_{D=6}\,, \hspace{1cm}&\underbrace{\left<\bar{q} \Gamma_1q\,\bar{q}\Gamma_2 q\right>}_{D=6}\,,&\hspace{1cm} \underbrace{\left<g_s^3f G^3\right>}_{D=6}\,,\nonumber
\end{eqnarray}
where $q=\{u,d,s\}$ is a light quark spinor and all indices are contracted.\footnote{The heavy quarks $c$, $b$ and $t$ do not form condensates because they are too massive to interact non-perturbatively with the QCD vacuum.}  We assume isospin symmetry  for $q=\{u,d\}$ and one must differentiate $q=s$ when $\rm SU(3)_F$-breaking effects are taken into account.  Higher dimensional condensates $D> 6$ are not very well determined and generally unknown. If required, however, they can be estimated by employing the \textit{vacuum saturation hypothesis} whereby the operator fields are simply split to form products of known condensates; for example, the quark-antiquark  $D=6$ operator can be simplified to the product of two $\bar{q} q$ operators \cite{PT:84,Shifman:1978by}. In practice, the OPE is truncated to a given order, and is usually justified by the stability of the resulting sum rule. The series, Eq.~(\ref{sr_eq4}), is then given in terms of a limited number of condensates allowing sum rules to be written in terms of a small set of parameters incorporating the general features of non-perturbative QCD, while retaining its predictive power.

The procedure works in reverse, of course, where the values of condensates are deduced from sum rules for which the hadronic parameters are known from other methods;  two-point correlation functions featuring $\bar{b} \gamma_\mu b$ or $\bar{c} \gamma_\mu c$ currents correspond to the $\Upsilon$ and $J/\Psi$ resonances respectively, of which the decay constants and masses are known. Values for the condensates are given, along with other input parameters, in Appendix~\ref{appendixB}. Uncertainties in the values of the condensates and other input parameters constitute part of the reducible theoretical uncertainty of the sum rule approach. 

\subsection{Dispersion Relation}
\begin{figure}[h]
$$ \epsfxsize=0.5\textwidth\epsffile{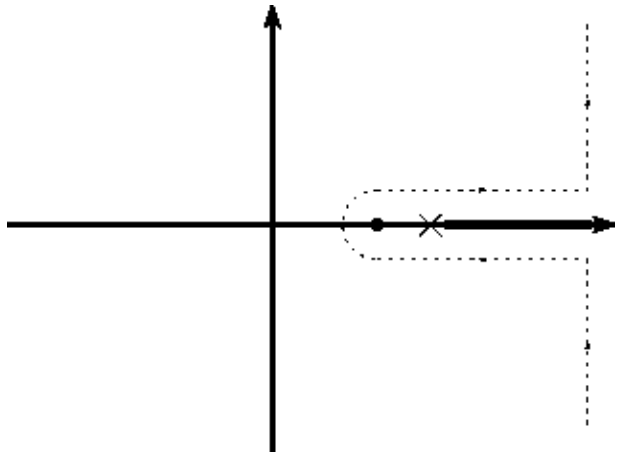}$$
  \caption[The spectral density function in the complex plane.]{\small The general features of a spectral density function $\rho^{\rm had}(s)$ in the complex plane. The blob represents the pole due to the ground-state, the cross possible poles due to  higher mass resonances, and the thick line the cut due to the continuum of multi-particle states. The dotted line is the integration contour.} 
  \label{sr_fig1}
\end{figure}

To proceed we need to relate the result of the OPE to a second representation of the correlation function which is obtained in terms of the spectrum of hadronic states in the physical region  $q^2>0$.   This is done via a \textit{dispersion relation}, which is derived from the analytic properties of the correlation function as follows. The function $\Pi(q^2)$ is analytic in all $q^2$ except on the real axis starting at a pole corresponding to the ground state particle. At higher energy  higher mass excited states and a continuum of many-particle states also feature. The higher mass resonances give poles above the ground state, the details of which depend on the physical spectrum of particles which possess the correct quantum numbers to couple to $\Pi$. The continuum of many-particle states, correspond to a continuous cut, see Fig.~\ref{sr_fig1}.

Using Cauchy's formula we can write
\begin{equation}
\Pi(q^2) =\frac{1}{2 \pi i} \oint\limits_{|z|=R} dz\, \frac{\Pi(z)}{z-q^2}+\frac{1}{2 \pi i} \int^R_0 dz\, \frac{\Pi(z+i \epsilon)-\Pi(z- i \epsilon)}{z-q^2}\,,
\end{equation}
where the region of integration is split into the parts just above and below the positive real axis and the circle of radius $R$. Provided that the correlation function vanishes at least as quickly as $q^{-2}$ as $|q^2| \sim R \to \infty$ then the integral over the circle at radius $R$ goes to zero.\footnote{If $\Pi$ does not vanish quickly enough we subtract the first few terms in its Taylor expansion as required. We shall see that this does not matter in the end, due to the Borel transformation.} The remaining integral can be simplified using the fact that below the first pole at $q^2=s_{\rm min}$,  $\Pi(q^2)$ is real and above this point, according to the Schwarz reflection principle, $\Pi(z+i \epsilon)-\Pi(z- i \epsilon)= 2 i\, \textrm{Im}\,\Pi(z+ i \epsilon)$. Hence
\begin{equation}\label{disp}
\Pi(q^2) = \int_{s_{\rm min}}^{\infty} ds\, \frac{\rho(s)}{s-q^2- i \epsilon}\,,
\end{equation}
where the function $\rho(s)=\frac{1}{\pi} \textrm{Im}\, \Pi(s)$ is the \textit{spectral density} and describes the physical particle spectrum as a function of energy $s$. 

\subsection{Unitarity Relation}
As we have seen, for large negative $q^2$ our correlation function is dominated by short-distance physics.  As $q^2$  becomes more positive the separation of the quarks increases. For large enough positive values of $q^2$ long-distance QCD interactions become more important and the correlation function then describes the creation of hadrons, which is the basis of its second representation. As discussed in the last section, $\Pi$ uncovers a very complicated spectrum of states for $q^2>0$. We describe this situation by using the \textit{unitarity relation}, which allows in insertion of a complete set of states into the correlation function
\begin{equation}\label{complete}
\textbf{1}= \sum_{n} \int d \Omega_n\, \ket{n(p)}\bra{n(p)}\,,
\end{equation}
where $d \Omega_n$ includes all phase-space factors and momentum conservation and the sum runs over all possible particles and polarisations, starting from the ground state $M$ of mass $m_M$. Inserting (\ref{complete}) between the currents of our original correlation function (\ref{sr_eq1}) yields an expression which we can relate to the hadronic spectral density
\begin{equation}\label{unitarity}
 \Pi^{\rm had}(q^2) =   \int \frac{d^4 p}{(2 \pi)^4} \frac{1}{m^2_{M}-p^2} \int d^4 x \,e^{i q \cdot x} \bra{0} J_{1}(x)   \ket{M(p)}\bra{M(p)}J_{2}(0) \ket{0} + \dots\,,
\end{equation}
where the dots denote higher mass states which contribute to the continuum. We are usually interested in the ground state, and can insert the expressions for the matrix elements on the right hand side. The local matrix elements considered here can be used to extract vacuum-meson decay constants, for example.  Using the unitarity relation (\ref{unitarity}) one can single out the ground state $M$ by comparing it to (\ref{disp}) and writing the hadronic spectral density  as:
\begin{equation}\label{gs}
\rho^{\rm had}(s)=f_{M} \,\delta(s-m_{M}^2)+\rho^{\textrm{cont}}(s),
\end{equation}
where $f_{M}$  is directly related to the matrix elements of the currents $J_1$ and $J_2$ in Eq.~(\ref{unitarity}). For example, one could choose $J_1=J_2^\dagger= \bar q \gamma_{z} s$ to extract $(f^\parallel_{K^*})^2$ c.f. Eq.~(\ref{das_eq22}). The exact form of the spectral density beyond the ground state is mostly unknown and the higher mass states and continuum contributions are usually lumped together in one function $\rho^{\textrm{cont}}(s)$. If the next highest particle above the ground state occurs at an energy not very much higher than $m_{M}$ then it is possible to explicitly include this particle as another delta-function term, analogously to the ground state. This procedure was used, for example, while investigating the leading-twist $K^*$ and $\rho$ DA parameters for which the relevant correlators couple to the  $K_1$ and $b_1$ resonances respectively \cite{Ball:1996tb,Ball:2005vx}.

\subsection{Quark-Hadron Duality}
It is possible to write the result of the OPE as a dispersion relation, with spectral density $\rho^{\textrm{OPE}}(s)$. As $\rho^{\textrm{cont}}(s)$ is mostly unknown  we replace it by $\rho^{\textrm{OPE}}(s)$ above a certain energy $s_0$
\begin{equation}\label{qhd}
\rho^{\textrm{cont}}(s) \to \rho^{\textrm{OPE}}(s)\,\Theta(s-s_0)\,.
\end{equation}
This assumption relies on the validity of  the hadronic  representation  being approximated by the partonic representation at higher energies. Inserting Eqs.~(\ref{gs}) and  (\ref{qhd}) into Eq.~(\ref{disp}) one finds
\begin{equation}
\Pi^{\rm had}(q^2)=\frac{f_M}{m_M^2-q^2}+\int_{s_{0}}^{\infty} ds\, \frac{\rho^{\rm OPE}(s)}{s-q^2- i \epsilon}\,.
\end{equation}
Now the assumption is not so strict because we only require a duality between the integrated spectral densities, not the spectral densities themselves. This is called \textit{semi-global quark-hadron duality}. The parameter $s_0$ is called the  \textit{continuum threshold} and its value  is specific for each particle spectrum being roughly equal to the energy of the next highest resonance above the ground state: $s_0 \sim (m_M + \Delta)^2$ where $\Delta \sim \mathcal{O}(\Lambda_{\textrm{QCD}})$.  Ultimately it must be determined from the sum rule itself by requiring the numerical value of the determined quantity to be largely insensitive to its variation and this introduces the first source of systematic uncertainty to the sum rule method. We are now in a position to equate both representations
\begin{equation}
\Pi^{\rm{had}}=\Pi^{\rm{OPE}}\,, 
\end{equation}
to derive our sought after sum rule, however, before we do so, there is one last procedure to discuss, which greatly improves the behaviour of the sum rule.

\subsection{Borel Transformation And The Sum Rule}
The sum rule can be improved by suppressing the continuum contribution, which we have assumed to be well described by $ \rho^{\textrm{OPE}}(s>s_0)$ and the possible detrimental impact of this assumption is thus reduced.  We do this by performing a \textit{Borel transformation} to both sides of the sum rule. The transformation is obtained by applying the operator
\begin{equation}\label{borel1}
\mathcal{\hat{B}} = 
\lim_{\stackrel{-q^2,n \to \infty}{-q^2/n=M^2}}\frac{(-q^2)^{(n+1)}}{n!}\left(\frac{d\phantom{q^2}}{dq^2}\right)^{n+1},
\end{equation}
which takes a function of $q^2$ and produces a new function of  the \textit{Borel parameter} $M^2$. One frequently encountered example is
\begin{equation}
\mathcal{\hat{B}} \frac{1}{(m^2-q^2)^k}= \frac{1}{(k-1)!}\frac{e^{-m^2/M^2}}{(M^2)^{k}}\,,
\end{equation}
providing an exponential suppression of the unknown continuum contributions, and a suppression of the power-corrections by factorials thus reducing the impact of neglected higher dimensional condensates. Also, as $\mathcal{\hat{B}} (q^2)^k=0$, any subtraction terms introduced to Eq.~(\ref{disp}), which can only appear as polynomials in $q^2$, are killed off. The Borel transformation improves the stability and accuracy of the sum rule.

The Borel parameter $M^2$ is the second and last sum rule specific parameter to be introduced; along with $s_0$ it is required to impact very little, when varied, on the numerical value of the quantity being determined. The variation of $M^2$ changes the relative impact of the power-corrections and perturbation theory contributions.  In evaluating sum rules one looks for a \textit{Borel window} which is  usually in the range $1\,\rm GeV^2\leqslant M^2\leqslant 2 \,\rm GeV^2$ for a typical mesonic DA parameter.  The sum rule should be reliable if a weak dependence (a plateau) is found, the contribution from the continuum is small, and there are no unnatural numerical cancellations.

We now equate Eqs.~(\ref{disp}) and (\ref{sr_eq4}) to reach the sum rule
\begin{equation}\label{sr1}
f_M\,e^{-m_M^2/M^2}= \int^{s_0}_0 ds\,e^{-s/M^2} \,\rho^{\textrm{OPE}}(s)\,,
\end{equation}
where the hadronic quantity $f_M$ is given as a function of the universal non-perturbative condensates, the perturbative short-distance coefficients as calculated from QCD, and the sum rule parameters $s_0$ and $M^2$. The sum rule is saturated by the ground state and higher mass states are suppressed. As the correlation function (\ref{sr_eq1}) does not depend on the renormalisation scale, the $\mu$ dependence of the condensates, when multiplied by their coefficient functions, must cancel in the sum of (\ref{sr_eq4}). The sum is always truncated, however, and the residual $\mu$ dependence will be a source of theoretical uncertainty.

\subsection{Non-local Formalism}
One way to gain access to parameters higher in conformal spin is to calculate sum rules involving operators  which are related to moments of DAs
\begin{equation}
\bra{0}\bar{q}(0) (\deriv \cdot z)^k \Gamma s(0)\ket{V}\sim \int^1_0 du\, (2u-1)^k \phi(u)\equiv \left<\xi^k\right>\,.
\end{equation}
For the $K^*$ for example the first few moments of both the leading-twist DAs are $\left<\xi^0\right>=1$, $\left<\xi^1\right>=\frac{3}{5} a_1(K^*)$, $\left<\xi^2\right>=\frac{1}{35}(7+12 a_2(K^*))$ and $\left<\xi^3\right>=\frac{1}{105}(27 a_1(K^*)+20 a_3(K^*))$. A more elegant method, enabling the  DA parameters to be extracted individually, relies on calculating a correlator of two currents, one of which is non-local, with fields at light-like separations ($z^2=0$) \cite{Ball:2003sc}. Consider the following
\begin{equation}
\Pi(q\cdot z) = i \int d^4 x \,e^{i q \cdot x} \bra{0} T J(x) \bar{s}(0) \gamma_{z} q(z) \ket{0}\,,
\label{nonlocalCF}
\end{equation}
where $J(x)$ is local, and the non-local current yields the leading-twist DA (\ref{das_eq16}). The sum rule (\ref{sr1}) then reads
\begin{equation}\label{sr2}
f_J f^\parallel_{K^*}\,e^{-m_M^2/M^2}\int^1_0 du\,e^{-i \bar{u}q\cdot z} \phi_{2;K^*}^\parallel= \int^{s_0}_0 ds\,e^{-s/M^2} \,\int^1_0 du\,e^{-i \bar{u}q\cdot z}\rho^{\textrm{OPE}}(s,u)\,.
\end{equation}
The integration over $u$ on the right hand side naturally arises via the Feynman parameterisation used in the calculation. At this point one can exploit the orthogonality of the Gegenbauer polynomials by replacing the exponential weight function $e^{-i\xi q\cdot z} \to C_n^{3/2}(\xi)$ on both sides to project out $a_n^\parallel(K^*)$ via Eqs.~(\ref{das_eq15}) and (\ref{das_eq17}). 
\begin{figure}[h]
$$ \epsfxsize=0.3\textwidth\epsffile{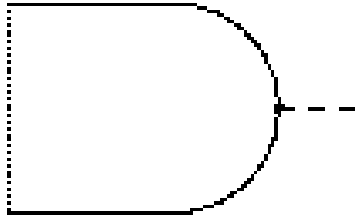}$$
  \caption[A generic diagram for a non-local sum rule.]{\small A generic non-local diagram. The dotted line denotes the path ordered gauge factor $[z,-z]$ between the two quark fields. The momentum $q$ is injected at point $y$ - the vertex on the right hand side.} 
  \label{sr_fig2}
\end{figure}
In Fig.~\ref{sr_fig2} we show the leading diagram of the non-local correlation function (\ref{nonlocalCF}). The dotted line denotes the path ordered gauge factor $[z,-z]$ between the two quark fields. The non-local formalism allows, in principle, an extraction of  parameters of arbitrary order $n$. In practice, however, only the parameters of the lowest few orders $n$ are accessible due to instability of the resulting sum rules. One finds that the power-corrections in $\rho^{\textrm{OPE}}$ grow with positive powers of $n$ compared to the perturbative contribution. For high enough $n$ this behaviour upsets the hierarchy of contributions to the OPE, where non-perturbative terms are expected to be moderately sized corrections to the leading term. Hence the method is justified for low-order coefficients $n\leq2$ where the non-perturbative coefficients describe the general features of the DA. It breaks down for higher-order coefficients $n>3$ because the local vacuum condensates appear with $\delta$-functions which cannot accommodate the information needed to describe the more detailed shape of the DA, see Refs.~\cite{Ball:1997rj,Ball:2003sc}. 

\section{QCD Sum Rules On The Light-Cone}\label{LCSR}
A modification of the QCD sum rule method known as QCD sum rules on the light-cone, or \textit{light-cone sum rules} (LCSRs)  \cite{Balitsky:1989ry,Braun:1988qv,Chernyak:1990ag}, was developed to overcome difficulties encountered when calculating transition and electromagnetic form factors in the SVZ method.\footnote{The term ``light-cone sum rules" first appears in Ref.~\cite{Ball:1991bs}.} The problems are related to the asymptotic scaling behaviour of the form factors in the heavy-quark limit $m_b \to \infty$. LCSRs rely on the use of DAs as their universal non-perturbative hadronic input and lead to the correct asymptotic behaviour in the heavy-quark limit. The DAs represent a  partial re-summation of the operators appearing in the condensates and appear ordered in contributions of increasing twist \cite{Ball:1997rj}. We can view LCSRs as a marriage of the SVZ technique and the theory of hard exclusive processes \cite{Chernyak:1977fk,Lepage:1980fj,Efremov:1979qk,Chernyak:1977as}. In the case of the ``heavy-to-light'' $B\to M$ transition form factors, LCSRs have been applied successfully to pseudoscalar transition form factors \cite{oldpseudo,Ball:1998tj,Ball:2001fp,Ball:2004ye} and vector transition form factors \cite{Ball:1998kk,Ball:2004rg}. 

For LCSRs to become competitive with the SVZ sum rules, a good knowledge of higher-twist DAs is required. This motivates the determination of the non-perturbative DA parameters via SVZ sum rules and via LCSR, the DAs themselves to determine other non-perturbative parameters, such as transition form factors. As with SVZ sum rules, the starting point of LCSRs is with a suitable correlation function. For the extraction of $B\to M$ transition form factors we require a two-point correlator, this time sandwiched between the vacuum and the meson state $M$, which is the example considered in this section. One employs much of the same methodology as in the last section, although now one requires the correlation function to be expanded in an OPE on the light-cone. In doing so one finds that the correlation function factorises and can be written in terms of a convolution of hard scattering kernels and the universal non-perturbative DAs of the light-meson.   To that end, consider a correlation function of two quark currents taken between the vacuum and an on-shell meson $M$
\begin{equation}\label{CFLCSR}
\Pi(q,p_B)=i\int d^4x \,e^{i q\cdot x} \bra{M(p)}T J_1(x) j_B^{\dagger}(0)\ket{0}\,,
\end{equation}
where $j_B = m_b\, \bar{q}i \gamma_5 b$ is the \textit{interpolating current} of the $B$ meson which defines the $B$ meson decay constant
\begin{equation}
\bra{B(p_B)}j_B\ket{0}=m^2_B f_B\,.
\label{bdecayconstant}
\end{equation}
The current $J_1(x)$ is chosen to project out the form factor of interest. The  momentum $q$ is  injected into the weak vertex, $p_B$ is the momentum of the $B$ meson and $p$ is the  momentum of $M$ with $q+p=p_B$. The correlation function is dominated by light-like distances for virtualities
\begin{equation}
m_b^2 - p_B^2 \geq \mathcal{O}(\Lambda_{\rm QCD} m_b)\,, \qquad m_b^2 - q^2 \geq \mathcal{O}(\Lambda_{\rm QCD} m_b)\,,
\end{equation}
which ensures the slow variation of the exponential in Eq.~(\ref{CFLCSR}) and its suitability for an expansion around the light-cone. The light-cone expansion results in the transverse and ``minus'' degrees of freedom being integrated out, leaving the longitudinal momenta of the partons as the relevant degrees of freedom. As a result a cutoff $\mu$ is introduced below which the transverse momenta are included in the resulting light-mesons DAs. The contributions from momenta above this cutoff are calculable in perturbation theory. The procedure yields the \textit{collinear factorisation} of the correlation function
\begin{equation}\label{LCSR:fac}
\Pi(q^2,p_B^2)=\sum_n \int_0^1 \,du\, T^{(n)}(u,q^2,p_B^2,\mu)\,\phi_{n;M}(u,\mu)\,,
\end{equation}
where $u$ $(1-u)$ denotes the momentum fraction of the outgoing quark (antiquark) and the sum is over all twist and possible polarisation contributions. The scale dependence of the hard scattering kernels $T^{(n)}$ must cancel that of the DAs $\phi_{n;M}$. The factorisation formula has to be verified by direct calculation and a proof to all orders in  $\alpha_s$ does not exist. The verification relies on the cancelation of divergences, of which there are two types: the IR and UV singluarities arising from loop calculations and so-called soft singularities which appear when the convolution over $u$ does not converge at the end-point regions ($u \sim 0~\textrm{or}~1$) i.e. when one of the quarks is soft. In terms of kinematics there are two main contributing processes: the hard-scattering mechanism and the soft contribution or Feynman mechanism. Both mechanisms are included in the LCSR approach for which there are no soft divergences and the IR/UV divergences can be treated in dimensional regularisation. 

One can write the result of the light-cone expansion (\ref{LCSR:fac}) as a dispersion relation in $p_B^2$
\begin{equation}
\Pi^{\rm  LC} (p_B^2, q^2) = \int_{m_b^2}^{\infty} ds\,\frac{\rho^{\rm  LC} (s, q^2)}{s-p_B^2}\,.
\end{equation}
Taking the imaginary part, to obtain $\rho^{\rm  LC} $, is straight forward after integration over the momentum fraction $u$ is performed. The correlation function has a cut in $p_B^2$ starting at $m_b^2$ over the physical region. One now matches this calculation to the hadronic representation of the correlation function, which can also be written as a dispersion relation
\begin{equation}
\Pi^{\rm  had} (p_B^2, q^2) = \int_{m_B^2}^{\infty} ds\,\frac{\rho^{\rm  had} (s, q^2)}{s-p_B^2}\,,
\end{equation}
where the physical spectral density is given by the ground state $B$ meson plus higher mass states forming a continuum as
\begin{equation}
\rho^{\rm had} (s, q^2)= F_M\, \delta(s-m_B^2) +  \rho^{\rm LC } (s, q^2) \,\Theta(s-s_0)\,.
\end{equation}
The quantity $F_M$ will contain the form factor we require. We perform the Borel transformation to arrive at the LCSR
\begin{equation}
F_M\, e^{-m_B^2/M^2}  =\int_{m_b^2}^{s_0} ds \, e^{-s/M^2} \rho^{\rm LC} (s, q^2)\,.
\end{equation}
To extract the form factor we need to find a sets of parameters $M^2$ and $s_0$ such that the form factor is largely insensitive to their variation. As with SVZ sum rules, there is no rigourous way to do this and so the procedure introduces the irreducible source of uncertainty to the method. 

\section{Example Calculation - The Gluon Condensate}\label{example}
\begin{figure}[h]
$$ \epsfxsize=0.8\textwidth\epsffile{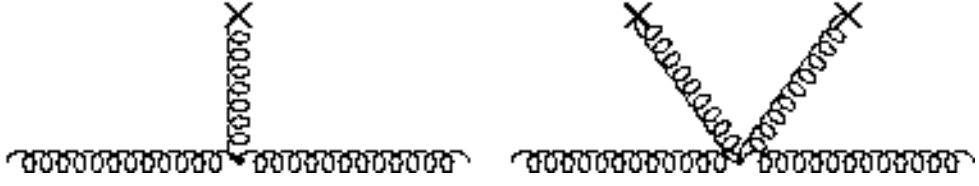}$$
\caption[Interactions of the background field $\mathcal{A}^a_\mu$ with the quantum field $a^a_\mu$.]{\small  The interactions of the background field $\mathcal{A}^a_\mu$ (denoted by a cross) with the quantum field $a^a_\mu$ corresponding to $\mathcal{L}^{\mathcal{A}aa}_{int}$ and $\mathcal{L}^{\mathcal{AA}aa}_{int}$ respectively.} 
\label{sr_fig3}
\end{figure}

\begin{figure}[h]
$$ \epsfxsize=0.7\textwidth\epsffile{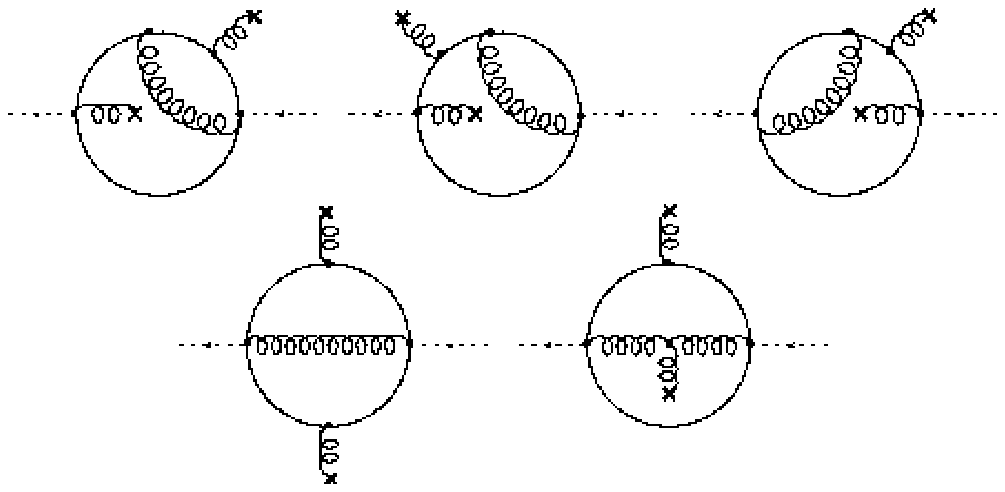}$$
$$\epsfxsize=0.5\textwidth\epsffile{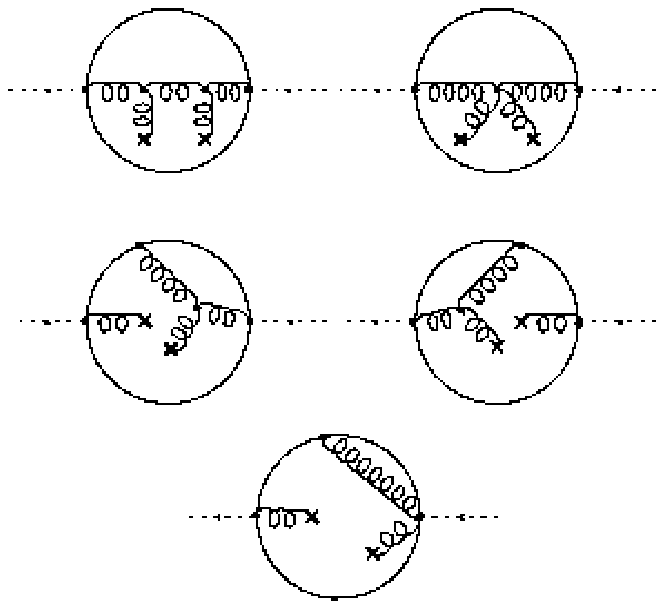}$$
\caption[Diagrams contributing to the gluon condensate at $\mathcal{O}(\alpha_s)$.]{\small The diagrams contributing to the gluon condensate at $\mathcal{O}(\alpha_s)$ for the SVZ sum rule of the $K$ twist-3 DA parameter $f_{3K}$ -- see Ref.~\cite{Ball:2006wn}. For each diagram the fixed-point $x_0=0$ is at the left most vertex and the right most is at $y$.} 
\label{sr_fig4}
\end{figure}

Here we present an example calculation within the SVZ sum rule framework. The result of the calculation is used in the sum rule for the G-even $K$ meson three-particle twist-3 DA parameter $f_{3K}$, see Ref.~\cite{Ball:2006wn}. We calculate the $\alpha_s$ correction to the gluon condensate $\left< \frac{\alpha_s}{\pi} G^2 \right>$ which proceeds from the following local correlation function
\begin{equation}\label{cf2}
\Pi^{(G^2)}   =  i \int d^4 y \, e^{i q\cdot y}\bra{0} T \bar q(0) \sigma^{\mu z}  g_sG_{\mu  z}(0) s(0) \bar s(y) 
\sigma^{\nu z} g_sG_{\nu  z}(y) q(y)\ket{0}\,,
\end{equation}
for which the leading-order contribution  vanishes. A convienient way of extracting the gluon condensate is to make use of the \textit{back-ground field technique} in which the fixed-point gauge allows the Taylor expansion of quark and gluon fields to be written in a gauge-covariant form, see  Ref.~\cite{Novikov:1983gd} for details. The gluon field in the QCD Lagrangian (\ref{basics_eq1}) is split into ``quantum''  and  ``classical''  (background) fields
\begin{equation}\label{split}
A^a_\mu\to a^a_\mu+\mathcal{A}^a_\mu\,,
\end{equation}
where the background field $\mathcal{A}^a_\mu$ is taken in the fixed-point gauge at $x_0=0$. The quantum field $a^a_\mu$ is taken to be in the Feynman gauge, thus requiring the gauge fixing term ($\xi=1$)
\begin{equation}
\mathcal{L}^{\rm fix}=-\frac{1}{2}(\partial^\mu a_\mu^a+g_s f^{abc}\mathcal{A}^{b \mu} a_\mu^c)^2\,,
\end{equation}
to be added to the QCD Lagrangian. The quantum field propagates perturbatively and we may use the standard expression
\begin{equation}
\begC1{a^a_\mu}\conC{(x)\,}\endC1{a^b_\nu}(y)=i \delta^{ab}\int \frac{d^4 l}{(2 \pi)^4}D_{\mu\nu}(l)e^{-i l\cdot (x-y)}\,,\qquad D_{\mu\nu}(l)=\frac{-g_{\mu\nu}}{l^2},
\end{equation}
The background field does not propagate perturbatively, and is the field that goes to form the condensate; it represents the low-energy, long distance modes of the gluon field that probe the non-perturbative structure of the QCD vacuum. The fixed-point gauge condition allows $\mathcal{A}_\mu^a(x)$ to be expressed in terms of the gluonic field strength tensor as
\begin{equation}
\mathcal{A}_\mu^a(x)=\sum^\infty_{n=0}\frac{1}{n!(n+2)} x^\omega x^{\omega_1}... x^{\omega_n}\left[ D_{\omega_1}(0),\left[D_{\omega_2}(0),\left[...\left[D_{\omega_n}(0),G_{\omega\mu}^a(0)\right]...\right]\right]\right]\,,
\end{equation}
and translating to momentum space one finds
\begin{equation}\label{condfield}
\mathcal{A}_\mu^a(k)=-\frac{i}{2} G^a_{\omega\mu}(0)\left[(2\pi)^4 \delta^{(4)}(k)\right]\frac{\partial\phantom{ k^\omega}}{\partial k^\omega}+\dots\,,
\end{equation}
where we only require the first term; higher order terms give rise to higher dimensional condensates which we do not consider. As we have to introduce two condensate gluons to construct $\left<G^2\right>$ we introduce two auxiliary vacuum momenta $k$ and $k^\prime$ for which the fixed-point $x_0=0$ is a sink. After integration over coordinates these momenta appear in the quark and gluon propagators. The two corresponding derivatives are then taken, and the vacuum momenta set to zero. The following expression proves very useful in managing derivatives of quark propagators
\begin{equation}\label{quarkderiv}
\frac{\partial}{\partial p_\mu} S^{(q)}(p)=-S^{(q)}(p)\gamma^\mu S^{(q)}(p)\,,\qquad S^{(q)}(p)=\frac{\slash{p}+m_q}{p^2-m_q^2}\,,
\end{equation}
for arbitrary quark flavour $q$. The gluon condensate is finally extracted using
\begin{equation}\label{vacav}
G^a_{\omega\mu}(0)G^b_{\omega^{\prime}\nu}(0)=\frac{1}{D(D-1)}\frac{\delta^{ab}}{N_c^2-1}\left(g_{\omega\omega^{\prime}}g_{\mu \nu}-g_{\omega\nu}g_{\omega^{\prime}\mu}\right)\left<G^2\right>\,,
\end{equation}
where $D$ is the spacetime dimension. Due to Eq.~(\ref{split}) the expansion of $\mathcal{L}_{\rm QCD}$  yields ``interaction'' terms in which background fields are radiated from the propagating gluons at single or double vertices, both of which contribute to the $\mathcal{O}(\alpha_s)$ correction to the gluon condensate. These vertices are shown in Fig.~\ref{sr_fig3} and the corresponding terms are 
\begin{eqnarray}\label{glueint}
 \mathcal{L}^{\mathcal{A}aa}_{int}&=&-\frac{1}{2}g_s f^{abc}\left[\left(\partial^\mu\mathcal{A}^{a\nu}-\partial^\nu\mathcal{A}^{a\mu}\right)a_\mu^b a_\nu^c\right.\nonumber\\
 &+&\left.(\partial^\mu a^{a\nu}-\partial^\nu a^{a\mu})(\mathcal{A}^b_\mu a^c_\nu+a^b_\mu\mathcal{A}^c_\nu)+2(\partial_\mu a^{a\mu})\mathcal{A}^{b\nu}a_\nu^c\right]\,,\nonumber\\
 \mathcal{L}^{\mathcal{AA}aa}_{int}&=&-\frac{1}{2} g_s^2 f^{abc} f^{ade}\left[\mathcal{A}^b_\mu \mathcal{A}^{d\mu} a_\nu^e a^{c\nu} +\mathcal{A}^b_\mu  a^{d\mu} \mathcal{A}^e_\nu a^{c\nu}+\mathcal{A}^b_\mu  a^{c\mu} \mathcal{A}^d_\nu a^{e\nu}\right]\,,
\end{eqnarray}
where terms which vanish eventually via Eq.~(\ref{vacav}) due to $f^{abc}\delta^{bc}=0$ are omitted.
Contributions also stem directly from the gluonic field strength tensors in Eq.~(\ref{cf2}) which give rise to gluon emission of either one or two fields from the vertices at co-ordinates $0$ and $y$. Due to the gauge condition there is no ``left-right'' symmetry and all diagrams with two gluons, of which at least one is a condensate gluon, emerging from the vertex at $x=0$ vanish due to $A_\mu(0)=0$.  Diagrams with two condensate gluons at point $y$, which originate from the non-linear part of the gluonic field strength tensor,  also give zero due to $f^{abc}\delta^{bc}=0$. There is an ``up-down'' symmetry where diagrams related by a  reflection in the central horizontal axis are equal. Overall we find there to be 10 distinct non-zero diagrams to be calculated which are shown in Fig.~\ref{sr_fig4}. 

Some of the diagrams are divergent, however, all divergences cancel in the sum of all diagrams.\footnote{We use dimensional regularisation and the $\overline{MS}$ renormalisation scheme throughout this thesis.} For an explicit example consider the last diagram in the second line of Fig.~\ref{sr_fig4}. It is evident that we require $\mathcal{L}^{\mathcal{A}aa}_{int}$ to be contracted in all possible ways with quantum fields originating from the linear part of the gluonic field strength tensors at points $0$ and $y$. This, multiplied by the condensate field originating from the quark loop yields the gluonic part of the calculation
\begin{equation}\label{allcon}
\sim\mathcal{A}^d_\delta(v) \left.(\partial_\mu a^a_z(0)-\partial_z a^a_\mu(0))\, \mathcal{L}^{\mathcal{A}aa}_{int}(w) \,(\partial_\nu a^b_z(y)-\partial_z a^b_\nu(y))\right|_{\rm all\,contractions}\,,
\end{equation}
which is eventually given in momentum space by (omitting Lorentz indices)
\begin{equation}\label{gluonicpart}
\sim \frac{\partial}{\partial k}\frac{\partial}{\partial k^\prime}\frac{f(l,k^\prime)}{l^2 (l-k^\prime)^2} \left<G^2\right>\,,
\end{equation}
where the condensate gluon within $\mathcal{L}^{\mathcal{A}aa}_{int}(w)$ is expressed by Eq.~(\ref{condfield}) with momentum $k^\prime$ and $f(l,k^\prime)$ is a function of the loop momentum $l$ and the vacuum momentum $k^\prime$. The  quark loop yields a usual trace 
\begin{equation}\label{quarkpart}
\sim \frac{\textrm{tr}\left[(\slash{p}+\slash{q}-\slash{l})\sigma^{\mu z}(\slash{p}+\slash{k})\gamma^\delta\slash{p}\,\sigma^{\nu z}\right]}{(p+q-l)^2(p+k)^2 p^2}\,,
\end{equation}
and after multiplying together Eqs.~(\ref{gluonicpart}) and (\ref{quarkpart}), performing the derivatives in $k$ and $k^\prime$ and integrating over the momenta $p$ and $l$ we find
\begin{equation}
\Pi^{(G^2)}_{\rm example}  = \frac{1}{384}\frac{\alpha_s}{\pi}  \left\langle\frac{\alpha_s}{\pi}\,G^2\right\rangle \frac{(q\cdot z)^4}{q^2}\,.
\end{equation}
In this way we can include all the other diagrams shown in Fig.~\ref{sr_fig4} to obtain the contribution to the sum rule
\begin{equation}
\Pi^{(G^2)}  = 
-\frac{89}{5184}\,\frac{\alpha_s}{\pi}\,\left\langle\frac{\alpha_s}{\pi}\,G^2\right\rangle \,\frac{(q \cdot z)^4}{q^2}\,,
\label{last}
\end{equation}
which differs from the result obtained in Ref.~\cite{Zhitnitsky:1985dd}; the logarithmic term is not reproduced:
\begin{equation}
\sim \log \left(\frac{M^2}{\mu^2}\right) \left< \frac{\alpha_s}{\pi} \,G^2 \right>\,.
\label{least}
\end{equation}

\chapter{The Determination Of Vector Meson Twist-2 And Twist-3  Parameters}\label{chapter4_det}
In this chapter we determine the leading twist-2 and twist-3 two- and three-particle vector meson DA parameters using the non-local modification of SVZ sum rules. The parameters are defined in Chapter~\ref{chapter2_DAs} and the sum rule method is outlined in Chapter~\ref{chapter3_SR}. We express the relevant correlation functions, via the OPE, in terms of the perturbative and condensate  contributions.  Key to the analysis is the inclusion of all G-parity and  $\rm SU(3)_F$-breaking effects which, as discussed in Chapter~\ref{chapter2_DAs}, come from a variety of sources, and allow a consistent determination of the parameters for the $\rho$, $K^*$, and $\phi$. Motivation for the present analysis comes from various sources, including:
\begin{itemize}
\item{values for the decay constants and leading-twist DA Gegenbauer moments are required as input for QCD factorisation frameworks which provide a systematic method for the calculation of $B$ decay matrix elements. We discuss one such framework in Chapter~\ref{chapter6_QCDF}.}
\item{Twist-2 and twist-3 DAs provide the leading non-perturbative input within the method of LCSR, as discussed in Chapter~\ref{chapter3_SR}, and as such are applied to many problems in heavy-flavour physics, such as the calculation of $B$ transition form factors and the estimation of $B$ decay matrix elements including power-suppressed contributions to QCD factorisation frameworks, see Chapter~\ref{chapter7_rad}.}
\item{A full determination of the twist-3 DA parameters, including $\rm SU(3)_{F}$-breaking and G-parity violating effects, and the inclusion of $\mathcal{O}(\alpha_s)$ and $\mathcal{O}(m_s^2)$ corrections to the quark condensate contributions to the twist-2 DA parameter sum rules are new to the present analysis, allowing $a_2^{\parallel,\perp}(\phi)$ to be determined, for the first time, to the same accuracy as $a_2^{\parallel,\perp}(\rho,K^*)$.}
\end{itemize}
All input parameters for the sum rules, and useful formulas, such as those required to take the imaginary parts of intermediate results, and various relevant integrals, are given in Appendix~\ref{appendixB}. In performing the calculations we find Refs.~\cite{PT:84,Borodulin:1995xd} very useful. The material covered in this chapter partially follows that of Ref.~\cite{Ball:2007rt}.

\section{Twist-2}
In this section we focus on the determination of the twist-2 DA Gegenbauer coefficients $a_{2}^{\parallel,\perp}$ defined by Eqs.~(\ref{das_eq16}) and (\ref{das_eq19}).  The sum rules for $f_{K^*}^{\parallel,\perp}$, including $\rm SU(3)_F$-breaking corrections, were calculated in Refs.~\cite{Govaerts:1986ua,Ball:2005vx,Ball:2006fz}. Those for the G-parity violating $a_1^{\parallel,\perp}(K^*)$ in Refs.\cite{Ball:2005vx,Ball:2006fz} and those for $a_2^{\parallel,\perp}(K^*)$ in \cite{Ball:2003sc} apart from perturbative terms in $m_s^2$ and the $\mathcal{O}(\alpha_s)$ and $\mathcal{O}(m_s^2)$ corrections to the quark condensate, which are new to the present analysis.  Motivation for including these corrections is found by examining the individual contributions to the sum rules for $a_2^{\parallel,\perp}(K^*)$ given in Ref.~\cite{Ball:2003sc}. They are found to be dominated by $\left<\bar s s\right>$ as we can see from the following explicit break down of contributions:
\begin{eqnarray}
a_2^\parallel(K^*)&=&\overbrace{0.05}^{\textrm{PT}}+\overbrace{0.08}^{\left<\frac{\alpha_s}{\pi} G^2\right>}+\overbrace{0.11}^{\left<\bar s g_s \sigma G s\right>}+\overbrace{0.04}^{\left<\bar q q\right>^2}-\overbrace{0.16}^{\left<\bar s s\right>}+\overbrace{0.02}^{\left<\bar s s\right>^2}-\overbrace{0.05}^{\left<\bar q q\right>\left<\bar s s\right>}\nonumber\\
a_2^\perp(K^*)&=&0.06+\,\,0.10\,\,+\,\,\,0.25\,\,\,+0.03-0.27+0.02\,-0\,,
\end{eqnarray}
for the reference point $s_0=1.2\,\textrm{GeV}^2$, $M^2=1\,\textrm{GeV}^2$ and $\mu=1\,\textrm{GeV}$. Moreover, for the  $\phi$ the impact of a finite strange quark mass may be even more pronounced with respect to perturbation theory and the gluon condensate. 

Firstly, we give an overview of the calculation of the $\mathcal{O}(\alpha_s)$ and $\mathcal{O}(m_{s}^2)$ corrections to the quark condensate $\left<\bar s s\right>$; the calculations for $\left<\bar q q\right>$ are analogous. We only need extract terms proportional to $m_s$  as the contributions proportional to $m_q$ are identical; we can find the contributions for $\phi$ by simply replacing $\left<\bar q q\right>\to\left<\bar s s\right>$ and doubling the terms in $m_{s} \left<\bar s s\right>$, $m_{s} \left<\bar q q\right>$ and $m_{s} \left<\bar s g_s G s\right>$. Contributions for $\rho$ are found by setting $m_s\to0$. Secondly, we go on to analyse the sum rules for $a_2^{\parallel,\perp}(\phi)$. We end this section by presenting the results.

\subsection{Calculation}
For both polarisations we begin from the diagonal correlation function
\begin{equation}
\Pi_{2;K^*}(q\cdot z) = i \int d^4y \,e^{-iq\cdot y} \bra{0} T \bar q(y) \Gamma s(y) \bar s(0) \Gamma [0,z]q(z) \ket{0}\,
\label{C.0}
\end{equation}
where $\Gamma^{\parallel}=\gamma_z$ and $\Gamma^{\perp}=\sigma_{\mu z}$. For the longitudinal parameters the sum rule is exactly that given by Eq.~(\ref{sr2}) with $f_J\to f^\parallel_{K^*}$ and for the transverse parameters the sum rule is analogous. Both polarisations have the same projections onto the DA parameters
\begin{eqnarray}
\left(f_{K^*}\right)^2e^{-m_{K^*}^2/M^2}\left[1\right]
 & = & \int_0^{s_0}ds\, e^{-s/M^2} \int^1_0 du\,\left[1\right]
      \frac{1}{\pi}\, {\rm Im}_{s}
\pi_{2;K^*}(u)\,,\label{C0}
\\
\left(f_{K^*}\right)^2  e^{-m_{K^*}^2/M^2}
      \,\left[\frac{9}{5}\,a_{1}(K^*)\right]
& = & \int_0^{s_0}ds\, e^{-s/M^2} \int^1_0 du\,
     \left[ 3 \xi \right]\frac{1}{\pi}\, {\rm Im}_{s}
 \pi_{2;K^*}(u)\,,
\nonumber\\
\left(f_{K^*}\right)^2 e^{-m_{K^*}^2/M^2}
      \,\left[\frac{18}{7}\,a_{2}(K^*)\right]
& = & \int_0^{s_0}ds\, e^{-s/M^2} \int^1_0 du\,
    \left[ \frac{1}{2}(  15 \xi^2-3)\right]\frac{1}{\pi}\, {\rm Im}_{s}
    \pi_{2;K^*}(u)\,, \nonumber
\end{eqnarray}
where ${\rm Im}_s$ denotes taking the imaginary part with respect to $s$.  The fact that we are dealing with non-local correlation functions means that we do not integrate over the co-ordinate $z$. The resulting residual exponential function remains throughout the calculation and can contribute to the momentum integrals yielding powers of $i c (q\cdot z)$, where $c$ is a constant. Ultimately the exponential functions can be cast into  the ``canonical form'' set by the exponential appearing in front of the leading-twist DA i.e. $e^{- i \bar{u}q\cdot z}$ -- see Eq.~(\ref{sr2}). 
\subsection*{Quark Condensate}
\begin{figure}[h]
$$ \epsfxsize=0.25\textwidth\epsffile{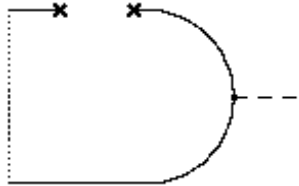}$$
\caption[Diagram contributing to the quark condensate $\left<\bar s s \right>$ at leading-order.]{\small The leading-order diagram contributing to the quark condensate $\left<\bar s s \right>$.} 
\label{det_fig1}
\end{figure}
The tree-level diagram is shown in Fig.~\ref{det_fig1}. To extract the quark condensates to $\mathcal{O}(m_s^2)$ we use the following expansion of the quark fields (for general quark flavour $q$)
\begin{eqnarray}
\bra{0}\!:\! \bar q^i_\alpha(x_1)\, q^j_\beta(x_2)\!:\!\ket{0}&=&\delta^{ij}\frac{\left<\bar q q\right>}{12}\left\{\delta_{\beta\alpha}\left(1-\frac{\Delta^2}{2D}m_q^2\right)\right.\nonumber\\
&-&\left.m_q\frac{i }{D}(\gamma_\lambda)_{\beta\alpha}\Delta^\lambda\left(1-\frac{\Delta^2 }{2(2+D)}m_q^2\right)\right\}\,,
\label{quarkextract}
\end{eqnarray}
where  $\Delta_\mu=(x_2-x_1)_\mu$ and $i,j$ are colour and $\alpha,\beta$ spinor indices. One can deal with the co-ordinate $\Delta_\mu$ by trading it, via partial integration (PI), for a derivative of the trace that arises from the quark loop. A convenient way to do so is via an auxiliary momentum $Q$
\begin{equation}
\Delta_\kappa \stackrel{\textrm{PI}}{\longrightarrow} ie^{-i\Delta\cdot Q}\frac{\partial}{\partial Q_\kappa}\Big|_{Q\to 0}\,.
\label{deriv1}
\end{equation}
\begin{figure}[h]
$$ \epsfxsize=0.8\textwidth\epsffile{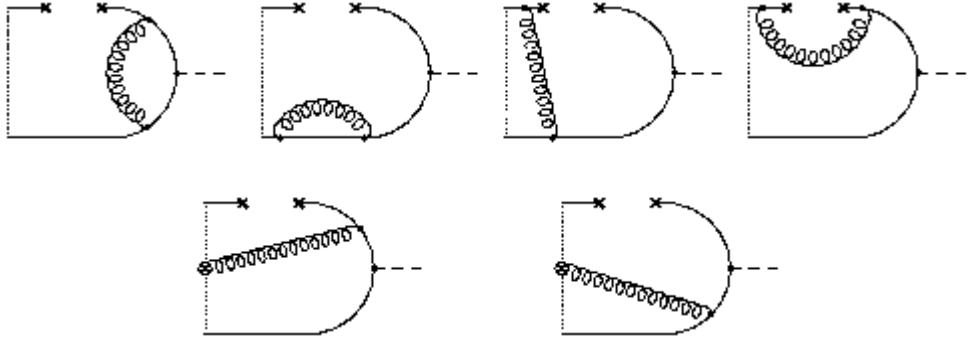}$$
\caption[Diagrams contributing to the quark condensate $\left<\bar s s \right>$ at $\mathcal{O}(\alpha_s)$.]{\small Diagrams contributing to the quark condensate $\left<\bar s s \right>$ at $\mathcal{O}(\alpha_s)$. The crossed circle $\otimes$ depicts the emission of a gluon from the non-local gauge factor -- see Eq.~(\ref{emissiongf}). The corresponding diagrams for $\left<\bar q q\right>$ are identical but reflected top to bottom.} 
\label{det_fig2}
\end{figure}
Diagrams for the $\mathcal{O}(\alpha_s)$ corrections to the strange quark condensate are shown in Fig.~\ref{det_fig2}.  Importantly there are contributions from the gauge-factor which need to be included
\begin{equation}
[0,z]=\textrm{P} \exp{\left\{-i g_s \int^1_0 dt\,z^\mu A_\mu(\bar{t}z)\right\}}=1-i g_s \int^1_0 dt\,z^\mu A_\mu(\bar{t}z)+\dots\,.
\label{emissiongf}
\end{equation}
Calculating $\mathcal{O}(\alpha_s)$ corrections leads to divergent diagrams and the dependence of the condensate on the spacetime dimension $D$ leads to $\mathcal{O}(\epsilon)$ contributions at tree level, that then cause finite counter-terms upon renormalisation. Also, the derivative with respect to $Q_\kappa$ in Eq.~(\ref{deriv1}) yields $\gamma_\kappa$ in the trace via Eq.~(\ref{quarkderiv}) which can also give a finite counter-term. This happens for the vertex correction diagrams.

\subsection{Evaluation of The Sum Rules}
\begin{figure}[h]
$$ \epsfxsize=0.5\textwidth\epsffile{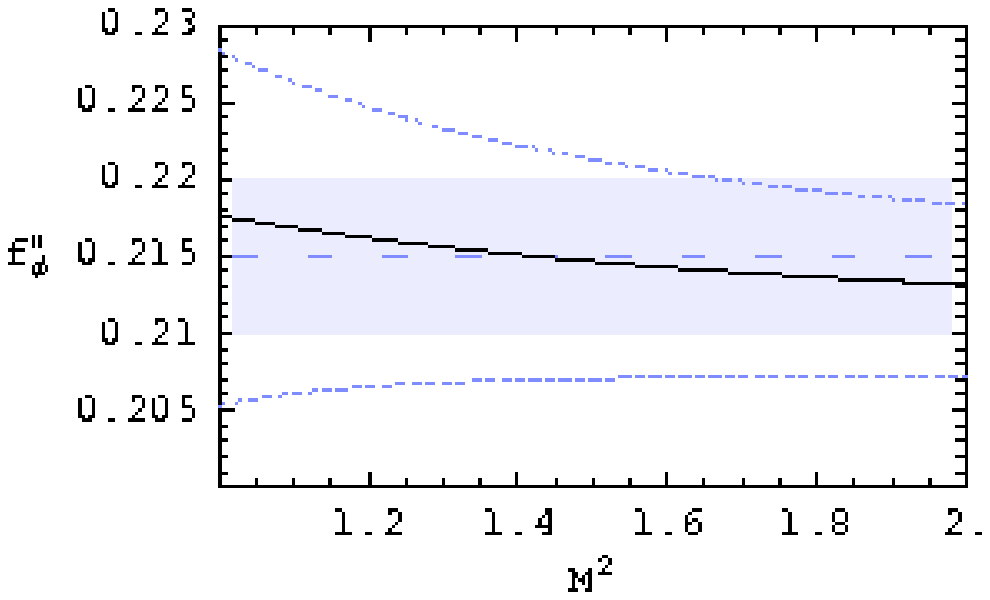}\quad \epsfxsize=0.5\textwidth\epsffile{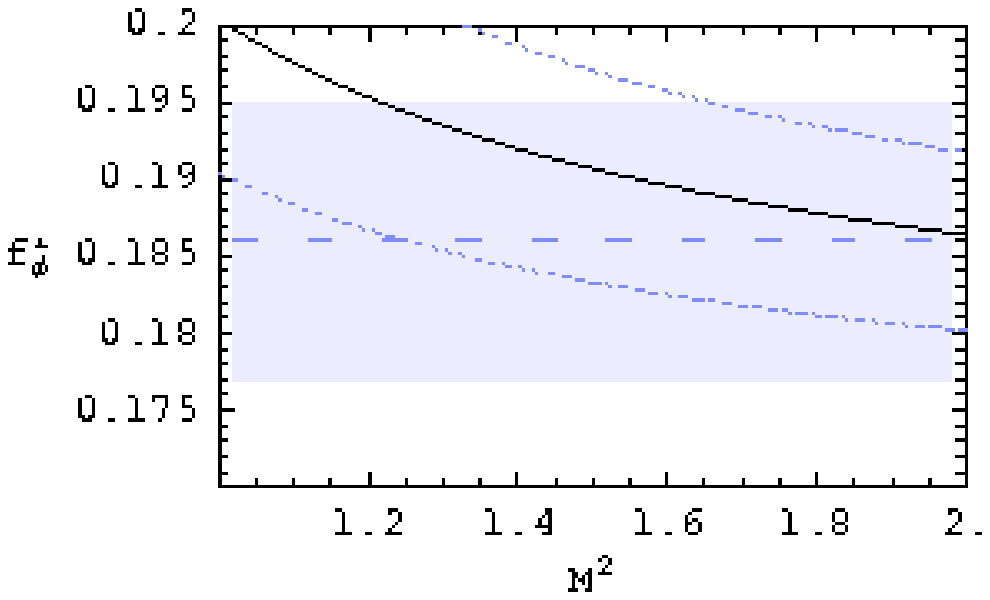}$$
$$ \epsfxsize=0.5\textwidth\epsffile{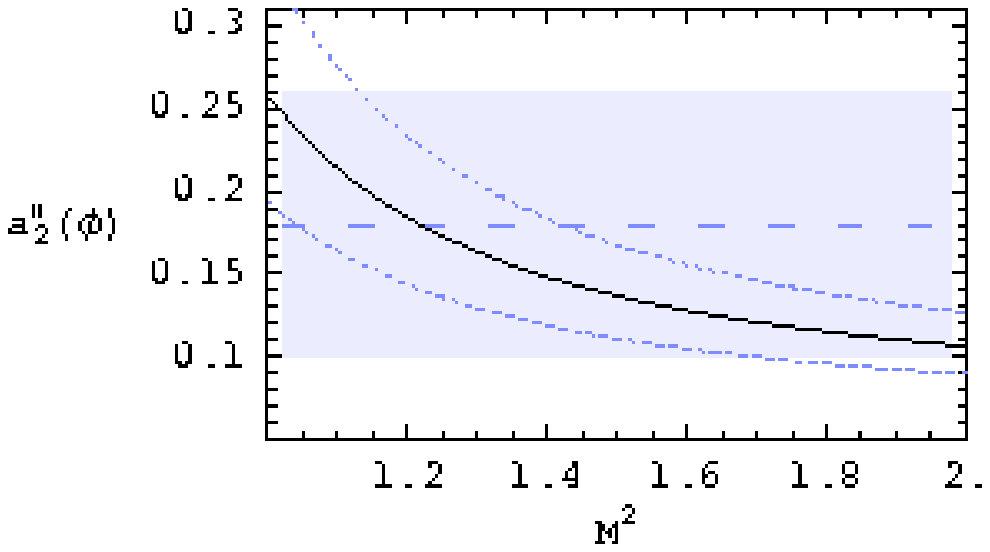}\quad \epsfxsize=0.5\textwidth\epsffile{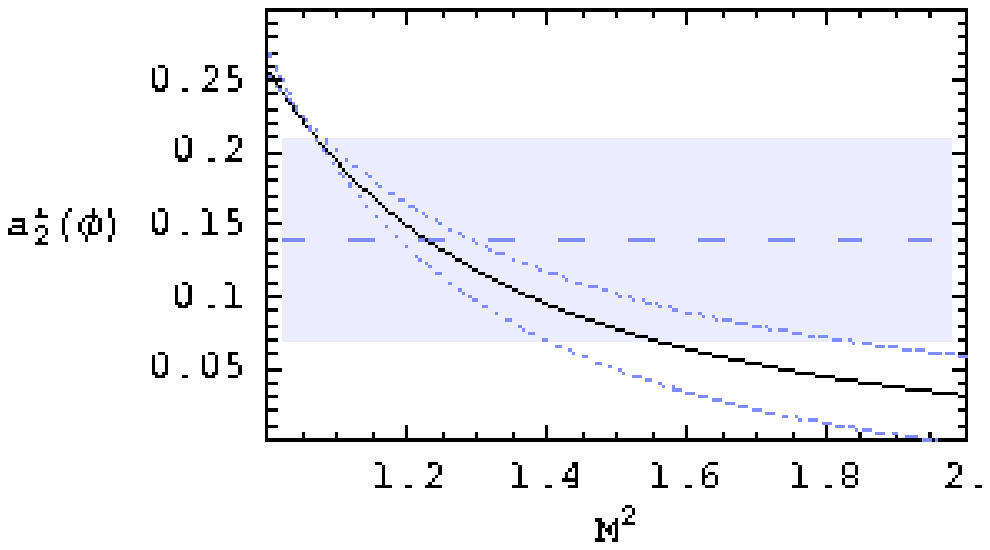}$$
  \caption[The hadronic parameters $f^{\parallel,\perp}_\phi$ and $a_2^{\parallel,\perp}(\phi)$ as a function of $M^2$.]{\small  The decay constants $f^\parallel_\phi$ (upper left) and $f^\perp_\phi$ (upper right) and the  Gegenbauer coefficients $a_2^\parallel(\phi)$ (lower left) and $a_2^\perp(\phi)$ (lower right) plotted as a function of $M^2$. The continuum thresholds are $s_0^\parallel =1.85\pm 0.05\,{\rm GeV}^2$ and $s_0^\perp =1.40\pm 0.05\,{\rm GeV}^2$ -- see text.   Solid line: central input parameters of Tab.~\ref{QCDSRinput}. Dashed lines: variation due to the uncertainties of $m_s$ and the gluon condensate. All quantities are evaluated at $\mu=1\,{\rm GeV}$.} 
  \label{aaa}
\end{figure}

The new quark condensate contributions are added to the results presented in the literature, see Refs.~\cite{Ball:2005vx,Ball:2007rt}.  For $f_{K^*}^{\parallel,\perp}$ the sum rules read
\begin{eqnarray}
\lefteqn{(f_{K^*}^\parallel)^2 e^{-m_{K^*}^2/M^2}  = 
\frac{1}{4\pi^2}\int\limits_{m_s^2}^{s_0}
ds\,e^{-s/M^2} \,\frac{(s-m_s^2)^2 (s+2m_s^2)}{s^3} +
\frac{\alpha_s}{\pi}\, \frac{M^2}{4\pi^2}\left( 1 -
e^{-s_0/M^2}\right)}\nonumber\\[5pt]
&&{} +\frac{m_s\langle \bar s s\rangle}{M^2}\left(1 +
\frac{m_s^2}{3M^2} - 
\frac{13}{9}\,\frac{\alpha_s}{\pi}\right)+
\frac{4}{3}\,\frac{\alpha_s}{\pi} \, \frac{m_s\langle \bar q
  q\rangle}{M^2}+\frac{1}{12M^2}\,
\langle\frac{\alpha_s}{\pi}\,G^2\rangle \left(1+\frac{1}{3}\frac{m_s^2}{M^2}\right)\label{tw2sr1}
\nonumber\\[5pt]
&&{}  -\frac{16\pi\alpha_s}{9M^4}\,
\langle \bar q q\rangle\langle \bar s s\rangle +
\frac{16\pi\alpha_s}{81M^4}\,\left( \langle \bar q q\rangle^2 +
\langle \bar s s\rangle^2 \right),\label{eq:fKP}\\ \nonumber \\
\lefteqn{(f_{K^*}^\perp)^2 e^{-m_{K^*}^2/M^2}
= \frac{1}{4\pi^2}\int\limits_{m_s^2}^{s_0}
ds\,e^{-s/M^2} \,\frac{(s-m_s^2)^2 (s+2m_s^2)}{s^3} }\nonumber\\[5pt]
&&{}+ \frac{1}{4\pi^2}\int\limits_{0}^{s_0}
ds\,e^{-s/M^2} \,\frac{\alpha_s}{\pi}\left( \frac{7}{9} +
\frac{2}{3}\,\ln \frac{s}{\mu^2}\right) 
-\frac{1}{12M^2}\,\langle\frac{\alpha_s}{\pi}\,G^2\rangle 
\nonumber\\[5pt]
&&{}
\times\left\{ 1-\frac{2m_s^2}{M^2}\left( \frac{7}{6}-\gamma_E + {\rm   Ei}\left(
-\frac{s_0}{M^2}\right) - \ln\,\frac{\mu^2}{M^2} +
\frac{M^2}{s_0}\left( 1 - \frac{M^2}{s_0}\right) e^{-s_0/M^2}
\right)\right\} \nonumber\\[5pt]
&&{} +\frac{m_s\langle \bar s
  s\rangle}{M^2}\left\{1+\frac{m_s^2}{3M^2}+
\frac{\alpha_s}{\pi}\left(-\frac{22}{9} + \frac{2}{3}
\left[ 1-\gamma_E + \ln\,\frac{M^2}{\mu^2} +
  \frac{M^2}{s_0}\,e^{-s_0/M^2} + {\rm Ei}\left(-\frac{s_0}{M^2}\right)\right]
\right)\right\}\nonumber\\[5pt]
&&{} 
-\frac{1}{3M^4}\,m_s\langle \bar s\sigma gGs\rangle -
\frac{32\pi\alpha_s}{81M^4}\,\left( \langle \bar q q\rangle^2 +
\langle \bar s s\rangle^2 \right)\,,\label{eq:fKT}
\end{eqnarray}
and for $a_2^{\parallel,\perp}(K^*)$
\begin{eqnarray}
\lefteqn{
a_2^\parallel(K^*) (f_{K^*}^\parallel)^2 e^{-m_{K^*}^2/M^2} = }\nonumber\\
&&{}\frac{7}{4\pi^2}\,m_s^4\int\limits_{m_s^2}^{s_0}
ds\,e^{-s/M^2} \,\frac{(s-m_s^2)^2(2m_s^2-s)}{s^5} +
\frac{7}{72\pi^2}\,\frac{\alpha_s}{\pi} \,M^2 (1-e^{-s_0/M^2})
+\frac{7}{36M^2}\,\left\langle\frac{\alpha_s}{\pi}\,G^2\right\rangle 
\nonumber\\[5pt]
&&{}+\frac{7}{3}\,\frac{m_s\langle \bar s
  s\rangle}{M^2}\left\{1+ 
\frac{\alpha_s}{\pi}\left[ -\frac{184}{27} +
\frac{25}{18} \left(1-\gamma_E + \ln\,\frac{M^2}{\mu^2} +
\frac{M^2}{s_0}\,e^{-s_0/M^2} + {\rm Ei}\left(-\frac{s_0}{M^2}\right)
\right)\right]\right\}  \nonumber\\[5pt]
&&{}+\frac{49}{27}\,\frac{\alpha_s}{\pi} \, \frac{m_s\langle \bar q
 q\rangle}{M^2} - \frac{35}{18}\, 
\frac{m_s\langle\bar s \sigma g Gs\rangle}{M^4}
+\frac{224\pi\alpha_s}{81M^4}\,\left( \langle \bar q q\rangle^2 +
\langle \bar s s\rangle^2 \right) -
\frac{112\pi\alpha_s}{27M^4}\,\langle \bar q q\rangle \langle \bar s
s\rangle\,,\label{tw2sr3}\\ \nonumber \\[5pt]
\lefteqn{
a_2^\perp(K^*) (f_{K^*}^\perp)^2 e^{-m_{K^*}^2/M^2} = }\nonumber\\[5pt]
&&{}\frac{7}{4\pi^2}\,m_s^4\int\limits_{m_s^2}^{s_0}
ds\,e^{-s/M^2} \,\frac{(s-m_s^2)^2(2m_s^2-s)}{s^5} +
\frac{7}{90\pi^2}\,\frac{\alpha_s}{\pi} \,M^2 (1-e^{-s_0/M^2})
+\frac{7}{54M^2}\,\left\langle\frac{\alpha_s}{\pi}\,G^2\right\rangle 
\nonumber\\[5pt]
&&{}+\frac{7}{3}\,\frac{m_s\langle \bar s
  s\rangle}{M^2}\left\{1+ 
\frac{\alpha_s}{\pi}\left[ -\frac{206}{27} +
\frac{16}{9} \left(1-\gamma_E + \ln\,\frac{M^2}{\mu^2} +
\frac{M^2}{s_0}\,e^{-s_0/M^2} + {\rm Ei}\left(-\frac{s_0}{M^2}\right)
\right)\right]\right\}  \nonumber\\[5pt]
&& - \frac{49}{18}\, 
\frac{m_s\langle\bar s \sigma g Gs\rangle}{M^4}
+\frac{112\pi\alpha_s}{81M^4}\,\left( \langle \bar q q\rangle^2 +
\langle \bar s s\rangle^2 \right)\,.\label{tw2sr4}
\end{eqnarray}
To obtain the sum rules for $f_{\phi}^{\parallel,\perp}$ and
$a_2^{\parallel,\perp}(\phi)$, one has to substitute $\langle \bar qq\rangle\to \langle
\bar s s\rangle$ and to double the terms in $m_s\langle
\bar s s\rangle$, $m_s\langle \bar q q\rangle$ and $m_s \langle \bar s
\sigma g G s \rangle$, and replace the perturbative  contribution by
\begin{eqnarray}
\mbox{for $(f_{\phi}^{\parallel,\perp})^2$:~~}&&
 \frac{1}{4\pi^2}\int_{4m_s^2}^{s_0} ds \,e^{-s/M^2} \frac{(s+2 m_s^2)
 \sqrt{ 1-4 m_s^2/s}}{s}\,,\nonumber\\
\mbox{for $a_2^{\parallel,\perp}(\phi)(f_{\phi}^{\parallel,\perp})^2$:~~}&&
 -\frac{7}{2\pi^2}\int_{4m_s^2}^{s_0} ds \,e^{-s/M^2} \frac{m_s^4
 \sqrt{ 1-4 m_s^2/s}}{s^2}\,.
\end{eqnarray}
We have derived sum rules for the decay constants $f^{\parallel,\perp}_V$, however, numerical values can be extracted from experiment for the longitudinal decay constants.  The perpendicular decay constants, on the other hand, must be determined from non-perturbative methods; results are available from Lattice QCD calculations and previous QCD sum rule determinations.  A detailed discussion of the latest numerical values of the decay constants can be found in Ref.~\cite{Ball:2006eu} from which we just quote the following
\begin{eqnarray}
f_{\phi}^\parallel=(215\pm5)\,\textrm{MeV}\,,\qquad f_{\phi}^\perp =(186\pm9)\,\textrm{MeV}\,,
\label{decayresults}
\end{eqnarray}
where $f_{\phi}^\parallel$ is an experimental result, and $f_{\phi}^\perp$ is from Lattice QCD \cite{Becirevic:2003pn}. We can compare these results to the sum rules of Eqs.~(\ref{eq:fKP}) and (\ref{eq:fKT}) which are plotted in the upper row of Fig.~\ref{aaa}. The sum rule determinations of  $a_2^{\parallel,\perp}(\phi)$ are plotted in the lower row.  

In all the plots the dashed line and shaded region represent the central value and uncertainty of the parameter in question. To evaluate the sum rules we use the input parameters of Tab.~\ref{QCDSRinput}. For the continuum threshold we note that for the sum rule determination of $f_{K^*}^\parallel$ in Ref.~\cite{Ball:2005vx} it is taken to be $s_0^\parallel(K^*)=1.7\,{\rm GeV}^2$, and we expect for $\phi$ it to be slightly larger. Indeed, by taking $s_0^\parallel(\phi)=1.85\pm 0.05\,{\rm GeV}^2$ we find a stable  plateau and excellent agreement with the experimental result for $f_{\phi}^{\parallel}$ (upper left plot). Likewise, guided by $s_0^\perp(K^*)=1.3\,{\rm GeV}^2$ \cite{Ball:2005vx} we find $s_0^\perp(\phi)=1.40\pm 0.05\,{\rm GeV}^2$ yields a result consistent with that from Lattice QCD (upper right plot). We use these thresholds in evaluating the sum rules for $a_2^{\parallel,\perp}(\phi)$ and also replace the decay constants by their sum rules, which helps reduce dependence on the Borel parameters. The results are plotted for $a_2^{\parallel}(\phi)$ (lower left plot) and $a_2^{\perp}(\phi)$ (lower right plot). It is found that the longitudinal parameters exhibit a stronger continuum threshold dependence, which is reflected in the larger uncertainty of the determined value of $a_2^\parallel(\phi)$.  The sum rule determinations of the other particle DA parameters follow analogously and all the numerical results  are given in Tab.~\ref{det_tab1}.
\begin{table}[h]
\renewcommand{\arraystretch}{1.3}
\addtolength{\arraycolsep}{3pt}
$$
\begin{array}{| l || c | c ||  l | l || c | c |}
\hline
& \multicolumn{2}{c||}{\rho} & \multicolumn{2}{c||}{K^*}  &  
\multicolumn{2}{c|}{\phi}\\
\cline{2-7}
& \mu = 1\,{\rm GeV} & \mu = 2\,{\rm GeV} & \mu = 1\,{\rm GeV} & \mu =
2\,{\rm GeV} & \mu = 1\,{\rm GeV} & \mu = 2\,{\rm GeV}\\
\hline
a_1^\parallel & 0 & 0 & \phantom{-}0.03(2) & \phantom{-}0.02(2) & 0 & 0 
\\
a_1^\perp & 0 & 0 & \phantom{-}0.04(3) & \phantom{-}0.03(3) & 0 & 0
\\
a_2^\parallel & 0.15(7) & 0.10(5) & \phantom{-}0.11(9) & 
\phantom{-}0.08(6) & 0.18(8) & 0.13(6)
\\
a_2^\perp & 0.14(6) & 0.11(5) & \phantom{-}0.10(8) &
\phantom{-}0.08(6) & 0.14(7) & 0.11(5)
\\\hline
\end{array}
$$
\renewcommand{\arraystretch}{1}
\addtolength{\arraycolsep}{-3pt}
\caption[Results for the leading twist-2 distribution amplitude parameters.]{\small Results for the twist-2 hadronic DA parameters  at the scale  $\mu= 1\,{\rm GeV}$ and scaled up to $\mu= 2\,{\rm GeV}$
using the evolution equations (\ref{evo}). Note that  $a_1^{\parallel,\perp}({K^*})$ refers to a $(s\bar q)$ bound state; for a $(q\bar  s)$ state it changes sign.}
\label{det_tab1}
\end{table}

\section{Twist-3}
In this section we determine the twist-3 three-particle parameters of the DAs $\Phi_{3;K^*}^\perp$, $\Phi_{3;K^*}^\parallel$ and $\widetilde\Phi_{3;K^*}^\parallel$ as defined by Eq.~(\ref{das_eq27}).  Previous determinations of these parameters are rather few and far between, thus motivating the present analysis. The chiral-even $\rho$ parameters $\zeta^\parallel_{3\rho}$, $\omega^\parallel_{3\rho}$, and $\widetilde{\omega}^\parallel_{3\rho}$ were obtained in Ref.~\cite{Zhitnitsky:1985dd}, and  $\omega^\perp_{3\rho}$ was obtained in Ref.~\cite{Ball:1998sk}. We make a comparison with these results in Section~\ref{section_evaluation}.

Firstly, we outline the calculation of the three functions $\pi_{3;K^*}$ which all proceed in a similar manner, and secondly we explicitly discuss the sum rules for $\widetilde\Phi_{3;K^*}^\parallel$ and present the results. In the diagrams that follow, $q$ is the upper line and $s$ is the lower line.

\subsection{Calculation}
Each DA is accessed via a correlation function featuring its defining current. The chiral-even twist-3 parameters $\zeta_{3K^*}^\parallel$, $\widetilde\omega_{3K^*}^\parallel$, $\widetilde\lambda_{3K^*}^\parallel$
can be determined from
\begin{equation}
\widetilde\Pi^\parallel_{3;K^*}(v,q\cdot z) = \frac{i g_{\alpha\mu}^{\perp}}{(q\cdot z)^2 (2-D)} \int d^4y \,e^{-iq\cdot y} \bra{0} T \bar q(z) g_s
\widetilde G^{\alpha z} (vz) \gamma_z \gamma_5 s(0) \bar s(y)
\gamma^\mu q(y) \ket{0}\,,
\label{C.1}
\end{equation}
where the definition of $g^\perp_{\mu\nu}$ is given in Appendix~\ref{appendixA}.\footnote{We also make use of the relation $\gamma_\mu\gamma_5=\frac{i}{6}\epsilon_{\mu\lambda\nu\pi}\gamma^\lambda\gamma^\nu\gamma^\pi$ defined in $D$ dimensions.} The parameters $\kappa_{3K^*}^\parallel$, $\omega_{3K^*}^\parallel$ and $\lambda_{3K^*}^\parallel$ can be obtained from the correlation
function $\Pi_{3;K^*}^\parallel$ obtained from $\widetilde\Pi^\parallel_{3;K^*}$ by making the replacement
\begin{equation}
g_s\widetilde G_{\alpha z} \gamma_z\gamma_5 \to g_s G_{\alpha z} i \gamma_z\,.
\end{equation}
Lastly for the chiral-odd operator 
\begin{equation}
\Pi^\perp_{3;K^*}(v,q\cdot z)= \frac{1}{ (q\cdot z)^3 }\int d^4y e^{-iq\cdot y} \bra{0}T \bar q(z) \sigma_{z\mu} g_s
G_{z\mu}(vz) s(0) \bar s(y) \sigma_{qz} q(y) \ket{0}\,.
\label{C.3} 
\end{equation}
All three correlation functions $\Pi$ can be written as
\begin{equation}
\Pi_{3;K^*}(v,q\cdot z) = \int {\cal D}\underline{\alpha}\,e^{-i q\cdot z (\alpha_2 + v \alpha_3)}\pi_{3;K^*}(\underline{\alpha})\,,
\label{C.4}
\end{equation}
where the exponential function is due to the fact that we keep the correlation functions non-local. The calculation proceeds for each correlation function analogously.  Considering Eq.~(\ref{C.1}) for instance, firstly we express it in terms of hadronic contributions
\begin{equation}
\widetilde{\Pi}_{3;K^*}^\parallel(v,q\cdot z) =
\frac{(f_{K^*}^\parallel)^2
  m_{K^*}^2}{m_{K^*}^2-q^2} \int{\cal
  D}(\underline{\alpha})\,e^{-iq\cdot z(\alpha_2+v\alpha_3)}\,
\widetilde{\Phi}_{3;K^*}^ \parallel(\underline{\alpha}) + \dots\, ,
\label{C.5}
\end{equation}
where the dots denote contributions from higher-mass states. To derive the sum rule we tread down a well worn path;  express Eq.~(\ref{C.4}) as a dispersion relation and equate to
Eq.~(\ref{C.5}), subtract the continuum contribution for $s>s_0$, perform the Borel transformation and project out the desired DA parameter by substitution of the relevant polynomial.  The three hadronic parameters $\zeta_{3K^*}^\parallel$, $\widetilde\omega_{3K^*}^\parallel$, $\widetilde\lambda_{3K^*}^\parallel$ are projected out like so:
\begin{eqnarray}
\left(f_{K^*}^ \parallel\right)^2 m_{K^*}^2 e^{-m_{K^*}^2/M^2}
      \left[  \zeta_{3K^*}^\parallel\right]  
& = & \int_0^{s_0}ds\, e^{-s/M^2} \int {\cal D}\underline{\alpha}\,
    \left[1\right]  \frac{1}{\pi}\, {\rm Im}_{s}
    \widetilde{\pi}^\parallel_{3;K^*}(\underline{\alpha})\,,
\\
\left(f_{K^*}^ \parallel\right)^2 m_{K^*}^2 e^{-m_{K^*}^2/M^2}
      \,    \left[\frac{1}{14}\,\widetilde\lambda_{3K^*}^\parallel\right]  
& = & \int_0^{s_0}ds\, e^{-s/M^2} \int {\cal D}\underline{\alpha}\,
         \left[ \alpha_1-\alpha_2\right]  \frac{1}{\pi}\, {\rm Im}_{s}
      \widetilde{\pi}_{3;K^*}^\parallel(\underline{\alpha})\,,
\nonumber\\
\left(f_{K^*}^ \parallel\right)^2 m_{K^*}^2 e^{-m_{K^*}^2/M^2}
      \,    \left[\frac{3}{28}\,\widetilde\omega_{3K^*}^\parallel\right]  
& = & \int_0^{s_0}ds\, e^{-s/M^2} \int {\cal D}\underline{\alpha}\,
         \left[\alpha_3-\frac{3}{7}\right]  \frac{1}{\pi}\, {\rm Im}_{s}
      \widetilde{\pi}_{3;K^*}^\parallel(\underline{\alpha})\,.\nonumber
      \label{C.6}
\end{eqnarray}
The formulas for the other parameters are analogous. In calculating  the functions  $\pi_{3;K^*}$ we keep explicit mass corrections $\mathcal{O}(m_s^2,m_q^2,m_s m_q)$ and all operators up to $D=6$ except  the triple gluon condensate $\left<g_s^3 f G^3\right>$ which is expected to yield a negligible contribution. By retaining all mass terms the resulting formulas for $\pi_{3;K^*}$ can be used to derive sum rules for all the DA parameters for $K^*$, $\rho$ and $\phi$ by setting  $m_q=0$, $m_q= m_s=0$ and $m_q= m_s$ respectively. For $\rho$ and $\phi$ expressions for the three-particle twist-3 DAs are analogous to Eq.~(\ref{das_eq27}), except that the G-parity violating parameters $\kappa$ and $\lambda$ vanish.

\subsection*{Perturbation Theory}
\begin{figure}[h]
$$ \epsfxsize=0.4\textwidth\epsffile{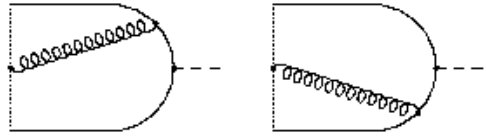}$$
  \caption[Diagrams contributing to perturbation theory.]{\small Diagrams contributing to perturbation theory.} 
  \label{a}
\end{figure}

The perturbation theory calculation is given by the two diagrams shown in Fig.~\ref{a}. 
As an example, consider the first diagram, which up to an overall factor can be written generally as
\begin{eqnarray}
\Pi^{(\rm{PT}_1)}&=&g_s^2\int\frac{d^D p}{(4 \pi)^D}\int\frac{d^D l}{(4 \pi)^D}\,\textrm{Tr}[\Gamma_1 S^{(s)}(\slash{p}+\slash{q})\Gamma_2 S^{(q)}(\slash{p})\gamma^{\beta}S^{(q)}(\slash{p}+\slash{l})] \nonumber\\
&&\cdot\,[l_\mu D_{\nu\beta}(l)-l_\nu D_{\mu\beta}(l)]\,e^{iz \cdot(l \bar{v}+p)}\,.\label{part1}
\end{eqnarray}
where the Dirac matrices $\Gamma_{1,2}$ depend on the correlation function. In performing the two successive integrations over $l$ and $p$, Feynman parameterisation leads to shifting the variables $l\to l-p\bar x$ and $p\to p-q\bar y$ respectively. Each time the exponential in (\ref{part1}) is also shifted. In expanding the part of the exponential that contributes to the integral, for example, for $l$ we have $e^{i  l  \cdot z \bar{v}}=1+i (l  \cdot z)\bar{v} +\dots$, only the first two terms contribute; higher order terms are killed off either via $z^2=0$ or because integrals with odd numbers of open indices, for example $l^{\mu_1}l^{\mu_2}l^{\mu_3}$, in the numerator vanish due to symmetry. After the integrations any terms $(\mathcal{T})$ including factors of $i( q  \cdot z)\bar{v}$ are dealt with by trading them for derivatives of $\mathcal{T}$ by using partial integration of the final exponential
\begin{equation}
i (q\cdot z) \bar{v}=\frac{1}{\bar{y}}\frac{\partial}{\partial\bar{x}}e^{-i q  \cdot z \bar{y}(1-\bar{x}\bar{v})} \qquad\Rightarrow\qquad (q\cdot z) \bar{v}\, \mathcal{T} \stackrel{\textrm{PI}}{\longrightarrow} \frac{i}{\bar{y}}\frac{\partial }{\partial\bar{x}}\mathcal{T} \,,\label{PI}
\end{equation}
where surface terms do not contribute as they vanish for $x=\{1,0\}$. The exponential can be matched to the ``canonical form'' by writing
\begin{equation}
\int^1_0dx\,\int^1_0dy\, e^{-i q  \cdot z \bar{y}(1-\bar{x}\bar{v})} = \int^1_0dx\,\int^1_0dy\,  \int {\cal D}\underline{\alpha}\, \delta(\alpha_1-y)\delta(\alpha_2-\bar{x}\bar{y})\delta(\alpha_3- x\bar{y})\,e^{-iq\cdot z(\bar\alpha_1-\bar{v}\alpha_3)}\,.
\end{equation}
Performing the $x$ and $y$ integration of the whole expression gives the desired result
\begin{equation}
\int {\cal D}\underline{\alpha}\, e^{-iq\cdot z(\alpha_2+v\alpha_3)}\pi^{(\textrm{PT}_1)}(\underline{\alpha})\,.
\end{equation}
The second diagram follows analogously. Both diagrams are divergent and need to be renormalised separately. We find finite counter terms which are proportional to the quark masses. 

\subsection*{Gluon Condensate}
\begin{figure}[h]
$$ \epsfxsize=0.4\textwidth\epsffile{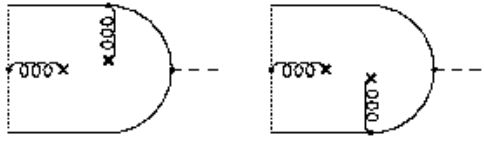}$$
  \caption[Diagrams contributing to the gluon condensate $\left<\frac{\alpha_s}{\pi} G^2\right>$.]{\small Diagrams contributing to the gluon condensate $\left<\frac{\alpha_s}{\pi} G^2\right>$.}
  \label{b} 
\end{figure}

The leading order contribution to the gluon condensate $\left<\frac{\alpha_s}{\pi} G^2\right>$ is found using the background field method as outlined in Section~\ref{example}. There are only two diagrams contributing  as depicted in Fig.~\ref{b}. One vacuum momentum $k$, from the gluon attached to the quark line, is introduced and hence one derivative is taken. As the gluon emerging from the non-local vertex $G(vz)$ carries no momentum these diagrams are proportional to $\delta(\alpha_3)$ and the remaining momentum fractions are related by $1-\alpha_1=\alpha_2$; the identification of the momentum fractions with the Feynman parameters is therefore straightforward. 
The calculation requires the integration over one momentum $p$ and the result can simply be written unexpanded in the quark masses.

\subsection*{Mixed Condensate}
\begin{figure}[h]
$$ \epsfxsize=0.4\textwidth\epsffile{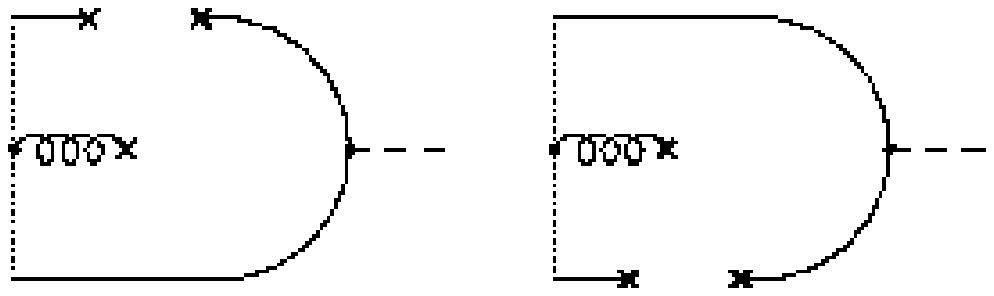}$$
  \caption[Diagrams contributing to the mixed condensates $\left<\bar q \sigma g_s G q \right>$ and $\left<\bar s \sigma g_s G s \right>$.]{\small Diagrams contributing to the mixed condensates $\left<\bar q \sigma g_s G q \right>$ and $\left<\bar s \sigma g_s G s \right>$.}
  \label{f} 
\end{figure}

The mixed condensates $\left<\bar q \sigma g_s G q \right>$ and $\left<\bar s \sigma g_s G s \right>$ originate from the diagrams shown in Fig.~\ref{f}. 
To extract the mixed condensates one uses the first non-local term in the expansion  $(D=4)$ \cite{PT:84}
\begin{eqnarray}
\lefteqn{\bra{0}\!:\!\bar{q}^i_\alpha(x_1) g_s (G_{\mu\nu})_{ij}(y)q^j_\beta(x_2)\!:\!\ket{0}=}\hspace{1.2in}\\
&&\delta^{ij}\left[\frac{\left<\bar{q} g_s \sigma G q\right>}{144}\left\{\sigma_{\mu\nu}+\frac{m_q}{2}\left[\Delta_\mu \gamma_\nu-\Delta_\nu \gamma_\mu -i(\Delta^\lambda \gamma_\lambda)\sigma_{\mu\nu}\right]\right\}\right.\nonumber\\
&&\left.+g_s^2 \left<6\right>\left\{\frac{i}{288}(x_2^\xi \sigma_{\mu\nu}\gamma_\xi-x_1^\xi \gamma_\xi \sigma_{\mu\nu})-\frac{1}{216}(y_\mu\gamma_\nu-y_\nu\gamma_\mu)\right\}\right]_{\beta\alpha}\,. \nonumber
\label{mixed}
\end{eqnarray}
The first $\sigma_{\mu\nu}$ does not contribute, but the term $\sim m_q$ does. The $\Delta_\mu$s can be expressed as derivatives of the trace via partial integration which is dealt with simply by using Eq.~(\ref{quarkderiv}). Along with the condensate gluon, the quark condensate lines carry no momentum. There is therefore no loop integration to perform and the results are proportional to $\delta(\alpha_3)\delta(\alpha_{1,2})$.

\subsection*{Quark Condensates}
\begin{figure}[h]
$$ \epsfxsize=0.8\textwidth\epsffile{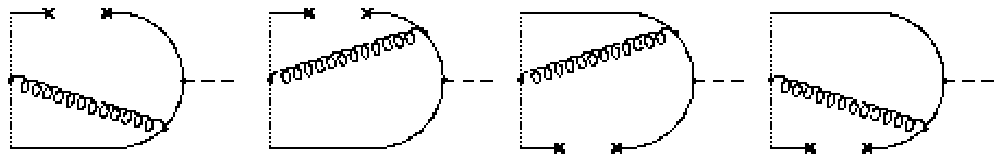}$$
  \caption[Diagrams contributing to the quark  condensates $\left<\bar q q\right>$ and $\left<\bar s s \right>$.]{\small Diagrams contributing to the quark  condensates $\left<\bar q q\right>$ and $\left<\bar s s \right>$.}
  \label{c} 
\end{figure}

\begin{figure}[h]
$$ \epsfxsize=0.4\textwidth\epsffile{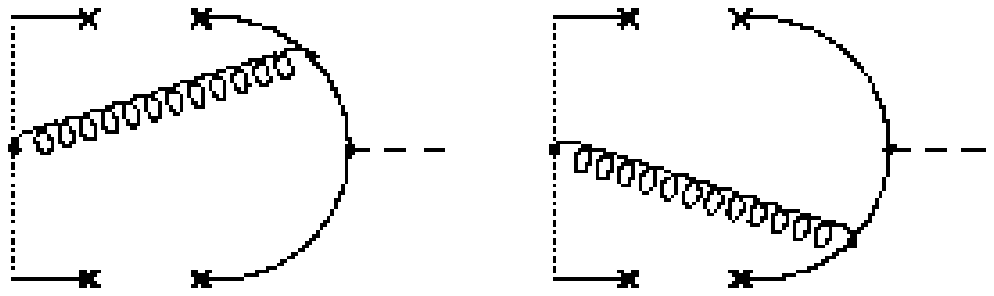}$$
  \caption[Diagrams contributing to the quark  condensate $\left<\bar q q\right>\left<\bar s s \right>$.]{\small  Diagrams contributing to the quark  condensate $\left<\bar q q\right>\left<\bar s s \right>$.}\label{d} 
\end{figure}

\begin{figure}[h]
$$ \epsfxsize=0.4\textwidth\epsffile{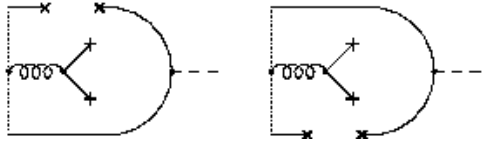}$$
  \caption[Diagrams contributing to the quark  condensates $\left<\bar q q\right>^2$ and $\left<\bar s s \right>^2$.]{\small Diagrams contributing to the quark  condensates $\left<\bar q q\right>^2$ and $\left<\bar s s \right>^2$ from the expansion of the mixed condensate -- see Eq.~(\ref{mixed}).}
  \label{e} 
\end{figure}

The diagrams of Fig.~\ref{c} generate the condensates $m_{q,s}\left<\bar q q\right>$ and $m_{q,s}\left<\bar s s \right>$. 
We do not consider $\mathcal{O}(m_{q,s}^2)$ corrections, which are however of dimension six, as they are very well suppressed with respect to the other contributions. To extract all $\mathcal{O}(m_{q,s})$ mass corrections the first non-local term in the expansion of the quark fields, given by Eq.~(\ref{quarkextract}), is needed. There is one loop momentum to integrate over and one finds contributions from the exponential which can be dealt with via partial integration in the same way as with the perturbation theory calculation, see Eq.~(\ref{PI}). The results are proportional to $\delta(\alpha_{1,2})$.  The diagrams in Fig.~\ref{d} generate the condensate $\left<\bar q q\right>\left<\bar s s \right>$ which is already of dimension six, so we do not require mass corrections. 
The two diagrams are of equal magnitude and cancel, however only for $\widetilde{\pi}^{\parallel}_{3;K^{*}}$ they add. There is no loop integral to perform and the result is proportional to $\delta(\alpha_1)\delta(\alpha_2)$. The four quark condensate is simplified via the vacuum saturation hypothesis (VSH) \cite{PT:84,Shifman:1978by}
\begin{equation}
\bra{0}\!:\!\bar{q}_\alpha^i (x_1) q_\beta^j(x_2)\bar{s}_\gamma^k (x_3) s_\delta^l(x_4)\!:\!\ket{0} \stackrel{\textrm{VSH}}{\longrightarrow}\bra{0}\!:\!\bar{q}_\alpha^i (x_1) q_\beta^j(x_2)\!:\!\ket{0}\bra{0}\!:\!\bar{s}_\gamma^k (x_3) s_\delta^l(x_4)\!:\!\ket{0}\,.
\end{equation}
The diagrams in Fig.~\ref{e} generate the condensates $\left<\bar q q\right>^2$ and $\left<\bar s s \right>^2$. They stem from the operator $\left<6\right>$ appearing in the expansion of the mixed condensate, Eq.~(\ref{mixed}), which simplifies as
\begin{equation}
 \left<6\right>=\langle\bar q \gamma_\kappa t^a q \sum_{u,d,s}\bar{q} \gamma^\kappa t^a q\rangle\stackrel{\textrm{VSH}}{\longrightarrow} -\frac{4}{9}\left<\bar q q\right>^2\,;
 \end{equation}
thus at higher order the mixed condensate also contributes to the quark condensates. The light-like co-ordinate of the gluonic field strength tensor $v z_\mu$ simplifies the resulting trace via $z^2=0$ from Eq.~(\ref{mixed}) and the other co-ordinates are dealt with as before. 
\subsection*{Results}
For the functions $\pi_{3;K^*}$, given by Eq.~(\ref{C.6}), we find (dropping all terms that vanish upon taking the imaginary part):
\begin{eqnarray}
\pi^{\perp }_{3;K^{*}}\left(\underline{\alpha}\right)
&=&
\frac{\alpha_{s}}{2\pi^{3}}\ln\frac{-q^2}{\mu^{2}}
\left[q^2\alpha_{1}\alpha_{2}\alpha_{3}\left(\frac{1}{\bar{\alpha}_{2}}-
\frac{1}{\bar{\alpha}_{1}}\right)\right.
\nonumber \\
&+&m_{s}m_{q}\frac{\alpha_{3}^{2}}{\bar{\alpha}_{1}\bar{\alpha}_{2}}
\left[\bar{\alpha}_{2}\left(\ln\frac{\alpha_{2}\alpha_{3}}{\bar{\alpha}_{1}}+
\frac{1}{2}\ln\frac{-q^2}{\mu^{2}}\right)-\left\{\alpha_{1}
\leftrightarrow\alpha_{2}\right\}\right]
\nonumber \\
&+&m_{s}^{2}\left\{-\alpha_{2}\alpha_{3}\left(\frac{1}{\bar{\alpha}_{2}}-
\frac{1}{\bar{\alpha}_{1}}\right)-\frac{\alpha_{2}\alpha_{3}^{2}}{
\bar{\alpha}_{2}^{2}}\left(\ln\frac{\alpha_{1}\alpha_{3}}{\bar{\alpha}_{2}}+
\frac{1}{2}\ln\frac{-q^2}{\mu^{2}}\right)\right\}
-m_{q}^{2}\left\{\alpha_{1}\leftrightarrow\alpha_{2}\right\}] 
\nonumber \\
&+&\frac{1}{12}\langle\frac{\alpha_{s}}{\pi}G^{2}\rangle
\frac{\alpha_{1}\alpha_{2}\left(\alpha_{1}-\alpha_{2}\right)\delta
\left(\alpha_{3}\right)}{\alpha_{1}m_{q}^{2}+\alpha_{2}m_{s}^{2}-
\alpha_{1}\alpha_{2}q^2}
\nonumber \\
&+&\frac{2}{3q^2}\frac{\alpha_{s}}{\pi}\left\{\right.
\frac{\bar{\alpha}_{3}}{2}\left(1+\alpha_{3}\right)\left(m_{q}
\langle\bar{q}q\rangle\delta\!\left(\alpha_{2}\right)-m_{s}\langle\bar{s}s
\rangle\delta\!\left(\alpha_{1}\right)\right) 
\nonumber \\
&+&\alpha_{3}\left[1+\alpha_{3}\left(1+\ln\left(\alpha_{3}
\bar{\alpha}_{3}\right)+\ln\frac{-q^2}{\mu^{2}}\right)\right]\left(m_{s}
\langle\bar{q}q\rangle\delta\!\left(\alpha_{2}\right)-m_{q}\langle\bar{s}s
\rangle\delta\!\left(\alpha_{1}\right)\right)\left.\right\} 
\nonumber \\
&+&\frac{1}{6q^4}\delta\!\left(\alpha_{3}\right)\left\{m_{q}\langle\bar{q}
\sigma g_s G q\rangle\delta\!\left(\alpha_{2}\right)-m_{s}\langle\bar{s}
\sigma g_s G s\rangle\delta\!\left(\alpha_{1}\right)\right\}
\nonumber \\
&+&\frac{16}{27q^4} \pi \alpha_{s} \delta\!\left( \alpha_{3}\right) 
\left\{\langle\bar{q}
q\rangle^{2}\delta\!\left(\alpha_{2}\right)-\langle\bar{s}s\rangle^{2}
\delta\!\left(\alpha_{1}\right)\right\},\label{corresult1}
\end{eqnarray}
\begin{eqnarray}
\pi^{\parallel}_{3;K^{*}}\left(\underline{\alpha}\right)
&=&
\frac{\alpha_{s}}{4\pi^{3}}\ln\frac{-q^2}{\mu^{2}}\left[\right.q^2
\alpha_{1}\alpha_{2}\alpha_{3}\left(\frac{1}{\bar{\alpha}_{2}}-
\frac{1}{\bar{\alpha}_{1}}\right)
\nonumber \\
&+&m_{s}m_{q}\frac{\alpha_{3}^{2}}{\bar{\alpha}_{1}\bar{\alpha}_{2}}
\left\{\bar{\alpha}_{2}\left(\ln\frac{\alpha_{2}\alpha_{3}}{\bar{\alpha}_{1}}+
\frac{1}{2}\ln\frac{-q^2}{\mu^{2}}\right)-\left\{\alpha_{1}
\leftrightarrow\alpha_{2}\right\}\right\}
\nonumber \\
&+&m_{s}^{2}\left\{-\alpha_{2}\alpha_{3}\left(\frac{1}{\bar{\alpha}_{2}}-
\frac{1}{\bar{\alpha}_{1}}\right)-\frac{\alpha_{2}\alpha_{3}^{2}}{
\bar{\alpha}_{2}^{2}}\left(\ln\frac{\alpha_{1}\alpha_{3}}{\bar{\alpha}_{2}}+
\frac{1}{2}\ln\frac{-q^2}{\mu^{2}}\right)\right\}
-m_{q}^{2}\left\{\alpha_{1}\leftrightarrow\alpha_{2}\right\}\left.\right]
\nonumber \\
&+&\frac{1}{24}\langle\frac{\alpha_{s}}{\pi}G^{2}\rangle\frac{\alpha_{1}
\alpha_{2}\left(\alpha_{1}-\alpha_{2}\right)\delta\!\left(\alpha_{3}\right)}{
\alpha_{2}m_{s}^{2}+\alpha_{1}m_{q}^{2}-\alpha_{1}\alpha_{2}q^2}
\nonumber\\
&+&\frac{1}{3q^2}\frac{\alpha_{s}}{\pi}\left\{\right.\frac{
\bar{\alpha}_{3}}{2}\left(1+\alpha_{3}\right)\left(m_{q}\langle\bar{q}q
\rangle\delta\!\left(\alpha_{2}\right)-m_{s}\langle\bar{s}s\rangle\delta
\left(\alpha_{1}\right)\right) 
\nonumber \\
&+&\alpha_{3}\left[1+\alpha_{3}\left(\ln\left(\alpha_{3}\bar{\alpha}_{3}
\right)+\ln\frac{-q^2}{\mu^{2}}\right)\right]\left(m_{s}\langle\bar{q}q
\rangle\delta\!\left(\alpha_{2}\right)-m_{q}\langle\bar{s}s\rangle\delta
\left(\alpha_{1}\right)\right)\left.\right\}
\nonumber \\
&+&\frac{1}{12 q^4}\delta\!\left(\alpha_{3}\right)\left\{
m_{q}\langle\bar{q}
\sigma g_s G q\rangle\delta\!\left(\alpha_{2}\right)-m_{s}\langle\bar{s}
\sigma g_s G s\rangle\delta\!\left(\alpha_{1}\right)\right\}
\nonumber \\
&+&\frac{8}{27 q^4}\alpha_{s}\pi \delta\!\left( \alpha_{3}\right)
\left(\langle\bar{q}q\rangle^{2}\delta\!\left(\alpha_{2}\right) -
\langle\bar{s}s\rangle^{2}\delta\!\left(\alpha_{1}\right) \right),\label{corresult2}
\end{eqnarray}
\begin{eqnarray}
\widetilde{\pi}^{\parallel}_{3;K^{*}}\left(\underline{\alpha}\right)
&=&\frac{\alpha_{s}}{4\pi^{3}}\ln\frac{-q^2}{\mu^{2}}\left[\right.-
q^2\alpha_{1}\alpha_{2}\alpha_{3}\left(\frac{1}{\bar{\alpha}_{1}}+
\frac{1}{\bar{\alpha}_{2}}\right)
\nonumber \\
&+&m_{s}m_{q}\frac{\alpha_{3}^{2}}{\bar{\alpha}_{1}\bar{\alpha}_{2}}
\left\{\bar{\alpha}_{1}\left(\ln\frac{\alpha_{1}\alpha_{3}}{\bar{\alpha}_{2}}-
\frac{1}{2}\ln\frac{-q^2}{\mu^{2}}\right)+\left\{\alpha_{1}
\leftrightarrow\alpha_{2}\right\}\right\}
\nonumber \\
&+&m_{s}^{2}\left\{\alpha_{2}\alpha_{3}\left(\frac{1}{\bar{\alpha}_{1}}+
\frac{1}{\bar{\alpha}_{2}}\right)+\frac{\alpha_{2}\alpha_{3}^{2}}{
\bar{\alpha}_{2}^{2}}\left(\ln\frac{\alpha_{1}\alpha_{3}}{\bar{\alpha}_{2}}+
\frac{1}{2}\ln\frac{-q^2}{\mu^{2}}\right)\right\}
+m_{q}^{2}\left\{\alpha_{1}\leftrightarrow\alpha_{2}\right\}\left.\right]
\nonumber \\
&+&\frac{1}{24}\langle\frac{\alpha_{s}}{\pi}G^{2}\rangle\frac{\alpha_{1}
\alpha_{2}\delta\!\left(\alpha_{3}\right)}{\alpha_{2}m_{s}^{2}+\alpha_{1}
m_{q}^{2}-\alpha_{1}\alpha_{2}q^2}
\nonumber\\
&+&\frac{1}{3q^2}\frac{\alpha_{s}}{\pi}\left\{\right.\frac{
\bar{\alpha}_{3}^{2}}{2}\left(m_{s}\langle\bar{s}s\rangle\delta\!\left(
\alpha_{1}\right)+m_{q}\langle\bar{q}q\rangle\delta\!\left(\alpha_{2}\right)
\right) 
\nonumber \\
&+&\alpha_{3}\left[1-\alpha_{3}\left(2+\ln\left(\alpha_{3}\bar{\alpha}_{3}
\right)+\ln\frac{-q^2}{\mu^{2}}\right)\right]\left(m_{s}\langle\bar{q}q
\rangle\delta\!\left(\alpha_{2}\right)+m_{q}\langle\bar{s}s\rangle\delta
\left(\alpha_{1}\right)\right)\left.\right\}
\nonumber \\
&+&\frac{1}{12 q^4}\delta\!\left(\alpha_{3}\right)\left\{ m_{q}\langle
\bar{q}\sigma g_s
Gq\rangle\delta\!\left(\alpha_{2}\right)+m_{s}\langle\bar{s}
\sigma g_s G s\rangle\delta\!\left(\alpha_{1}\right)\right\}
\nonumber \\
&+&\frac{8}{27 q^4}\alpha_{s}\pi \delta\!\left( \alpha_{3}\right)
\left(\langle\bar{q}q\rangle^{2}\delta\!\left(\alpha_{2}\right)+
\langle\bar{s}s\rangle^{2}\delta\!\left(\alpha_{1}\right) \right)
\nonumber\\
&+&\frac{2}{3 q^4}\alpha_{s}\pi \delta\!\left( \alpha_{1}\right) 
\delta\!\left( \alpha_{2}\right)\langle\bar{q}q\rangle\langle\bar{s}s
\rangle.\label{corresult3}
\end{eqnarray}

\subsection{Evaluation of The Sum Rules}\label{section_evaluation}
In the following we consider $\widetilde{\pi}_{3;K^*}^\parallel$; the sum rules for the other DA parameters and particles $\rho$ and $\phi$ follow similarly. The values of the input parameters  and the continuum thresholds used for all sum rules are given in Appendix~\ref{appendixB}. 

One subtlety must be noted: upon integration over $\alpha_i$ and subsequent
expansion in powers of the quark masses, the gluon condensate contribution yields
terms in $m_{q,s}^2 \ln (m_{q,s}^2/(-q^2))$, which are long-distance
effects and must not appear in the short-distance OPE of the correlation functions of Eqs~(\ref{C.1}) and (\ref{C.3}). The appearance of these 
logarithmic terms is due to the fact that the expressions of Eqs.~(\ref{corresult1}-\ref{corresult3}) are
obtained using Wick's theorem which implies that the condensates are normal-ordered: 
$\langle O \rangle =\bra{0}\!:\!O\!:\! \ket{0}$ \cite{logms}. Rewriting the OPE in terms of
non-normal-ordered operators, all infrared sensitive terms can be
absorbed into the corresponding condensates. Indeed, using, 
\begin{equation}
\bra{0}\bar s g_s G s \ket{0} = \bra{0}\!:\!\bar s g_s G s\!:\!\ket{0} +
\frac{m_s}{2}\, \ln\,\frac{m_s^2}{\mu^2} \,\bra{0}\! :\!
\frac{\alpha_s}{\pi}\, G^2\!:\!\ket{0}\,,
\end{equation}
and the corresponding formula for $q$ quarks, all terms in $\ln m_{q,s}^2$ can be absorbed into the mixed quark-quark-gluon condensate and the resulting short-distance coefficients
can be expanded in powers of $m_{q,s}^2$. 
\begin{figure}[h]
$$
\epsfxsize=0.6\textwidth\epsffile{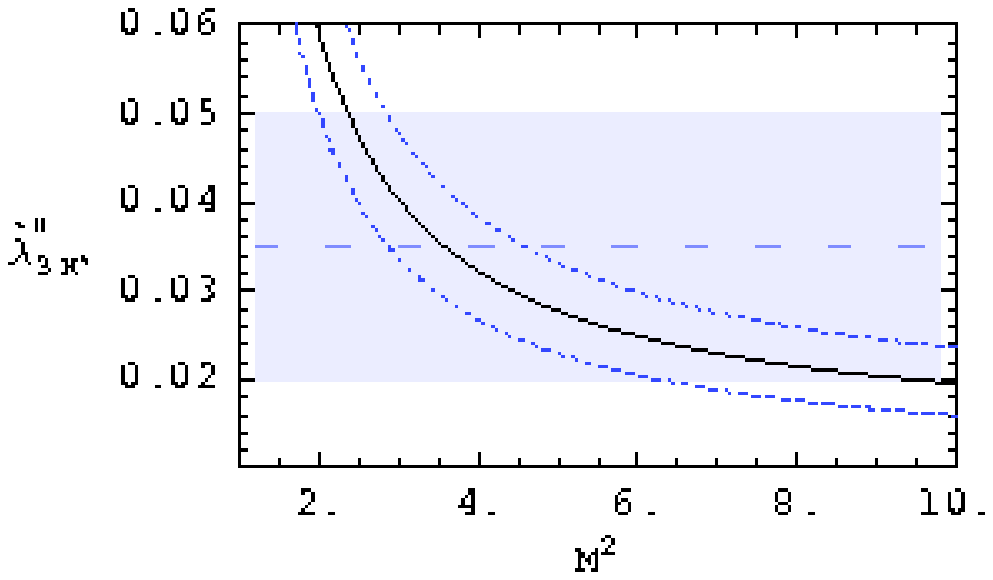}
$$
$$
\epsfxsize=0.6\textwidth\epsffile{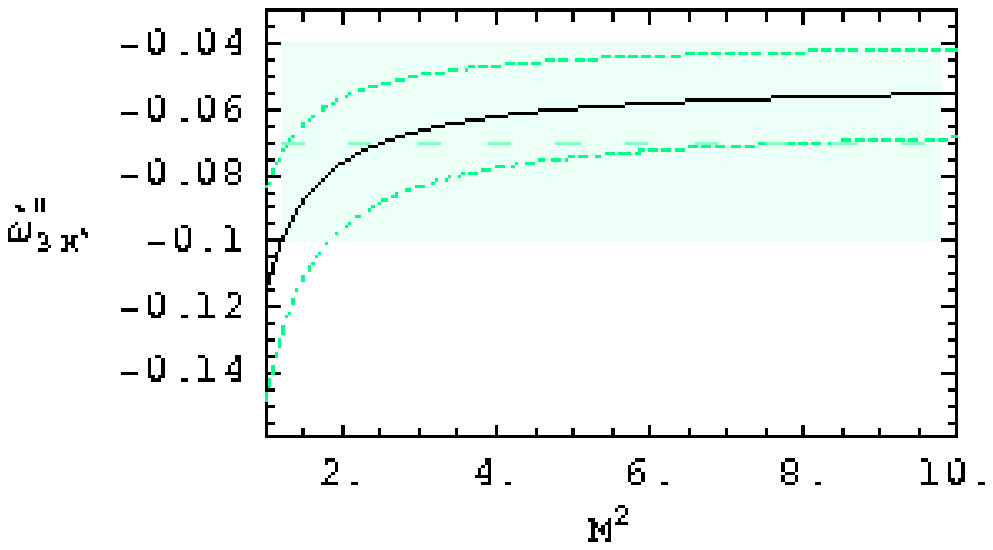}
$$
$$
\epsfxsize=0.6\textwidth\epsffile{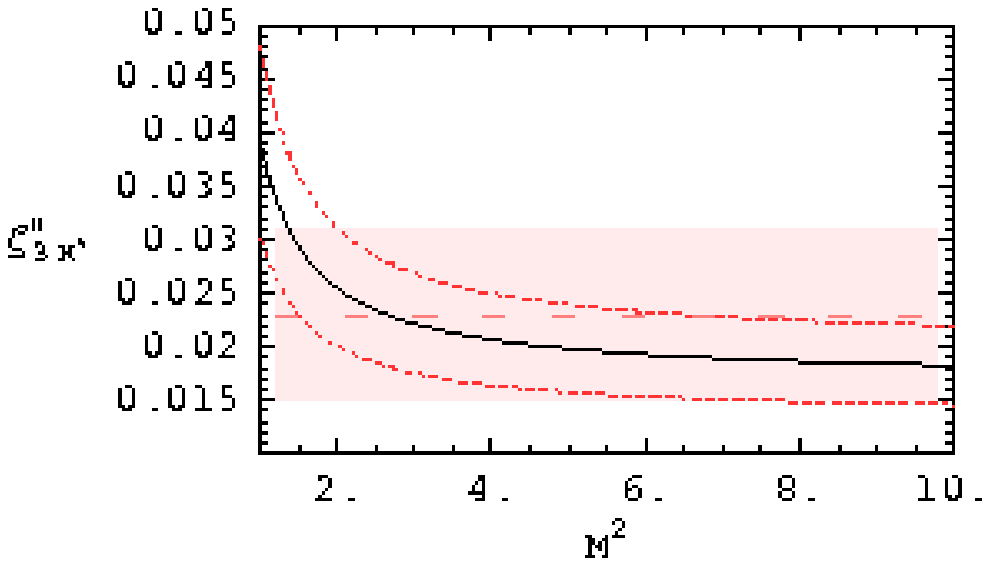}
$$
\caption[Hadronic parameters of $\widetilde\Phi_{3;K^*}^\parallel$ as functions of $M^2$.]{\small  Hadronic parameters of the twist-3 distribution amplitude $\widetilde\Phi_{3;K^*}^\parallel$ as functions of $M^2$.  Upper: $\widetilde\lambda_{3K^*}^\parallel$, middle:  $\widetilde\omega_{3K^*}^\parallel$, and lower:  $\zeta_{3K^*}^\parallel$. The solid curve is for central input values for $\mu=1\,{\rm GeV}$ and outer curves take into consideration their uncertainties -- see Tab.~\ref{QCDSRinput}. Horizontal dashed line is the extracted DA parameter value and shaded region its uncertainty -- see Tab.~\ref{det_tab2}.}
  \label{g} 
\end{figure}

In Fig.~\ref{g} we plot the sum rules  for $\widetilde\lambda_{3K^*}^\parallel$,  $\widetilde\omega_{3K^*}^\parallel$ and $\zeta_{3K^*}^\parallel$, given by Eqs.~(\ref{C.6}), which are evaluated for the central input parameters of Tab.~\ref{QCDSRinput} and at a scale $\mu=1\,{\rm GeV}$. The parameters unfortunately exhibit  very strong $M^2$ dependence, which leads to increased uncertainty of their values; we do not find a stable plateau in the region $ 1\,\textrm{GeV}^2\leqslant M^2\leqslant  2.5\,\textrm{GeV}^2$. On the other hand, there is only a very small $s_0$ dependence $\approx 1 \%$ over the range $s_0^\parallel (K^*) = (1.3\pm 0.3)\,{\rm GeV}^2$. The curves flatten at high $M^2$ which is expected, as the power corrections become negligible compared to the perturbative contribution.\footnote{The quark condensates survive as $M^2\to\infty$ as $\hat{\mathcal{B}}\left[q^{-2}\right]=-1$ but perturbation theory $\sim M^4$ -- see Appendix~\ref{appendixB}.} The sum rules for the other parameters and particles show the same general behaviour which is fairly typical of non-diagonal correlation functions. If one were to use diagonal correlation functions then it is possible that the sum rules would be better behaved and thus the uncertainties would be reduced somewhat. The calculation of diagonal correlation functions of three-particle operators, as we saw with the gluon condensate in Chapter~\ref{chapter3_SR}, is rather more involved, especially when calculating radiative corrections, which may very well be necessary in this case. 

All the numerical results, including the uncertainties from the variation of $M^2$, $s_0$, and input parameters,  are given in Tab.~\ref{det_tab2}.  The results are presented at the scale  $\mu= 1\,{\rm GeV}$ and scaled up to $\mu= 2\,{\rm GeV}$, 
using the evolution equations, Eq.~(\ref{evo}).  The only previous determination for  comparison is for the chiral-even $\rho$ parameters, $\zeta^\parallel_{3\rho}(1\,\rm{GeV})=0.033\pm0.003$, $\omega^\parallel_{3\rho}(1\,\rm{GeV})=0.2$, and 
$\widetilde{\omega}^\parallel_{3\rho}(1\,\rm{GeV})=-0.1$ \cite{Zhitnitsky:1985dd}
and  $\omega^\perp_{3\rho}(1\,\rm{GeV})=0.3\pm0.3$ \cite{Ball:1998sk}. These results agree with ours, although we consider the uncertainty of $\zeta^\parallel_{3\rho}$ to be optimistic.

\begin{table}[h]
\renewcommand{\arraystretch}{1.3}
\addtolength{\arraycolsep}{3pt}
$$
\begin{array}{| l || c | c ||  l | l || c | c |}
\hline
& \multicolumn{2}{c||}{\rho} & \multicolumn{2}{c||}{K^*}  &  
\multicolumn{2}{c|}{\phi}\\
\cline{2-7}
& \mu = 1\,{\rm GeV} & \mu = 2\,{\rm GeV} & \mu = 1\,{\rm GeV} & \mu =
2\,{\rm GeV} & \mu = 1\,{\rm GeV} & \mu = 2\,{\rm GeV}\\
\hline
\zeta_{3V}^\parallel 
& 0.030(10) & 0.020(9) & \phantom{-}0.023(8) & \phantom{-}0.015(6) & 
0.024(8) & 0.017(6)
\\
\widetilde\lambda_{3V}^\parallel 
& 0 & 0 & \phantom{-}0.035(15)& \phantom{-}0.017(8) & 0 & 0
\\
\widetilde\omega_{3V}^\parallel 
& -0.09(3) & -0.04(2) & -0.07(3) &  -0.03(2) & -0.045(15) & -0.022(8)
\\
\kappa_{3V}^\parallel 
& 0 & 0 & \phantom{-}0.000(1) & -0.001(2) & 0 & 0
\\
\omega_{3V}^\parallel 
& 0.15(5) & 0.09(3) & \phantom{-}0.10(4) & \phantom{-}0.06(3) &
0.09(3) & 0.06(2)
\\
\lambda_{3V}^\parallel 
& 0 & 0 & -0.008(4) & -0.004(2) & 0 & 0
\\
\kappa_{3V}^\perp 
& 0 & 0 & \phantom{-}0.003(3) & -0.001(2) & 0 & 0 
\\
\omega_{3V}^\perp 
& 0.55(25) & 0.37(19) & \phantom{-}0.3(1) & \phantom{-}0.2(1) &  
0.20(8) & 0.15(7)
\\
\lambda_{3V}^\perp 
& 0 & 0 & -0.025(20) & -0.015(10) & 0 & 0\\\hline 
\end{array}
$$
\renewcommand{\arraystretch}{1}
\addtolength{\arraycolsep}{-3pt}
\caption[Results for the leading twist-3 distribution amplitude parameters.]{\small Results for the leading three-particle twist-3 hadronic parameters of the DAs of Eq.~(\ref{das_eq27}). The results are presented at the scale  $\mu= 1\,{\rm GeV}$ and scaled up to $\mu= 2\,{\rm GeV}$ using the evolution equations (\ref{evo}). The sign of the  parameters corresponds to the sign convention for the strong coupling defined by the covariant derivative $D_\mu = \partial_\mu - i g_s A^a_\mu t^a$; they change sign if $g_s$ is fixed by $D_\mu = \partial_\mu + i g_s A^a_\mu t^a$.}
\label{det_tab2}
\end{table}

\begin{figure}[t]
$$
\epsfxsize=0.5\textwidth\epsffile{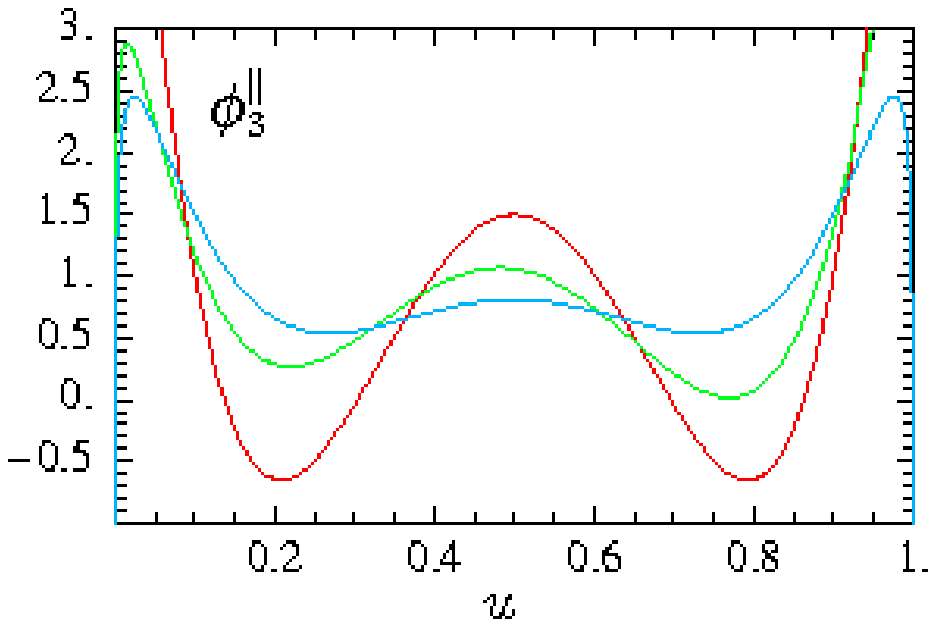}\quad
\epsfxsize=0.5\textwidth\epsffile{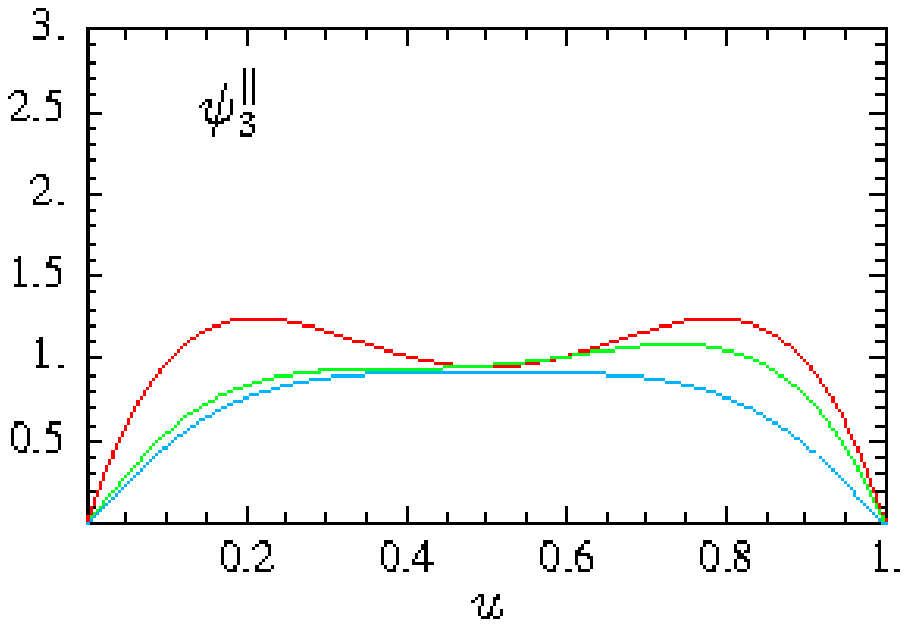}
$$
\caption[The  distribution amplitudes $\phi^\parallel_{3;V}$ and $\psi^\parallel_{3;V}$ as a function of  $u$.]{\small Left: $\phi^\parallel_{3}$ as a function of  $u$ for the central values of hadronic parameters, for $\mu=1\,$GeV.   Red line: $\phi_{3;\rho}^\parallel$, green:   $\phi_{3;K^*}^\parallel$, blue: $\phi_{3;\phi}^\parallel$. Right: same for $\psi^\parallel_{3}$.}
$$
\epsfxsize=0.5\textwidth\epsffile{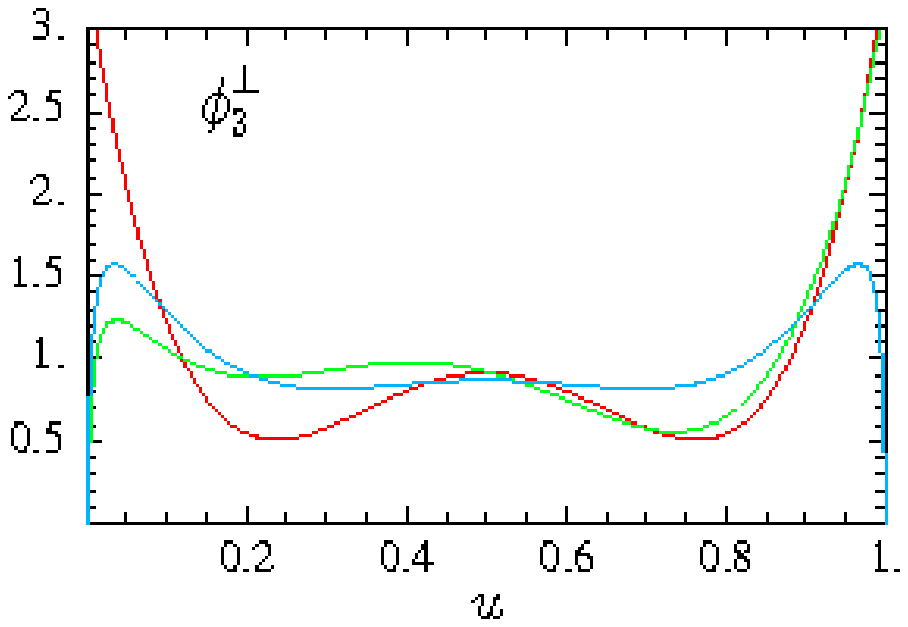}\quad
\epsfxsize=0.5\textwidth\epsffile{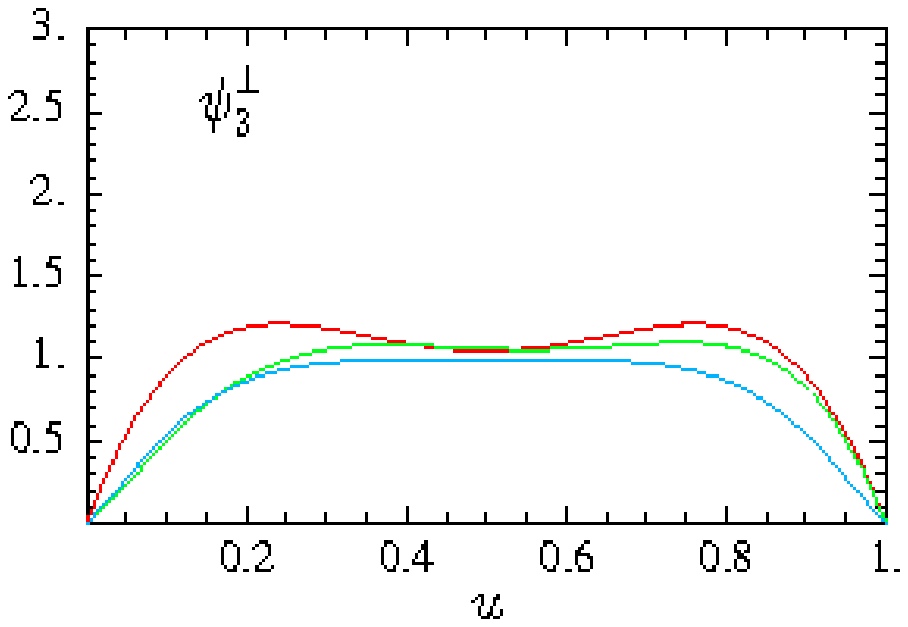}
$$
\caption[The distribution amplitudes $\phi^\perp_{3;V}$ and $\psi^\perp_{3;V}$ as a function of  $u$.]{\small Left: $\phi^\perp_{3}$ as a function of $u$ for the central values of hadronic parameters, for $\mu=1\,$GeV.   Red line: $\phi_{3;\rho}^\perp$, green:   $\phi_{3;K^*}^\perp$, blue: $\phi_{3;\phi}^\perp$. Right: same for $\psi^\perp_{3}$.}
  \label{graphs}
\end{figure}

In Fig.~\ref{graphs} we plot the two-particle twist-3 DAs as defined by Eqs.~(\ref{das_eq30}- \ref{das_eq33}). G-parity violating effects cause the small asymmetry of the $K^*$ curves. The effects of $\rm SU(3)_F$-breaking are larger and cause the pronounced difference between $\phi_{3}^\parallel$ and $\phi_{3}^\perp$  for the $\rho$ and $\phi$. We notice in particular the end-point behaviour of the DAs is greatly modified. The fact that both $\phi_{3;\rho}^{\parallel\perp}$ and $\phi_{3;K^*}^{\parallel\perp}$ diverge as $u\to1$ and $\phi_{3;\rho}^{\parallel\perp}$ for $u\to0$ is in itself not  a problem. It is only the leading-twist DA that can be considered a probability distribution and likewise there is no cause for concern that $\phi_{3;\rho}^{\parallel}$ takes negative values. Moreover, in practical calculations we are only interested in convolutions of the DAs with hard scattering kernels, which are generally finite. If not, this signals a problem with the hard scattering kernel, rather than the DA, as happens with end-point divergences within the QCD factorisation framework for non-leptonic $B$ decays, see Chapter~\ref{chapter6_QCDF}. 

\chapter{ $B \to \etapb$ Form Factors in QCD}\label{chapter5_eta}
In this chapter we discuss the semileptonic $B\to\etapb$ form factors $f_+^{B\to\etapb}$ in the LCSR approach. The previous LCSR determination of the $B\to\etapb$ form factors  presented in Ref.~\cite{Ball:2004ye} is completed by calculating the gluonic contribution, the mechanism for which involves the annihilation of the $B$ meson to two gluons. The $\etapb$ particles undergo pronounced mixing with each other due to the $\rm U(1)_A$ anomaly of QCD and the $\eta$-$\etap$ system, after many years of investigation, has succumbed to the phenomenologically motivated mixing scheme proposed by Feldmann, Kroll and Stech \cite{Feldmann:1998vh,Feldmann:1998sh}. The consideration of this mixing scheme is central to the correct description of the $B\to\etapb$ form factors.

Motivation to complete the calculation of  $f_+^{B\to\etapb}$ comes from a variety of sources, with probably the most prominent being:
\begin{itemize}
\item{the flavour-singlet contributions to the QCD factorisation framework to be discussed in Chapter~\ref{chapter6_QCDF} were added by Beneke and Neubert in Ref.~\cite{Beneke:2002jn}. It is found that the branching ratios of $B \to \etap (V,P)$ are very sensitive to $f_+^{B\to\etapb}$  as the leading-order annihilation diagrams can be interpreted as a  gluon contribution to the $B \to \etapb$ form factors \cite{Beneke:2003zv}.  Therefore a consistent estimation of the annihilation diagrams necessitates the inclusion of the gluonic contributions to the form factor.}
\item{There exists a ``tension'' in the determinations of $|V_{ub}|$ from  inclusive semileptonic decays $B\to X_u l\nu$ and their exclusive counterparts, namely from $B\to\pi l \nu$. The former have led to larger values than the latter, and the reason for the discrepancy is unclear. $B\to \etapb$ transitions are at leading order a $b\to u$ transition and so sensitive to $|V_{ub}|$ which can, in principle, be extracted from $B\to \etapb l \nu$.  An improved calculation of $f_+^{B\to\etapb}$ would reduce the theoretical uncertainty of the result.}
\item{Finally, the observation that exclusive $B\to \etap K$ and inclusive $B\to \etap X$ decays have shown unexpectedly large branching ratios with respect to $B\to\pi$ transitions, for example, is an unresolved issue which an improved calculation of $f_+^{B\to\etapb}$ may help clarify.}
\end{itemize}
We begin by introducing the $\etapb$ system and define two closely related $\eta$-$\etap$ mixing schemes. We then discuss the calculation of the flavour-singlet contribution to the form factor before lastly we discuss the results of the LCSR analysis, the framework for which was covered in Chapter~\ref{chapter3_SR}.  The material presented in this chapter follows that of Ref.~\cite{Ball:2007hb}.

\section{The $\eta$-$\etap$ System}
The approximate chiral symmetry of light quarks $u,d$ and $s$ in QCD seems to be broken by Nature to reveal the pseudoscalar mesons $(\pi^0,\pi^+,\pi^-, K^+ ,K^-, K^0 ,\bar{K}^0, \eta)$ as the corresponding octet of Goldstone bosons (all massless in the \textit{chiral limit} $m_{u,d,s}\to 0$) of the broken $\rm SU(3)\otimes SU(3)$ symmetry. There is another symmetry of the QCD Lagrangian (\ref{basics_eq1}); a global $\rm U(1)_A$ symmetry which exists at the classical level in the chiral limit. Due to non-vanishing quark masses,  the broken $\rm U(1)_A$ symmetry creates a Goldstone boson, but such a light particle does not appear in the physical spectrum and this embodies the \textit{$\rm U(1)_A$ problem}. At the quantum level, however, the $\rm U(1)_A$ symmetry in the massless limit is broken due to the QCD anomaly and so was not present in the first place; thus a ninth state, the  $\etap$, exists as a singlet and only becomes massless in the chiral limit \textit{and} as $N_c\to\infty$, causing the effects of anomaly to vanish. The situation is complicated by instanton effects, but was ultimately resolved by 't Hooft  with the same conclusion \cite{Hooft:1976up,Hooft:1986nc}.  It has been known for a while that the $\rm U(1)_A$ anomaly plays a decisive role in the $\etapb$ system with the $\etap$ consisting of a large gluonic component \cite{Witten:1978bc,Ball:1995zv}. The large mass of the $\etap$ is mostly generated by the anomaly and $\rm SU(3)_F$-breaking effects.\footnote{The particles $\etapb$ have masses $m_{\eta}=547.51 \pm 0.18 ~\textrm{MeV}$ and $m_{\etap}=957.78\pm 0.14 ~\textrm{MeV}$  and quantum numbers  $J^{PC}=0^{-+}$ \cite{Yao:2006px}.}

The $\eta$-$\etap$ system has been of considerable interest for a number of years \cite{Fritzsch:1976qc,Isgur:1976qg,Novikov:1979ux}. Vast simplifications can be made in studying the low-energy particle spectrum of QCD by employing  \textit{Chiral Perturbation Theory} (ChPT) which is an effective theory in which the heavy quarks are integrated out and the dynamically relevant  light quarks remain at a scale $\mu\sim\Lambda_{\rm QCD}$ after an expansion in powers of energies, momenta and quark masses. Alongside the $1/N_c$ expansion, ChPT is the method of choice for analysing the light pseudoscalar mesons.\footnote{Another interesting approach to understanding the $\etapb$ system was given in Ref.~\cite{Katz:2007tf}.} We do not discuss ChPT in any detail although we do quote a few of its constraints; for more details see for example \cite{Weinberg:1978kz,Gasser:1984gg,Leutwyler:2001hn} 

\begin{figure}[h]
$$ \epsfxsize=0.6\textwidth\epsffile{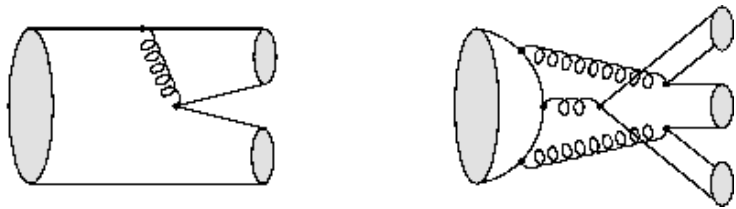}$$
\caption[Examples of an OZI-rule suppressed and allowed strong decays.]{\small Examples of strong interaction decays. Left: $\phi \to K^+ K^-$, right: $\phi \to \pi^+\pi^-\pi^0$. The former occurs preferentially over the latter due to the fact that the annihilation of the $\phi$ requires all gluons to be hard, yielding a suppression via a small $\alpha_s$ which need not be the case for the first decay. This forms the basis of the OZI-rule.}
\label{eta_ozi}
\end{figure}

Concerning $\eta$-$\etap$  mixing, ChPT requires a description in terms of two mixing angles  beyond leading-order \cite{Leutwyler:1997yr,Feldmann:1997vc}. How this is implemented in practice has caused some confusion in the past but a consistent picture has emerged \cite{Feldmann:1998vh,Feldmann:1998sh}. Key to the phenomenological picture of the $\eta$-$\etap$ system is the understanding that the main contributions to the mixing are due to the $\rm U(1)_A$ anomaly of QCD, and so-called \textit{OZI-rule violating} processes. Named after Okubo, Zweig and Iizuka the OZI-rule states that strong interaction processes that must proceed via the annihilation of all initial state quarks to gluons are suppressed \cite{Okubo:1963fa,Iizuka:1966fk,Zweig:1964wu}. In Fig.~\ref{eta_ozi} we show the unsuppressed process $\phi \to K^+ K^-$ (left) alongside the suppressed process $\phi \to \pi^+\pi^-\pi^0$ (right) for which the rule was originally formulated.  Such processes are shown to be $\mathcal{O}(1/N_c)$ in a $1/N_c$ expansion and phenomenologically they are found to be small $\approx 10\%$; they can be safely neglected, leaving the $\rm U(1)_A$ anomaly as the only mixing mechanism. For the mixing schemes we discuss in the next section, this assumption has been confronted with experimental data and holds to the expected accuracy. 

\begin{figure}[h]
$$ \epsfxsize=0.3\textwidth\epsffile{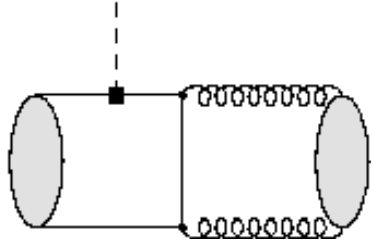}$$
\caption[$B\to \etapb$ via the $\rm U(1)_A$ anomaly.]{\small $B\to \etapb$ via the $\rm U(1)_A$ anomaly. The $b\to u$ transition allows for an annihilation of the $B$ meson's quarks to two gluons, thus probing the gluonic content of $\etapb$.}
\label{eta_u1a}
\end{figure}

A schematic picture of the $\rm U(1)_A$ anomaly at work for $B\to\etapb$ is shown in Fig.~\ref{eta_u1a}., where the flavour-singlet contribution is defined as the amplitude for producing either a quark-antiquark pair in a singlet state which does not contain the $B$'s spectator quark, or two gluons, which then hadronise into an $\etapb$.

What about mixing between other pseudoscalar mesons? In $\eta$ - $\etap$ - $\pi^{0}$ mixing the  gluonic component present in the $\pi^0$ is found to be at the level of a few percent  and so can be neglected \cite{Feldmann:1998sh,Feldmann:1999uf,Kroll:2004rs}. There also exists a $c \bar{c}$ component to $\etapb$ ($\eta_{c}$) which is considered in Ref.~\cite{Feldmann:1998vh} and found to be small with the conclusion that it is not the solution to the abnormally large $B \to K \etap$ branching ratio. Sometimes other particles are included as possible glueball candidates produced via OZI-rule suppressed processes in $J/\psi$ decay, see for example Refs.~\cite{Ball:1995zv,Li:2007ky}.  Although it is unclear whether pseudoscalar mesons contain pure glueball properties, Ref.~\cite{Kroll:2003yi}  concludes that it is unlikely. Thus the $\eta$-$\etap$  system stands out on its own. 

Phenomenologically, the semileptonic decay $B \to \etapb l \nu_l$ can be used to determine the size of  the CKM matrix element $|V_{ub}|$ from the spectrum
\begin{equation}\label{eq:spectrum}
\frac{d\Gamma}{dq^2}(B \to P l \nu_l) = \frac{G_F^2 |V_{ub}|^2}{
192\pi^3m_B^3}\lambda^{3/2}_P(q^2) |f^P_+(q^2)|^2 \,,
\end{equation}
where $P=\{\eta,\etap\}$ and $\lambda_P(x) = (m_B^2+m_{P}^2-x)^2-4m_B^2m_{P}^2$. Alternatively,
as we shall see, the ratio of branching ratios ${\cal B}(B\to\etap
\ell\nu)/{\cal B}(B\to \eta\ell\nu)$ can be used to constrain the
gluonic Gegenbauer moment $B_2^g$.

\section{State Mixing}
The first step in describing $\eta$-$\etap$ mixing is to decompose the two physical states $\ket{\etapb}$ into other, more convenient orthogonal states. As proposed in Refs.~\cite{Feldmann:1998vh,Feldmann:1998sh}  one can proceed in two ways; either by employing the singlet-octet scheme (SO) or the quark-flavour scheme (QF). The SO axial-vector currents are respectively
\begin{equation}
 J^{0}_{\mu 5}=\frac{1}{\sqrt{3}}\left(\bar u \gamma_\mu\gamma_5 u+\bar d \gamma_\mu\gamma_5  d +\bar s \gamma_\mu\gamma_5  s\right)\,,\quad  J^{8}_{\mu 5}=\frac{1}{\sqrt{6}}\left(\bar u \gamma_\mu\gamma_5  u+\bar d \gamma_\mu\gamma_5  d -2\bar s \gamma_\mu\gamma_5  s\right)\,,
\end{equation}
and their couplings are given by
\begin{equation}
\bra{0} J^{i}_{\mu 5}\ket{P(p)} = i f^{i}_{P} p_{\mu} \qquad (i=0,8)\,,
\label{SOdc}
\end{equation}
where $J^{8}_{\mu 5}$ denotes the $\rm SU(3)_F$-octet and $J^{0}_{\mu 5}$ the $\rm SU(3)_F$-singlet axial-vector current. The four quantities are related to the decay constants of a pure singlet or octet state $\ket{\eta_i}$ by two mixing angles  $\theta_i$
\begin{equation}
\left(
\begin{array}{cc}
f_\eta ^8 & f_\eta^0 \\
f_{\etap} ^8 & f_{\etap}^0
\end{array}\right)
= 
\left(
\begin{array}{cc}
\cos\theta_8 & -\sin\theta_0 \\
\sin\theta_8 & \phantom{-}\cos\theta_0 
\end{array}\right)
\left(
\begin{array}{cc}
f_8 & 0 \\
0 & f_0
\end{array}
\right).
\label{corr}
\end{equation}
Evidently $\rm SU(3)_F$-breaking effects cause $\theta_i\neq 0$ and $f_8\neq f_\pi$, and as such the SO scheme is very natural. In fact, at leading-order in ChPT an expansion in quark masses  and $1/N_{c}$ gives \cite{Leutwyler:1997yr}
\begin{equation}
\sin (\theta_{0}-\theta_{8})=\frac{2 \sqrt{2} (f_{K}^{2}-f_{\pi}^{2})}{4f_{K}^{2}-f_{\pi}^{2}}+\dots\,,
\label{su3}
\end{equation}
where the dots denote neglected higher-order terms which are required to match phenomenology \cite{Kaiser:1998ds}. The impact of the U(1)$_{\rm A}$ anomaly is plainly localised in $f_0$ via the divergence of the singlet current $J_{\mu 5}^0$ which can be written
\begin{equation}
\partial^\mu J^{a}_{\mu 5} = 2\,\bar{q} \left[ t^a \hat{m} i \gamma_5 \right]q + \delta^{a 0}\,\frac{\alpha_{s}}{4 \pi} G \widetilde{G}\,,
\label{anomaly}
\end{equation} 
where $a=\{0,1,\dots,8\}$, $\textrm{Tr}[t^a t^b ] = \frac{1}{2}\delta^{a b}$, $t^0 = \textbf{1}/\sqrt{3}$ and the mass matrix $\hat{m}=\textrm{diag}[m_u,m_d,m_s]$. The SO scheme diagonalises the renormalisation-scale dependence of parameters; $f_8$ and $\theta_i$ are
scale-independent, whereas $f_0$ renormalises multiplicatively
\begin{equation}
\mu\,\frac{d f_0}{d \mu} = - N_f \left(\frac{\alpha_s}{\pi}\right)^2 f_0 + O(\alpha_s^3)\,.
\label{scaledep}
\end{equation}
In the QF mixing scheme, on the other hand, the basic axial-vector currents are
\begin{equation}
J^q_{\mu 5} = \frac{1}{\sqrt{2}} \left(\bar u \gamma_\mu \gamma_5 u + 
\bar d \gamma_\mu \gamma_5 d \right),\qquad
J^s_{\mu 5} = \bar s \gamma_\mu \gamma_5 s\,,
\end{equation}
and the corresponding couplings are
\begin{equation}\label{6}
\bra{0} J^r_{\mu 5}\ket{P(p)} = i f_P^r p_\mu \quad (r=q,s)\,.
\end{equation}
The mixing is analogous to (\ref{corr}) with
\begin{equation}
\left(
\begin{array}{cc}
f_\eta ^q & f_\eta^s \\
f_{\etap} ^q & f_{\etap}^s 
\end{array}\right)
= 
\left(
\begin{array}{cc}
\cos\phi_q & -\sin\phi_s \\
\sin\phi_q & \phantom{-}\cos\phi_s 
\end{array}\right)
\left(
\begin{array}{cc}
f_q & 0 \\
0 & f_s
\end{array}
\right).
\label{corr2}
\end{equation}
Both quark flavour states $\ket{\eta_{q,s}}$ have vanishing vacuum-particle matrix elements with the opposite currents
\begin{equation}
\bra{0}J^s_{\mu 5}\ket{\eta_q}=\bra{0}J^q_{\mu 5}\ket{\eta_s}=0\,,
\end{equation}
which is an assumption that has been tested. It is in part motivated by the observation  of near ideal mixing in vector and tensor mesons. It implies that the mixing of states is the same as that of the decay constants and moreover leads to the diagonalisation of the mass matrix, which we come back to shortly. This hypothesis does not hold for the SO basis. It is found by Refs.~\cite{Feldmann:1997vc,Feldmann:1999uf} that the difference  between the two mixing angles of the QF scheme $\phi_q-\phi_s$ is generated by OZI-rule suppressed processes and is not caused by SU(3)$_{\rm F}$-breaking effects, as for the SO scheme (\ref{su3}). While the numerical values of $\theta_i$ differ largely, with typical values $\theta_8\approx -20^\circ$ and $\theta_0\approx - 5^\circ$, one finds $\phi_s-\phi_q\, \lesssim\, 5^\circ$, with $\phi_q\approx
\phi_s \approx 40^\circ$ \cite{Feldmann:1998vh,Feldmann:1998sh,Feldmann:1997vc}. This observation led the authors of Refs.~\cite{Feldmann:1998vh,Feldmann:1998sh} to suggest the QF scheme as an approximation to
describe $\eta$-$\etap$ mixing, based on neglecting the difference $\phi_q-\phi_s$ (and all other OZI-breaking effects):
\begin{equation}
\phi\equiv \phi_{q,s},\qquad \phi_q-\phi_s\equiv 0\,.
\end{equation}
The state mixing is then given by
\begin{equation}
\left(
\begin{array}{c}
\ket{\eta} \\ \ket{\etap}
\end{array}
\right) 
= 
\left(
\begin{array}{ll}
\cos\phi & -\sin\phi\\
\sin\phi & \phantom{-}\cos\phi
\end{array}
\right)
\left(
\begin{array}{c}
\ket{\eta_q}\\ \ket{\eta_s}
\end{array}
\right)\,.
\label{8}
\end{equation}
The re\-nor\-ma\-li\-sa\-tion-scale dependence of $f_0$ given by Eq.~(\ref{scaledep}) is not reproduced as it is induced precisely by neglected OZI-breaking terms \cite{Feldmann:1999uf}. Numerically, this is not a problem as the scale-dependence of $f_0$ is a two-loop effect. In the case of non-local matrix elements, the DAs, this lack of scale dependence of the QF scheme is somewhat problematic. We come back to this point in the next section.

Returning to the diagonalisation of the mass matrix; from Eq.~(\ref{SOdc}) one finds the quadratic diagonal mass matrix, for example
\begin{equation}\label{div}
\bra{0} \partial^{\mu}J^{s}_{\mu 5}\ket{\eta (p)} = M^{2}_{\eta} f^{s}_{\eta}\,,
\end{equation}
which, via Eq.~(\ref{anomaly}), gives the mass matrix in QF basis
\begin{equation}
\mathcal{M}_{\rm QF}^2=\left(
\begin{array}{cc}
m_{qq}^2 +\frac{\sqrt{2}}{f_q}\bra{0}\frac{\alpha_s}{4\pi}G\widetilde{G}\ket{\eta_q} & \frac{1}{f_s}\bra{0}\frac{\alpha_s}{4\pi}G\widetilde{G}\ket{\eta_q} \\
\frac{\sqrt{2}}{f_q}\bra{0}\frac{\alpha_s}{4\pi}G\widetilde{G}\ket{\eta_s}&m_{ss}^2+\frac{1}{f_s}\bra{0}\frac{\alpha_s}{4\pi}G\widetilde{G}\ket{\eta_s}
\end{array}\right)\,,
\label{mass}
\end{equation}
with the short-hand notation
\begin{equation}
m_{qq}^2=\frac{\sqrt{2}}{f_q}\bra{0}m_u \bar{u}i\gamma_5 u+m_d \bar{d} i \gamma_5 d\ket{\eta_q}\,,\qquad m_{ss}^2=\frac{2}{f_s}\bra{0}m_s \bar{s} i \gamma_5 s\ket{\eta_s}\,.
\end{equation}
From Eq.~(\ref{mass}) the crucial impact of the anomaly, as the only term in the off-diagonal elements, is evident. To first order in $\rm SU(3)_F$-breaking, the decay constants and quantities $m_{qq,ss}^2$ are fixed giving the theoretical estimate
\begin{eqnarray}
f_q=f_\pi\,,&\quad& f_s =\sqrt{2 f_K^2-f^2_\pi}\,,\nonumber\\
 m_{qq}^2=M_\pi^2\,,&\quad& m_{ss}^2=2 M_K^2-M_\pi^2\,,
\end{eqnarray}
which also leads to a fixed value of $\phi$; there is no free parameter left and thus the QF scheme is totally determined \cite{Feldmann:1998vh}. We do not work in this limit, however, and take numerical values of the decay constants and mixing angle from phenomenology. Given enough data to fix all independent parameters, there is no reason to prefer the QF over the SO scheme. The QF scheme is beneficial when considering DAs as the SO scheme leads to a proliferation of unknown parameters. For this reason we decide to use the QF scheme for the analysis. Its basic parameters have been determined as \cite{Feldmann:1998vh,Feldmann:1998sh}
\begin{equation}
f_q  =  (1.07\pm 0.02)f_\pi,\qquad f_s = (1.34\pm
0.06)f_\pi\,,\qquad
\phi =  39.3^\circ\pm 1.0^\circ\,.
\end{equation}
This can be translated into values for the SO parameters as
\begin{eqnarray}
f_8 & = & \sqrt{\frac{1}{3}\,f_q^2 + \frac{2}{3} f_s^2} = (1.26\pm
0.04) f_\pi\,,\nonumber\\
f_0  &=&  \sqrt{\frac{2}{3}\,f_q^2 + \frac{1}{3} f_s^2} = (1.17\pm
0.03) f_\pi\,,\nonumber\\
\theta_8 & = & \phi-{\rm arctan}[\sqrt{2} f_s/f_q] = (-21.2 \pm
1.6)^\circ\,,\nonumber\\
\theta_0  &=&  \phi-{\rm arctan}[\sqrt{2} f_q/f_s] = (-9.2 \pm
1.7)^\circ\,,
\end{eqnarray}
Note that in the QF scheme $f_{q,s}$ are scale-independent parameters, and so is $f_0$ as obtained from the above relations. The SO decay constants are related to those of the QF scheme by a change of basis
\begin{equation}\label{11}
\left(
\begin{array}{cc}
f_\eta ^8 & f_\eta^0 \\
f_{\etap} ^8 & f_{\etap}^0 
\end{array}\right)
= 
\left(
\begin{array}{cc}
\cos\phi & -\sin\phi \\
\sin\phi & \phantom{-}\cos\phi 
\end{array}\right)
\left(
\begin{array}{cc}
f_q & 0 \\
0 & f_s
\end{array}\right)
\left(
\begin{array}{cc}
\phantom{-}\sqrt{\frac{1}{3}} & \sqrt{\frac{2}{3}}\\
-\sqrt{\frac{2}{3}} & \sqrt{\frac{1}{3}}
\end{array}\right).
\end{equation}
The last matrix originates from the ideal mixing angle $\theta_{\textrm{ideal}}=\arctan{\sqrt{2}}$ which rotates from the QF basis to the SO basis.

\section{Pseudoscalar Meson Distribution Amplitudes}
As discussed in Chapter~\ref{chapter3_SR}, the method of LCSRs relies on the non-perturbative universal light-cone DAs; specifically here we require pseudoscalar meson DAs including the two-gluon DA. At leading-twist both these DAs contribute and indeed mix with each other under renormalisation. The quark-antiquark DAs are extensions of the matrix elements given by Eqs.~(\ref{SOdc}) and (\ref{6}) to those of non-local operators on the light-cone. Pseudoscalar mesons' quark-antiquark DAs have been investigated previously in  Refs.~\cite{Ball:1998je,Ball:2006wn,Braun:1989iv}. The two-gluon DAs of leading and higher twist have been investigated in Ref.~\cite{AP03}. In this analysis we only include the effects of the leading-twist two-gluon DA, which is justified as its effects turn out to be fairly small and higher-twist DAs are estimated to have even smaller impact. Following Ref.~\cite{Kroll:2002nt}, the twist-2 two-quark DAs of $\etapb$ are defined as
\begin{equation}
\bra{0} \bar\Psi(z) {\cal C}^i\gamma_z \gamma_5 [z,-z] \Psi(-z) \ket{P(p)} = i (p\cdot z) f_P^i \int_0^1 du\, e^{i \xi p \cdot  z} \phi_{2;P}^i(u) \,.
\end{equation}
$\phi_{2;P}^i(u)$ is the twist-2 DA of the meson $P$ with respect to the current whose flavour content is given  by ${\cal C}^i$, with $\Psi = (u,d,s)$ the triplet of light-quark fields in flavour space. For the SO currents, one has ${\cal C}^0 = \mbox{\boldmath $1$}/\sqrt{3}$ and ${\cal C}^8 = \sqrt{2}\, t^8$, while for the QF currents ${\cal C}^q = (\sqrt{2} {\cal C}^0 + {\cal C}^8)/\sqrt{3}$ and  ${\cal C}^s = ({\cal C}^0 - \sqrt{2} {\cal C}^8)/\sqrt{3}$. Due to the positive G-parity of $\eta$ and $\etap$, the two-quark DAs are symmetric under $u\leftrightarrow 1-u$, and hence all odd Gegenbauer moments vanish:
\begin{equation}
\phi_{2;P}^i(u) = \phi_{2;P}^i(1-u)\,,
\end{equation}
and the DAs are expanded in terms of Gegenbauer polynomials in exactly the same way as for the vector mesons
\begin{equation}
\phi_{2;P}^i(u) = 6 u (1-u) \left( 1 + \sum_{n=2,4,\dots} a_n^{P,i}(\mu)
C^{3/2}_n(\xi) \right) \,\quad (i=1,8,q,s)\,,
\label{expansion}
\end{equation}
where $a_n^{P,i}$ are the quark Gegenbauer moments. The gluonic twist-2 DA is defined as\footnote{This definition refers to the ``$\sigma$-rescaled'' DA $\phi^\sigma_g$ in Ref.~\cite{Kroll:2002nt} with  $\sigma = \sqrt{3}/C_F$. It agrees with that used in Refs.~\cite{AP03,Charng:2006zj}, which means that we can use their results for the two-gluon Gegenbauer moment
$B^g_2$ without rescaling.}
\begin{equation}
\bra{0} G_{\mu z}(z) [z,-z] \widetilde G^{\mu z}(-z) \ket{P(p)} = \frac{1}{2}\,(p\cdot z)^2 \frac{C_F}{\sqrt{3}} f_P^0 \int_0^1 du\, e^{i\xi p\cdot z} \psi_{2;P}^g(u)\,.
\end{equation}
In order to perform the calculation of the correlation function defined in the next section, we also need the matrix element of the meson $P$ over two gluon fields. Dropping the gauge factor $[z,-z]$ one has
\begin{equation}
\bra{0} A^a_\alpha(z) A^b_\beta(-z)\ket{P(p)} =
\frac{1}{4}\,\epsilon_{\alpha\beta\rho\sigma} \,\frac{z^\rho
  p^\sigma}{p\cdot z} \,\frac{C_F}{\sqrt{3}}\, f_P^0 \,\frac{\delta^{ab}}{8}
\int_0^1 du\, e^{i\xi p\cdot z}\,\frac{\psi_{2;P}^g(u)}{u(1-u)}\,.
\label{gluefields}
\end{equation}
The two-gluon asymptotic DA is $u^{2j-1}(1-u)^{2j-1}$ with $j=3/2$ the lowest conformal spin of the operator $G_{\mu z}$ and the expansion goes in terms of Gegenbauer polynomials $C^{5/2}_n$, see Eq.~(\ref{basics_eq18}). One can show that $\psi_{2;P}^g$ is antisymmetric:
\begin{equation}
\psi_{2;P}^g(u) = - \psi_{2:P}^g(1-u)\,,
\end{equation}
and in particular $\int_0^1 du\, \psi_{2;P}^g(u) = 0$ and the local twist-2 matrix element $\bra{0} G_{\mu z} \widetilde G^{\mu z}\ket{P}$ vanishes. The non-vanishing coupling $\bra{0}G_{\alpha\beta}  \widetilde G^{\alpha\beta}\ket{P}$ induced by the U(1)$_{\rm A}$ anomaly is a twist-4 effect. The corresponding matrix elements are discussed in Refs.~\cite{Feldmann:1998vh,Feldmann:1998sh} and are given, in the QF scheme, by:
\begin{eqnarray}
\bra{0}\frac{\alpha_s}{4\pi} G\widetilde{G} \ket{\eta_q} & = & 
f_s (m_\eta^2-m_{\etap}^2) \sin\phi \cos\phi\,,\nonumber\\
\bra{0}\frac{ \alpha_s}{4\pi} G\widetilde{G} \ket{\eta_s} & = & 
f_q (m_\eta^2-m_{\etap}^2)/\sqrt{2} \sin\phi \cos\phi\,.
\label{extra}
\end{eqnarray}
In taking the ratios of both sides of the above relations one can see that $\rm SU(3)_F$-breaking in the decay constants $f_q/f_s$ is driven by the anomaly. There are no twist-3 two-gluon DAs and the remaining twist-4 DAs also have vanishing normalisation \cite{AP03}.  The conformal expansion of the twist-2 two-gluon DA reads
\begin{equation}
\psi_{2;P}^g(u,\mu) = u^2 (1-u)^2 \sum_{n=2,4,\dots} B^{P,g}_n(\mu)
C^{5/2}_{n-1}(\xi)\,,
\end{equation}
with the gluonic Gegenbauer moments $B^{P,g}_n$. In this analysis, we truncate both $\phi^i_{2;P}$ and $\psi^g_{2;P}$ at $n=2$. An estimate of the effect of higher Gegenbauer moments in $\phi_{2;\pi}$ on the $B\to\pi$ form factor $f_+^\pi$ has been given in Ref.~\cite{Ball:2005ei}, based on a
certain class of models for the full DA beyond conformal expansion. The effect of neglecting $a_{n\geq 4}^\pi$ was found to be very small $\approx 2\%$ hence we expect the truncation error from neglecing $B^g_{n\geq 4}$ to be of similar size.

\begin{figure}[h]
$$\epsfxsize=0.45\textwidth\epsffile{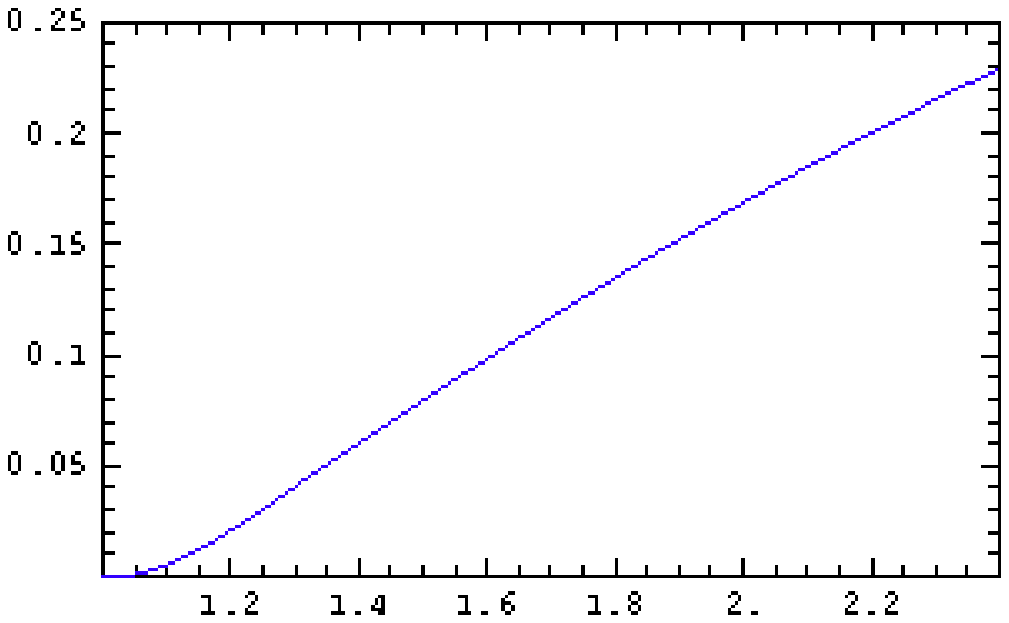}\qquad \qquad\epsfxsize=0.45\textwidth\epsffile{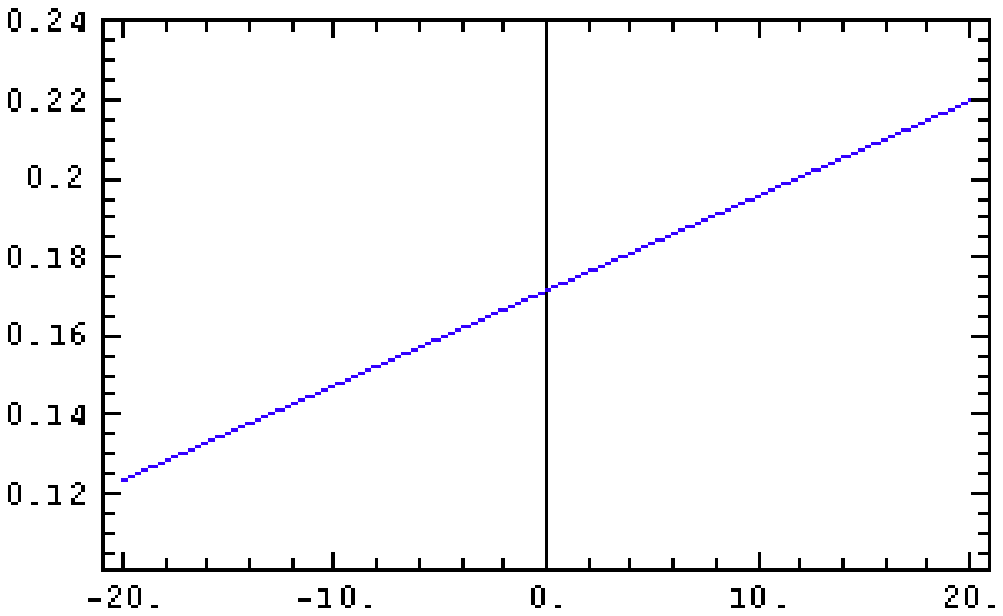}$$
\caption[Scale dependence of the twist-2 distribution amplitude parameters.]{\small  Left: $\Delta=100\,| (a^0_2(\mu)-a^8_2(\mu))/a^0_2(\mu)|$ as a function of scale $\mu$, according to Eqs.~(\ref{21}) and (\ref{22}) with $B_2^g=0$. Right: dependence of $a^0_2(2.4\,{\rm GeV})$ on $B^g_2(1\,{\rm GeV})$ for  $a^0_2(1\,{\rm GeV})=0.25$ according to Eq.\,(\ref{21})}
\label{eta_fig2}
\end{figure}

$\phi_{2;P}^0$ and $\psi_{2;P}^g$ mix upon a change of scale $\mu$ and as discussed in  Refs.~\cite{Baier:1981pm,Kroll:2002nt} this amounts to a mixing of $a_2^{P,0}$ and $B^{P,g}_2$, resulting in the renormalisation-group equation to LO accuracy
\begin{equation}
\mu\,\frac{d}{d\mu}\left(\begin{array}{c} a_2^0\\ B^g_2
\end{array}\right)
=
-\frac{\alpha_s}{4\pi} \left(\begin{array}{cc} \displaystyle\frac{100}{9} &
\displaystyle  -\frac{10}{81}\\\vphantom{\displaystyle\frac{100}{9}}
 -36 & 22\end{array}\right)
\left(\begin{array}{c} a_2^0\\ B^g_2
\end{array}\right),
\label{20}
\end{equation}
where for simplicity we have dropped the superscript $P$. The solution for $a_2^0$ reads
\begin{eqnarray}
a_2^0(\mu^2) & = & \left[ \left(\frac{1}{2} -
  \frac{49}{2\sqrt{2761}}\right) L^{\gamma_2^+/(2\beta_0)} + 
\left(\frac{1}{2} +
  \frac{49}{2\sqrt{2761}}\right) L^{\gamma_2^-/(2\beta_0)}\right]
  a_2^0(\mu_0^2) \nonumber\\
&&{}+ \frac{5}{9\sqrt{2761}}\left[L^{\gamma_2^-/(2\beta_0)}-
L^{\gamma_2^+/(2\beta_0)}\right] B_2^g(\mu_0^2)
\label{21}
\end{eqnarray}
with the anomalous dimensions $\gamma_2^\pm = (149\pm \sqrt{2761})/9$. The octet Gegenbauer moment does not have another DA with which it can mix and so its evolution is simpler
\begin{equation}
a_2^8(\mu^2) = L^{50/(9\beta_0)} a_2^8(\mu_0^2)\,.
\label{22}
\end{equation}
The mixing amongst the DAs complicates matters; as the scale dependence of the decay constants is lost in the QF scheme, one expects to have to lose scale dependence in the DAs too, and we must be careful to be consistent. The verification of the anomalous dimensions in Eq.~(\ref{20}) from the singlet and octet parts of the form factor calculations is a crucial test of the LCSR analysis. For this reason, we discuss the implications of mixing on the twist-2 DA parameters, and only briefly cover higher-twist quark DAs which are included in the octet part; for a detailed discussion one is referred to Ref.~\cite{Ball:2007hb}. Following Ref.~\cite{Kroll:2002nt}, for the DAs introduced by Eq.~(\ref{expansion}) we have, in terms of the quark valence Fock states $\ket{q\bar q}$ and $\ket{s\bar s}$
\begin{equation}
\ket{\eta_q} \sim \phi_2^q (u) \ket{q\bar q} + \phi_2^{\rm
 OZI}(u) \ket{s\bar s}\,,\quad
\ket{\eta_s} \sim  \phi_2^{\rm OZI} (u) \ket{q\bar q} + 
\phi_2^s(u) \ket{s\bar s}\,,
\end{equation}
where $q\bar q$ is shorthand for $(u\bar u + d\bar d)/\sqrt{2}$ and 
\begin{equation}
\phi_2^q = \frac{1}{3}\,(\phi_2^8 + 2\phi_2^0)\,,\quad
\phi_2^s = \frac{1}{3}\,(2\phi_2^8 + \phi_2^0)\,,\quad
\phi_2^{\rm OZI} = \frac{\sqrt{2}}{3} (\phi_2^0-\phi_2^8)\,.
\label{darel}
\end{equation}
In the QF scheme, the ``wrong-flavour'' DA  $\phi_2^{\rm OZI}$, which is generated by OZI-violating interactions,  is set to 0. Once this is done at a certain scale, however, the different evolution of $a_n^0$ and $a_n^8$  will generate a non-zero $\phi_2^{\rm OZI}$ already to LO accuracy. A consistent implementation of the QF scheme hence requires one to either set $a_n^{0,8}\equiv 0$ and also $B^g_n\equiv 0$, or to set $a_n^8\equiv a_n^0$ and neglect the different 
scale-dependence of these parameters. The induced non-zero DA $\phi_2^{\rm OZI}$ is numerically very small for the scales relevant for our calculation, $\mu=1\,$GeV and $2.4\,$GeV.\footnote{$2.4\,$GeV is  a typical scale in the calculation of form factors from LCSRs: $\mu=\sqrt{m_B^2-m_b^2}$ is chosen as an intermediate scale between $m_b$ and the typical hadronic scale $1\,{\rm GeV}$.} The left panel of Fig.~\ref{eta_fig2} shows a plot of $\Delta=100\,| (a^0_2(\mu)-a^8_2(\mu))/a^0_2(\mu)|$ as a function of scale $\mu$, according to Eqs.~(\ref{21}) and (\ref{22}),  for $a_2^8(1\,{\rm GeV})\equiv a_2^0(1\,{\rm GeV})$ and $B_2^g=0$.  We see that $\Delta$ is less than $0.25\,\%$ over the range $1\,\textrm{GeV}<\mu<2.4\,\textrm{GeV}$.  Choosing $a_2^8(1\,{\rm GeV})=0.25\pm0.15$, guided by our knowledge of twist-2 DAs of the $\pi$; we have $a_2^8(2.4\,{\rm   GeV}) = 0.171$ from Eq.~(\ref{22}), and $a_2^0(2.4\,{\rm GeV}) = 0.171$ for $B^g_2=0$, from Eq.~(\ref{21}). Evidently, the impact of the different anomalous dimensions of $a_2^0$ and $a_2^8$ is negligible.
Also, the evolution of $a_2^0$ is not hugely different to that of $a_2^8$, for a wide range of values of $B^2_g$.  The right panel of Fig.~\ref{eta_fig2} shows the evolution of the singlet Gegenbauer moment $a_2^0$ from $\mu=1\,{\rm GeV}$ - $2.4\, {\rm GeV}$, from  Eq.~(\ref{21}), for the range of gluon Gegenbauer moments $|B_2^g(1\,{\rm  GeV})|<20$, which is a \textit{very} conservative estimated range, as discussed below. The mixing of $B_2^g$ into $a_2^0$ is up to $20\%$ for $B_2^g=20$ and $40\%$ for $B_2^g=-20$.

From the conclusions of the above discussion we are justified in implementing the QF scheme for DAs as follows: we set $\phi_2^0\equiv \phi_2^8$ at the scale $\mu=1\,$GeV, which, by virtue of Eq.~(\ref{darel}), implies $\phi_2^q\equiv\phi_2^s$ at the same scale. We then evolve $a_2$ according to the scaling-law for the octet Gegenbauer moment (\ref{22}).\footnote{This is equivalent to imposing the QF-scheme  relation $a_2^0=a_2^8$ as the scale $\mu=2.4\,$GeV and defining $B^g_2$ as $B^g_2(2.4\,{\rm GeV})$.}  We also set $\psi_{2;\eta}^g=\psi_{2;\etap}^g$; again any SU(3)$_{\rm F}$-breaking of this relation is expected to have only very small impact on $f_+^{B\to\etapb}$. The twist-2 parameters used in our calculation are then reduced to two: $a_2$ and $B^g_2$.  

Concerning numerical values, we assume that the bulk of  SU(3)$_{\rm F}$-breaking effects is described by the decay constants via $f_q\neq f_\pi$, and that SU(3)$_{\rm F}$-breaking in Gegenbauer moments is sub-leading \cite{Ball:2006wn}. Sum rules for $a_2^\pi$ and $a_2^q$ would essentially be the same, with $f_\pi\neq f_q$ driving the SU(3)$_{\rm F}$-breaking and any small differences in $s_0$ and $M^2$ being negligible. This motivates setting $a_2^q = a_2^\pi$,  with $a_2^\pi(1\,{\rm GeV}) = 0.25\pm 0.15$ as an average over a large number of  calculations and fits to experimental data \cite{Ball:2006wn}. 

For $B_2^g$, however, no direct calculation is available.  Results from fits to data have been obtained from the $\etap\gamma$ transition form factor, yielding $B_2^g(1\,{\rm GeV}) = 9\pm 12$ \cite{Kroll:2002nt}, and the combined analysis of this form factor and the inclusive decay $\Upsilon(1S)\to \etap X$ yielding $B_2^g(1.4\,{\rm GeV}) = 4.6\pm 2.5$ \cite{AP03}.  Caution must be taken when considering these results as they are highly correlated with the simultaneous determination of $a_2^0$ and $a_2^8$ from the same data, yielding $a_2^0(1\,{\rm GeV}) = -0.08\pm 0.04$, $a_2^8(1\,{\rm GeV}) =
-0.04\pm 0.04$ and $a_2^0(1.4\,{\rm GeV}) = a_2^8(1.4\,{\rm GeV}) = -0.054\pm 0.029$, respectively. The same analysis, applied to the $\pi\gamma$ form factor, returns $a_2^\pi (1\,{\rm GeV}) =-0.06\pm 0.03$ \cite{vogt}. These results are not really compatible with those from the direct calculation of $a_2^\pi$ from Lattice QCD and QCD sum rules; in particular the sign of $a_2^\pi$ is unambiguously fixed as being positive. A possible reason for this discrepancy is the neglection of higher-order terms in the light-cone expansion and that, in addition, as one of the photons in the process is nearly real with virtuality $q^2\approx 0$, one also has to take into account long-distance photon interactions, of order $1 \sqrt{q^2}$, as discussed in Ref.~\cite{rady}. For this reason, we assume the very conservative range $B_2^g(2.4\,{\rm GeV}) = 0\pm20$ in the analysis.

As far as higher-twist quark DAs are concerned, we only need those involving currents with flavour content $\bar q q = (\bar u u + \bar d d)/\sqrt{2}$. In line with the implementation of the QF scheme
for twist-2 DAs, we include SU(3)$_{\rm F}$-breaking only via the decay constants. The precise definitions of all twist-3 and 4 DAs, as well as up-to-date numerical values of the $\pi$'s hadronic parameters can be found in Ref.~\cite{Ball:2006wn}. A discussion of the correct treatment of these DAs
within LCSR, as modified to describe $\etapb$, can be found in Ref.~\cite{Ball:2007hb}. 

\section{Calculation}
We define the $B\to P$ form factors analogously to those of other pseudoscalar mesons as  \cite{Ball:2004ye}
\begin{equation}
\bra{P(p)} \bar u \gamma_\mu b \ket{B(p+q)} = \left\{
(2p+q)_\mu - \frac{m_B^2-m_P^2}{q^2}\,q_\mu\right\} \frac{f_+^P(q^2)}{\sqrt{2}} + \frac{m_B^2-m_P^2}{q^2}\,q_\mu\,\frac{f_0^P(q^2)}{\sqrt{2}}\,.
\label{FF}
\end{equation}
where the factor of $1/\sqrt{2}$ on the right-hand side is to ensure that in the SU(3)$_{\rm F}$ symmetry limit, without $\eta$-$\etap$ mixing, $f_+^\eta = f_+^\pi$. For semileptonic decays $B\to \etapb l \nu_l$ the form factor $f_0^P$ appears proportional to $q^2\approx m_l^2$ which is negligible for light leptons $l=\{e,\mu\}$ for which only $f_+^P$ is required.  Using the LCSR method outlined in Chapter~\ref{chapter3_SR} we extract the semileptonic form factor $f_{+}^P$  from the following correlation function
\begin{eqnarray}
\Pi^P_{\mu}(p,q) &=& i\int d^4x\,e^{i q\cdot x} \bra{P(p)}  T [\bar u \gamma_\mu b](x) j_B^{\dagger}(0)\ket{0}\\
&=& \Pi_+^P(q^2,p_B^2) (2p+q)_\mu + \dots\,,\nonumber
\label{eq:corr}
\end{eqnarray}
where $j_B= m_b \bar u i\gamma_5 b$ is the interpolating current for the $B$ meson and $p_B^2=(p+q)^2$ its virtuality. In calculating the correlation function, we use Eq.~(\ref{8}) which relates the physical states $\ket{\etapb}$ and the QF basis states  $\ket{\eta_{q,s}}$ so that
\begin{equation}
\Pi^\eta_\mu = \frac{1}{\sqrt{2}}\left(\Pi^q_{\mu} \cos\phi - \Pi^s_{\mu} \sin\phi\right),\quad
\Pi^{\etap}_\mu = \frac{1}{\sqrt{2}}\left(\Pi^q_{\mu} \sin\phi + \Pi^s_{\mu} \cos\phi\right)\,.
\label{32}
\end{equation}
The interpolating current $\bar u \gamma_\mu b$ only probes the $\bar u u $ quark component of the $\etapb$ so $\Pi^s_\mu$ vanishes to leading order in $\alpha_s$ and at $O(\alpha_s)$ is due only to gluonic Fock states of the meson. $\Pi^q_\mu$, on the other hand, receives contributions from both quark and gluon states. The final LCSR for $f_+^P$ then reads
\begin{equation}\label{35}
e^{-m_B^2/M^2}\,m_B^2 f_B\, \frac{f_+^P(q^2)}{\sqrt{2}} = \int_{m_b^2}^{s_0}
ds\,e^{-s/M^2}\, \frac{1}{\pi}\,{\rm Im}_s\,\Pi^P_+(s,q^2)\,,
\end{equation}
with the usual sum rule specific parameters $M^2$, the Borel parameter, and $s_0$, the continuum threshold.
\subsection*{Quark Contribution}
\begin{figure}[h]
$$ \epsfxsize=0.85\textwidth\epsffile{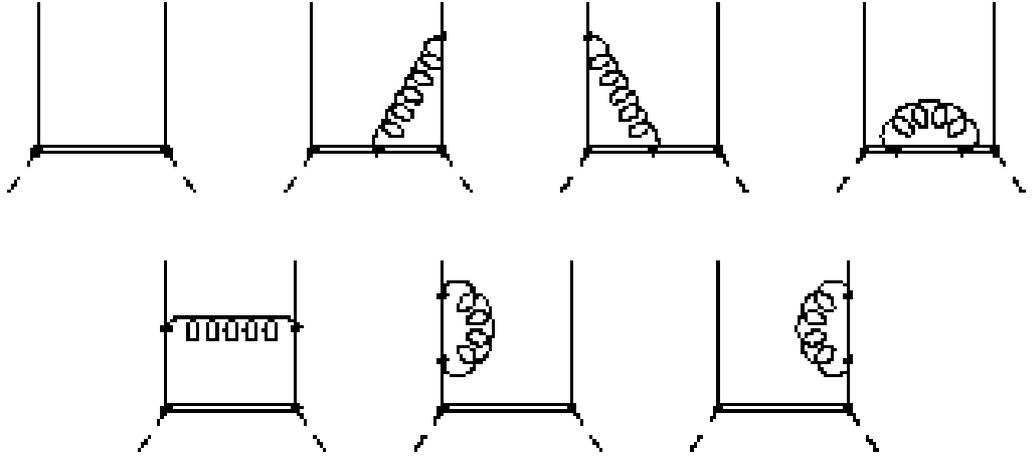}$$
\caption[The quark contributions to $f_+^{\etapb}(q^2)$ to $\mathcal{O}(\alpha_s)$.]{\small The quark-antiquark contributions to the semileptonic $B\to\etapb$ form factors $f_+^{\etapb}(q^2)$ from light-cone sum rules. The top left diagram is the leading one, the others are $\mathcal{O}(\alpha_s)$. The double line corresponds to the $b$ quark and the dashed lines the injection of the weak vertex momentum $q$, and the momentum of the $B$ meson $p_B$.} 
\label{eta_fig1}
\end{figure}

The quark contributions follow from the studies already undertaken for the $\pi$, for more details see Ref.~\cite{Ball:2004ye}. We briefly cover the general features of the calculation to put the singlet contribution in context. The leading quark contributions to $\Pi_+^P$ originate from the diagrams of Fig.~\ref{eta_fig1}, where first order $\mathcal{O}(\alpha_s)$ corrections are shown. The external quarks have momentum fractions $u p$ and $(1-u)p$ and are on-shell; $p^2=m_P^2$.
The two-particle DAs are projected out by using the general spinor decomposition of quark fields
\begin{eqnarray}
\bar{q}_a q_b^\prime &=& \frac{1}{4} (\textbf{1})_{ba}(\bar q q^\prime)-\frac{1}{4} (i \gamma_5)_{ba}(\bar q i \gamma_5  q^\prime)+\frac{1}{4} (\gamma_\mu)_{ba}(\bar q \gamma^\mu q^\prime)-\frac{1}{4} (\gamma_\mu \gamma_5)_{ba}(\bar q\gamma^\mu \gamma_5  q^\prime)\nonumber\\
&-&\frac{1}{8} (i \sigma_{\mu\nu}\gamma_5)_{ba}(\bar q i \sigma^{\mu\nu} \gamma_5 q^\prime)\,.
\end{eqnarray}
The vacuum-meson matrix elements of each term above either vanish or yield a DA depending on the quantum numbers of the meson in question. For pseudoscalar mesons the leading-twist contribution comes from $\gamma_\mu\gamma_5$, whereas $i\gamma_5$ and $i\sigma_{\mu\nu}\gamma_5$ give two-particle twist-3 contributions, and although the two-particle twist-3 contributions appear in the sum rules as formally $1/m_b$, they are \textit{chirally enhanced} by numerically large factors \cite{Ball:1998je} and so are included in typical LCSR analyses \cite{Ball:2004ye}. Three-particle twist-3 and two- and three-particle twist-4 DAs are also included; all twist-2 and -3 contributions include $O(\alpha_s)$ corrections twist-4 contributions are to tree level accuracy. The corresponding expressions yield $\Pi^q_+$, with the replacement $f_\pi\to f_q$.  

\subsection*{Gluonic Contribution}
\begin{figure}[h]
$$ \epsfxsize=0.85\textwidth\epsffile{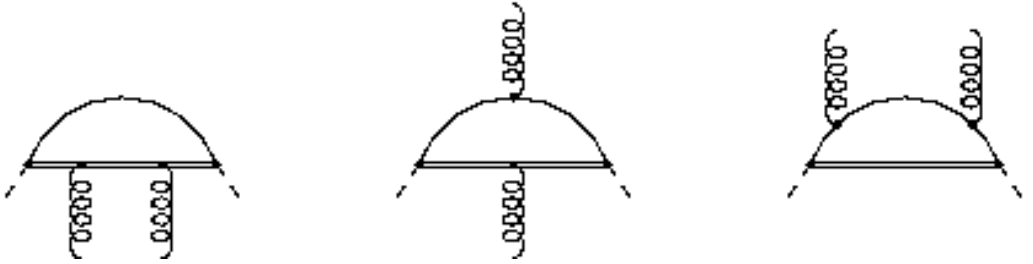}$$
\caption[The leading diagrams for the flavour-singlet contribution to $f_+^{\etapb}(q^2)$.]{\small The leading diagrams for the flavour-singlet contribution to the semileptonic $B\to\etapb$ form factors  from light-cone sum rules. The double line corresponds to the $b$ quark. The dashed lines the injection of weak vertex momentum $q$, and momentum of the $B$ meson  interpolating current $p_B$.} 
\label{eta_diags}
\end{figure}

In order to obtain the gluonic contribution to $\Pi_+^P$, one needs to calculate the diagrams shown in Fig.~\ref{eta_diags}. The last diagram is divergent and the other two are finite.
The gluon fields are introduced in the standard way
\begin{eqnarray}
\left.\Pi^{P}_{\mu}\right|_{\textrm{gluon}}=i\int d^4[x,w,y] \, e^{i q\cdot x} \bra{P(p)} T [\bar{u}\gamma_{\mu} b](x) [m_b \bar{b} i\gamma_5 u](0) \, S \mathcal{L}^{q_1}_{g}(w)\mathcal{L}^{q_2}_{g}(y)\ket{0}\,,\nonumber
\end{eqnarray}
with the usual interaction Lagrangian $\mathcal{L}^{q_{i}}_{g}(x) = ig_{s} [\bar{q_i} \gamma^{\alpha} A_{\alpha}^a t^a q_i](x)$ with $q_{i} = \{u,b\}$ and the statistical factor $S$ takes values $1$ if $q_1\neq q_2$ and  $1/2$ if $q_1=q_2$. The integral is over each co-ordinate separately. To extract the gluon contribution we need the projection onto the twist-2 two-gluon DA, which can be read off Eq.~(\ref{gluefields}), which amounts to the following replacement of the gluon fields (up to the numerical factor)
\begin{equation}
A_{\alpha}^a(w) A_{\beta}^b(y) \stackrel{\textrm{twist-2}}{\longrightarrow}\delta^{ab} \epsilon_{\alpha \beta \rho\sigma} \,\frac{\tilde{z}^\rho p^\sigma }{p \cdot \tilde{z}}\,\,\int^1_0 du\,\frac{\psi^g_{2;P}(u)}{u \bar{u}} \,e^{i p \cdot (u w +  \bar{u} y)}\,,
\end{equation}
where the separation $\tilde{z}$ is light-like i.e. $\tilde{z}^2 = (w-y)^2=0$. Via partial integration we can simplify the resulting expression for $\left.\Pi^{P}_{+}\right|_{\textrm{gluon}}$; the co-ordinate $\tilde{z}$ is traded for a derivative of the hard scattering kernel with respect to the momentum of one of the emitted gluons; and the dot product $1/ (p\cdot \tilde{z})$ can be traded for an integral with respect to the DA momentum fraction. As the boundary terms vanish due to the leading-twist gluon DA being antisymmetric, the calculation takes a rather simple form:
\begin{equation}
\left.\Pi^{P}_{+}\right|_{\textrm{gluon}}=\left.\int^1_0 du  \, \left[\frac{\partial \,T_{\mu }^\rho(u p) }{\partial(u p)^\rho}\right] \int^u_0 dv \,\frac{\psi_{2;P}^g(v)}{v\bar{v}}\right|_{p_\mu \to \frac{1}{2},\,q_\mu \to 0}\,,
\end{equation}
where $T_{\mu }^\rho(u p) $ is the hard scattering kernel.  Both the gluonic and quark contributions are renormalisation scale dependent. The relevant term concerning the quark Gegenbauer moment $a_2$ is
\begin{equation}
\Pi^q_+ \sim 18 f_q  a_2 \left( 1 + \frac{\alpha_s}{4\pi}
\,\frac{50}{9}\,\ln\,\frac{\mu^2}{m_b^2}\right)F(p_B^2,q^2)\,,
\label{33}
\end{equation}
where $F(p_B^2,q^2)$ is a function of $p_B^2$ and $q^2$. The logarithmic terms in the convolution of the gluonic diagrams of Fig.~\ref{eta_diags} with $\psi_{2;P}^g$ are
\begin{equation}
\Pi^P_+ \sim -\frac{10}{9\sqrt{3}}\,\frac{\alpha_s}{4\pi}\, B_2^g f^0_P
\ln\,\frac{\mu^2}{m_b^2}\, F(p_B^2,q^2) \,.
\end{equation}
By expressing $f_q$ via Eq.~(\ref{11}) in terms of $f^0_\eta$ and $f^0_{\etap}$, respectively, and inserting Eq.~(\ref{33}) into Eq.~(\ref{32}), one verifies that the renormalisation-group equation, Eq.~(\ref{20}), is fulfilled.  The twist-2 two-gluon contribution to the correlation functions $\Pi_+^P$, Eq.~(\ref{32}), is given in terms of a spectral density as
\begin{equation}
\left.\Pi_{+}^{P}\right|_{\textrm{gluon}} = \int_{m_b^2}^\infty ds\,\frac{\rho^P_{\textrm{gluon}}(s)}{s-p_B^2}
\end{equation}
with the result being
\begin{eqnarray}
\rho^P_{\textrm{gluon}}(s) 
& = & B_2^g \alpha_s f_0^P m_b\, \frac{5}{36\sqrt{3}}\, \frac{m_b^2-s}{(s-q^2)^5} \, \left\{ 59 m_b^6 + 21 q^6 - 63 q^4 s - 19 q^2 s^2 + 2 s^3\right. \nonumber\\
&& \hspace*{3cm}\left. + \,m_b^2 s (164 q^2 + 13 s) - m_b^4 (82 q^2 + 95s)\right\}
\nonumber\\
&& {} + B_2^g \alpha_s f_0^P m_b\, \frac{5}{6 \sqrt{3}}\, \frac{(m_b^2-q^2)(s-m_b^2)}{(s-q^2)^5} \,\{ 5 m_b^4 + q^4 + 3 q^2 s + s^2 - 5m_b^2 (q^2+s)\} \nonumber\\
&& \hspace*{3cm} \times\left\{ 2 \ln\,\frac{s-m_b^2}{m_b^2} - \ln\,\frac{\mu^2}{m_b^2} \right\}.
\end{eqnarray}

\section{Discussion}
\begin{figure}[h]
$$ \epsfxsize=0.45\textwidth\epsffile{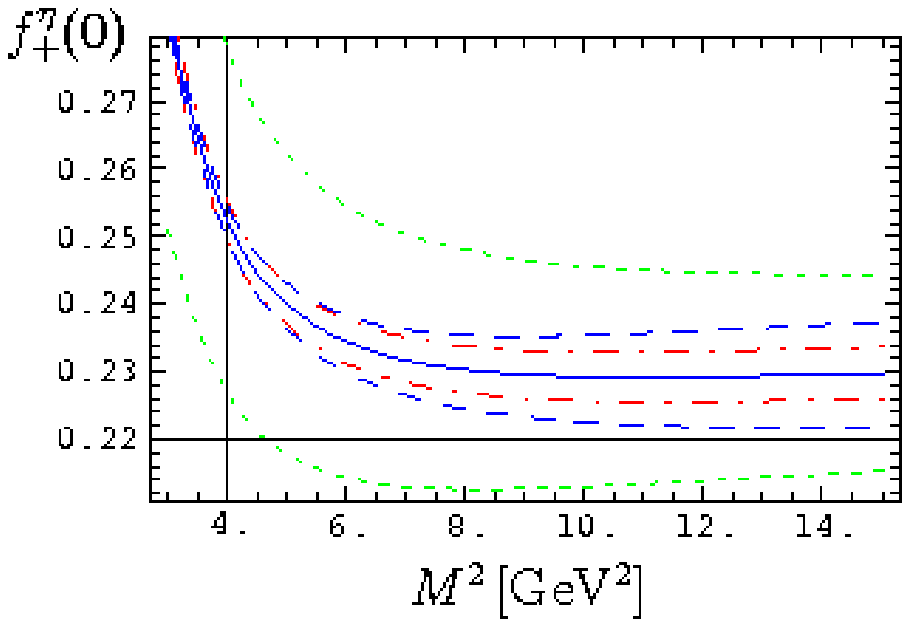}\qquad\qquad \epsfxsize=0.45\textwidth\epsffile{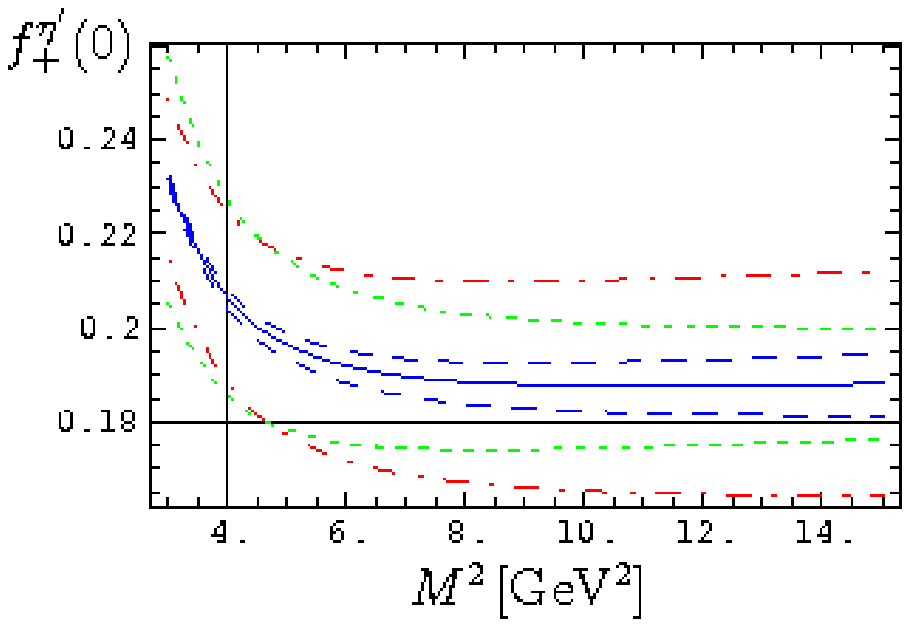}$$
\caption[$f_+^\eta(0)$  and $f_+^{\etap}(0)$ as a
  function of the Borel parameter $M^2$.]{\small $f_+^\eta(0)$ (left) and $f_+^{\etap}(0)$ (right) as a
  function of the Borel parameter $M^2$ and various choices of input
  parameters. Solid curves: central values of input parameters and
  $s_0=34.2\,{\rm GeV}^2$. Long-dashed curves: $s_0$ varied by
  $\pm 0.7\,{\rm GeV}^2$. Short-dashed curves: $a_2(1\,{\rm
  GeV})$ varied by $\pm 0.15$. Dash-dotted curves: $B_2^g$ varied
  by $\pm 10$.} 
\label{eta_diags3}
\end{figure}

For the evaluation of the LCSR, Eq.~(\ref{35}), as with any sum rule, optimum values of  $M^2$ and $s_0$ need to be found. The standard procedure \cite{Ball:2004ye} is to replace $f_B$ by its sum rule, derived via SVZ sum rules, thus reducing the dependence of the LCSR on $m_b$ for which we use the one-loop pole mass $m_b=4.80\pm0.05\,\rm{GeV}$ \cite{Colangelo:2000dp}. From the $f_B$ sum rule the optimum threshold parameter  $s_0=34.2 \pm 0.7\,{\rm GeV}^2$ is found, and this value is taken over to the LCSR. As mentioned before $\mu=2.4\,{\rm GeV}$ is chosen as an intermediate scale between $m_b$ and $1\,{\rm GeV}$. The Borel parameter is taken to be $M^2>6\,{\rm GeV}^2$ and is varied in the range $6\,{\rm GeV}^2<M^2<14\,{\rm GeV}^2$ to reflect the corresponding uncertainty. In Fig.~\ref{eta_diags3} we plot $f^{B\to \eta}_+(0)$ and $f^{B\to \etap}_+(0)$ respectively as functions of $M^2$, making explicit the result of varying $s_0$ by $\pm0.7\,{\rm GeV}^2$, $a_2$ by $\pm0.15$ and $B_2^g$ by $\pm10$.
As expected, $f_+^\eta(0)$ is not very sensitive to the gluonic twist-2 DA parameter $B^g_2$ (dashed-dotted curves), but is quite sensitive to the Gegenbauer moment $a_2$ (short-dashed
curves). For  $f_+^{\etap}(0)$, on the other hand, the dependence on $B^g_2$ is more pronounced than that of $a_2$. Varying all relevant parameters within their respective ranges, i.e.\ $\Delta m_b = \pm
0.05\,{\rm GeV}$, $\Delta a_2(1\,{\rm GeV})=\pm0.15$ and $\Delta B^g_2=\pm 20$, as well as all twist-3 and twist-4 parameters within the ranges given in Ref.~\cite{Ball:2006wn}, we find
\begin{eqnarray}\label{fp1}
f_+^\eta(0) & = & 0.229\pm \overbrace{0.005}^{M^2}\pm \overbrace{0.006}^{s_0}\pm
\overbrace{0.016}^{a_2^\eta}\pm \overbrace{0.007}^{B^g_2}\pm \overbrace{0.005}^{f_q,\phi}\pm \overbrace{0.011}^{{\rm T3}}\pm \overbrace{0.001}^{{\rm T4}} \pm \overbrace{0.007}^{f_B,m_b}\nonumber\\
& = &
0.229\pm \underbrace{0.024}_{{\rm param.}}\pm \underbrace{0.011}_{{\rm syst.}}\,,\\
f_+^{\etap}(0) & = & 0.188\pm \overbrace{0.004}^{M^2}\pm \overbrace{0.005}^{s_0}\pm
\overbrace{0.013}^{a_2^{\etap}}\pm \overbrace{0.043}^{B^g_2}\pm \overbrace{0.005}^{f_q,\phi}
\pm \overbrace{0.009}^{{\rm T3}}\pm \overbrace{0.005}^{{\rm T4}} \pm \overbrace{0.006}^{f_B,m_b}\nonumber\\
& =&
0.188\pm \underbrace{0.043}_{B^g_2} \pm \underbrace{0.019}_{{\rm param.}}\pm \underbrace{0.009}_{{\rm syst.}}\,.
\label{fp0}
\end{eqnarray}
The entry labelled ``T4'' also contains an estimate of the possible impact
of the local twist-4 two-gluon matrix elements (\ref{extra}). For
this estimate, we exploit the fact that the asymptotic DA of the
non-local generalisation of Eq.~(\ref{extra}) is the same as for the
twist-2 two-quark DA: $6 u (1-u)$.\footnote{This follows from Eq.~(\ref{das_eq18}). For
$G_{\perp\perp}$, one has $l=2$ and $s=0$.} We then
assume that the corresponding correlation function is the same as that
for the leading conformal wave in the two-quark twist-2 contribution,
i.e.\ the coefficient in the Gegenbauer moment $a_0=1$, and replace
$a_0$ by $\bra{0} \alpha_s
G\tilde{G}/(4\pi)\ket{\eta_{q,s}}/(f_{q,s}m_b^2)$. The factor
$1/m_b^2$ comes from the fact that this is a twist-4 effect and hence
suppressed by two powers of $m_b$ with respect to the twist-2
contribution. This is only a
rough estimate, of course, as the true spectral density will be
different. The results (\ref{fp0}) show that for small $B^g_2\approx 2$ both twist-2 and -4
two-gluon effects can indeed be of similar size. In this case,
however, the total flavour-singlet contribution to $f_+^{\etap}$ will
also be small, $\sim 0.008$.
In the third lines, we have added all uncertainties from the input
parameters (param.) in quadrature and the sum-rule specific
uncertainties from $M^2$ and $s_0$ (syst.) linearly. For
$f_+^{\etap}(0)$, we have displayed the dependence on $B^g_2$
separately. The new result for $f_+^\eta(0)$ is, within errors, in agreement with
the previous one from LCSR, $f_+^\eta(0) = 0.275\pm 0.036$, obtained in
Ref.~\cite{Ball:2004ye}. That for $f_+^{\etap}(0)$ is new to the present analysis.
The results agree well with those obtained in Ref.~\cite{Charng:2006zj},
from pQCD,
$f_+^\eta(0) = 0.208$ and $f_+^{\etap}(0) = 0.171$, including a
rescaling by a factor $\sqrt{2}$ to bring their definition of the form
factors into agreement with Eq.~(\ref{FF}). We confirm the finding of
Ref.~\cite{Charng:2006zj} that the range of the singlet contribution to the form
factor estimated in Ref.~\cite{Beneke:2002jn} is likely to be too large, unless
$B_2^g$ assumes extreme values $\sim 40$.
\begin{figure}[h]
$$ \epsfxsize=0.55\textwidth\epsffile{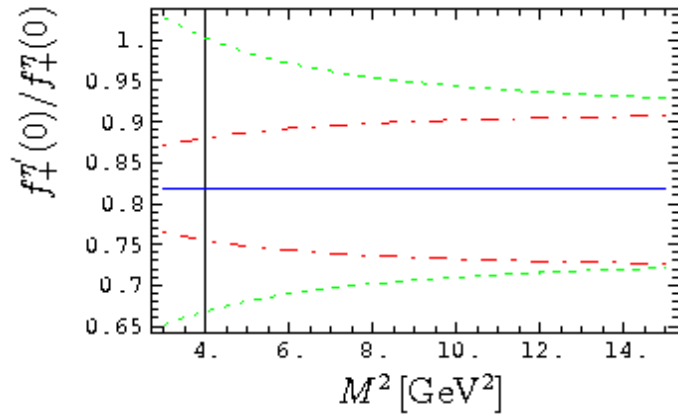}$$
\caption[$f_+^{\etap}(0)/f_+^\eta(0)$ as a
  function of the Borel parameter $M^2$.]{\small $f_+^{\etap}(0)/f_+^\eta(0)$ as a
  function of the Borel parameter $M^2$ and various choices of input
  parameters. Solid line: central values of input parameters,
  which corresponds to $f_+^{\etap}(0)/f_+^\eta(0)\equiv \tan\phi =
  0.814$. Dot-dashed curves: $B_2^g$ varied
  by $\pm 10$. Dotted curves: $a_2^{\eta,\etap}(1\,{\rm GeV})$
  varied independently: $a_2^\eta=0.1$, $a_2^{\etap}=0.4$ and
  $a_2^{\etap}=0.4$, $a_2^\eta=0.1$.} 
\label{eta_fig5}
\end{figure}

In Fig.~\ref{eta_fig5} we plot the ratio $f^{\etap}_+(0)/f^{\eta}_+(0)$
as a function of the Borel parameter. In the ratio, many uncertainties
cancel, in particular that on $f_B$. As we have chosen $B_2^g=0$ as
central value, $f^{\etap}_+(0)/f^{\eta}_+(0)\equiv \tan\phi = 0.814$ exactly,
see Eq.~(\ref{32}). The figure also illustrates the change of the
result upon inclusion of a non-zero $B^g_2$ (dot-dashed curves). The ratio is actually rather sensitive to that parameter. While the dependence on $a_2$ largely cancels when $a_2^\eta$ and
$a_2^{\etap}$ are set equal, there is a considerable residual dependence on
$a_2^\eta - a_2^{\etap}\neq 0$ (dotted curves). While
$|a_2^\eta - a_2^{\etap}|=0.3$ as illustrated by these curves is
rather unlikely, and would signal very large OZI-breaking
contributions (recall that $a_2^\eta\neq a_2^{\etap}$ or,
equivalently, $a_2^1\neq a_2^8$ signals the presence of
``wrong-flavour'' contributions to the $\eta_{q,s}$ DAs and is set to 0 in
the QF mixing scheme), one should nonetheless keep in mind that moderate
corrections of this type are not excluded and compete with the
OZI-allowed corrections in $B^g_2$.
\begin{figure}
$$\epsfxsize=0.55\textwidth\epsffile{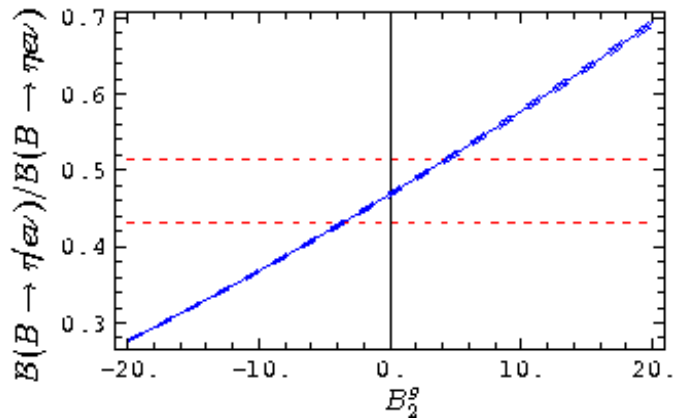}$$
\caption[${\cal    B}(B\to\etap e\nu)/{\cal B}(B\to\eta e\nu)$ as a function of the    singlet-parameter $B_2^g$.]{\small The ratio of branching ratios    $R_{\eta\etap} = {\cal    B}(B\to\etap e\nu)/{\cal B}(B\to\eta e\nu)$ as a function of the    singlet-parameter $B_2^g$. Solid curve: central values of    input parameters; Long-dashed curves: including variation of extrapolation model parameters, See Ref.~\cite{Ball:2007hb}. Short-dashed curves: theoretical uncertainty of $R_{\eta\etap}$ for $B^g_2=0$, for $a_2^{\eta,\etap}(1\,{\rm GeV})$ varied independently, as in Fig.~\ref{eta_fig5}.
    }\label{eta_fig8}
\end{figure}

Finally, in Fig.~\ref{eta_fig8} we show the dependence of the ratio of
branching ratios $R_{\eta\etap} = {\cal
    B}(B\to\etap e\nu)/{\cal B}(B\to\eta e\nu)$ on
$B_2^g$. The advantage of this observable is that all hadronic effects
are encoded in the form factors and that $|V_{ub}|$ cancels.
The solid curve corresponds to the branching ratios obtained
from the central values of input parameters; the long-dashed curves illustrate the dependence on parameters originating from the model  used to extrapolate the $q^2$ dependence of the form factor from beyond the limit of the LCSR approach, in this case $q^2=16\,{\rm GeV}^2$, to the maximum possible value $q_{\rm max}^2=(m_B-m_{\etapb})^2$. It may be noted that the dependence on these parameters is very small. We do not go into detail about the extrapolation procedure and refer the reader to Ref.~\cite{Ball:2007hb}.  On the other hand, $R_{\eta\etap}$ also depends on $a_2^\eta\neq
a_2^{\etap}$. This dependence is shown by the short-dashed
curves. The conclusion is that large values of $B^g_2$, $|B^g_2|>5$,
can be distinguished from the OZI-breaking parameter
$|a_2^\eta-a_2^{\etap}|$, once an accurate experimental value of
$R_{\eta\etap}$ is available, but that for smallish $B^g_2$ and
unknown $|a_2^\eta-a_2^{\etap}|$ only mutual constraints on these
parameters can be extracted from the data. In this case  also twist-4 gluonic DAs can become important.

\chapter{QCD Factorisation}\label{chapter6_QCDF}
In this chapter we discuss the framework of QCD factorisation which was introduced  in the context of exclusive two-body non-leptonic $B$ decays by Beneke, Buchalla, Neubert and Sachrajda in Refs.~\cite{Beneke:1999br, Beneke:2000ry}. We shall refer to the the original implementation of the framework as the BBNS approach.  We also focus on its application to the radiative $B$ decays $B \to V \gamma$, as presented by Bosch and Buchalla in Refs.~\cite{Bosch:2001gv,Bosch:2002bw}.  

QCD factorisation allows a rigourous determination of the $B$ decay matrix elements of the weak effective Hamiltonian (\ref{basics_eq20}) to leading order in the heavy-quark limit of QCD  $m_b\gg \Lambda_{\rm QCD}$, and yields a neat factorisation formula. It relies on the factorisation of hadronic matrix elements into universal non-perturbative hadronic parameters, given by transition form factors and meson light-cone DAs,  and process dependent hard-scattering kernels, calculable in perturbation theory. The validity of the QCD factorisation formula, to all orders in $\alpha_s$, and the impact of generally unknown power corrections, formally suppressed by powers of $1/m_b$, must be addressed case by case. The introduction of the QCD factorisation framework has made more discerning phenomenological studies of exclusive $B$ decays possible whereby key observables, such as branching ratios, CP and isospin asymmetries, can be calculated and confronted with experimental data.

The dependence of the factorisation formula on meson DAs, either directly or via LCSR calculations of the transition form factors, greatly motivates their study, with their better determination reducing the theoretical uncertainty of the QCD factorisation predictions, and aiding the quest to discover new physics effects from decay observables. 

We begin with a short introduction, in the context of $B\to M_1 M_2$ decays, of the general features of QCD factorisation, and in particular, discuss the appearance of meson DAs. We then discuss the framework as applied to the radiative $B$ decays $B \to V \gamma$. We postpone all discussions of phenomenology to Chapter~\ref{chapter7_rad} in which we perform an analysis of the decays $B_{u,d} \to (\rho, \omega, K^*)\gamma$ and $B_{s} \to (\bar{K}^*,\phi)\gamma$ using QCD factorisation, augmented by the inclusion of the dominant power-suppressed corrections.

\section{Introduction}
\textit{QCD factorisation} (QCDF) \cite{Beneke:1999br, Beneke:2000ry} was introduced in the context of the ``heavy-to-light'' decays $B \to \pi\pi$ where the factorisation of the relevant QCD matrix elements was shown to apply, to leading order in a $1/m_b$ expansion, to a large class of non-leptonic $B$ decays.   Consequently, QCDF has opened up the rich and varied landscape of $B$ decays to a more complete quantitative analysis. The existence of factorisation in non-leptonic decays is non-trivial and complicated by the possible gluonic interactions amongst the initial and final states. Conversly, leptonic and semi-leptonic decays factorise much more easily into the product of a quark current and a leptonic current, which cannot interact via gluon exchange.

Phenomenologically,  QCDF has been remarkably successful, especially given the range of processes for which the method holds. After its introduction, it was swiftly generalised to encompass $\pi K$ final states \cite{Beneke:2001ev}, pseudoscalar-vector final states \cite{Beneke:2003zv} and vector-vector meson final states \cite{Kagan:2004uw}. The gluonic flavour-singlet contributions to $B \to K^{(*)} \etapb $ decays were added by Ref.~\cite{Beneke:2002jn}. To date, the framework has been extended to many other processes, including for example, (double) radiative $B$ decays $B \to \gamma (\gamma, V)$ \cite{Bosch:2002bv,Bosch:2002bw} and $B \to \gamma l \nu$ \cite{Descotes-Genon:2002mw}.  Also,  other factorisation frameworks have since been developed and applied to the same problems:
\begin{itemize}
\item{\textit{Soft Collinear Effective Theory} (SCET) \cite{SCET1, SCET2, SCET3, SCET4}  makes a careful distinction between a hierarchy of ``hard'' $(m_b)$, ``hard-collinear'' ($\sqrt{\Lambda_{\rm QCD} m_b}$) and ``collinear'' ($\Lambda_{\rm QCD}$) scales via contributions of internal quark and gluon lines. Details of the differences between the SCET and BBNS approaches to QCD factorisation can be found in Refs.~\cite{Bauer:2004tj,Beneke:2004bn,Bauer:2005wb}.}
\item{The \textit{Perturbative QCD}  (pQCD) approach \cite{PQCD}, which yields a factorisation formula that depends on the mesons' transverse momenta.}
\item{The method of LCSRs, although having existed before the advent of QCDF, was applied to $B \to \pi \pi$, both to the matrix elements which exhibit factorisation and also a class of power corrections, providing some useful complementary insights, see Refs.~\cite{Khodjamirian:2000mi, Khodjamirian:2005wn}.}
\end{itemize}
We now go on to discuss the general features of QCDF. 
\section{General Structure}
Consider the case of non-leptonic decays where the $B$ meson decays into two mesons. The simplest way of dealing with the resulting matrix elements is to employ \textit{naive factorisation} \cite{Fakirov:1977ta,Cabibbo:1977zv}. Simply put, naive factorisation splits each local operator $Q_i$ of the effective Hamiltonian into two colour-singlet currents, whose matrix elements are proportional to a decay constant and a transition form factor respectively. For example, consider the four-quark operator $Q_2^U =(\bar{D} U)_{\rm V-A}(\bar U b)_{\rm V-A}$ then
 \begin{equation}
\bra{M_1 M_2}(\bar{D} U)_{\rm V-A}(\bar U b)_{\rm V-A}\ket{B} \,\stackrel{\rm{NF}}{\longrightarrow}\,\underbrace{\bra{M_2}(\bar{D} U)_{\rm V-A}\ket{0}}_{f_{M_2}}\underbrace{\bra{M_1}(\bar U b)_{\rm V-A}\ket{B}}_{F^{B \to M_1}}\,.
\label{qcdf_1}
\end{equation}
The motivation for factorising in this way comes from the \textit{colour transparency} argument \cite{Bjorken:1988kk}. It follows that a major shortcoming of naive factorisation is that it assumes the exchange of gluons of virtualites $\mu \lesssim m_b$ to be negligible and hence rescattering between the decay products is not considered; there is then no mechanism for the generation of strong phase effects between different amplitudes. Also, the matrix elements (\ref{qcdf_1})  do not display the correct renormalisation-scale dependence.

The framework of QCDF allows the calculation of $\mathcal{O}(\alpha_s)$ corrections to naive factorisation, which occur at scales $\mu\lesssim m_b$. It is constructed by observing the cancelation of infrared (IR) and collinear divergences, via consistent power-counting arguments, allowing the use of perturbation theory to describe the hard-gluon exchanges. The resulting intuitive factorisation formula thus presents a massive simplification of the long-distance QCD effects, with QCDF recovering naive factorisation in the limit $m_b\to \infty$. In terms of two-body non-leptonic $B$ decays to light pseudoscalar mesons $B\to M_1 M_2$ the factorisation formula, as presented in Ref.~\cite{Beneke:1999br}, reads schematically as
\begin{eqnarray}
\bra{M_1 M_2} Q_i\ket{B}&=& F^{B \to M_1}\int^1_0 du\,T^{I}_i (u)\, \phi_{2;M_2}(u) + (M_1 \leftrightarrow M_2)\nonumber \\
&+&\int^1_0 d\xi\,du\,dv\,T^{II}_i (\xi,u,v)\,\phi_B(\xi) \,\phi_{2;M_1}(v)\,\phi_{2;M_2}(u) \nonumber \\
&+&\mathcal{O}(\Lambda_{\rm QCD}/m_b)\,
\label{qcdf_2}
\end{eqnarray}
where $F^{B \to M_1}$ is the relevant form factor, $T^{I,II}_i$ are the hard-scattering kernels, $\phi_{B}$ is one of the leading-twist DAs of the $B$ meson and $\phi_{2;P}$ the leading-twist DA of the final state meson $P$, and the $Q_i$ are the operators of the effective Hamiltonian. The matrix elements are given as the convolution of the universal DAs and the process dependent hard-scattering kernels, with respect to the meson momentum fractions. Since the transition form factor and the DAs are real functions, all strong phases are generated by the hard-scattering kernels and are suppressed by powers of $\alpha_s$. Factorisation has be proven to one-loop for ``light-light'' final states and two-loop for ``heavy-light'' final states \cite{Beneke:2000ry}. It has be proven to all orders in $\alpha_s$ for $B\to D \pi $ using SCET \cite{SCET2}.

The ability of QCDF to accurately describe $B$ decay processes is limited by two main considerations; firstly, by the nature of the factorisation formula itself, which is valid up to power corrections $\mathcal{O}(1/m_b)$ and to a given order in $\alpha_s$; and secondly by uncertainties of the necessary input parameters, such as the DAs, the transition form factors, the strange quark mass, the $B$ meson decay constant $f_B$ etc.  Whether a discrepancy between experiment and QCDF predictions can be put down to new physics, or not, requires an estimation of neglected power corrections; certainly the $b$ quark mass is not asymptotically large $m_b\sim 5 \,{\rm GeV}$ and power corrections are therefore expected to feature at the level of $\mathcal{O}(\Lambda_{\rm QCD}/m_b) \sim 10\%$. The size and nature of power corrections can be probed via phenomenology, however, the task is not straight forward; even the initial focus of the approach, the decays $B \to \pi (K,\pi)$, which stands as a crucial test, has not been resolved satisfactorily, see for example Ref.~\cite{Feldmann:2004mg} and Refs.~\cite{Fleischer:2005vz,Fleischer:2007wd}. Better determined input parameters will nevertheless shed light, case by case, on whether power corrections are important, and the QCDF predictions must be used to determine or constrain CKM matrix elements (UT angles), or detect signs of new physics, with that in mind.

\section{Light-Cone Distribution Amplitudes}
To leading-order in the heavy-quark limit the leading-twist final state meson DAs contribute to the factorisation formula and can be safely truncated after the second Gegenbauer moment $a_2$. For pseudoscalar meson final states the two-particle twist-3 DAs come with large normalisation factors $r^P_\chi$ and are said to be \textit{chirally enhanced}, and are therefore included even though they are formally $1/m_b$ suppressed. The vector mesons  do not have the same large normalisation factors but their two-particle twist-3 DAs are included in the BBNS approach for consistency. For a pseudoscalar  or vector meson, with valence quark content $\bar q q^{\prime}$, the normalisation factors are respectively
\begin{equation}
r_\chi^P(\mu) = \frac{2 m_P^2}{m_b(\mu)(m_q+m_{q^\prime})(\mu)} \sim\frac{\Lambda_{\rm QCD}}{m_b}\,,\qquad r^V_\chi(\mu)=\frac{2 m_V}{m_b(\mu)}\frac{f_V^\perp(\mu)}{f_V^\parallel}\,.
\label{qcdf_3}
\end{equation}
Three-particle twist-3 DAs are neglected because they do not come with large normalisations.  The inclusion of the chirally enhanced DAs leads to end-point divergences from the convolutions of the two-particle twist-3 pseudoscalar DAs with the corresponding hard-scattering kernels originating from both the hard-spectator scattering and annihilation contributions. The resulting divergent integrals signal the breakdown of factorisation and are parameterised by two universal unknown parameters $X_{H,A}$, introducing a source of theoretical uncertainty to the BBNS approach \cite{Beneke:1999br}. 

At leading-twist the $B$ meson is described by two DAs, only one of which is required as input for Eq.~(\ref{qcdf_2}) and appears in the hard-spectator diagrams contributing to $T^{II}_i$. The DAs of the $B$ mesons are complicated by the fact that the momentum of the meson is shared in a highly antisymmetric way: the $b$ quark has most of it. The $B$ meson DAs are given, at leading-order in $1/m_b$, by
\begin{equation}
\bra{0}\bar{q}_\alpha(0) b_{\beta}(z)\ket{B(p_B)}=i \frac{f_B}{4} \left[(\slash{p}_B+m_b) \gamma^5\right]_{\beta\gamma}\int^1_0 d\xi\,e^{-i\xi (p_B)_+  z_-}\left[\Phi_{B1}(\xi)+\slash{n}_- \Phi_{B2}(\xi)\right]_{\gamma \alpha}\,,
\label{qcdf_4}
\end{equation}
with the decay constant $f_B$ given by Eq.~(\ref{bdecayconstant}). With a careful choice of $n_-=(1,0,0,-1)$ only the following normalisation conditions are required
\begin{equation}
\int^1_0d\xi\, \Phi_{B1}(\xi)=1\,,\qquad\int^1_0d\xi\, \Phi_{B2}(\xi)=0\,,
\label{qcdf_5}
\end{equation}
along with the first inverse moment of $ \Phi_{B1}$ which is parameterised as
\begin{equation}
\int^1_0 d\xi\,\frac{\Phi_{B1}(\xi)}{\xi}\equiv \frac{m_B}{\lambda_B}\,,
\label{qcdf_6}
\end{equation}
and the numerical value of $\lambda_B$ is a source of uncertainty in the QCDF framework for both $B\to M_1 M_2$ and $B\to V \gamma$. We now discuss the radiative decays $B\to V \gamma$ within QCDF.

\section{Radiative $B$ decays to Vector Mesons}\label{qcdf_rad}
We consider the leading contributions to the $B\to V \gamma$ QCDF factorisation formula  as of Refs.~\cite{Bosch:2001gv,Bosch:2002bw,Bosch:2004nd,Beneke:2001at} in which a model independent framework is presented.  Contributions that are power-suppressed by one power of $1/m_b$ or more \textit{and} are $\mathcal{O}(\alpha_s)$ are not considered. At the quark level the decays are $b\to D \gamma$ transitions, where $D=\{s,d\}$. If otherwise not stated, in the following we refer to $\bar{B}\to V\gamma$ decays where $\bar{B}$ ($V$) denotes a $b \bar{q}$ ($D \bar{q}$)  bound state. For $B\to V \gamma$ decays the matrix element of each relevant local operator in the effective Hamiltonian factorises as
\begin{equation}
\bra{V\gamma}Q_i\ket{B}=e^* \cdot \left[ T_1^{B\to V} (0) \,T^{I}_i+\int^1_0 d\xi du\,T^{II}_i(\xi,u)\phi_B(\xi) \phi_{2;V}^{\perp}(u)\right]+\mathcal{O}(1/m_b)\,,
\label{qcdf_7}
\end{equation}
where $e_\mu$ is the photon polarisation vector and $T_1^{B \to V}(0)$ is the relevant form factor. $\phi_{2;V}^{\perp}$ the leading-twist DA of the perpendicularly polarised final state vector meson (\ref{das_eq19});  contributions from $\phi^{\parallel}_{2;V}$ are power-suppressed in the heavy-quark limit. Problems of end-point divergences are not encountered in $B\to V\gamma$ decays and the twist-3 vector meson DA does not feature -- the $B$ meson DAs (\ref{qcdf_6}) do however. The factorisation formula is accurate up to corrections suppressed by powers of $1/m_b$, as shown, and was proven to hold to all orders in $\alpha_s$ in SCET \cite{Becher:2005fg}.  The form factor $T_1^{B \to V}(0)$ has been calculated, for example, from LCSR in Ref.~\cite{Ball:2004rg}. 

\begin{figure}[h]
$$\epsfxsize=0.2\textwidth\epsffile{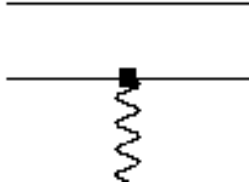}$$
\caption[The leading contribution to $B \to V\gamma$.]{\small The leading contribution to $B \to V\gamma$ due to the  electromagnetic dipole operator $Q_{7\gamma}$. }\label{qcdf_fig1}
\end{figure}

\begin{figure}[h]
$$\epsfxsize=0.5\textwidth\epsffile{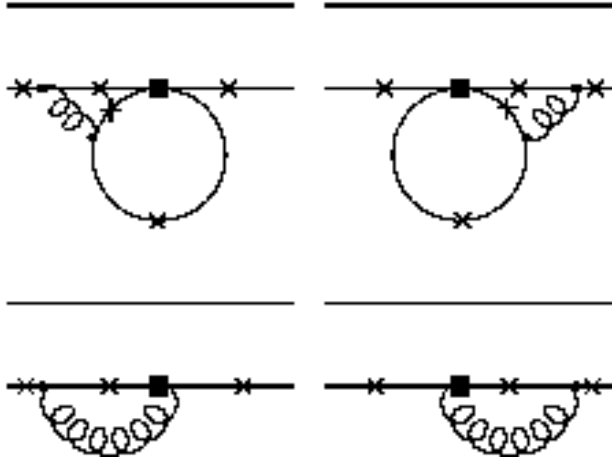}$$
\caption[Contributions to the hard-scattering kernel $T^{I}_i$ for $B\to V \gamma$ decays.]{\small Penguin contractions of $Q_{1,\dots,6}$ (top line) and the chromomagnetic dipole operator $Q_8$ (bottom line) contributing to the hard-vertex corrections of $T^{I}_i$ at $\mathcal{O}(\alpha_s)$. Crosses denote possible photon emission vertices.}
\label{qcdf_fig2}
\end{figure}

\begin{figure}[h]
$$\epsfxsize=0.5\textwidth\epsffile{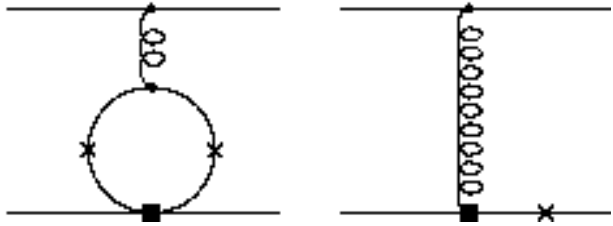}$$
\caption[Contributions to the hard-scattering kernel $T^{II}_i$ for $B\to V \gamma$ decays.]{\small Penguin contractions of $Q_{1,\dots,6}$ (left) and the chromomagnetic dipole operator $Q_{8g}$ (right) contributing to the hard-scattering kernel  $T^{II}_i$ at $\mathcal{O}(\alpha_s)$. Crosses denote possible photon emission vertices at leading order. Photon emission from the other quark lines power-suppressed. Photon emission from the final state meson for $Q_{8g}$ breaks factorisation.}
\label{qcdf_fig3}
\end{figure}

\begin{figure}[h]
$$\epsfxsize=0.2\textwidth\epsffile{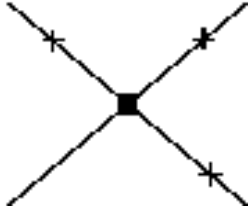}$$
\caption[Weak annihilation contributions to $B\to V \gamma$.]{\small Weak annihilation contributions, which are suppressed by one power of $1/m_b$. Crosses denote possible photon emission vertices at leading order. The dominant mechanism for $Q_{1,\dots,4}$ is the emission of the photon from the light quark in the $B$ meson and for $Q_{5,6}$ it is the emission from the final state vector meson quarks. Other possible emissions are either vanishing or more strongly suppressed.}
\label{qcdf_fig4}
\end{figure}

The $B\to V \gamma$ decay produces either left- or right-handed photons, which therefore constitute, in principle,  two separate observable processes. In practise the direct measurement of the photon's helicity is very difficult; indirectly, however, it can be accessed by measurement of the time-dependent CP asymmetry in $\bar B^0\to V^0\gamma$, which vanishes if one of them is absent, see Chapter~\ref{chapter7_rad}.  We define the two amplitudes as
\begin{equation}
\bar{\cal A}_{L(R)} = {\cal A}(\bar B\to V\gamma_{L(R)})\,,  \qquad
{\cal A}_{L(R)} = {\cal A}(B\to \bar V \gamma_{L(R)})\,.
\label{qcdf_8}
\end{equation}
For ($B$) $\bar B$  decays the production of the (left-) right-handed photon is suppressed by $1/m_b$ with respect to the opposite helicity. The decays are dominated by the electromagnetic dipole operator $Q_{7\gamma}$, and as such are penguin mediated and so loop-suppressed. The operators $Q_{7\gamma}^{L(R)}$ are given by
\begin{equation}
Q_{7\gamma}^{L(R)} = \frac{e}{8\pi^2}\, m_b \bar D \sigma_{\mu\nu}\left(1 \pm \gamma_5\right)b F^{\mu\nu}\,,
\label{qcdf_9}
\end{equation} 
and generate left- (right-) handed photons.  Their matrix elements can be parameterised in terms of the form factor $T_1^{B\to V}$ as
\begin{eqnarray}
\lefteqn{\bra{V(p,\eta) \gamma_{L(R)}(q,e)} Q_{7\gamma}^{L(R)} \ket{\bar
B}}\hspace*{1cm}\nonumber\\
&=& -\frac{e}{2\pi^2}\, m_b T_1^{B\to V}(0) \left[
\epsilon^{\mu\nu\rho\sigma} e_\mu^* \eta_\nu^* p_\rho q_\sigma \pm i
\{ (e^* \cdot \eta^*) (p \cdot  q) - (e^* \cdot p)(\eta^* \cdot q)\}\right]
\nonumber\\
&\equiv& -\frac{e}{2\pi^2}\, m_b T_1^{B\to V}(0) S_{L(R)}\,,
\label{qcdf_10}
\end{eqnarray}
where $S_{L,R}$ are the helicity amplitudes corresponding to left- and right-handed photons, respectively, and $e_\mu$ $(\eta_\mu)$ is the polarisation four-vector of the photon (vector meson).   The leading-order diagram is given in Fig.~\ref{qcdf_fig1} which is also the leading diagram for the form factor $T_1^{B\to V}$.
The factorisation formula (\ref{qcdf_10}) is therefore trivial to leading order in $\alpha_s$ and the heavy-quark limit; the matrix element given by the standard form factor, the scattering kernel $T^I_7$ by a purely kinematical function and $T^{II}_7$ does not feature.  The electroweak penguin operators $Q_{7,\dots,10}$ appear at higher-order and safely neglected in the analysis. All other operators begin to contribute at $\mathcal{O}(\alpha_s)$. The hard-vertex corrections contribute to $T^I_i$ yielding functions of $m_{u,c}^2/m_b^2$ and originate from penguin contractions of the operators $Q_{1,\dots,6}$ and the chromomagnetic operator $Q_{8g}$ as shown in Fig.~\ref{qcdf_fig2}. 
The hard-spectator scattering diagrams of Fig.~\ref{qcdf_fig3}, in which the spectator quark of the $B$ meson participates, contribute to  $T^{II}_i$ and involve the same operators as the hard-vertex corrections. The hard-gluon exchange probes the momentum distribution of the $B$ and vector mesons and so requires the introduction of the mesons' light-cone DAs, as suggested by the factorisation formula; it is in these contributions that the $B$ meson DA parameter $\lambda_B$ and decay constants $f_B$ and $f_V^\perp$ appear. 
Also, the dominant power-suppressed  \textit{weak annihilation} (WA) contributions, shown in Fig.~\ref{qcdf_fig4}, are calculable in the QCDF approach, and involve the operators $Q_{1,\dots,6}$. WA contributions are
$O(1/m_b)$; photon emission from the $b$ quark and the quarks in the vector meson is further suppressed and $O(1/m_b^2)$ -- unless the weak interaction operator is $Q_{5,6}$, which can be Fierz transformed into $(\bar D (1+\gamma_5) q) (\bar q (1-\gamma_5) b)$ and picks up an additional factor $m_B$ from the projection onto the $B$ meson DA thus resulting in this contribution being $O(1/m_b)$. Consequently, due to the large Wilson coefficients $C_{1,2}$ these contributions are sizeable and important phenomenologically, see Chapter~\ref{chapter7_rad}. 

The decay amplitude is then given by
\begin{equation}
\mathcal{A}(\bar{B}\to V \gamma_{L(R)})=\frac{G_F}{\sqrt{2}}\left(\lambda_u^D a_{7 L(R)}^u(V)+ \lambda_c^D a_{7 L(R)}^c(V)\right)\bra{V\gamma_{L(R)}}Q_{7\gamma}^{L(R)}\ket{\bar{B}}\,,
\label{qcdf_11}
\end{equation}
where the left-handed coefficients are given, to leading order in QCDF, by
\begin{equation}
a_{7L}^{U,{\rm QCDF}}(V)=C_7+\mathcal{O}(\alpha_s,1/m_b)\,,
\label{qcdf_12}
\end{equation} 
and the right-handed parameters, for a $b\to D$ transition, by \cite{Ball:2006cv}
\begin{equation}
a_{7R}^{U,{\rm QCDF}} = C_7\,\frac{m_D}{m_b}+\mathcal{O}(1/m_b,\alpha_s/m_b)\,.
\label{qcdf_13}
\end{equation}
Explicit expressions for the $\mathcal{O}(\alpha_s)$ corrections to the left-handed coefficients can be found in Refs.~\cite{Bosch:2001gv,Bosch:2002bw} and will be considered in Chapter~\ref{chapter7_rad}, alongside the dominant power-suppressed corrections.
\chapter{$B \to V \gamma$ Beyond QCD Factorisation}\label{chapter7_rad}
In this chapter we perform a phenomenological analysis of the exclusive radiative $B$ decays to vector mesons. We make use of the QCDF framework outlined in Chapter~\ref{chapter6_QCDF} and investigate the impact of the leading power-corrections on the branching ratios, CP asymmetries and isospin asymmetries for all $b\to D$ transitions; $B_{u,d} \to (\rho, \omega, K^*)\gamma$ and  $B_{s} \to (\phi, \bar{K}^*)\gamma$.  Weak annihilation effects, although power-suppressed, are calculable in QCDF, and are included for all decay modes in this analysis.  The other power-suppressed contributions ``beyond QCDF''  considered are; soft photon emission from the soft $B$ spectator quark \cite{Ball:2003fq}; and long-distance contributions from heavy quark loops \cite{Ball:2006cv} and light quark loops \cite{Ball:2006eu} which have been estimated from LCSR.  The estimation of the light quark loop contribution is new to the present analysis. Whereas the branching ratios are generally dominated by the leading contributions, and power-suppressed contributions play a minor role, the same cannot be said for the CP and isospin asymmetries for which the impact of power-corrections is in fact crucial.

The motivation to study radiative $B$ decays stems from a variety of sources:
\begin{itemize}
\item{as loop-induced, penguin mediated decays, they allow the extraction of the CKM matrix element $|V_{t,(d,s)}|$ complimentarily to the determination from $B$ mixing and also that from the SM UT analysis based on the tree-level observables $|V_{ub}/V_{cb}|$ and the angle $\gamma$.}
\item{They are sensitive to new physics contributions, which may occur within the penguin loops, with the time-dependent CP asymmetry a very promising avenue of investigation. They are also subject to large short-distance QCD corrections, which now approach next-to-next-to-leading-order accuracy, see Refs.~\cite{misiak,NNLObsgamma}.}
\item{The decay rates are of order $G_F^2 \alpha_{\rm QED}$ and are enhanced with respect to other loop-induced non-radiative rare decays which are of order $G_F^2 \alpha_{\rm QED}^2$. Also, the $b\to s$ modes are CKM-favoured. Consequently there exist good experimental results for the exclusive branching ratios; $B\to K^* \gamma$ is known to 5\%, but the $b\to d$ transitions are not so well known.}
\end{itemize}

As discussed in Chapter~\ref{chapter6_QCDF}, the QCDF framework for $B\to V\gamma$ relies on the leading-twist vector meson DA $\phi_{2;V}^\perp$. Moreover, the LCSR calculations of the form factors $T_1^{B\to V}$ and the parameters entering expressions for the soft-quark contributions rely also on the higher-twist DAs of the vector mesons and thus we find immediate use for the results of the twist-2 and twist-3 DA parameters of Chapter~\ref{chapter4_det}, as presented in Tab.~\ref{det_tab1} and Tab.~\ref{det_tab2}.\footnote{The analysis presented in Ref.~\cite{Ball:2006eu} used preliminary input for the DA parameters, values for which were later finalised in Ref.~\cite{Ball:2007rt}. The conclusions and numerics of the analysis are unaffected, due somewhat to the large errors attributed to the soft quark loop calculations in which the twist-3 DA parameters feature.}

We begin with an introduction, and then go on to discuss the power-suppressed contributions and investigate their impact on the decay observables. We extract the CKM parameter $|V_{t,d}/V_{ts}|$ from the branching ratio results, assuming no new physics contributions, and discuss possible new physics contributions to the CP and isospin asymmetries. The material covered in this chapter follows that of Ref.~\cite{Ball:2006eu}.

\section{Introduction}
$B \to V \gamma$ decays are a very rich and promising probe of flavour physics. Both the inclusive decay $B\to X_s \gamma$ and the exclusive decays $B\to (K^*,\rho)\gamma$ have been under scrutiny for many years, see for example Refs.~\cite{Neubert:2002ku,Kagan:1998bh}.  The experimental results for $B\to (\rho,\omega,K^*)\gamma$ are shown in Tab.~\ref{rad_tab1}.  For $B_s\to\phi\gamma$  only an upper bound ${\cal B}(B_s\to\phi\gamma)<120\times 10^{-6}$ exists and no experimental information is available for $B_s\to \bar K^*\gamma$ \cite{Yao:2006px}. 
\begin{table}[h]
\renewcommand{\arraystretch}{1.4}\addtolength{\arraycolsep}{3pt}
$$
\begin{array}{l||c|c||l|c}
\hline
{\cal B} \times 10^6 & \mbox{{\sc BaBar} \cite{babar_rad}} & \mbox{Belle
  \cite{belle_rad}} & {\cal B} \times 10^6 & \mbox{HFAG \cite{Barberio:2007cr}}
\\\hline
B\to (\rho,\omega)\gamma & 1.25^{+0.25}_{-0.24}\pm 0.09 &
                           1.32^{+0.34}_{-0.31}{}^{+0.10}_{-0.09} 
                         &
B^+\to K^{*+}\gamma & 40.3\pm 2.6
\\
B^+\to \rho^+\gamma &      1.10^{+0.37}_{-0.33}\pm 0.09 &
                           0.55^{+0.42}_{-0.36}{}^{+0.09}_{-0.08} 
                    &
B^0\to K^{*0}\gamma &  40.1\pm 2.0
\\
B^0\to\rho^0\gamma  &      0.79^{+0.22}_{-0.20}\pm0.06 &
                           1.25^{+0.37}_{-0.33}{}^{+0.07}_{-0.06} 
\\
B^0\to\omega\gamma  &      <0.78 &
                           0.96^{+0.34}_{-0.27}{}^{+0.05}_{-0.10}
\\\hline
\end{array}
$$
\caption[Experimental branching ratios of exclusive $b\to (d,s)\gamma$
  transitions.]{\small Experimental branching ratios of exclusive $b\to (d,s)\gamma$
  transitions. All entries are CP averaged. The first error is statistical, the second
  systematic. $B\to (\rho,\omega)\gamma$ is the CP   average of the
  isospin average over $\rho$ and $\omega$ channels:\\  $\overline{\cal B}(B\to
  (\rho,\omega)\gamma) = \frac{1}{2}  \left\{ \overline{\cal B}(B^\pm\to \rho^\pm\gamma) +
  \frac{\tau_{B^\pm}}{\tau_{B^0}} \left[ \overline{\cal B}(B^0\to 
  \rho^0\gamma) +  \overline{\cal B}(B^0\to \omega \gamma)\right]\right\}$.}
  \label{rad_tab1}
\end{table}

In the SM the decays are flavour-changing-neutral-current (FCNC) $b\to D\gamma$ transitions, mediated by penguin diagrams; they are therefore loop-suppressed and potentially very sensitive to new physics. To determine the relative sizes of contributions to the decays one must consider the following points:
\begin{itemize}
\item{the leading term is loop-suppressed $\sim 1/(4 \pi)^2$ and proportional to $C_7\sim-0.3$.}
\item{Evidently from Eq.~(\ref{qcdf_11}) for each mode there are two amplitudes proportional to different CKM factors $\lambda^{(D)}_{u,c}$. For $b\to d$ transitions both $\lambda^{(d)}_u$ and $\lambda^{(d)}_c$ are $\sim\lambda^3$, however, for $b\to s$ transitions $\lambda^{(s)}_u\sim\lambda^4$ and $\lambda^{(s)}_c\sim\lambda^2$; there is a relative CKM suppression of the up-quark contribution.}
\item{Power suppressed corrections from WA are formally $\sim 1/m_b$ although come with  large Wilson coefficients $C_1\sim-0.3$ and $C_2\sim 1$ and are not loop suppressed. The WA contributions drive the isospin asymmetries.}
\item{The production of ``wrong'' helicity photons is suppressed by $m_D/m_b$ (\ref{qcdf_13}). The interplay of both helicity amplitudes generates the time-dependent CP asymmetries, which are small in the SM due to this suppression.}
\end{itemize}

\section{Wilson Coefficients}
Considerable effort has gone into calculating the Wilson coefficients to NLO accuracy. Using the expressions for the NLO anomalous dimension matrices available in the literature we employ the renormalisation techniques of Eqs.~(\ref{basics_eq22}-\ref{basics_eq29}) to calculate the Wilson coefficients at the required scales. Numerical values of all the NLO Wilson coefficients $C_i$ used in the analysis are given in Tab.~\ref{rad_tab2}. The situation is complicated by the fact that the QCDF results of Ref.~\cite{Bosch:2002bw} are given in terms of two bases. The first, the so-called BBL basis named after the authors of Ref.~\cite{Buchalla:1995vs}, is that of Eqs.~(\ref{basics_eq20}) and (\ref{basics_eq21}) except with $Q_1$ and $Q_2$ exchanged with respect to the basis of Ref.~\cite{Bosch:2001gv}. The second is the so-called CMM basis of Ref.~\cite{munz, buras}. The two bases differ except for $Q_{7(8)}^{\rm BBL}=Q_{7(8)}^{\rm CMM}$. Following Ref.~\cite{Bosch:2002bw}, the CMM set is used for calculating hard-vertex  corrections to the QCDF formulas and the BBL set at the lower scale $\mu_h\sim \sqrt{\Lambda_{h} \,\mu}$ (with $\lambda_h\sim 0.5\,{\rm GeV}$ and $\mu= \mathcal{O}(m_b)$)  is used to calculate hard-spectator corrections. Power corrections are calculated from the BBL set at scale $m_b$. 

NLO accuracy is mandatory only for $C_7$, as it is for this term only that the hadronic matrix element is also known to NLO accuracy. We evaluate all $\mathcal{O}(\alpha_s)$ and power-suppressed corrections using both LO and NLO scaling for Wilson coefficients and hadronic matrix elements and include the resulting scale dependence in the theoretical uncertainty.
\begin{table}[h]
\renewcommand{\arraystretch}{1.3}
\addtolength{\arraycolsep}{2pt}
$$
\begin{array}{c|c|c|c|c|c|c}
C^{\rm CMM}_1(m_b) & C^{\rm CMM}_2(m_b) & C^{\rm CMM}_3(m_b) & 
C^{\rm CMM}_4(m_b) & C^{\rm CMM}_5(m_b) & C^{\rm CMM}_6(m_b) &
C^{\rm CMM}_7(m_b)
\\\hline
-0.322 & 1.009 & -0.005 & -0.087 & 0.0004 & -0.001 & -0.309 
\\\hline\hline
C^{\rm BBL}_1(m_b) & C^{\rm BBL}_2(m_b) & C^{\rm BBL}_3(m_b) & 
C^{\rm BBL}_4(m_b) & C^{\rm BBL}_5(m_b) & C^{\rm BBL}_6(m_b) 
& C^{\rm CMM}_8(m_b) 
\\\hline
-0.189 & 1.081  & 0.014 & -0.036 & 0.009  & -0.042 & -0.170
\\\hline\hline
C^{\rm BBL}_1(\mu_h) & C^{\rm BBL}_2(\mu_h) & C^{\rm BBL}_3(\mu_h) & 
C^{\rm BBL}_4(\mu_h) & C^{\rm BBL}_5(\mu_h) & C^{\rm BBL}_6(\mu_h) &
C^{\rm CMM}_8(\mu_h)
\\\hline
-0.288 & 1.133 & 0.021 & -0.051 & 0.010 & -0.065 & -0.191
\end{array}
$$
\caption[Numerical values of the next-to-leading-order Wilson coefficients.]{\small NLO Wilson coefficients to be used in the analysis, at the scales $m_b=4.2\,$GeV and $\mu_h=2.2\,$GeV. The coefficients labelled BBL correspond to the operator  basis of Ref.~\cite{Buchalla:1995vs} and given in Eq.~(\ref{basics_eq21}),   whereas CMM denotes the basis of Ref.~\cite{munz}. We use   $\alpha_s(m_Z) = 0.1176$ \cite{Yao:2006px} and ${m}_t({m}_t) =   163.6\,$GeV \cite{mt}. Note that $C_1^{\rm BBL}$ and $C_2^{\rm BBL}$   are exchanged with respect to the basis of Ref.~\cite{Bosch:2001gv} and  that $C_{7(8)}^{\rm BBL}=C_{7(8)}^{\rm CMM}$. Following Ref.~\cite{ Bosch:2002bw}, the CMM set is used for calculating hard-vertex  corrections to the QCDF formulas and the BBL set at the lower scale  $\mu_h$ is used to calculate hard-spectator corrections. The BBL set at  scale $m_b$ is used for the calculation of power-corrections.}
\label{rad_tab2}
\end{table}

\section{Leading and Power Suppressed Contributions}
It proves convenient to split to the coefficients in Eq.~(\ref{qcdf_11}) into three contributions which we will investigate separately:
\begin{eqnarray}
a_{7L}^U( V) &=& a_{7L}^{U,{\rm QCDF}}( V) + a_{7L}^{U,{\rm ann}}(  V) + a_{7L}^{U,{\rm soft}}( V)+\dots\,,\nonumber\\
a_{7R}^U( V) &=& a_{7R}^{U,{\rm QCDF}}( V) + a_{7R}^{U,{\rm ann}}(V)+  a_{7R}^{U,{\rm soft}}( V)+\dots\,,
\label{asplit}
\end{eqnarray} 
where the leading term in the $1/m_b$ expansion is given by Eq.~(\ref{qcdf_12}) and all other terms are suppressed by at least one power of $m_b$. The dots denote terms of higher order in $\alpha_s$ and further $1/m_b$ corrections to QCDF, most of which are incalculable. We only include those power-suppressed terms that are either numerically large or relevant for isospin and CP asymmetries.

\subsection{Leading Contributions}
The diagrams giving the leading QCDF contributions are given in Chapter~\ref{chapter6_QCDF}. It turns out that, at the level of two decimal places, all $a_{7L}^{c,{\rm QCDF}}$ are equal and so are
$a_{7L}^{u,{\rm QCDF}}$.\footnote{Explicit formulas for  $a_{7L}^{U,{\rm QCDF}}$, complete to $\mathcal{O}(\alpha_s)$, can be found in Ref.~\cite{Bosch:2002bw}.} For central values of the input parameters of Tab.~\ref{rad_tab8} we obtain
\begin{eqnarray}
a^{c,{\rm QCDF}}_{7L}(V) & = & -\overbrace{(0.41+0.03i)}^{\shortstack{{\rm \footnotesize Vertex}\\{\rm \footnotesize Corrections}}} - 
\overbrace{(0.01+0.01i)}^{\shortstack{{\rm \footnotesize Hard-Spectator}\\{\rm \footnotesize Corrections}}}\,,\nonumber\\
a^{u,{\rm QCDF}}_{7L}(V) & = & -(0.45+0.07i) + (0.02-0i)\,.
\label{10}
\end{eqnarray}
The size of the hard-spectator corrections is set by the factor
\begin{equation}
h_V = \frac{2 \pi^2}{9}\,\frac{f_B f_V^\perp}{m_B  
T_1^{B\to V}(0)\lambda_B}\,.
\end{equation}
For $B_s$ decays one has to set $f_B\to f_{B_s}$ and correspondingly for the other $B$ meson parameters. We estimate the value of  $\lambda_{B_s}$, the first inverse moment of the twist-2 $B$-meson light-cone DA, from $\lambda_{B_d}$ by a simple scaling argument:
\begin{equation}
\frac{m_{B_s}}{\lambda_{B_s}}\,(\Lambda_{\rm QCD}+m_s) = 
\frac{m_{B_q}}{\lambda_{B_q}}\,\Lambda_{\rm QCD}\,,
\label{rad_bs}
\end{equation}
which follows from the assumption that the $B_q$ DA peaks at the spectator momentum $k_+ = \Lambda_{\rm QCD}$, whereas that of $B_s$ peaks at $\Lambda_{\rm QCD}+m_s$. Its numerical value is given, along with all the other input parameters, in Tab.~\ref{rad_tab8}.

\subsection{Weak Annihilation}
\begin{figure}[h]
$$\epsfxsize=0.6\textwidth\epsffile{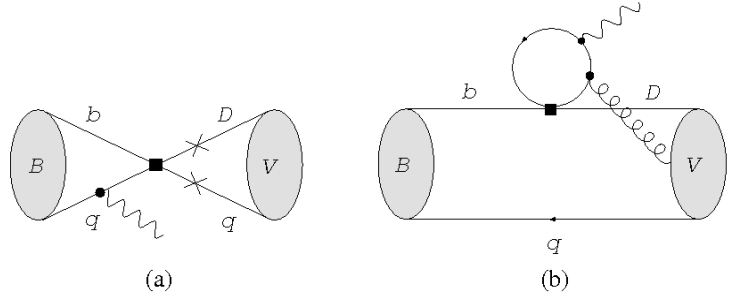}$$
\caption[Diagrams for weak annihilation and soft-gluon emission from a quark loop.]{\small (a) Weak annihilation diagram where photon emission from the $B$ meson spectator quark is power-suppressed. The crosses denote possible photon emission vertices for $Q_{5,6}$ only. (b) soft-gluon emission from a quark loop, where there is also a second diagram in which the gluon is picked up by the $B$ meson.}
\label{rad_fig1}
\end{figure}

 $a_{7L}^{U,{\rm ann}}$ encodes the $\mathcal{O}(1/m_b)$ contribution of the WA diagram of Fig.~\ref{rad_fig1}(a) which drives  the isospin asymmetries and has been  calculated in QCDF in  Ref.~\cite{Bosch:2002bw} with $\alpha_s$ corrections given in Ref.~\cite{Kagan:2001zk} for $\rho$ and $K^*$  and in Ref.~\cite{Bosch:2004nd} for $\omega$. WA receives contributions from the current-current operator $Q_2^u$, which for $b\to s$ transitions is doubly CKM suppressed, and QCD penguin operators $Q_{3,\dots,6}$, which are not CKM suppressed. Formulas for $a_{7L}^{U,{\rm ann}}(\rho,K^*)$ in QCDF can be found in Refs.~\cite{Bosch:2002bw,Bosch:2004nd};  in this approximation, there is no contribution to  $a_{7R}^{U,{\rm ann}}$. 

Preliminary results for the $\mathcal{O}(\alpha_s)$ corrections to WA in $B\to\rho \gamma$ were presented in Ref.~\cite{chamonix}.   In QCDF,  the $a_{7L}^{U,{\rm ann}}$  are expressed in terms of the hadronic quantities
\begin{equation}
b^V = \frac{2\pi^2}{T_1^{B\to V}(0)} \,\frac{f_B m_V f_V}{m_B m_b
  \lambda_B}\,, \qquad
d^V_{v} = -\frac{4\pi^2}{T_1^{B\to V}(0)} \,\frac{f_B f_V^\perp}{m_B
  m_b} \,\int_0^1 dv\,\frac{\phi_{2;V}^\perp(v)}{v}
\end{equation}
and $d^V_{\bar v}$, obtained by replacing $1/v\to 1/\bar v$ in the integrand; $\phi_{2;V}^\perp$ is the twist-2 DA of a transversely polarised vector meson, (\ref{das_eq19}). Numerically, one finds, for instance for the $\rho$, $b^\rho = 0.22$ and $d^\rho = -0.59$, at the scale $\mu = 4.2\,$GeV. As $T_1\sim 1/m_b^{3/2}$ and $f_B\sim m_b^{-1/2}$ in the heavy-quark limit, these terms are $\mathcal{O}(1/m_b)$, but not numerically small because of the tree-enhancement factors of $\pi^2$.
\begin{table}
\renewcommand{\arraystretch}{1.3}
\addtolength{\arraycolsep}{3pt}
$$
\begin{array}{l||c|c|c|c|c}
\mbox{WA} &  
B^-\to K^{*-} & \bar B^0\to K^{*0} & B\to (\rho,\omega) & B_s\to\phi & 
B_s\to \bar  K^*\\\hline
\mbox{induced by} & \mbox{C (and P)} & \mbox{P} & \mbox{C and P} &
\mbox{P} & \mbox{P}\\
\mbox{CKM} & \lambda^2 \mbox{~(and 1)} & 1 & 1 & 1 & 1
\end{array}
$$
\caption[Parametric size of the weak annihilation contributions.]{\small Parametric size of WA  contributions to $B\to V\gamma$. C denotes the charged-current operators $Q_{1,2}$, P  the penguin operators $Q_{3,\dots,6}$; their Wilson coefficients are   small -- see Tab.~\ref{rad_tab2}. CKM denotes the order in  the Wolfenstein parameter $\lambda$ with respect to the dominant amplitude induced by $Q_7$.}
\label{rad_tab3}
\end{table}

For $\omega$, $\bar K^*$ and $\phi$ we obtain
\begin{eqnarray}
\left. a_{7L}^{u,{\rm ann}}(\omega)\right|_{\rm QCDF}
 & = & Q_d b^\omega (a_1 + 2 (a_3+a_5)
+ a_4) + Q_d (d^\omega_v + d^\omega_{\bar v}) a_6\,,\nonumber\\
\left. a_{7L}^{c,{\rm ann}}(\omega)\right|_{\rm QCDF} 
& = & Q_d b^\omega (2 (a_3+a_5)
+ a_4) + Q_d (d^\omega_v + d^\omega_{\bar v}) a_6\,,\nonumber\\
\left. a_{7L}^{U,{\rm ann}}(\phi)\right|_{\rm QCDF} 
& = & Q_s b^\phi (a_3+a_5) + 
         Q_s (d^\phi_v + d^\phi_{\bar v}) a_6\,,\nonumber\\
\left. a_{7L}^{U,{\rm ann}}(\bar K^*)\right|_{\rm QCDF} 
& = & Q_s b^{\bar K^*} a_4 + 
         Q_s (d^{\bar K^*}_v Q_d/Q_s + d^{\bar K^*}_{\bar v}) a_6\,,
\label{15}
\end{eqnarray}
with $a_1 = C_1+C_2/3$, $a_3 = C_3+C_4/3$, $a_4 = C_4+C_3/3$, $a_5 = C_5+C_6/3$, $a_6 = C_6+C_5/3$.\footnote{Note that $a_1\leftrightarrow a_2$ as compared to \cite{Bosch:2002bw} as in our operator basis  (i.e.\ the BBL basis) $Q_1$  and $Q_2$ are exchanged.} The expressions for $\phi$ and $\bar K^*$ are new; for $\omega$, we do not agree with \cite{Bosch:2004nd}.  Apart from for $\rho$ and $\omega$, all the WA coefficients are numerically small and do not change the branching ratio significantly; the terms in $a_6$, however, are relevant  for the isospin asymmetries. 

In Tab.~\ref{rad_tab3} we show the relative weights of these diagrams in terms of CKM factors and Wilson coefficients. The numerically largest contribution occurs for $B^\pm\to \rho^\pm\gamma$: it comes with the large combination of  Wilson coefficients $C_2+C_1/3=1.02$ and is not CKM suppressed. For $B^0\to (\rho^0,\omega)\gamma$ it comes with the factor $C_1+C_2/3 = 0.17$
instead and an additional suppression factor $1/2$ from the electric charge of the spectator quark ($d$ instead of $u$). For all other decays, WA is suppressed by small (penguin) Wilson coefficients. Apart from $B\to(\rho,\omega)\gamma$, WA is not relevant so much for the total values of $a_{7L}$, but rather for isospin breaking, which is set by photon emission from the spectator quark. WA is the only mechanism to contribute to isospin asymmetries at tree-level; see Ref.~\cite{Kagan:2001zk} for $\mathcal{O}(\alpha_s)$ contributions.

\begin{figure}
$$\epsfxsize=\textwidth\epsffile{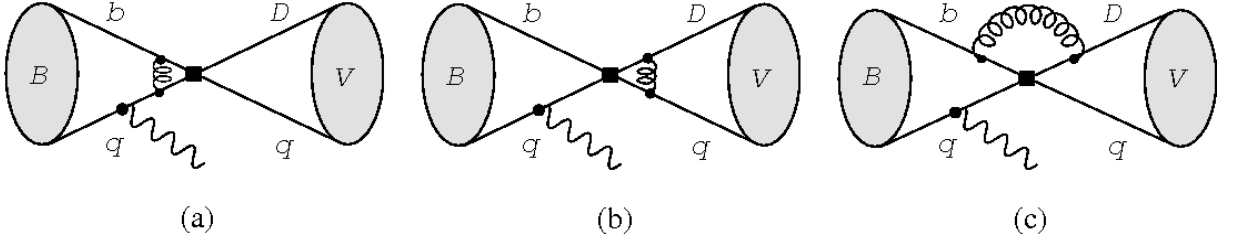}$$
\caption[Example radiative corrections to 
weak annihilation.]{\small Example radiative corrections to 
weak annihilation. The corrections to the $B$ vertex in (a) are known 
\cite{bellnu,Descotes-Genon:2002mw} and those to the $V$ vertex in (b) are included in $f_V$. 
For the non-factorisable corrections in (c) only preliminary results
are available \cite{chamonix}.}
\label{rad_fig2}
\end{figure}

In view of the large size of $a_{7L}^{u,{\rm ann}}(\rho)$ it is appropriate to have a look at further  corrections. The most  obvious ones are $\mathcal{O}(\alpha_s)$ corrections to the QCDF expressions, shown in Fig.~\ref{rad_fig2}.
As it turns out, the corrections to the $B$ vertex in Fig.~\ref{rad_fig2}(a) are known: they also enter the decay $B\to\gamma \ell\nu$ and were calculated in Ref.~\cite{bellnu,Descotes-Genon:2002mw}. Numerically, they are at the level of 10\%.  Fig.~\ref{rad_fig2}(b) shows the vertex corrections to the $V$ vertex, which are actually included in the decay constant $f_V$. For the non-factorisable corrections shown in Fig.~\ref{rad_fig2}(c) preliminary results have been reported in Ref.~\cite{chamonix} according to which these corrections are of a size similar to the $B$ vertex corrections. 

In Ref.~\cite{Kagan:2001zk} also another class of $1/m_b$ corrections to $B\to K^*\gamma$ was calculated, namely $\mathcal{O}(\alpha_s)$  corrections to the isospin asymmetry in this decay. As these corrections break factorisation (require an infra-red cut-off in the  momentum  distribution of the valence quarks in the $K^*$ meson) and are numerically small, we do not include them in our analysis. 

\subsection{Long-Distance Photon Emission}
Another class of corrections is suppressed by one power of $m_b$ with respect to the QCDF contributions and is  due to  long-distance photon emission from the soft $B$ spectator quark. A first calculation of this effect was attempted in Ref.~\cite{WA} and was corrected and extended in Ref.~\cite{Ball:2003fq}. The long-distance photon emission from a soft-quark line requires the inclusion of higher-twist terms in the expansion of the quark propagator in a photon background field, beyond the leading-twist (perturbative) contribution; a comprehensive discussion of this topic can be found in Ref.~\cite{Ball:2002ps}. The quantity calculated in Ref.~\cite{Ball:2003fq} is
\begin{eqnarray}
\lefteqn{
\bra{\rho^-(p)\gamma(q)} (\bar d u)_{V-A} (\bar u b)_{V-A}\ket{B^-(p+q)} =}\hspace*{3cm}\nonumber\\
& = & e\,\frac{m_\rho f_\rho}{m_B} \eta^*_\mu
\left\{F_V \epsilon^{\mu\nu\rho\sigma} e^*_\nu p_{\rho}
  q_\sigma - i F_A [e^{*\mu} (p \cdot q) - q^\mu
    (e^* \cdot p)]\right\}\nonumber\\
& = & -e \,\frac{m_\rho f_\rho}{m_B} \left\{ \frac{1}{2}\,
  F_V (S_L+S_R) + \frac{1}{2}\, F_A (S_L-S_R)\right\}
  \label{problem}
\end{eqnarray}
in terms of the photon-helicity amplitudes $S_{L,R}$.\footnote{Eq.~(\ref{problem}) differs from the one given in \cite{Ball:2003fq} by an overall sign, which is due to the different convention used in \cite{Ball:2003fq} (and in \cite{Ball:2002ps}) for the covariant derivative: $D_\mu = \partial_\mu
- i e Q_f A_\mu$ instead of $D_\mu = \partial_\mu + i e Q_f A_\mu$ as in this analysis.}
In QCDF, $F_{A,V}$ are given by $Q_u f_B/\lambda_B$ and induce a term $Q_u a_2 b^\rho$ in  $a_{7L}^{u,{\rm    ann}}(\rho^-)$. The long-distance photon contribution to $F_{V,A}$ was found to be \cite{Ball:2003fq}
\begin{equation}
F^{\rm soft}_A = -0.07\pm 0.02 \equiv Q_u G_A\,,\qquad 
F^{\rm soft}_V = -0.09\pm
0.02 \equiv Q_u G_V\,.
\label{Fsoft}
\end{equation}
with $G_A+G_V = -0.24\pm 0.06$ and $G_V-G_A = -0.030\pm 0.015$.\footnote{Again, there is a relative sign with respect to the results in \cite{Ball:2003fq}. This comes from the fact that the product $e F_{A,V}^{\rm soft}$ is independent of the sign convention for $e$, and as we have changed the overall sign of (\ref{problem}) with respect to \cite{Ball:2003fq}, we also have to change the sign of $F_{A,V}^{\rm soft}$. Stated differently: the relative sign between $F_{A,V}^{\rm soft}$ and $F_{A,V}^{\rm hard}$ in \cite{Ball:2003fq} is wrong because of a mismatch in sign conventions for $e$ in the covariant derivative.}
In order to obtain concise expressions for $a_{7L(R)}^{U,{\rm ann}}$, it proves convenient to define one more hadronic quantity:
\begin{equation}
g^\rho_{L,R} = \frac{\pi^2}{T_1^\rho}\,\frac{m_\rho f_\rho}{m_b m_B}\,
(G_V\pm G_A)
\end{equation}
and correspondingly for other mesons. $g_L$ is $\mathcal{O}(1/m_b^2)$ as $G_V+G_A$ has the same power scaling in $m_b$ as $T_1$, i.e.\ $\sim m_b^{-3/2}$, as one can read off from the explicit expressions in \cite{WA}. The difference $G_V-G_A$, on the other hand, is a twist-3 effect due to three-particle light-cone DAs of the photon and is suppressed by one more power of $m_b$, i.e.\ $g_R\sim
1/m_b^3$. This quantity will enter the CP asymmetry. Our final expressions for $a_{7L(R)}^{U,{\rm ann}}$ then read:
\begin{eqnarray}
a_{7L}^{U,{\rm ann}}(V) & = & \left. a_{7L}^{U,{\rm
    ann}}(V)\right|_{\rm QCDF} (b^V\to b^V + g^V_L)\,,\nonumber\\
a_{7R}^{U,{\rm ann}}(V) & = & \left. a_{7L}^{U,{\rm
    ann}}(V)\right|_{\rm QCDF} (b^V\to g^V_R, d^V\to 0)\,.
    \label{20A}
\end{eqnarray}
Numerically, one has $g^{\rho}_L/b^\rho = -0.3$, so these corrections, despite being suppressed by one more power in $1/m_b$, are not small numerically and larger than the known $\mathcal{O}(\alpha_s)$
corrections to QCDF from $B\to\gamma\ell\nu$. Based on this, we feel justified in including these long-distance corrections in our analysis, while dropping the radiative ones of Figs.~\ref{rad_fig2}(a) and (c). For central values of the input parameters we find the following numerical values for the various WA and long-distance photon contributions, including in particular those to which $Q_{1,2}$ contribute (with no Cabibbo suppression):
\begin{eqnarray}
a_{7L}^{c,{\rm ann}}(K^{*0})  &=&  -0.013-0.001\, {\rm LD}\,,
\qquad  a_{7L}^{c,{\rm ann}}(K^{*-}) =  0.004+0.001\, {\rm
  LD}\,,\nonumber\\
a_{7L}^{u,{\rm ann}}(\rho^0) &=&  -0.001-0.004\, {\rm LD}\,,
 \qquad ~~ a_{7L}^{u,{\rm ann}}(\rho^-) =  0.149-0.043\, {\rm
  LD}\,,\nonumber\\
 a_{7L}^{u,{\rm ann}}(\omega)  &=&  -0.024+0.003\, {\rm LD}\,.
 \label{LDcont}
\end{eqnarray}
The contribution from the long-distance photon emission is labelled ``LD'' (LD$\to 1$ at the end). 
The unexpectedly small $a_{7L}^{u,{\rm ann}}(\rho^0)$ is due to a numerical cancellation between the charged-current and  penguin-operator contributions. Comparing these results with those from QCDF,
Eq.~(\ref{10}), it is evident that WA is, as expected,  largely irrelevant for the branching ratios, except for $B^\pm\to \rho^\pm\gamma$.

\subsection{Soft Quark Loops}
$a_{7L(R)}^{U,{\rm soft}}$ encodes soft-gluon emission from a (light or heavy quark) loop as shown in Fig.~\ref{rad_fig1}(b). Soft-gluon emission from a charm loop was first considered in  Ref.~\cite{KRSW97} as a potentially relevant long-distance contribution to the branching ratio of $B\to K^*\gamma$, however, the same diagram also contributes dominantly to the time-dependent CP asymmetry in $B^0\to K^{*0}\gamma$ \cite{grin05}.  As for $a_{7R}^U$, the dominant contributions to $a_{7R}^c(K^*)$ were calculated in Ref.~\cite{Ball:2006cv} and new to this analysis is their generalisation to the other vector mesons and the inclusion of contributions from light-quark loops. Motivation to include light quark loops stems from the fact that they are doubly CKM-suppressed for $b\to s\gamma$ transitions, but not for $b\to d\gamma$, for which they are on an equal footing as the heavy quark loops. The quark loop contributions are suppressed by one power of $m_b$ with respect to $a_{7L}^{U,{\rm QCDF}}$, but they also induce a right-handed photon amplitude which is of the same order in $1/m_b$ as $a_{7R}^{U,{\rm QCDF}}$ (\ref{qcdf_13}), and this amplitude induces the time-dependent CP asymmetry.  The asymmetry is expected to be very small in the SM and $\propto m_D/m_b$ due to the chiral suppression of the leading transition (\ref{qcdf_13}), but could be drastically enhanced by new physics contributions -- thus constituting an excellent ``null test'' of the SM \cite{Gershon:2006mt,Ball:2006cv}.  It was noticed in Refs.~\cite{grin04,grin05}  that the chiral suppression is relaxed by emission of a gluon from the quark loop and contributes dominantly to the time-dependent CP asymmetry in $B^0\to K^{*0}\gamma$, which motivates the inclusion of these contributions. The task of the present analysis, however, is not so much to calculate these contributions to high accuracy, but to exclude the possibility of  {\em  large} contributions to the CP asymmetry.  With this in mind, the theoretical uncertainties of the results are very generously estimated  --- which is somewhat unavoidable due to the current uncertainties of the relevant hadronic input parameters.

Potentially the most important contribution to the soft-gluon emission  diagram in Fig.~\ref{rad_fig1}(b) 
comes from the charged-current operator $Q_2^U$ with the  large Wilson coefficient $C_2\sim 1$; it vanishes for $Q_1^U$  by gauge invariance. In addition, the penguin operators $Q_{3,4,6}$ give a non-zero contribution. Details of the derivation of $a_7^{U,{\rm soft}}$ can be found in Ref.~\cite{Ball:2006eu} in which the following expression is obtained:
\begin{eqnarray}
a_{7L(R)}^{U,{\rm soft}}(V) & = &  \frac{\pi^2}{m_b T_1^{B\to V}(0)} \left\{ Q_U C_2 (l_U\pm \tilde l_U)(V) + Q_D C_3 (l_D\pm \tilde l_D)(V)\right.\nonumber\\
&&\left. + \sum_q Q_q (C_4-C_6) (l_q\pm \tilde l_q)(V)\right\}.
\label{24}
\end{eqnarray}
Here the sum over $q$ runs over all five active quarks $u,d,s,c,b$. The contribution from $Q_5$ is proportional to $m_D$ and hence helicity suppressed and neglected.  Assuming $\rm SU(3)_F$-flavour symmetry for the light quark loops, one has $l_u=l_d=l_s$, and ditto for $\tilde l_{u,d,s}$, which causes a cancellation of these contributions in the last term of Eq.~(\ref{24}). ${\rm SU(3)_F}$-breaking effects are estimated to be around 10\% \cite{Ball:2006eu}.  The parameters  $l_c(K^*)$ and $\tilde{l}_c(K^*)$ were first calculated from three-point sum rules in Ref.~\cite{KRSW97} and were re-calculated in the more suitable method of LCSR via a local OPE in Ref.~\cite{Ball:2006cv}. The analysis therein as been updated and extended to $l_b,\, \tilde{l}_b$ and the other particles $\rho,\,\omega ,\,\bar{K}^*,\,\phi$ for the present analysis \cite{Ball:2006eu}. The results for $l_c$ and $\tilde l_c$ are given in the upper table of Tab.~\ref{rad_tab5}. Those for $l_b$ and $\tilde l_b$ are obtained as
\begin{equation}
l_b = \frac{m_c^2}{m_b^2}\, l_c\,,\qquad \tilde l_b = \frac{m_c^2}{m_b^2}\, \tilde l_c\,.
\end{equation}
For light-quark loops the photon is almost at threshold and the local OPE does not apply. In Ref.~\cite{Ball:2006eu} a method was developed for calculating these contributions via LCSRs.  Similar to the method of Ref.~\cite{Khodjamirian:2000mi}  used for the calculation of soft-gluon contributions to $B\to\pi\pi$, a dispersion relation approach is used to connect the off-shell matrix element to the physical regime $q^2=0$.  The results are presented in the lower table of Tab.~\ref{rad_tab5}.
\begin{table}[h]
\renewcommand{\arraystretch}{1.3}
\addtolength{\arraycolsep}{3pt}
$$
\begin{array}{l ||r|r|r|r}
& \multicolumn{1}{c|}{l_c} & \multicolumn{1}{c|}{\tilde l_c} 
& \multicolumn{1}{c|}{l_c-\tilde l_c} & \multicolumn{1}{c}{l_c +
    \tilde l_c} \\
 \hline
 B \to K^*   & -355 \pm 280 &  -596 \pm 520 & 242 \pm 370 & -952 \pm 800  \\
 B \to (\rho,\omega)   & -382 \pm 300 & -502\pm  430 & 120 \pm 390 & 
-884\pm 660  \\
 B_s \to \bar K^*   &  -347\pm 260 & -342\pm 400 & -4\pm 300 & -689\pm 600 \\
 B_s \to \phi   & -312 \pm  240 & -618 \pm 500 & 306 \pm 320 & -930 \pm
750 
\end{array}
$$\\
\renewcommand{\arraystretch}{1.3}
\addtolength{\arraycolsep}{3pt}
$$
\begin{array}{l ||r|r|r|r}
& \multicolumn{1}{c|}{l_u} & \multicolumn{1}{c|}{\tilde l_u} &   
\multicolumn{1}{c|}{l_u-\tilde l_u} & \multicolumn{1}{c}{l_u + \tilde l_u} \\
 \hline
 B \to K^*   & 536 \pm 70\% &  635 \pm 70\% & -99 \pm 300 & 1172 \pm 70 \% \\
 B \to (\rho,\omega)   & 827 \pm 70\% & 828\pm 70\% & -1\pm 300&
 1655\pm 70\%  \\
 B_s \to \bar K^*   &  454\pm 70\% & 572\pm 70\% & -118\pm 300 & 1025\pm
70\% \\
 B_s \to \phi   & 156\pm 70\% & 737\pm 70\% & -581\pm 300 & 893\pm 70\% \\
\end{array}
$$
\caption[Soft-gluon contributions from $c$-quark and $u$-quark  loops in units KeV. ]{\small Soft-gluon contributions from $c$-quark (upper table) and $u$-quark (lower table) loops in units KeV. The quantities $l_{c,u}$ and $\tilde l_{c,u}$ are defined in Ref.~\cite{Ball:2006eu}. We assume equal parameters for $\rho$ and $\omega$. $l_b$ is obtained as $l_b = l_c m_c^2/m_b^2$ and correspondingly for $\tilde l_b$. The uncertainty for $l_u-\tilde l_u$  is given in absolute numbers because of cancellations. In the $\rm SU(3)_F$-flavour limit assumed in this calculation one has  $l_u = l_d = l_s  \equiv l_q$}
\label{rad_tab5}
\end{table}

\section{Phenomenological Results}
In this section we combine the different contributions to the factorisation coefficients $a_{7L(R)}^U$ and give results for the observables, namely the branching ratios, the isospin asymmetries and the time-dependent CP asymmetries. 
\subsection{Branching Ratios}\label{rad_brs}
The (non-CP-averaged) branching ratio of the $b\to D\gamma$ decay 
$\bar B\to V\gamma$ is given by
\begin{eqnarray}
{\cal B}(\bar B\to V\gamma) & = & \frac{\tau_B}{c_V^2}\,
\frac{G_F^2\alpha_{\rm QED} m_B^3  m_b^2}{32 \pi^4} \left(1-\frac{m_V^2}{m_B^2}\right)^3 \left[T_1^{B\to V}(0)\right]^2\nonumber\\
&&\times \left\{ \left| \sum_{U=u,c} \lambda_U^{(D)} a_{7L}^U(V)\right|^2 +  \left| \sum_{U=u,c} \lambda_U^{(D)}
  a_{7R}^U(V)\right|^2\right\}
  \label{BR}
\end{eqnarray}
with the isospin factors $c_{\rho^\pm,K^*,\phi}=1$  and $c_{\rho^0,\omega} = \sqrt{2}$. The branching ratio for the CP-conjugated channel $B\to \bar V\gamma$ ($\bar b\to \bar D\gamma$ at parton level) is obtained by replacing $\lambda_U^{(D)}\to (\lambda_U^{(D)})^*$.  With the input parameters from Tab.~\ref{rad_tab8} and the lifetimes given in Tab.~\ref{rad_tab9} we find the following CP-averaged branching ratios  for $B\to K^*\gamma$, making explicit various sources of uncertainty:
\begin{eqnarray}
\overline{\cal B}(B^- \to K^{*-}\gamma) & = & (53.3\pm \overbrace{13.5}^{T_1}
\pm \overbrace{4.8}^{\mu}
\pm \overbrace{1.8}^{V_{cb}}\pm \overbrace{1.9}^{l_{u,c}} \pm \overbrace{1.3}^{\mbox{other}})\times
10^{-6}\nonumber\\
& =& (53.3\pm \underbrace{13.5}_{T_1}\pm 5.8)\times 10^{-6}\,,
\nonumber\\
\overline{\cal B}(\bar B^0 \to K^{*0}\gamma) & = & (54.2\pm \overbrace{13.2}^{T_1}
\pm \overbrace{6.0}^{\mu}
\pm \overbrace{1.8}^{V_{cb}}\pm \overbrace{1.8}^{l_{u,c}} \pm \overbrace{1.4}^{\mbox{other}})\times
10^{-6}\nonumber\\
& = & (54.2\pm \underbrace{13.2}_{T_1}\pm 6.7)\times 10^{-6}\,.
\label{50}
\end{eqnarray}
We have added all individual uncertainties in quadrature, except for that induced by the form factor. The uncertainty in $\mu$ is that induced by the renormalisation-scale dependence, with  $\mu= m_b(m_b)\pm 1\,$GeV.  ``Other'' sources of uncertainty include the dependence on the parameters in Tab.~\ref{rad_tab7}, on the size of LD WA contributions and the replacement of NLO by LO Wilson coefficients.  The above results agree, within errors, with the experimental ones given in Tab.~\ref{rad_tab1}, within the large theoretical uncertainty induced by the form factor.

As the uncertainties of all form factors in Tab.~\ref{rad_tab8} are of roughly the same size, one might conclude that the predictions for all branching ratios will carry uncertainties similar to those in
(\ref{50}). This is, however, not the case: the accuracy of the theoretical predictions can be improved by making use of the fact that the {\em ratio} of form factors is known much better than the individual form factors themselves. The reason is that the values given in Tab.~\ref{rad_tab8}, which were calculated using the same method, LCSRs, and with a common set of input parameters,  include common systematic uncertainties (dependence on $f_B$, $m_b$ etc.) which partially cancel in the ratio. In Ref.~\cite{Ball:2006nr}  the ratio of the $K^*$ and $\rho$ form factors was found to be
\begin{equation}
\xi_\rho \equiv \frac{T_1^{B\to K^*}(0)}{T_1^{B\to \rho}(0)} = 1.17\pm
0.09\,.
\label{xirho}
\end{equation}
The uncertainty is by a factor 2 smaller than if we had calculated $\xi_\rho$
from the entries in Tab.~\ref{rad_tab8}; analogously for $\omega$ one finds
\begin{equation}
\xi_\omega \equiv \frac{T_1^{B\to K^*}(0)}{T_1^{B\to \omega}(0)} = 1.30\pm
0.10\,.
\label{xiomega}
\end{equation}
The difference between $\xi_\rho$ and $\xi_\omega$ is mainly due to the difference between $f_\omega^\perp$ and $f_\rho^\perp$.  For the $B_s$ form factors, we also need the ratio of decay constants $f_{B_s}/f_{B_d}$. The status of $f_B$ from Lattice QCD was reviewed in  Ref.~\cite{Onogi}; the present state-of-the-art calculations are unquenched with $N_f=2+1$ active flavours \cite{unquenchedfB},  whose average is $f_{B_s}/f_{B_d}=1.23\pm 0.07$. Again, this ratio is fully consistent with that quoted in Tab.~\ref{rad_tab8}, but has a smaller uncertainty. One then finds the following ratios for $B_s$ form factors:
\begin{equation}
\xi_\phi \equiv \frac{T_1^{B\to K^*}(0)}{T_1^{B_s\to\phi}(0)} = 1.01\pm 0.13
\,,\qquad
\xi_{\bar K^*} \equiv \frac{T_1^{B\to K^*}(0)}{T_1^{B_s\to\bar K^*}(0)} 
= 1.09\pm 0.09\,.
\label{xis}
\end{equation}
The uncertainty of $\xi_{\bar K^*}$ is smaller than that of $\xi_\phi$ because the input parameters for $K^*$ and $\bar K^*$ are the same (except for G-odd parameters like $a_1^\perp$) and cancel in the ratio; the uncertainty is dominated by that of $f_{B_s}/f_{B_d}$. To benefit from this reduced theoretical uncertainty in predicting branching ratios, one has to calculate ratios of branching ratios, which mainly depend on $\xi_V$ and only mildly on  $T_1$  itself: in addition to the  overall normalisation, $T_1$ also enters hard-spectator interactions and power-suppressed corrections, whose size is set by hadronic quantities $\propto 1/T_1$. As these corrections are subleading (in  $\alpha_s$ or $1/m_b$),  however, a small shift in $T_1$ has only very minor impact on the  branching ratios. The absolute scale for the branching ratios is set by the CP- and isospin-averaged  branching ratio with the smallest experimental uncertainty, i.e.\ $B\to K^*\gamma$; from  Tab.~\ref{rad_tab1}, one finds:
\begin{equation}
\overline{\cal B}(B\to K^*\gamma) = \frac{1}{2}\left\{ \overline{\cal
  B}(B^\pm \to K^{*\pm}\gamma) + \frac{\tau_{B^\pm}}{\tau_{B^0}}\, 
  \overline{\cal B}(\bar B^0 \to K^{*0}\gamma)\right\} = (41.6\pm
  1.7)\times 10^{-6}\,.
  \label{x}
\end{equation}
That is, we obtain a theoretical prediction for $\overline{\cal B}(B\to
V\gamma)$ as 
\begin{equation}
\left.\overline{\cal B}(B\to V\gamma)\right|_{\rm th}= \left[ \frac{\overline{\cal B}(B\to V\gamma)}{
\overline{\cal B}(B\to K^*\gamma)}\right]_{{\rm th}} \, \left.
\overline{\cal B}(B\to K^*\gamma)\right|_{\rm exp}\,,
\end{equation}
where $\left[\dots\right]_{\rm th}$ depends mainly on $\xi_V$ and only in subleading terms on the individual form factors $T_1^{B\to K^*}$ and $T_1^{B\to V}$. It is obvious that, except for these subleading terms, this procedure is equivalent to extracting an {\em effective form factor} $\left.T_1^{B\to  K^*}(0)\right|_{\rm eff}$ from $B\to K^*\gamma$ and using $\left.T_1^{B\to V}(0)\right|_{\rm eff} = \left.T_1^{B\to   K^*}(0)\right|_{\rm  eff}/\xi_{V}$ for calculating the branching ratios for $B\to V\gamma$. From (\ref{x}) we find
\begin{equation}
\left. T_1^{B\to K^*}(0)\right|_{\rm eff} = 0.267\pm \overbrace{0.017}^{{\rm th}} \pm
\overbrace{0.006}^{{\rm exp}} = 0.267\pm 0.018\,,
\end{equation}
where the theoretical uncertainty  follows from the second uncertainty given in (\ref{50}). Eqs.~(\ref{xirho}), (\ref{xiomega}) and (\ref{xis}) then yield
\begin{eqnarray}
\left. T_1^{B\to \rho}(0)\right|_{\rm eff} &=& 0.228 \pm 0.023\,, \qquad
\left. T_1^{B\to \omega}(0)\right|_{\rm eff} = 0.205 \pm
0.021\,,\nonumber\\
\left. T_1^{B_s\to \bar K^*}(0)\right|_{\rm eff} &=& 0.245 \pm
0.024\,, 
\qquad
\left. T_1^{B_s\to \phi}(0)\right|_{\rm eff} = 0.260 \pm
0.036\,.
\label{56}
\end{eqnarray}
Note that all effective form factors agree, within errors, with the results from LCSRs given in Tab.~\ref{rad_tab8}, which confirms the results obtained from this method; the crucial point, however, is that the uncertainties are reduced by a factor of 2 (except for $T_1^{B_s\to \phi}$). We would like to stress that the motivation for this procedure is to achieve a reduction of the theoretical uncertainty of the
predicted branching fractions in $B\to (\rho,\omega)\gamma$ and $B_s$ decays.  The effective form factors do {\em not} constitute a new and independent theoretical determination, but are derived from the experimental results for $B\to K^*\gamma$ under the following assumptions:
\begin{itemize}
\item there is no new physics in $B\to K^*\gamma$;\footnote{Which is motivated by the
  results from inclusive $B\to X_s \gamma$ decays \cite{misiak}.}
\item QCDF is valid with no systematic uncertainties;
\item LCSRs can reliably predict the ratio of form factors at zero
  momentum transfer.
\end{itemize}
 From (\ref{BR}) and (\ref{56}), we then predict the following CP-averaged branching ratios:
\begin{eqnarray}
\overline{\cal B}(B^-\to \rho^-\gamma) & = & (1.16\pm \overbrace{0.22}^{T_1}\pm \overbrace{0.13}^{\rm Other})\times 10^{-6}\,, \nonumber\\
\overline{\cal B}(B^0\to \rho^0\gamma) & = & (0.55\pm 0.11\pm 
0.07)\times 10^{-6}\,, \nonumber\\
\overline{\cal B}(B^0\to \omega\gamma) & = & (0.44\pm 0.09 \
\pm 0.05)\times 10^{-6}\,,\nonumber\\
\overline{\cal B}(B_s\to \bar K^*\gamma) & = & (1.26\pm 0.25\pm 
0.18)\times 10^{-6}\,, \nonumber\\
\overline{\cal B}(B_s\to \phi\gamma) & = & (39.4\pm 10.7 \ \pm 5.3)\times 10^{-6}\,,
\label{57}
\end{eqnarray}
where the first uncertainty is induced by the effective form factors and  the second includes the variation of all inputs from Tab.~\ref{rad_tab8}  except for the angle $\gamma$ of the UT, which is fixed at
$\gamma=53^\circ$.\footnote{The value of the UT angle
$\gamma$ in Tab.~\ref{rad_tab8} comes from Belle's 
Dalitz-plot analysis of the CP asymmetry in $B^-\to (K_S^0 \pi^+\pi^-)
K^-$, with $K_S^0 \pi^+\pi^-$ \cite{Bellegamma} being a three-body final state common
to both $D^0$ and $\bar D^0$. Other determinations all come with theoretical uncertainties and/or possible contamination by unresolved new physics, so we take this result as a reference point.} The total uncertainty in each channel is $\sim 20\%$, except for $B_s\to \phi\gamma$, where it is 30\%. The results for $\rho$ and $\omega$ agree very well with those of {\sc  BaBar}, Tab.~\ref{rad_tab1}, but less so with the Belle results, although present experimental and theoretical uncertainties preclude a firm conclusion. Our prediction for $B_s\to \phi\gamma$ is well below the current experimental bound $120\times 10^{-6}$ \cite{Yao:2006px}. A branching ratio of the size given in (\ref{57}) implies that $\mathcal{O}(10^3)$ $B_s\to\phi\gamma$  events will be seen within the first few years of the LHC.

In Tab.~\ref{rad_tab6} we detail the contributions of individual terms to the branching ratios. In all cases ${\cal B}$ is dominated by the QCDF contribution, with WA most relevant for $B^-\to \rho^-\gamma$. This is expected as WA enters with the large Wilson coefficient $C_2\sim 1$. The effect is extenuated by long-distance (LD) photon emission, which itself is compensated by soft-gluon emission. The other channels follow a similar pattern, although the size of the effects is smaller.
\begin{table}[h]
\renewcommand{\arraystretch}{1.3}
\addtolength{\arraycolsep}{3pt}
$$
\begin{array}{l||c|c|c|c}
\mathcal{B}\times 10^{6}& \mbox{QCDF} & \mbox{+ WA (no LD)} & \mbox{+ WA (inc.\ LD)}& 
\mbox{+ soft gluons}\\\hline
B^-\to \rho^-\gamma & 1.05&1.17 & 1.11&1.16
\\
B^0\to \rho^0\gamma & 0.49&0.53& 0.53&0.55
\\
B^0\to \omega\gamma & 0.40& 0.42&0.42 &0.44
\\
B^-\to K^{*-}\gamma &39.7&38.4& 38.3&39.4
\\
B^0\to K^{*0}\gamma & 37.1&39.7 &39.9 &41.0
\\
B_s^0\to \bar K^{*0}\gamma & 1.12& 1.22& 1.23&1.26
\\
B_s^0\to \phi\gamma &34.6 & 38.2& 38.3&39.4
\end{array}
$$
\caption[Contributions to CP-averaged branching ratios.]{\small Contributions to CP-averaged branching ratios, using effective form factors and  central values of all other input parameters, Tab.~\ref{rad_tab8} (in particular $\gamma=53^\circ$). LD stands for long-distance photon-emission contributions. Each column  labelled ``+X'' includes the contributions listed in the previous  column plus the contribution induced by X. The entries in the last  column are our total central values.}
  \label{rad_tab6}
\end{table}
\begin{figure}[h]
$$ \epsfxsize=0.48\textwidth\epsffile{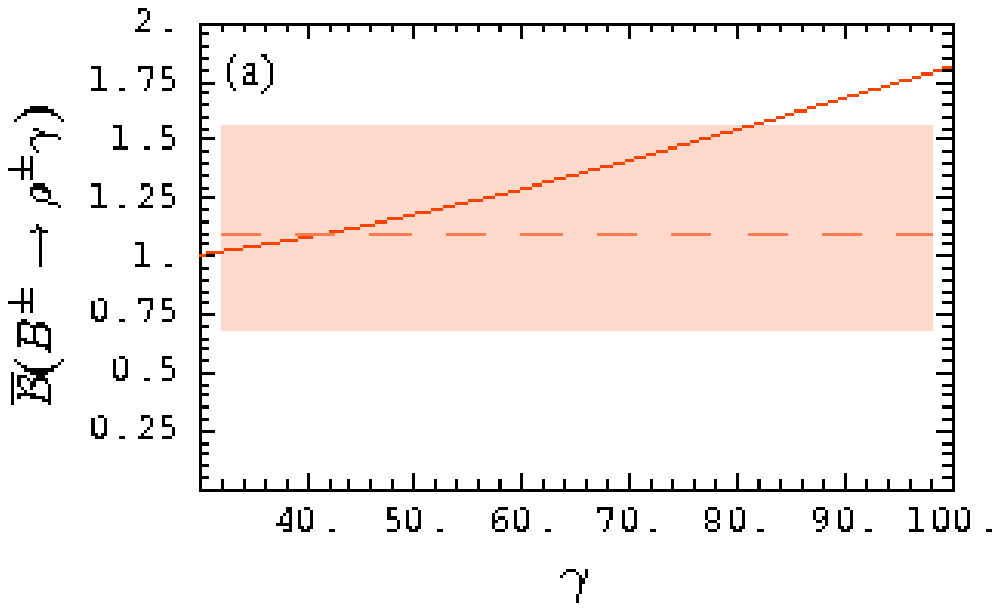}\quad\epsfxsize=0.48\textwidth\epsffile{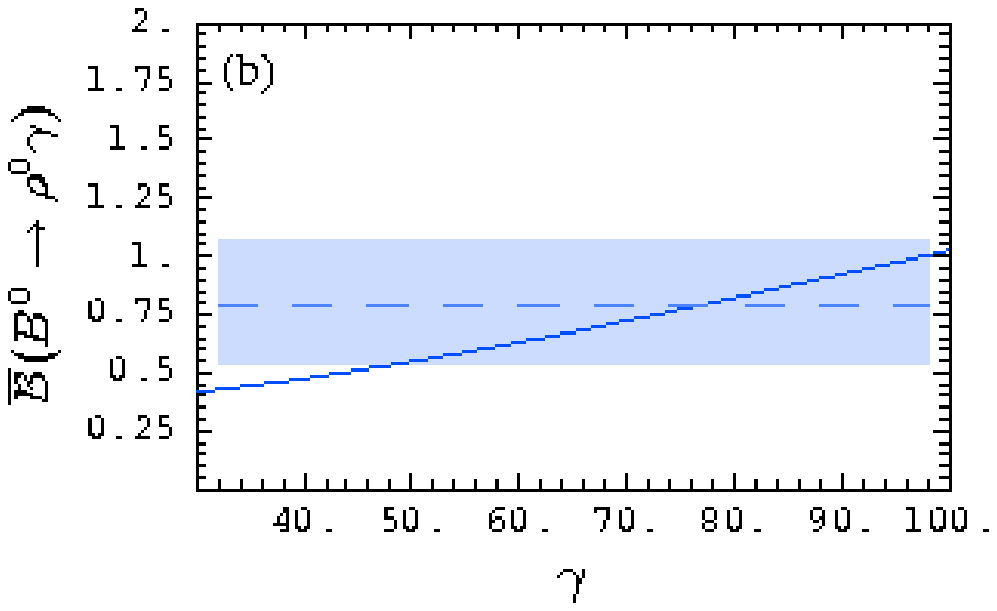}$$
$$ \epsfxsize=0.48\textwidth\epsffile{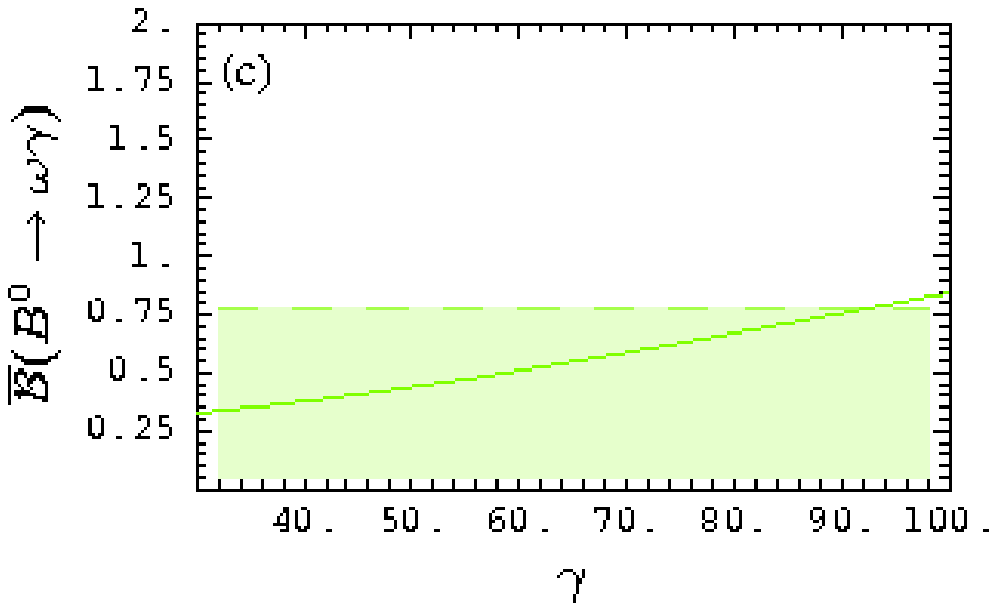}$$
  \caption[CP-averaged branching ratios of $B\to(\rho,\omega)\gamma$ as function   of angle $\gamma$.]{\small CP-averaged branching ratios of $B\to(\rho,\omega)\gamma$ as function   of UT angle $\gamma$, using the effective form factors and central values of   other input parameters. (a): $B^\pm \to \rho^\pm\gamma$, (b):   $B^0\to \rho^0\gamma$, (c): $B^0\to\omega\gamma$. The boxes indicate the 1$\sigma$ experimental results from {\sc BaBar} \cite{babar_rad}, Tab.~\ref{rad_tab1}.  Note that the resulting value of $\gamma$   from the average of all three channels is $\gamma =  (61.0^{+13.5}_{-16.0}({\rm exp})^{+8.9}_{-9.2})^\circ$ -- see Section~\ref{ckmextract}.} 
  \label{rad_fig3}
\end{figure}

We would like to close this section by making explicit the  dependence of the three $B\to (\rho,\omega)\gamma$ branching ratios on $\gamma$.  In Fig.~\ref{rad_fig3}  we plot these branching ratios, for central values of the input parameters, as functions of $\gamma$. We also indicate the present experimental results from {\sc BaBar} \cite{babar_rad}, Tab.~\ref{rad_tab1}, within their 1$\sigma$ uncertainty. 

\subsection{Isospin Asymmetries}\label{isosec}
\begin{figure}[tb]
$$\epsfxsize=0.45\textwidth\epsffile{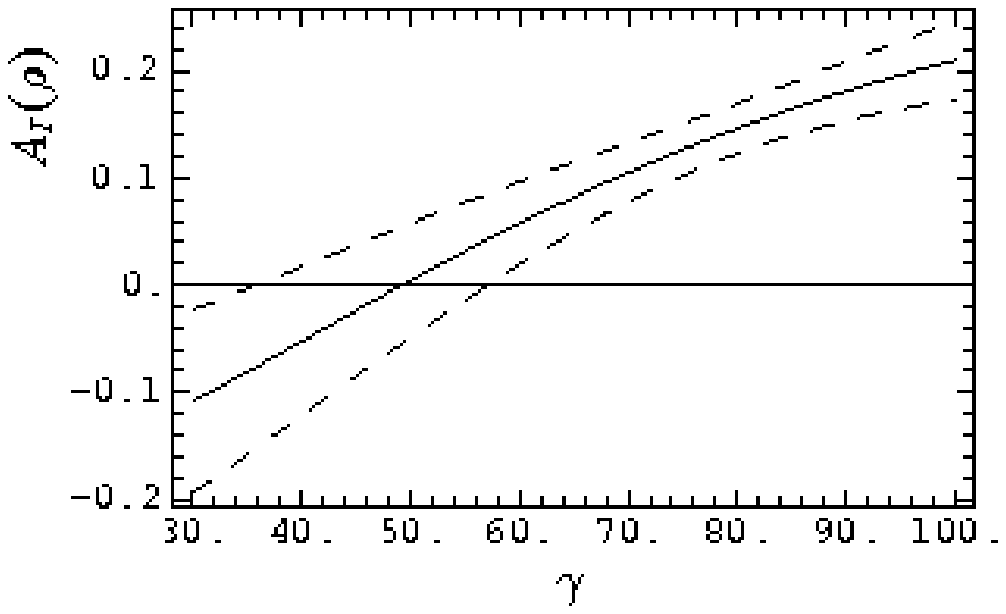}\quad
  \epsfxsize=0.45\textwidth\epsffile{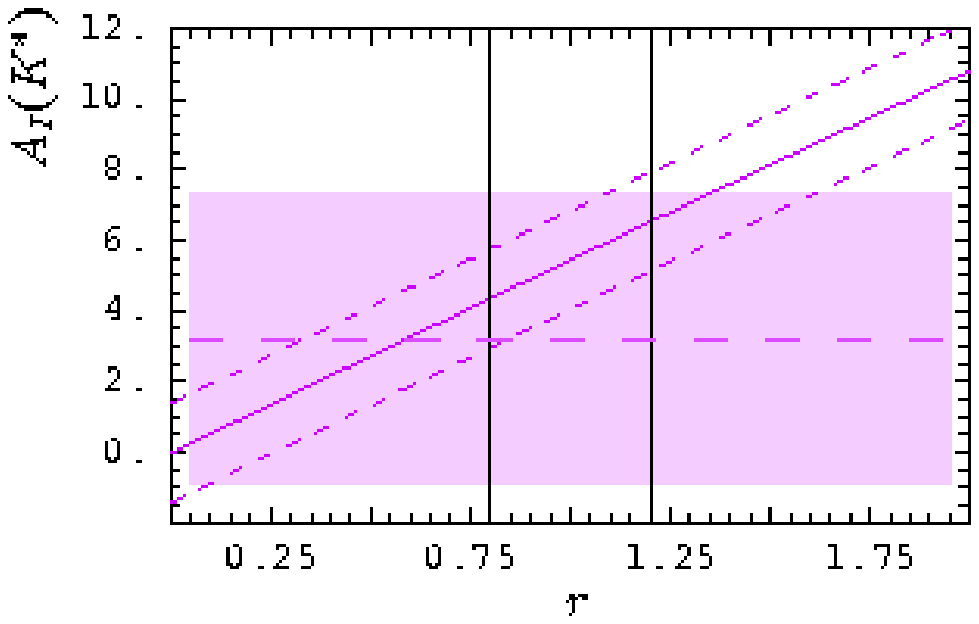}$$
\caption[$A_I(\rho)$  as function of the angle $\gamma$ and $A_I(K^*)$ as function of $r\equiv a_6/a_6^{\rm     SM}$.]{\small Left panel: isospin asymmetry $A_I(\rho)$, Eq.~(\ref{rad_air}),  as function of the UT angle $\gamma$. Solid line: central values of input parameters; dashed lines: theoretical uncertainty. Right panel: $A_I(K^*)$, Eq.~(\ref{rad_aik}), in percent,  as function of the ratio $r\equiv a_6/a_6^{\rm     SM}$ of the combination of penguin Wilson coefficients   $a_6\equiv C_6+C_5/3$. Solid line: central value of input parameters,  dashed lines: theoretical uncertainty. The box indicates the  present experimental uncertainty and the straight black lines the  theory uncertainty in $r$.}
    \label{rad_fig4}
\end{figure}

The asymmetries $A_I(\rho)$, $A_I(K^*)$, and   $A(\rho,\omega) $ are given by
\begin{eqnarray}
A(\rho,\omega) & = &  \frac{\overline{\Gamma}(B^0\to \omega   \gamma)}{\overline{\Gamma}(B^0\to \rho^0   \gamma)}-1\,, 
  \label{rad_arw} \\
A_{I}(\rho) & = & \frac{2\overline{\Gamma}(\bar B^0\to    \rho^0 \gamma)}{\overline{\Gamma}(\bar B^\pm\to    \rho^\pm \gamma)} - 1\,,
  \label{rad_air}\\
A_{I}(K^*) & = & \frac{\overline{\Gamma}(\bar B^0\to    K^{*0} \gamma) - \overline{\Gamma}(B^\pm\to 
  K^{*\pm} \gamma)}{\overline{\Gamma}(\bar B^0\to    K^{*0} \gamma) + \overline{\Gamma}(B^\pm\to 
  K^{*\pm} \gamma)}\,,
  \label{rad_aik}
\end{eqnarray}
where the partial decay rates are CP-averaged. Let us first discuss $A(\rho,\omega)$ and $A_{I}(\rho)$ which are  relevant for the experimental determination of $\overline{\cal  B}(B\to(\rho,\omega)\gamma)$,  which in turn is used for the determination of $|V_{td}/V_{ts}|$ (or $\gamma$), see Section~\ref{ckmextract}. The present experimental statistics for $b\to d\gamma$ transitions is rather low, so the experimental value of $\overline{\cal B}(B\to(\rho,\omega)\gamma)$ is obtained under the explicit assumption of perfect symmetry, i.e.\ $\overline{\Gamma}(B^\pm\to \rho^\pm \gamma) = 2 \overline{\Gamma}(B^0\to \rho^0 \gamma) = 2 \overline{\Gamma}(B^0\to \omega \gamma)$. In reality, the symmetry between $\rho^0$ and $\omega$ is broken by different values of the form factors, and isospin symmetry between neutral and charged $\rho$ is broken by photon emission from the spectator quark, the dominant mechanism of which is WA. From the formulas for individual branching ratios, Eq.~(\ref{BR}),  and the various contributions to the factorisation coefficients $a_{7L(R)}^U$, we find
\begin{equation}
A(\rho,\omega) = -0.20\pm \overbrace{0.09}^{{\rm th.}}\,.
\label{eq:AIrw}
\end{equation}
The uncertainty is dominated by that of the  form factor ratio  $T_1^{B\to\omega}(0)/T_1^{B\to\rho}(0)=0.90\pm 0.05$.\footnote{Note that this result is dominated by the ratio of  decay constants given in Tab.~\ref{rad_tab8} and discussed in Ref.~\cite{Ball:2006eu}. The experimental results entering these averages have a  large spread which may cast a shadow of doubt on the averaged final  branching ratios for $(\rho^0,\omega)\to e^+ e^-$ quoted by PDG  \cite{Yao:2006px}.} The dependence on all other input parameters is marginal. The result (\ref{eq:AIrw}) is not compatible with the strict isospin limit $A(\rho,\omega) =0$.  $A_{I}(\rho)$, on the other hand, is very sensitive to $\gamma$, whereas the form factors drop out. It is driven by the WA contribution and, in the QCDF framework, vanishes if WA is set to zero. In Fig.~\ref{rad_fig4}(a) we plot $A_{I}(\rho)$ as  function of $\gamma$, including the theoretical uncertainties. 
\begin{table}
\renewcommand{\arraystretch}{1.3}
\addtolength{\arraycolsep}{3pt}
$$
\begin{array}{c||c|c|c|c}
\gamma & 40^\circ & 50^\circ & 60^\circ & 70^\circ\\\hline
A_I(\rho) & -(5.3\pm 6.9)\% & (0.4\pm 5.3)\% & (5.7\pm 3.9)\% &
(10.5\pm 2.7)\%
\end{array}
$$
\caption[Isospin asymmetry $A_I(\rho)$ for different values  of $\gamma$.]{\small Isospin asymmetry $A_I(\rho)$  for different values  of $\gamma$.}
  \label{rad_tab7}
\end{table}
As suggested by the findings of Ref.~\cite{chamonix}, these results are  not expected to change considerably upon inclusion of  the non-factorisable radiative corrections of  Fig.~\ref{rad_fig2}(c). In Tab.~\ref{rad_tab7}, we give the corresponding  results for several values of $\gamma$, together with the theoretical  uncertainty. Our result agrees very well with that  obtained by the {\sc BaBar} collaboration: $A_I(\rho)_{\rm BaBar} =  0.56\pm 0.66$ \cite{babar_rad}.

$A_I(K^*)$ was first discussed in Ref.~\cite{Kagan:2001zk}, including power-suppressed $\mathcal{O}(\alpha_s)$ corrections which unfortunately violate QCDF, i.e.\ are divergent. It is for this reason that we decide to drop these corrections and include only leading-order terms in $\alpha_s$. We then find
\begin{eqnarray}
A_I(K^*) &=& (5.4\pm \overbrace{1.0}^{\mu} \pm \overbrace{0.6}^{{\rm NLO}\leftrightarrow{\rm
  LO}} \pm \overbrace{0.6}^{f_B} \pm \overbrace{0.6}^{{\rm other}})\%\nonumber\\
&=& (5.4\pm 1.4)\%\,,
\label{62}
\end{eqnarray}
where ${\rm NLO}\leftrightarrow{\rm LO}$ denotes the uncertainty induced by switching from NLO to LO accuracy in the Wilson coefficients and ``other'' summarises all other sources of theoretical uncertainty. 
As can be inferred from the entries in Tab.~\ref{rad_tab1}, the  present experimental result is $A_I(K^*)_{\rm exp}=(3.2\pm 4.1)\%$.  In Ref.~\cite{Kagan:2001zk} it was pointed out that $A_I(K^*)$
is very sensitive to the values of the Wilson coefficients $C_{5,6}^{\rm BBL}$ in the combination $a_6\equiv C_{5}^{\rm BBL}+C_6^{\rm   BBL}/3$. In the SM, varying the renormalisation scale as $\mu=m_b(m_b)\pm 1\,{\rm GeV}$ and switching between LO and NLO accuracy for the Wilson coefficients, one has $a_6= -0.039\pm 0.008$, which actually induces the bulk of the uncertainty in Eq.~(\ref{62}). In Fig.~\ref{rad_fig4}(b) we plot $A_I(K^*)$ as function of $a_6/a_6^{\rm SM}$, with $a_6^{\rm SM}=-0.039$.
The figure clearly indicates that, although there is presently no
discrepancy between theoretical prediction and experimental result,
a reduction of the experimental uncertainty
of $A_I(K^*)$ may well reveal some footprints of new physics
in this observable.

\subsection{CP Asymmetries}\label{cpsec}
The time-dependent CP asymmetry in $\bar B^0\to V^0\gamma$ is given analogously to Eq.~(\ref{basics_eq13}) as
\begin{equation}
A_{CP}(t)  = S(V\gamma) \sin(\Delta m_D\, t ) - C(V\gamma) 
\cos(\Delta m_D\, t)\,.
\label{rad_cpa}
\end{equation}
The above equation is technically only valid for $\Delta \Gamma =0$ and while this is a good assumption for $B^0_d$ decays, it is not so for $B_s^0$ decays. Although Eq.~(\ref{rad_cpa})  can easily be adapted to non-zero $\Delta \Gamma_s$ we refrain from doing so: the whole point in calculating the CP asymmetry is not so much to give precise predictions for $S$ and $C$, but rather to exclude the possibility of large corrections to the naive expectation $S\sim m_D/m_b$. With this is mind, small corrections from a non-zero $\Delta\Gamma_s$ are irrelevant. The time-dependent CP asymmetries are given in terms of the left- and right-handed photon amplitudes (\ref{qcdf_8}) by
\begin{equation}
S(V\gamma) 
 =  \frac{2 \,{\rm Im}\,\left(\frac{q}{p}({\cal A}_L^* \bar{\cal A}_L + 
                                       {\cal A}_R^* \bar{\cal A}_R)\right)}{
        |{\cal A}_L|^2 + |{\cal A}_R|^2 + |\bar{\cal A}_L|^2 + |\bar{\cal
                                       A}_R|^2}\,,
\quad
C(V\gamma)  =   \frac{|{\cal A}_L|^2 + |{\cal A}_R|^2 - |\bar{\cal A}_L|^2 - 
               |\bar{\cal A}_R|^2}{
        |{\cal A}_L|^2 + |{\cal A}_R|^2 + |\bar{\cal A}_L|^2 + |\bar{\cal
                                       A}_R|^2}\,.
\label{54}
\end{equation}
With ${\cal A}_{L,R}$ and $\bar{\cal A}_{L,R}$ as given in (\ref{qcdf_11}). The indirect CP asymmetry $S(V\gamma)$ relies on the interference of both left- and right-helicity amplitudes and vanishes if one of them is absent; it thus probes indirectly the photon helicity. The direct CP asymmetry
$C(V\gamma)$ is less sensitive to $\bar{\cal A}_R$, but very sensitive to the    strong phase of $\bar{\cal A}_L$ and vanishes if the radiative    corrections to $a_{7L}^{U,{\rm QCDF}}$, Eq.~(\ref{10}), are
     neglected. As the accuracy of the prediction of strong phases in      QCDF is subject to discussion, and in any case $C(V\gamma)$ is      less sensitive to new physics than $S(V\gamma)$, we shall
  not consider direct CP asymmetries in this analysis.

Let us briefly discuss the reason for the expected smallness of $S$. In the process $b\to D\gamma$, in the SM, the emitted photon is  predominantly  left-handed in $b$, and right-handed in $\bar b$ decays. This  is due to the fact that the dominant contribution to the  amplitude comes from the chiral-odd dipole operator $Q_7$. As only left-handed quarks participate in the weak  interaction, an effective operator of this type necessitates, in the SM, a helicity flip on one of the external quark lines, which results in a factor $m_b$ (and a left-handed photon) in $b_R\to D_L\gamma_L$ and a factor $m_D$ (and a right-handed photon) in $b_L\to D_R\gamma_R$. Hence, the emission of right-handed photons is suppressed by  a factor $m_D/m_b$, which leads to the QCDF prediction (\ref{qcdf_13}) for $a_{7R}^U$.  The interesting point is not the smallness of the CP asymmetry {\em   per se}, but the fact that the helicity suppression can easily be alleviated  in a large number of new physics scenarios where the spin flip occurs on an internal line, resulting in a factor $m_i/m_b$ instead of $m_D/m_b$. A prime example is left-right symmetric models \cite{LRS}, whose impact on the photon polarisation was discussed in
Refs.~\cite{alt, grin04,grin05}.  These models also come in a supersymmetric version whose effect on $b\to s\gamma$ was investigated in Ref.~\cite{frank}. Supersymmetry with no left-right symmetry can also provide large  contributions to $b\to D\gamma_R$, see Ref.~\cite{susy} for recent studies. Other
potential sources of large effects  are warped extra dimensions \cite{warped} or anomalous right-handed top couplings \cite{anomalous}.  Unless the amplitude for $b\to D\gamma_R$ is of the same order as the SM prediction for $b\to D \gamma_L$, or the enhancement of $b\to D \gamma_R$ goes along with a suppression of $b\to D \gamma_L$, the impact on the branching ratio is small,  as the two helicity amplitudes add incoherently. This implies there can be a substantial contribution of new physics to $b\to D\gamma$ escaping detection when only branching ratios are measured.

We can calculate $S$ directly from (\ref{54}) and obtain, making explicit the contributions from  different sources:
\begin{equation}
\renewcommand{\arraystretch}{1.3}
\begin{array}[b]{rll}
S(\rho\gamma) = 
& \phantom{-}(\overbrace{~0.01~}^{m_D/m_b}+\overbrace{~0.02~}^{\rm
  LD~WA}+\overbrace{~0.20~}^{\shortstack{\footnotesize{soft} \\  \footnotesize{gluons}}}~\pm~ 1.6)\%
& = \phantom{-}(0.2\pm 1.6)\%\,,\\
S(\omega\gamma) =
& \phantom{-}(0.01-0.08+0.22\pm 1.7)\% 
& = \phantom{-}(0.1\pm 1.7)\%\,,\\
S(K^*\gamma) =
& -(2.9-0+0.6\pm 1.6)\%
& = -(2.3\pm 1.6)\%\,,\\
S(\bar K^*\gamma) =
& \phantom{-}(0.12+0.03+0.11\pm 1.3)\%
& = \phantom{-}(0.3\pm 1.3)\%\,,\\
S(\phi\gamma) =
& \phantom{-}(0+0+5.3\pm8.2)\times 10^{-2}\,\%
& =  \phantom{-}(0.1\pm 0.1)\%\,.
\label{eq:SVgamma}
\end{array}
\end{equation}
Including only the helicity-suppressed contribution, one expects, for $B\to K^*\gamma$, neglecting the doubly Cabibbo suppressed amplitude in $\lambda_u^{(s)}$
\begin{equation}
\left.S(K^*\gamma)\right|_{\mbox{\footnotesize no soft gluons}} 
 =  -2\, \frac{m_s}{m_b}\,\sin\,\phi_d\approx -2.7\%\,.\label{75}
\end{equation}
For $B_s\to\phi\gamma$, one expects the CP asymmetry to vanish if the decay amplitude is proportional to $\lambda_t^{(s)}$, which, at tree-level, precludes any contributions of type $\sin(\phi_s) m_s/m_b$ and also any contribution from WA. This is     because the mixing angle $\phi_s$ is given by ${\rm
 arg}[(\lambda_t^{(s)})^2]$, Eq.~(\ref{basics_eq12}),  and the interference of amplitudes in    (\ref{54}) also yields a factor $(\lambda_t^{(s)})^2$,  if the individual amplitudes are proportional to  $\lambda_t^{(s)}$ or 
$(\lambda_t^{(s)})^*$, respectively; this is indeed the case for the helicity-suppressed term $m_s/m_b$ induced by the operator $Q_7$ and the WA contributions to $a_{7R}^U(\phi)$,
Eqs.~(\ref{15}) and (\ref{20A}), so that the phases cancel in (\ref{54}).

The actual results in (\ref{eq:SVgamma}) disagree with the above
expectations because of the contributions from soft-gluon emission,
which enter $a_{7R}^U$. Moreover, for $S(\phi\gamma)$ this is because the
soft-gluon emission from quark loops is different for $u$ and $c$ loops so that $a_{7R}^c\neq a_{7R}^u$ and hence  $\bar {\cal A}_{R}$ (${\cal A}_{L}$) is not proportional to   $\lambda_t^{(s)}$ ($(\lambda_t^{(s)})^*$). Note that a substantial enhancement of $S(\phi\gamma)$ by new physics requires
not only an enhancement of $|\bar{\cal A}_R|$ (and $|{\cal
  A}_L|$), but also the presence of a large phase in (\ref{54}); 
this could be  either
a large $B_s$ mixing phase which will also manifest itself in
a sizable CP violation in, for instance, $B_s\to J/\psi \phi$, see
Ref.~\cite{BF06,Ball:2006xx}; or it could be a new weak phase in $\bar{\cal A}_{R}$
(and ${\cal A}_L$); or it could be a non-zero strong phase in
one of the $a_{7R}^{c,u}$ coefficients. Based on the light quark loop results there is not much scope for
a large phase in $a_{7R}^{u}$ (whose contribution is, in addition,
doubly Cabibbo suppressed), but the situation could be different for
$a_{7R}^{c,{\rm soft}}$, where only the
leading-order term in a $1/m_c$ expansion is included, which does not carry a
complex phase \cite{Ball:2006eu}. It is not excluded that a
resummation of higher-order terms in this expansion will generate a
non-negligible strong phase --- which is not really relevant for our
results in Eq.~(\ref{eq:SVgamma}), but could be relevant for the
interpretation of any new physics to be found in that observable. For
$S(K^*\gamma)$, on the other hand, no new phases are required, and
any enhancement of $ |\bar{\cal A}_R|$ (and $|{\cal A}_L|$) by new physics will
result in a larger value of $S(K^*\gamma)$.

For all $S$ except $S(K^*\gamma)$,
the uncertainty is entirely dominated by that of the soft-gluon emission
terms $l_{u,c}-\tilde l_{u,c}$, whose uncertainties have been doubled with
respect to those given in Tab.~\ref{rad_tab5}. The smallness of
$S((\rho,\omega)\gamma)$ is due to the fact that the helicity
factor is given by $m_d/m_b$ (we use $m_{u,d}/m_s = 1/24.4$ from
ChPT). For $\bar K^*$,
the suppression from the small mixing
angle is relieved by the fact that both weak amplitudes in
$\lambda_U^{(d)}$ contribute
so that the CP asymmetry is comparable
with that of $\rho$ and $\omega$. Despite the generous uncertainties, it is
obvious that none of these CP symmetries is larger than
4\% in the SM, which makes these observables very interesting
for new physics searches. The present experimental result from the $B$
factories, $S(K^*\gamma)=-0.28\pm 0.26$ \cite{Barberio:2007cr}, certainly encourages
the hope that new physics may manifest itself in that
observable. While a measurement of the $b\to d$ CP asymmetries is
probably very difficult even at a super-flavour factory,
$S(K^*\gamma)$ is a promising observable for $B$ factories \cite{superB}, but
not for the LHC.\footnote{$K^*$ has to be traced via its decay into a CP
eigenstate, i.e.\ $K_S\pi^0$. Neutrals in the final state are not
really LHC's favourites.} $B_s\to \phi(\to K^+K^-)\gamma$, on the
other hand, will be studied in detail at the LHC, and in particular at
LHCb, and any largely enhanced value of $S(\phi\gamma)$ 
will be measured within the first years of running. 

\section{Extraction Of CKM Parameters}\label{ckmextract}

Let us now turn to the determination of CKM parameters from the
branching ratios determined in Section~\ref{rad_brs}. In this context, two particularly interesting
observables are 
\begin{equation}\label{58}
R_{\rho/\omega}\equiv\frac{\overline{\cal B}(B\to (\rho,\omega)\gamma)}{
\overline{\cal B}(B\to K^*\gamma)}\,,\qquad
R_{\rho}\equiv\frac{\overline{\cal B}(B\to \rho\gamma)}{
\overline{\cal B}(B\to K^*\gamma)}\,,\qquad
\end{equation}
given in terms of the CP- and isospin-averaged branching ratios of
$B\to(\rho,\omega)\gamma$ and $B\to \rho\gamma$, respectively, 
and $B\to K^*\gamma$ decays. $R_{\rho/\omega}$
has been measured by both {\sc BaBar} and Belle \cite{babar_rad,belle_rad}, a first
value of $R_\rho$ has been given by {\sc BaBar} \cite{babar_rad}. 
The experimental determinations
actually assume exact isospin symmetry, i.e.\
$\overline{\Gamma}(B^\pm\to \rho^\pm\gamma) \equiv
2 \overline{\Gamma}(B^0\to \rho^0\gamma)$, and also
$\overline{\Gamma}(B^0\to \rho^0\gamma) \equiv 
\overline{\Gamma}(B^0\to \omega\gamma)$; and as we have seen in Section~\ref{isosec}, these relations are not in fact exact. Hence, the present
experimental results for $R_{\rho/\omega}$ are theory-contaminated. 
As the isospin asymmetry between the charged and neutral $\rho$ decay 
rates turns out to be smaller than the asymmetry
 between $\rho^0$ and $\omega$, 
it would actually be preferable, from an experimental point of view,
to drop the $\omega$ channel and
measure $R_\rho$ instead of $R_{\rho/\omega}$, as done in the most
recent {\sc BaBar} analysis on that topic \cite{babar_rad}. 
We will give numerical results and
theory uncertainties for both $R_{\rho/\omega}$ and $R_{\rho}$.

One parametrisation of $R_{\rho/\omega}$ often quoted, 
in particular in experimental papers, is
\begin{equation}\label{Brat}
R_{\rho/\omega} = \left|\frac{V_{td}}{V_{ts}}\right|^2
\left(\frac{1-m_{\rho}^2/m_B^2}{1-m_{K^*}^2/m_B^2}\right)^3
\frac{1}{\xi^2_{\rho}} \left [ 1 +  \Delta R\right],
\end{equation}
with $\Delta R=0.1\pm 0.1$ \cite{BVga1} and again assuming isospin
symmetry for $\rho$ and $\omega$. This parametrisation creates the impression
that $\Delta R$ is a quantity completely unrelated to
and with a fixed value independent of
$|V_{td}/V_{ts}|$. We would like to point out here that this 
impression is {\em wrong}: $\Delta R$ contains both QCD
(factorisable and non-factorisable) effects and such from weak
interactions. In Ref.~\cite{Ball:2006nr}  $\Delta R$ is expressed in
terms of the factorisation coefficients $a_{7L}^U$, assuming isospin
symmetry for $\rho^0$ and $\omega$, as
\begin{eqnarray}
1+\Delta R & = & \left|
  \frac{a_{7L}^c(\rho)}{a_{7L}^c(K^*)}\right|^2 \left( 1 +
  {\rm Re}\,(\delta a_\pm + \delta a_0) \left[\frac{R_b^2 - R_b
  \cos\gamma}{1-2 R_b \cos\gamma + R_b^2}\right]\right.\nonumber\\
& & \left. + \frac{1}{2}\left( |\delta a_\pm|^2 + |\delta a_0|^2\right)
  \left\{ \frac{R_b^2}{1-2 R_b \cos\gamma + R_b^2}\right\} \right)
\label{delR}
\end{eqnarray}
with $\delta a_{0,\pm}=
a_{7L}^u(\rho^{0,\pm})/a_{7L}^c(\rho^{0,\pm})-1$. Eq.~(\ref{delR}) shows explicitly that $\Delta R$ depends both on QCD
 ($\delta a_{\pm,0}$) and CKM parameters ($R_b,\gamma$).
The point we would like to make is that the calculation of $\Delta R$
requires input values for $R_b$ and $\gamma$. Once these parameters
(and the Wolfenstein parameter $\lambda$)
are fixed, however, $|V_{td}/V_{ts}|$ is also fixed and given by
\begin{equation}\label{61}
\left|\frac{V_{td}}{V_{ts}}\right| = \lambda \sqrt{1-2 R_b \cos\gamma
  + R_b^2} \left[ 1 + \frac{1}{2}\,( 1 - 2 R_b \cos\gamma) \lambda^2 +
  \mathcal{O}(\lambda^4)\right]\,.
\end{equation} 
Hence, as $|V_{td}/V_{ts}|$ and $(R_b,\gamma)$ are not independent
of each other, 
it is {\em impossible} to extract $|V_{td}/V_{ts}|$ from (\ref{Brat})
with a fixed value of $\Delta R$. Of course $R_{\rho/\omega}$ and $R_\rho$ of (\ref{58}) 
{\em can} be used to extract information
about CKM parameters, but in order to do so one has to settle for a set
of truly independent parameters. Based on (\ref{61}), one can
exchange, say, $\gamma$ for $|V_{td}/V_{ts}|$.\footnote{Strictly speaking, (\ref{61}) only
  fixes $\cos\gamma$ as function of $|V_{td}/V_{ts}|$, leaving
  a twofold degeneracy of $\gamma$. Eq.~(\ref{delR}), however, only
  depends on $\cos\gamma$, so that indeed one can unambiguously 
replace $\gamma$ by $|V_{td}/V_{ts}|$.}  So we can either consider $R_V$ as a
function of the CKM parameters $R_b$ and $\gamma$ (let us call this
the $\gamma$ set of parameters) or as a function of $R_b$ and
$|V_{td}/V_{ts}|$ (to be called the $|V_{tx}|$ set). Using the
$\gamma$ set, a measurement of $R_V(\gamma,R_b)$ allows a
determination of $\gamma$, whereas 
$R_V(|V_{td}/V_{ts}|,R_b)$ allows the
determination of $|V_{td}/V_{ts}|$. 
In either case, the simple quadratic relation
(\ref{Brat}) between $R_V$ and $|V_{td}/V_{ts}|$ becomes more
complicated.

\begin{figure}
$$
\epsfxsize=0.45\textwidth\epsffile{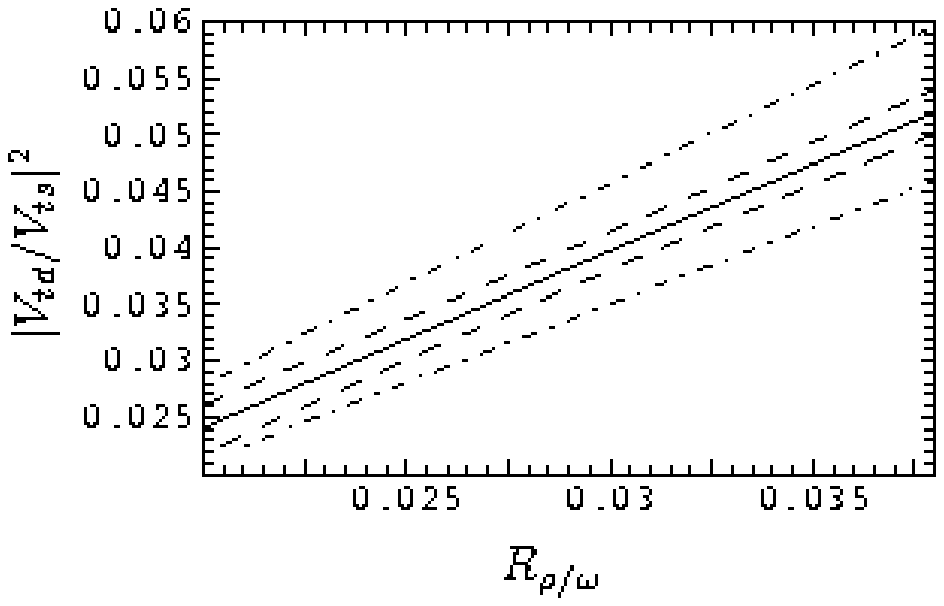}\qquad\epsfxsize=0.45\textwidth\epsffile{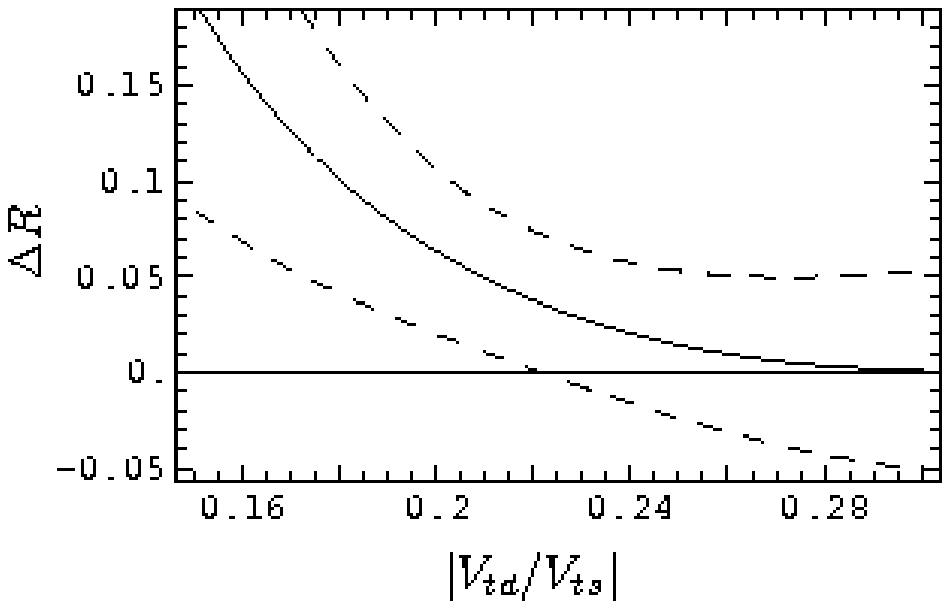}
$$
\caption[$|V_{td}/V_{ts}|^2$ as function of   $R_{\rho/\omega}$ and $\Delta R$ as function of   $|V_{td}/V_{ts}|$.]{\small Left panel: $|V_{td}/V_{ts}|^2$ as function of   $R_{\rho/\omega}$, Eq.~(\ref{58}),   in the $|V_{tx}|$ basis -- see   text.  Solid line: central values. Dash-dotted lines: theoretical   uncertainty induced by $\xi_\rho = 1.17\pm 0.09$, (\ref{xirho}).  Dashed lines: other   theoretical uncertainties, including those induced   by $|V_{ub}|$, $|V_{cb}|$ and the hadronic parameters of   Tab.~\ref{rad_tab8}. Right panel: $\Delta R$ from   Eq.~(\ref{delR}) as function of
  $|V_{td}/V_{ts}|$   in the $|V_{tx}|$ basis. Solid   line: central values. Dashed   lines: theoretical uncertainty.}\label{rad_fig5} 
  $$\epsfxsize=0.45\textwidth\epsffile{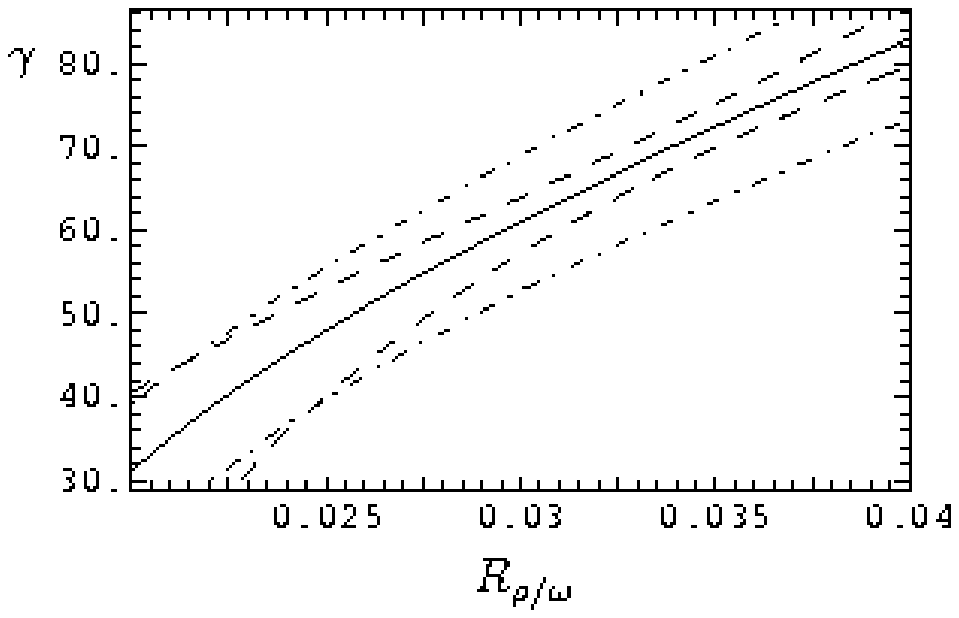}$$
\caption[The UT angle $\gamma$ as function of   $R_{\rho/\omega}$.]{\small The UTangle $\gamma$ as function of   $R_{\rho/\omega}$ in the $\gamma$ set of CKM parameters. Solid   lines: central values of input parameters. Dash-dotted lines: theoretical   uncertainty induced by $\xi_\rho = 1.17\pm 0.09$. Dashed lines: other   theoretical uncertainties.}\label{rad_fig6}
$$\epsfxsize=0.45\textwidth\epsffile{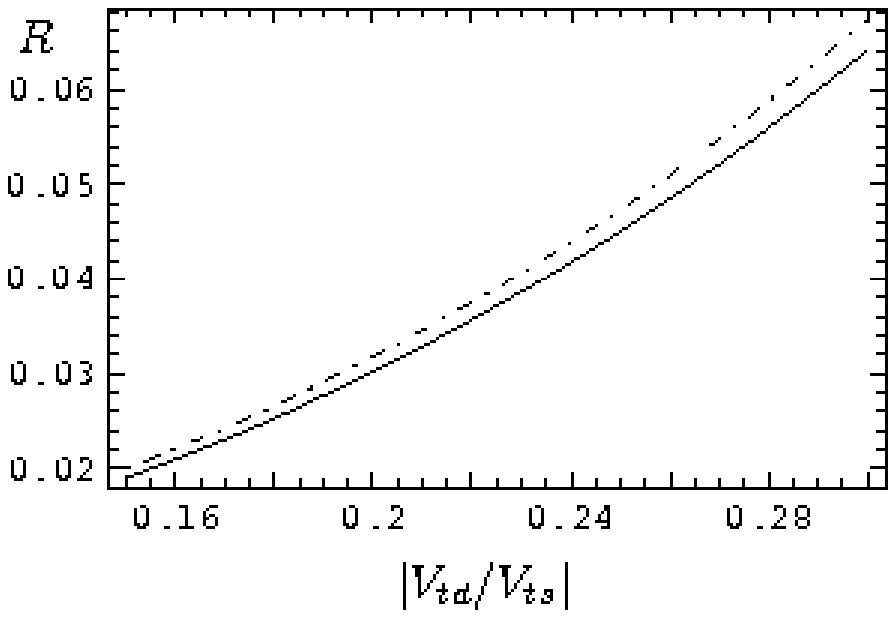}$$
\caption[Central values of  $R_{\rho/\omega}$ and $R_{\rho}$ as functions of $|V_{td}/V_{ts}|$]{\small Central values of  $R_{\rho/\omega}$ (solid line) and $R_{\rho}$   (dash-dotted line) as functions of $|V_{td}/V_{ts}|$.}\label{rad_fig7}
\end{figure}

 In Figs.~\ref{rad_fig5} and \ref{rad_fig6} we plot the resulting values of $|V_{td}/V_{ts}|^2$ and $\gamma$, respectively,  as a  function of $R_V$. Although the curve in Fig.~\ref{rad_fig5}(a) looks like a straight line, as naively expected from (\ref{Brat}), this is not exactly the case, because of the dependence of $\Delta R$ on $|V_{td}/V_{ts}|$. In  Fig.~\ref{rad_fig5}(b) we plot $\Delta R$ for the $|V_{tx}|$ set of parameters. The dependence of $\Delta R$ on $|V_{td}/V_{ts}|$ is rather strong. Apparently indeed $\Delta R=0.1\pm 0.1$ in the expected range $0.16<|V_{td}/V_{ts}|<0.24$, but this estimate does not reflect  the true theoretical uncertainty which is indicated by the dashed lines in the figure. 

It is now basically a matter of choice whether to use
$R_{\rho/\omega}$ to determine $|V_{td}/V_{ts}|$ or $\gamma$. Once one of
these parameters is known, the other one follows from Eq.~(\ref{61}). In
Fig.~\ref{rad_fig6} we plot $\gamma$ as a function of
$R_{\rho/\omega}$, together with the theoretical uncertainties. In
Fig.~\ref{rad_fig7} we also compare the central values of
$R_{\rho/\omega}$ with those of $R_{\rho}$, as a function of
$|V_{td}/V_{ts}|$. Although the difference is small, $R_{\rho}$ is
expected to be larger than $R_{\rho/\omega}$.  $R_{\rho/\omega}$ and $R_{\rho}$ are dominated by the  uncertainties of $\xi_{\rho}$ and as discussed in Ref.~\cite{Ball:2006nr}, a reduction of this uncertainty  would require a reduction of the uncertainty of the transverse decay constants $f_V^\perp$ of $\rho$ and $K^*$. With the most recent results from {\sc BaBar}, $R_{\rho/\omega} = 0.030\pm 0.006$ \cite{babar_rad}, and from Belle, $R_{\rho/\omega} = 0.032\pm 0.008$ \cite{belle_rad}, we then find
\begin{equation}
\renewcommand{\arraystretch}{1.3}
\begin{array}[b]{l@{\quad}l@{\quad\leftrightarrow\quad}l}
\mbox{{\sc BaBar}:} & \displaystyle
\left|\frac{V_{td}}{V_{ts}}\right| = 0.199\overbrace{^{+0.022}_{-0.025}}^{{\rm exp}}\pm
\overbrace{0.014}^{{\rm th}} &\displaystyle
\gamma = (61.0\overbrace{^{+13.5}_{-16.0}}^{{\rm exp}}\overbrace{^{+8.9}_{-9.3}}^{
{\rm th}})^\circ\,,\\[10pt]
\mbox{Belle:} & \displaystyle
\left|\frac{V_{td}}{V_{ts}}\right| = 0.207\,^{+0.028}_{-0.033}\,
^{+0.014}_{-0.015} &\displaystyle
\gamma = (65.7\,^{+17.3}_{-20.7}\,^{+8.9}_{-9.2})^\circ\,.
\end{array}
\label{63}
\end{equation}
These numbers  compare well with the Belle result  \cite{Bellegamma} 
from tree-level processes, $\gamma=(53\pm 20)^\circ$, quoted in
Tab.~\ref{rad_tab8}, and results from  global fits
\cite{global}. We also would like to point out that the above
determination of $\gamma$ is actually a determination of
$\cos\gamma$, via Eq.~(\ref{61}), and implies, in principle, a twofold
degeneracy $\gamma\leftrightarrow 2\pi-\gamma$. This is in contrast to the
determination from $B\to D^{(*)} K^{(*)}$ in \cite{Bellegamma}, which
carries a twofold degeneracy
$\gamma \leftrightarrow \pi+\gamma$. Obviously these two
determinations taken together remove the degeneracy and 
select $\gamma\approx 55^\circ<180^\circ$. If 
$\gamma\approx 55^\circ+180^\circ$ instead, one would have 
$|V_{td}/V_{ts}|\approx 0.29$ from
(\ref{61}), which is definitely ruled out by data. Hence, the result
(\ref{63}) confirms the SM interpretation of $\gamma$ from 
the tree-level CP asymmetries in $B\to D^{(*)} K^{(*)}$.

\begin{table}[ht]
$$
\begin{array}{c|c|c}
\tau_{B^0} & \tau_{B^\pm}/\tau_{B^0} & \tau_{B_s^0}/\tau_{B^0}
\\\hline
1.530(9)\,{\rm ps} & 1.071(9) & 0.958(39)
\end{array}
$$
\caption[$B$ lifetimes from HFAG.]{\small $B$ lifetimes from HFAG \cite{Barberio:2007cr}.}
\label{rad_tab9}
\addtolength{\arraycolsep}{3pt}
\renewcommand{\arraystretch}{1.3}
$$
\begin{array}{c|c|c|c|c|c}\hline\hline
\multicolumn{6}{c}{\mbox{CKM parameters and couplings}}\\\hline
\lambda \mbox{~\cite{Yao:2006px}} & |V_{cb}| \mbox{~\cite{inclmoments}} & 
|V_{ub}| & \gamma \mbox{~\cite{Bellegamma}} & \alpha_s(m_Z)
\mbox{~\cite{Yao:2006px}} & \alpha_{\rm QED}\\\hline
0.227(1) & 42.0(7)\times 10^{-3} & 4.0(7)\times
10^{-3} & (53\pm 20)^\circ & 0.1176(20) & 1/137\\\hline\hline
\multicolumn{6}{c}{\mbox{B parameters}}\\\hline
f_{B_q}\mbox{~\cite{Onogi}} & f_{B_s}\mbox{~\cite{Onogi}} &
    \lambda_{B_q}(\mu_h) \mbox{~\cite{Ball:2006nr}} & \lambda_{B_s}(\mu_h)
& \mu_h     \\\hline
200(25)\,{\rm MeV} & 240(30)\,{\rm MeV} & 0.51(12)\,{\rm GeV} &
0.6(2)\,{\rm GeV} & 2.2\,{\rm GeV}\\\hline\hline
\multicolumn{6}{c}{\mbox{$\rho$ parameters}}\\\hline
f_{\rho} & f_{\rho}^\perp & a_1^\perp({\rho}) &
a_2^\perp({\rho}) & T_1^{B\to\rho}(0)\\\hline
216(3)\,{\rm MeV} & 165(9)\,{\rm MeV} & 0 & 0.14(6) & 0.27(4)
\\\hline\hline
\multicolumn{6}{c}{\mbox{$\omega$  parameters}}\\\hline
f_{\omega} & f_{\omega}^\perp & a_1^\perp({\omega}) &
a_2^\perp({\omega}) & T_1^{B\to\omega}(0)\\\hline
187(5)\,{\rm MeV} & 151(9)\,{\rm MeV} & 0 & 0.15(7) & 0.25(4)
\\\hline\hline
\multicolumn{6}{c}{\mbox{$K^*$ parameters}}\\\hline
f_{K^*} & f_{K^*}^\perp & a_1^\perp({K^*})\mbox{~\cite{Ball:2005vx}} & 
a_2^\perp({K^*}) & T_1^{B_q\to K^*}(0) & 
T_1^{B_s\to \bar  K^*}(0)\\\hline
220(5)\,{\rm MeV} & 185(10)\,{\rm MeV} & 0.04(3) & 0.15(10) &
0.31(4) & 0.29(4)\\\hline\hline
\multicolumn{6}{c}{\mbox{$\phi$ parameters}}\\\hline
f_{\phi} & f_{\phi}^\perp & a_1^\perp({\phi}) &
a_2^\perp({\phi}) & T_1^{B_s\to\phi}(0)\\\hline
215(5)\,{\rm MeV} & 186(9)\,{\rm MeV} & 0 & 0.2(2) & 0.31(4) & \\\hline\hline
\multicolumn{6}{c}{\mbox{quark masses}}\\\hline
\multicolumn{2}{c|}{m_s(2\,{\rm GeV})\mbox{~\cite{ms}}} 
& m_b(m_b)\mbox{~\cite{inclmoments}} & 
m_c(m_c)\mbox{~\cite{czakon}} & \multicolumn{2}{c}{
m_t(m_t)\mbox{~\cite{mt}}}\\\hline
\multicolumn{2}{c|}{100(20)\,{\rm MeV}} 
& 4.20(4)\,{\rm GeV} & 1.30(2)\,{\rm GeV} & 
\multicolumn{2}{c}{163.6(2.0)\,{\rm GeV}} \\\hline\hline
\end{array}
$$
\caption[Summary of input parameters.]{\small Summary of input parameters. The  value of $|V_{ub}|$ is an average over inclusive and exclusive  determinations and the result from UTangles Refs.~\cite{Barberio:2007cr,global,Vub}. None of our results is very sensitive  to $|V_{ub}|$. For an explanation of our choice of the value of the UT angle  $\gamma$, see text. $\lambda_{B_s}$ is obtained from   $\lambda_{B_q}$, see Eq.~(\ref{rad_bs}).  The vector meson  decay constants $f_V$, $f_V^\perp$ are discussed in Ref.~\cite{Ball:2006eu}; the  values of the Gegenbauer moments $a_i^\perp$ are compiled from  various sources   \cite{Ball:2006nr,Ball:1996tb,Ball:1998sk,Ball:2003sc} and include only small ${\rm SU(3)_F}$-breaking,  in line with the findings for pseudoscalar mesons \cite{Ball:2006wn}. The form factors $T_1$  are 
updates of previous LCSR results  \cite{Ball:2004rg}, including the updated values of the decay constants
 $f_{\rho,\omega,\phi}$ and of $a_1^\perp({K^*})$ \cite{Ball:2005vx,Ball:2006fz}. 
 All scale-dependent quantities are given at the scale $\mu=1\,$GeV unless stated otherwise.}
\label{rad_tab8}
\end{table}
\chapter{Summary and Conclusions}\label{chapter8_conc}
This thesis has consisted of three main analyses centred on the investigations and determinations of meson light-cone distribution amplitudes.  We have seen how the determinations of decay observables in $B$ decays are reliant on the sound understanding of both theoretical and experimental uncertainties with the work presented in this thesis striving towards the former. To summarise:

We began, in Chapter~\ref{chapter1_basics}, with a brief introduction defining the QCD Lagrangian, discussing CP violation and the $\Delta B =1$ effective Hamiltonian.

In Chapter~\ref{chapter2_DAs} we investigated the structure of vector mesons distribution amplitudes to twist-3 accuracy. We included all $\rm SU(3)_F$-breaking and G-parity violating effects. The QCD equations of motion were implemented to unpick the interwoven relations between the distribution amplitudes ultimately expressing the two-particle twist-3 distribution amplitudes in terms of the three-particle twist-3 and two-particle twist-2 distribution amplitudes. The equations of motion result in integral equations which are readily solved order-by-order in conformal spin and to the order considered all the distribution amplitudes are then expressed by a small number of non-perturbative parameters. Finite quark mass effects appear in the equation of motion and therefore impact the two-particle twist-3 distribution amplitudes (\ref{das_eq31}-\ref{das_eq33}). Such effects also cause mixing between the twist-3 hadronic parameters under renormalisation scale evolution, see Eq.~(\ref{das_eq37}).

In Chapter~\ref{chapter3_SR} we discussed the methods of QCD sum rules (the SVZ method) and QCD sum rules on the light-cone. We outlined the procedures with example correlation functions and ended the chapter with an example calculation of the $\alpha_s$ corrections to the gluon condensate contribution to a $K$ meson sum rule. The calculation made use of the background field technique and served to illustrate the calculation of radiative corrections to -- and extraction of -- vacuum condensates in the SVZ method. The result of the calculation is in conflict with that in the literature, see Eqs.~(\ref{last}) and (\ref{least}).

In Chapter~\ref{chapter4_det}  we determined the leading hadronic parameters defined in Chapter~\ref{chapter2_DAs} via SVZ sum rules. We calculated the three-particle twist-3 parameters to NLO in conformal spin, also including all G-parity violating terms and finite strange quark mass effects. The determination of the twist-3 parameters is new for $K^*$ and $\phi$. The results for the $\rho$ agree within uncertainties  with previous determinations and are presented in Tabs.~\ref{det_tab1} and \ref{det_tab2}. We also calculate $\mathcal{O}(\alpha_s)$ and $\mathcal{O}(m_s^2)$ corrections to the quark condensate for the sum rules for $a_n^{\parallel,\perp}(V)$, which for $n=2$ is the first non-trivial Gegenbauer coefficient of the G-even particles $\rho$ and $\phi$. We add this contribution to the existing sum rules taken from the literature and update the value of $a_2^{\parallel,\perp}(\phi)$ which we find to be consistent with that found for $K^*$ and $\rho$;  $a^\perp_{1,2}(V)=a^\parallel_{1,2}(V)$ within uncertainties. The results find direct application in QCD factorisation descriptions of $B\to V$ decays, and the light-cone sum rule analyses of $B\to V$ transition form factors.

In Chapter~\ref{chapter5_eta} we calculated the form factors of $B\to \etap$
semileptonic transitions from light-cone sum rules,
including the gluonic singlet contributions. We built upon the previous light-cone sum rule determination of the $B\to \eta$ form factor by casting the calculation consistently within the phenomenologically motivated $\eta$-$\etap$ mixing scheme of Refs.~\cite{Feldmann:1998vh,Feldmann:1998sh}. We found that, as
expected, these contributions are more relevant for $f_+^{\etap}$ than
for $f_+^\eta$ and can amount up to 20\% in the former, 
depending on the only poorly
constrained leading Gegenbauer moment $B^g_2$ of the gluonic twist-2
distribution amplitude of $\etap$.  The numerical results, with each contribution listed separately, are given by Eqs.~(\ref{fp1}) and (\ref{fp0}). Consequently, it seems unlikely that the large exclusive $B\to \etap K$ and inclusive $B\to \etap X$ branching ratios can be explained by a large $B^g_2$, as it would have to assume a very extreme value. We also found that the form factors
are sensitive to the values of the twist-2 two-quark Gegenbauer
moments $a_2^{\eta,\etap}$ which, given the uncertainty of independent
determinations, we have set equal to $a_2^\pi$, see Fig.\ref{eta_diags3}.

The ratio
of branching ratios ${\cal B}(B\to\etap e\nu)/{\cal B}(B\to\eta e\nu)$
is sensitive to both $a_2$ and $B^g_2$ and may be used to constrain
these parameters, once it is measured with sufficient accuracy, see Fig.~\ref{eta_fig8}. The
extraction of $|V_{ub}|$ from these semileptonic decays, in particular
$B\to\eta e\nu$, with negligible singlet contribution, although
possible in principle, at the moment is obscured by the lack of
knowledge of $a_2$. We
would also like to stress that, in the framework of the quark-flavour
mixing scheme for the $\eta$-$\etap$ system as used in this analysis,
$B\to \etap$ transitions probe only the $\eta_q$ component of these
particles. The $\eta_s$ component could be probed directly for
instance in the $b\to s$ penguin transition $B_s\to \etap
\ell^+\ell^-$, although such a measurement would also be sensitive
to new physics in the penguin diagrams.

In Chapter~\ref{chapter6_QCDF} we discussed the QCD factorisation (QCDF) approach of Refs.~\cite{Beneke:1999br, Beneke:2000ry} and its application to the radiative $B$ decays $B \to V \gamma$ of Refs.~\cite{Bosch:2001gv,Bosch:2002bw}.   We discussed the appearance of distribution amplitudes in the factorisation formulas and focused on the leading contributions to the $B\to V \gamma$ decays.

In Chapter~\ref{chapter7_rad} we performed a phenomenological analysis of the radiative $B$ decays to vector mesons $B\to V \gamma$, using the framework discussed in Chapter~\ref{chapter6_QCDF}. We investigated the most relevant power-suppressed corrections to the QCDF predictions for the radiative decays $B_{u,d} \to (\rho, \omega, K^*)\gamma$ and  $B_{s} \to (\phi, \bar{K}^*)\gamma$. We use the QCDF framework presented in Refs.~\cite{Bosch:2001gv,Bosch:2002bw} in which we find use for the twist-2 DA parameters determined in Chapter~\ref{chapter4_det}. Besides the leading QCDF contributions we included long-distance photon emission and soft-gluon mission from quark loops. These effects, although formally $\sim 1/m_b$ with respect to the leading contributions, augment the QCDF predictions for the branching ratios, CP and isospin asymmetries. 

The impact of the power-suppressed corrections on the branching ratios is found to be very small, with the exception of the weak annihilation contributions to $B^\pm\to \rho^\pm \gamma$ which are large due to a large combination of Wilson coefficients $C_2+C_1/3=1.02$ and no CKM-suppression. Moreover, long-distance photon emission also impacts most here, see Eq.~(\ref{LDcont}). An explicit break down  of the results are given in Tab.~\ref{rad_tab6}. 

The isospin asymmetries $A(\rho,\omega)$, $A_I(\rho)$ and $A_I(K^*)$  are driven by weak annihilation and long-distance photon emission contributions. We found a non-zero asymmetry $A(\rho,\omega)=-0.20\pm0.09$ which suggests the  explicit assumption of perfect symmetry, i.e.\ $\overline{\Gamma}(B^\pm\to \rho^\pm \gamma) = 2 \overline{\Gamma}(B^0\to \rho^0 \gamma) = 2 \overline{\Gamma}(B^0\to \omega \gamma)$ used to obtain the experimental value of $\overline{\cal B}(B\to(\rho,\omega)\gamma)$ is not so well justified. We found $A_I(\rho)$ to depend strongly on the UT angle $\gamma$, as shown in Tab.~\ref{rad_tab7}. With our central value of $\gamma=53^\circ$ (see Tab~\ref{rad_tab8}) our result agrees very well with the {\sc BaBar} result $A_I(\rho)_{\rm BaBar} =  0.56\pm 0.66$ \cite{babar_rad}. For $A_I(K^*)$ we found a result consistent with the experimental result $A_I(K^*)_{\rm exp}=(3.2\pm4.1)\%$ and, via its sensitivity to the Wilson coefficient combination $C_5+C_6/3$ conclude that a reduction in the experimental uncertainty may uncover signs of new physics contributing to these Wilson coefficients, see Fig.~\ref{rad_fig4}.

The  indirect CP asymmetries $S(V\gamma)$ are caused by the interference between the amplitudes describing the production of left and right-handed photons, see Eqs.~(\ref{qcdf_8}) and (\ref{54}). The right-handed amplitude is suppressed by $m_D/m_b$ with respect to the left-handed one for $\bar B =b \bar q$ decays (and vice versa for $B$ decays). Due to this natural suppression in  the SM we expect the CP asymmetries to be small, and this suppression can be relieved by many new physics senarios. We investigated the soft-gluon effects arising from soft heavy and soft quark loops. The calculation of these contributions makes use of the three-particle twist-3 DA parameters determined in Chapter~\ref{chapter4_det}. They contribute to both the left and right-handed amplitudes, and so may also relieve to SM suppression. We found that although they do divert the results from the values naively expected, there is no scope for a large enhancement due to these power-suppressed contributions. The results are given in Eq.~(\ref{eq:SVgamma}).

Finally, using the most recent results from {\sc BaBar} and Belle,  we extracted the CKM parameter ratio $|V_{td}/V_{ts}|$ and equivalently the UT angle $\gamma$ from the  ratio of branching ratios $R_{\rho/\omega}$. The results are 
\begin{equation}
\renewcommand{\arraystretch}{1.3}
\begin{array}[b]{l@{\quad}l@{\quad\leftrightarrow\quad}l}
\mbox{{\sc BaBar}:} & \displaystyle
\left|\frac{V_{td}}{V_{ts}}\right| = 0.199\overbrace{^{+0.022}_{-0.025}}^{{\rm exp}}\pm
\overbrace{0.014}^{{\rm th}} &\displaystyle
\gamma = (61.0\overbrace{^{+13.5}_{-16.0}}^{{\rm exp}}\overbrace{^{+8.9}_{-9.3}}^{
{\rm th}})^\circ\,,\\[10pt]
\mbox{Belle:} & \displaystyle
\left|\frac{V_{td}}{V_{ts}}\right| = 0.207\,^{+0.028}_{-0.033}\,
^{+0.014}_{-0.015} &\displaystyle
\gamma = (65.7\,^{+17.3}_{-20.7}\,^{+8.9}_{-9.2})^\circ\,.
\end{array}
\end{equation}
and agree well with the Belle result $\gamma=(53\pm20)^\circ$ obtained from tree-level processes, and results from global fits \cite{global}. The result confirms the SM interpretation of $\gamma$ from 
the tree-level CP asymmetries in $B\to D^{(*)} K^{(*)}$.
\appendix
\chapter{Light-cone Co-ordinates}\label{appendixA}
To perform the light-cone expansion one relate the meson's 4-momentum $P_\mu$, polarisation vector $e^{(\lambda)}$ and the coordinate $x_\mu$ to two light-like vectors $p_\mu$ and $z_\mu$. We have the usual relations
\begin{equation}
p^2=0, \hspace{1in} z^2=0\,,
\end{equation}
and
\begin{equation}
P^2=m_{K^*}^2, \hspace{1in} e^{(\lambda)}\cdot e^{(\lambda)} =-1,\hspace{1in} P\cdot e^{(\lambda)} =0,
\end{equation}
so that the limit $m_{K^*}^2\to 0$ gives $p \to P$ and $x^2 \to 0$ gives $z \to x$. From this it follows that
\begin{eqnarray}\label{lccoords1}
  z_\mu &=& x_\mu-P_\mu\,\frac{1}{m_{K^*}^2}\left[x\cdot P -\sqrt{(x\cdot P)^2-x^2m^2_{K^*}}\,\right]
= x_\mu\left[1-\frac{x^2m_{{K^*}}^2}{4(z\cdot p)^2}\right]
-\frac{1}{2}p_\mu\,\frac{x^2}{p\cdot z}+ \mbox{\cal O}(x^4)\,,\nonumber\\
p_\mu &=& P_\mu-\frac{1}{2}\,z_\mu\, \frac{m^2_{K^*}}{p\cdot z}\,.
\end{eqnarray}
The meson's polarization vector $e^{(\lambda)}$ can be decomposed into projections onto the two light-like vectors and the orthogonal plane
\begin{eqnarray}\label{lccoords2}
 e^{(\lambda)}_\mu &=& \frac{e^{(\lambda)}z}{p\cdot z}\, p_\mu + \frac{e^{(\lambda)} p}{p\cdot z}\, z_\mu +
                     e^{(\lambda)}_{\perp\mu} =  \frac{e^{(\lambda)} z}{p\cdot z}\left( p_\mu -\frac{m^2_{K^*}}{2p\cdot z}\, z_\mu \right)+e^{(\lambda)}_{\perp\mu}\,,\nonumber\\
&=&(e^{(\lambda)} \cdot x)\frac{P_\mu (x\cdot P)-x_\mu m^2_{K^*}}{(x\cdot P)^2 -x^2 m^2_{K^*}}+ e^{(\lambda)}_{\perp\mu}\,.
\end{eqnarray}
We also need the projector $g_{\mu\nu}^\perp$ onto the directions orthogonal to $p$ and $z$
\begin{equation}
g^\perp_{\mu\nu} = g_{\mu\nu} -\frac{1}{p\cdot z}(p_\mu z_\nu+ p_\nu z_\mu)\,.
\end{equation}
Some useful scalar products are
\begin{eqnarray}
z\cdot P = z\cdot p &=& \sqrt{(x \cdot P)^2 - x^2 m^2_{K^*}}\,,\nonumber\\
p \cdot e^{(\lambda)}&=& -\frac{m^2_{K^*}}{2 pz} z \cdot e^{(\lambda)}\,,\nonumber\\
z \cdot e^{(\lambda)}&=&x \cdot e^{(\lambda)}\,.
\end{eqnarray}
Will use the notations
\begin{equation}\label{note1}
a_z\equiv a_\mu z^\mu, \qquad b_p\equiv b_\mu p^\mu,\qquad \slash{c}\equiv \gamma_\mu c^\mu,\qquad d_\mu^\perp\equiv g_{\mu \nu}^\perp d^\nu,
\end{equation}
for  arbitrary Lorentz vectors $a_\mu$, $b_\mu$, $c_\mu$ and $d_\mu$ and 
\begin{equation}
x^\mu = x_- n^\mu + x_+ \bar{n}^\mu +x^\mu_\perp\,,
\end{equation}
for null unit vectors $n^2=\bar{n}^2=0$ and $n \cdot \bar{n} =1$. The following notation is also used:\begin{equation}
a_+=a\cdot z\,,\qquad a_- =\frac{a\cdot p}{p\cdot z}\,,\qquad a^{\perp}_\mu
= a_\mu - \frac{a_- p_\mu}{p\cdot z}-a_+ z_\mu\,.
\end{equation}

\chapter{Useful formulas for sum rule determinations}\label{appendixB}
\section{Loop Integrals}
Here we summarise the loop integrals needed for calculating the twist-3 correlation
functions in Chapter~\ref{chapter4_det}.  At one loop, one has ($z^2=0$)\cite{Ball:2003sc}
\begin{eqnarray}
\int \left[d^L k\right]  e^{i f_k k\cdot z} \,\frac{(k\cdot z)^n}{(k^2)^a
  ((k-p)^2)^b} & = & (-1)^{a+b} \left(-p^2\right)^{D/2-a-b} (p\cdot z)^n
  \,\frac{\Gamma(a+b-D/2)}{\Gamma(a)\Gamma(b)}
\nonumber\\
&& \times\int_0^1 dw\,
  e^{i(1-w) f_k p\cdot z}\, w^{D/2-1-b} (1-w)^{D/2+n-1-a}\,,\nonumber\\ \label{E.1}
\end{eqnarray}
where the integration measure is defined as  $d^D k = i/(4\pi)^2 \left[d^L k\right]$ and $f_k$ is an arbitrary numerical factor, which in the cases considered in Chapter~\ref{chapter4_det} is either $v$ or $\bar v$. One also needs the integral
\begin{eqnarray}
\lefteqn{\int \left[d^L l\right]  e^{i f_l l\cdot z} \,\frac{(l\cdot p)(l\cdot z)^j}{(l^2)^c
  ((l-k)^2)^d}}
\nonumber\\
& = & (-1)^{\frac{D-4}{2}} \left(k^2\right)^{D/2-c-d} (k\cdot p) (k\cdot z)^j
  \,\frac{\Gamma(c+d-D/2)}{\Gamma(c)\Gamma(d)}\int_0^1 du\,
  e^{i(1-u) f_l k\cdot z}\, u^{D/2-1-d} (1-u)^{D/2+j-c}
\nonumber\\
&&{}+(-1)^{\frac{D-4}{2}} \left(k^2\right)^{D/2+1-c-d} (p\cdot z)(k\cdot z)^{j-1}
  \,\frac{\Gamma(c+d-D/2-1)}{2\Gamma(c)\Gamma(d)}
\nonumber\\
&&{}\times\int_0^1 du\,
  e^{i(1-u) f_l k\cdot z}\, u^{D/2-d} (1-u)^{D/2-1+j-c}\left( j + i f_l
  (1-u) (k\cdot z)\right)\,.\label{E.2}
\end{eqnarray}
Two-loop integrals are obtained by combining the above one-loop integrals.

\section{Borel Subtraction}
To derive the sum rules from $\widetilde{\pi}_{3;V}^\parallel$, $\pi_{3;V}^\parallel$ and $\pi_{3;V}^\perp$ we use the relation
\begin{equation}
\frac{1}{\pi}\textrm{Im}_s \left[-q^2-i0\right]^\alpha=\frac{s^\alpha}{\Gamma(-\alpha)\Gamma(1+\alpha)}\Theta(s)\,,
\end{equation}
where $s=-q^2$, to find the imaginary part. Using the following notation for the Borelisation and continuum subtraction procedure
 \begin{equation}
\hat{\mathcal{B}}_{sub}\left[X\right]=\int^{s_0}_0ds\,e^{-s/M^2}\frac{1}{\pi} \textrm{Im}_s X\,,
\end{equation}
and the definition of the Borel transform (\ref{borel1}) allows one to write the required results as
\begin{eqnarray}
\qquad\hat{\mathcal{B}}_{sub}\left[\frac{1}{(q^2)^\alpha}\right]&=&\frac{(-1)^\alpha}{(\alpha-1)! (M^2)^{\alpha-1}} \,,\qquad
\hat{\mathcal{B}}_{sub}\left[\ln(-q^2)\right]=-M^2+\int^\infty_{s_0}ds\,e^{-s/M^2}\,,\nonumber\\
\hat{\mathcal{B}}_{sub}\left[q^2\ln(-q^2)\right]&=&-M^4+\int^\infty_{s_0}ds\,e^{-s/M^2}s\,,\nonumber\\
\hat{\mathcal{B}}_{sub}\left[\frac{\ln(-q^2)}{q^2}\right]&=&\gamma_{E}-\ln M^2+\int^\infty_{s_0}ds\,e^{-s/M^2}\frac{1}{s}\,,\nonumber\\
\hat{\mathcal{B}}_{sub}\left[\frac{\ln(-q^2)}{q^4}\right]&=&\frac{1}{M^2}\left(1-\gamma_E +\ln M^2\right)+\int^\infty_{s_0}ds\,e^{-s/M^2}\frac{1}{s^2}\,,\nonumber\\
\hat{\mathcal{B}}_{sub}\left[\ln(-q^2)^2\right]&=&2M^2\left(\gamma_E-\ln M^2\right) +2\int^\infty_{s_0}ds\,e^{-s/M^2}\ln s\,,
\end{eqnarray}
where $\gamma_E$ is Euler's constant. 

\section{Input Parameters}
For the twist-2 and twist-3 DA parameter sum rule determinations of Chapter~\ref{chapter4_det} we use the following input parameters:
\begin{table}[ht]
\renewcommand{\arraystretch}{1.3}
\addtolength{\arraycolsep}{3pt}
$$
\begin{array}{|r@{\:=\:}l||r@{\:=\:}l|}
\hline
\quark & (-0.24\pm0.01)^3\,\mbox{GeV}^3 & \squark & (1-\delta_3)\,\quark\\
\mixed & m_0^2\,\quark &  \smixed & (1-\delta_5)\mixed\\[6pt]
\displaystyle \gluon & (0.012\pm 0.003)\, 
{\rm GeV}^4 & \multicolumn{2}{l|}{}\\[6pt]\hline
\multicolumn{4}{|c|}{m_0^2 = (0.8\pm 0.1)\,{\rm GeV}^2~,\qquad \delta_3
  = 0.2\pm 0.2, \qquad \delta_5 = 0.2\pm 0.2}\\\hline
\multicolumn{4}{|c|}{\overline{m}_s(2\,\mbox{GeV}) = (100\pm
20)\,\mbox{MeV}\qquad\longleftrightarrow\qquad\overline{m}_s(1\,\mbox{GeV})
= (133\pm 27)\,\mbox{MeV}}\\
\multicolumn{4}{|c|}{\overline{m}_q(\mu) = \overline{m}_s(\mu)/R\,,
  \qquad R = 24.6\pm 1.2}\\\hline
\multicolumn{4}{|c|}{\alpha_s(M_Z) = 0.1176\pm 0.002  ~\longleftrightarrow~ 
\alpha_s(1\,\mbox{GeV}) = 0.497\pm 0.005}\\\hline
\end{array}
$$
\renewcommand{\arraystretch}{1}
\addtolength{\arraycolsep}{-3pt}
\caption[Summary of input parameters for Chapter 4.]{\small Input parameters for sum rules at the renormalisation scale $\mu=1\,$GeV. The value of $m_s$ is obtained from unquenched lattice calculations with $N_f=2$ flavours  as summarised in \cite{mslatt}, which agrees with the results from QCD sum rule calculations \cite{jamin}. $\overline{m}_q$ is taken from chiral perturbation theory \cite{chPT}. $\alpha_s(M_Z)$ is the PDG  average \cite{Yao:2006px}.}
\label{QCDSRinput}
\end{table}

To evaluate the sum rules for the three-particle twist-3 DA parameters we use the following values of the continuum threshold $s_0$
\begin{eqnarray}
s_0^\parallel (\rho) &=& (1.3\pm 0.3)\,{\rm GeV}^2\,,\quad 
s_0^\parallel (K^*) = (1.3\pm 0.3)\,{\rm GeV}^2\,,\quad
s_0^\parallel (\phi) = (1.4\pm 0.3)\,{\rm GeV}^2\,,\quad \nonumber\\
s_0^\perp (\rho) &=& (1.5\pm 0.3)\,{\rm GeV}^2\,,\quad 
s_0^\perp (K^*) = (1.6\pm 0.3)\,{\rm GeV}^2\,,\quad
s_0^\perp (\phi) = (1.7\pm 0.3)\,{\rm GeV}^2\,. \nonumber\\
\end{eqnarray}
The threshold for the $\rho$ channel is from \cite{Shifman:1978bz}. 
\addcontentsline{toc}{chapter}{\protect\numberline{Bibliography}}

\clearpage
\end{document}